%% file: main.tex
\documentclass[12pt]{article}
\usepackage[pdftex]{graphicx}
\usepackage[margin=0.7in]{geometry}

\input{includes_macros}
\usepackage{enumitem}
\usepackage[bottom]{footmisc}
\usepackage[font=small, skip=8pt]{caption}

\def\figurespath{figures/}
\def\referencespath{references}

\theoremstyle{plain}
\newtheorem{theorem}{Theorem}
\AtAppendix{\counterwithin{theorem}{section}}
\newtheorem{proposition}[theorem]{Proposition}
\AtAppendix{\numberwithin{proposition}{section}}
\newtheorem{corollary}[theorem]{Corollary}
\AtAppendix{\numberwithin{corollary}{section}}
\newtheorem{lemma}[theorem]{Lemma}
\AtAppendix{\counterwithin{lemma}{section}}

\theoremstyle{definition}

\AtAppendix{\counterwithin{definition}{section}}
\newtheorem{assumption}{Assumption}
\AtAppendix{\counterwithin{assumption}{section}}
\newtheorem{example}{Example}
\AtAppendix{\counterwithin{example}{section}}
\renewcommand{\thmcontinues}[1]{continued}

\newenvironment{examplecont}[1]{%
  \par\medskip\noindent\textbf{Example~\ref{#1}} (continued)\textbf{.}\ %
}{\par\medskip}

\AtAppendix{\counterwithin{remark}{section}}

\AtAppendix{\counterwithin{figure}{section}}
\AtAppendix{\counterwithin{table}{section}}
\AtAppendix{\numberwithin{equation}{section}}

\input{\paperpath snippets}

\title{\papertitle\thanks{First version: December 11, 2023. \paperthanks}}
\author{Lihua Lei\thanks{Stanford Graduate School of Business and Department of Statistics (by courtesy); Email: \href{mailto:lihualei@stanford.edu}{\texttt{lihualei@stanford.edu}}} \and Brad Ross\thanks{National Bureau of Economic Research; Email: \href{mailto:rossb@nber.org}{\texttt{rossb@nber.org}}}}
\date{\today}

\begin{document}

\maketitle

\begin{abstract}
\paperabstract

~

\end{abstract}

\onehalfspacing

\input{\paperpath body}

\bibliography{\referencespath}

\appendix

\input{\paperpath appendix_body}

\input{\paperpath online_appendix_body}

\input{\paperpath supplement_body}

\typeout{get arXiv to do 4 passes: Label(s) may have changed. Rerun}

\end{document}

%% file: includes_macros.tex
\input{\rootpath general_includes}

\usepackage[backref=page]{hyperref}
\hypersetup{colorlinks,citecolor=blue,urlcolor=blue,linkcolor=blue}
\usepackage{bookmark}

\usepackage[natbibapa]{apacite}
\let \backreforig \backref
\renewcommand*{\backref}[1]{[\backreforig{#1}]}
\AtBeginDocument{%

  \renewcommand*{\PrintBackRefs}[1]{\unskip}%
}

\input{\rootpath defs}

%% file: general_includes.tex
\usepackage{amsmath}
\usepackage{amssymb}
\usepackage{amsthm}
\usepackage{mathtools}
\usepackage{thmtools}
\mathtoolsset{showonlyrefs}
\usepackage{mathdots}
\usepackage{bbm}
\usepackage{upgreek}
\usepackage{fancyhdr}
\usepackage{multicol}
\usepackage{bm}
\usepackage{listings}
\PassOptionsToPackage{usenames,dvipsnames}{color}  %
\usepackage{pdfpages}
\usepackage{tikz}
\usepackage[T1]{fontenc}
\usepackage{inconsolata}
\usepackage{framed}
\usepackage{wasysym}
\usepackage[thinlines]{easytable}
\usepackage[ruled]{algorithm2e}
\usepackage{makecell}

\usepackage[nodisplayskipstretch]{setspace}
\usepackage{float}
\usepackage{subcaption}
\usepackage{booktabs}
\allowdisplaybreaks
\usepackage{apptools}
\usepackage[section]{placeins}

%% file: defs.tex
\definecolor{shadecolor}{gray}{0.95}

\newcommand{\abs}[1]{\lvert #1 \rvert}
\newcommand{\absfit}[1]{\left\lvert #1 \right\rvert}
\newcommand{\norm}[1]{\left\lVert #1 \right\rVert}
\newcommand{\normnofit}[1]{\lVert #1 \rVert}

\newcommand{\inner}[1]{\left\langle#1\right\rangle}

\newcommand{\ind}[1]{\mathbbm{1}\{#1\}}
\newcommand{\pfrac}[2]{\frac{\partial #1}{\partial #2}}

\newcommand{\bmat}[1]{\begin{bmatrix}#1\end{bmatrix}}

\newcommand{\convin}[1]{\stackrel{#1}{\longrightarrow}}
\newcommand{\sconvin}[1]{\stackrel{#1}{\rightarrow}}
\newcommand{\bigo}[1]{\mathcal{O}\left(#1\right)}
\newcommand{\bigop}[1]{\mathcal{O}_{\Pr}\left(#1\right)}

\newcommand{\littleop}[1]{o_{\Pr}\left(#1\right)}

\newcommand\indep{\protect\mathpalette{\protect\independenT}{\perp}}
\def\independenT#1#2{\mathrel{\rlap{$#1#2$}\mkern2mu{#1#2}}}

\DeclareMathOperator{\Cov}{Cov}

\DeclareMathOperator{\E}{\mathbb{E}}

\DeclareMathOperator{\R}{\mathbb{R}}

\DeclareMathOperator*{\argmin}{arg\,min}

\let\Pr\relax
\DeclareMathOperator{\Pr}{\mathbb{P}}
\DeclareMathOperator{\distiid}{\stackrel{i.i.d.}{\sim}}

\DeclareMathOperator{\zeros}{\mathbf{0}}

\let\epsilon\relax
\DeclareMathOperator{\epsilon}{\varepsilon}
\DeclareMathOperator{\setst}{:}

%% file: snippets.tex
\newcommand*{\papertitle}{Estimating Counterfactual Matrix Means with Short Panel Data}
\newcommand*{\paperthanks}{Authors are listed alphabetically. We thank Alberto Abadie, Dmitry Arkhangelsky, Timothy Armstrong, Susan Athey, Mohsen Bayati, St\'{e}phane Bonhomme, Kirill Borusyak, Jiafeng Chen, Rebecca Diamond, Matthew Gentzkow, Bryan Graham, Christian Hansen, Guido Imbens, Koen Jochmans, Patrick Kline, Elena Manresa, Samuel Norris, David Ritzwoller, Jonathan Roth, Raffaele Saggio, Kevin Song, Jann Spiess, Vasilis Syrgkanis, Stefan Wager, Martin Weidner, Andrei Zeleneev, and the participants of several seminars and conferences for their thoughtful comments and valuable feedback. Brad acknowledges and appreciates financial support from the Bradley Graduate and Postgraduate Fellowship via grants to the Stanford Institute for Economic Policy
Research. Computational Support was provided by the Data, Analytics, and Research Computing (DARC) group at the Stanford Graduate School of Business (\href{https://scicrunch.org/resources/data/record/nlx_144509-1/SCR_022938/resolver?q=SCR_022938&l=SCR_022938&i=rrid:scr_022938}{RRID:SCR\_022938}). A software package implementing the methodology described in this paper can be found at \url{https://github.com/brad-ross/apm}; code implementing the empirical application in this paper can be found at \url{https://github.com/brad-ross/apm-paper}.}
\newcommand*{\paperabstract}{We develop a spectral approach for identifying and estimating average counterfactual outcomes under a low-rank factor model with short panel data and general outcome missingness patterns. Applications include event studies and studies of outcomes of ``matches'' between agents of two types, e.g. people and places, typically conducted using less-flexible Two-Way Fixed Effects (TWFE) models of outcomes. Given finite observed outcomes per unit, we show our approach identifies all counterfactual outcome means, including those not identified by existing methods, if a particular graph algorithm determines that units' sets of observed outcomes have sufficient overlap. Our analogous, computationally efficient estimation procedure yields consistent, asymptotically normal estimates of counterfactual outcome means under fixed-$T$ (number of outcomes), large-$N$ (sample size) asymptotics. When estimating province-level averages of held-out wages from an Italian matched employer-employee dataset, our estimator outperforms a TWFE-model-based estimator.}
\newcommand*{\paperkeywords}{panel data, missing not-at-random, factor model, interactive fixed effects, event study, bipartite network data}

%% file: body.tex
\section{Introduction}\label{sec:introduction}

\input{\paperpath sections/intro}

\section{Setup and Intuition}\label{sec:setup}

\input{\paperpath sections/intuition}

\section{Estimation and Inference Procedure}\label{sec:method}\label{sec:method:procedure}

\input{\paperpath sections/method}

\section{Theoretical Properties}\label{sec:theory}

\input{\paperpath sections/theory}

\section{Empirical Illustration}\label{sec:empirical_performance}

\input{\paperpath sections/empirical}

\section{Additional Discussion and Results}\label{sec:conclusion}

\input{\paperpath sections/conclusion}

%% file: sections/intro.tex
Researchers often aim to estimate average counterfactual outcomes in some population using ``short'' panel data with outcomes that are missing ``not-at-random'' \citep{rubin1976inference,little2019statistical}, i.e. sampled units for whom small, likely endogenous subsets of possible outcomes are observed.
For example, in event study settings, units receive a treatment at different times, and researchers observe units' pre-treatment control potential outcomes and post-treatment treated potential outcomes across several time periods. Estimating the average treatment effect on treated units requires estimating their average unobserved, post-treatment control potential outcomes, but often, units plausibly select into treatment times on characteristics that also affect their outcomes \citep{ashenfelter1985using, bertrand2004much, angrist2009mostly}.

As another example, several empirical literatures aim to characterize what determines the outcomes of counterfactual ``matches'' between pairs of agents of two different ``types,'' e.g. wages of individuals working at different firms \citep{abowd1999high,card2013workplace}, test scores of students taught by different teachers \citep{jackson2014teacher}, and earnings and health outcomes of people living in different places \citep{finkelstein2016sources,chetty2018impacts,card2023location}. It is reasonable to assume that units select into matches based on unobserved characteristics that also affect match outcomes, and exogenous match determinants are difficult to come by.

To estimate average counterfactual outcomes in these settings, researchers typically leverage multiple observations per unit to estimate a model of how low-dimensional, unobserved confounders affect outcomes.\footnote{Usually, these models also require ``strict exogeneity'' \citep{chamberlain1984panel}, namely that, conditional on confounders, the outcomes themselves are independent of which outcomes are observed. In keeping with much of applied practice, this paper does the same. However, \citet{ashenfelter1985using} and \citet{bonhomme2019distributional} discuss strict exogeneity's plausibility in event study and match outcome contexts, respectively.} A canonical model of this type is the Two-Way Fixed Effects (TWFE) model, which enables outcome means to be identified and estimated with short panel data under many outcome missingness patterns by ``differencing out'' unit fixed effects \citep{borusyak2021revisiting,jochmans2019fixed}. However, it severely restricts how unobserved confounders can affect outcomes, as discussed in the literatures on the ``parallel trends'' assumption implied by the TWFE model \citep{arkhangelsky2024causal,ghanem2022selection} and on estimating match effects (e.g. \citet{woodcock2015match,bonhomme2019distributional}).

In event study settings, a large literature has sought to relax these restrictions by positing a low-rank factor model of outcomes (see Section \ref{subsec:related_work}).\footnote{Linear factor models are also frequently called ``interactive fixed effects'' models since they assume outcomes are determined by the inner product of vectors of unit-specific and outcome-specific factors.} However, existing factor model-based methods cannot be applied generally: they require certain outcome missingness patterns, and unlike TWFE-based methods, many require a large number of observed outcomes per unit.

In this paper, we bridge the gap between the general applicability of TWFE-based methods and the expressivity of factor model-based methods. We develop a method for identifying, estimating, and conducting inference on counterfactual outcome means under factor models in short panels with general outcome missingness patterns, including those not identified by existing approaches.

\begin{figure}[t]
    \centering
    \begin{subfigure}{0.4\textwidth}
          \centering
          \includegraphics[width=\textwidth]{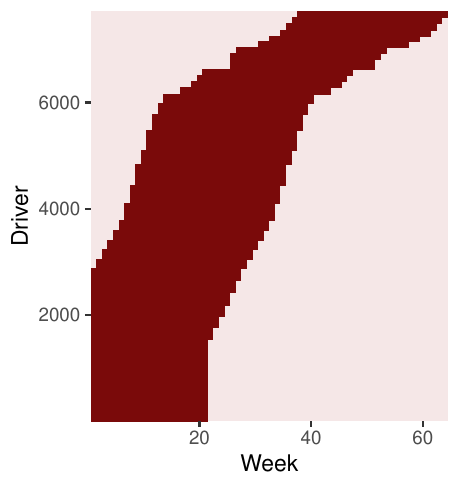}
          \caption{Event Study with Staggered Entry}
          \label{fig:control_outcomes_over_time_missingness_pattern}
      \end{subfigure}
      \hfill%
      \begin{subfigure}{0.59\textwidth}
          \centering
          \includegraphics[width=\textwidth]{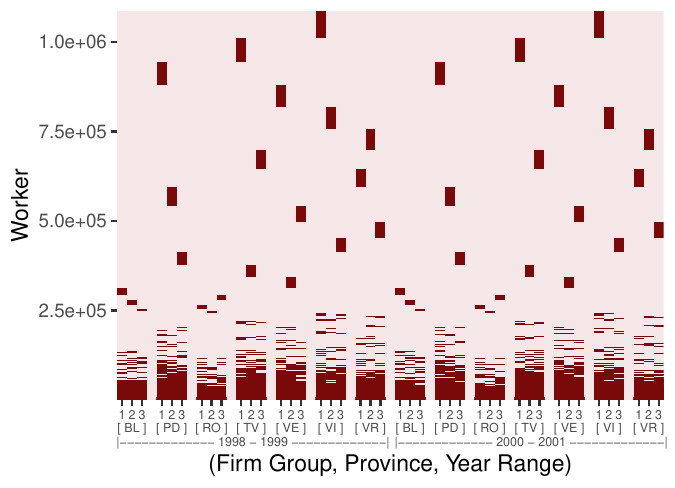}
          \caption{Bipartite Match Outcomes}
          \label{fig:match_outcome_missingness_pattern}
      \end{subfigure}
    \caption{Outcome observation patterns in panels based on two real-world panel datasets: \citet{ater2024can} in Figure \ref{fig:control_outcomes_over_time_missingness_pattern} and the VWH dataset in Figure \ref{fig:match_outcome_missingness_pattern}. Rows in each figure correspond to units and columns correspond to outcomes. Dark red indicates an observed outcome and light red indicates a missing outcome.}
    \label{fig:empirical_outcome_missingness_patterns}
\end{figure}

Figure \ref{fig:empirical_outcome_missingness_patterns} illustrates two empirical examples with complex outcome missingness patterns to which our methods uniquely apply. Figure \ref{fig:control_outcomes_over_time_missingness_pattern} is based on an event study from \citet{ater2024can}, evaluating the effect of a congestion pricing incentive on Israeli drivers' behavior; here, outcomes correspond to drivers' control potential outcomes summarizing driving behavior in the absence of the incentive. Drivers joined the study at staggered times and received treatment after a non-random 20 weeks of monitoring, generating the ``staircase'' pattern in Figure \ref{fig:control_outcomes_over_time_missingness_pattern} of observed control potential outcomes in dark red. Figure \ref{fig:match_outcome_missingness_pattern} is based on the Veneto Worker Histories (VWH) matched employer-employee dataset from the Veneto region of Italy, a bipartite matching setting that serves as the basis for our empirical application in Section \ref{sec:empirical_performance}. Outcomes are workers' average weekly wages at groups of similar firms in each Veneto province over two-year periods from 1998 - 2001; an outcome is observed and indicated in Figure \ref{fig:match_outcome_missingness_pattern} by dark red if a worker worked at any firm in that group during that period.

Both settings exhibit three features our approach is designed to accommodate. First, outcome missingness is plausibly determined by unobserved characteristics that also affect outcomes: in Figure \ref{fig:control_outcomes_over_time_missingness_pattern}'s setting, drivers who would benefit more from the treatment may have enrolled earlier; in Figure \ref{fig:match_outcome_missingness_pattern}'s setting, workers likely work at firms that value their skills most. Second, the TWFE model's requirement that unobserved confounders affect all outcomes identically is unlikely to hold: in Figure \ref{fig:control_outcomes_over_time_missingness_pattern}'s setting, remote workers likely responded to COVID-19 lockdowns differently than those in occupations requiring in-person work; in Figure \ref{fig:match_outcome_missingness_pattern}'s setting, workers whose skills are disproportionately valued by industries concentrated in certain provinces may earn higher wages there than elsewhere. Third, both are short panels: only a small number of outcomes are observed per unit relative to the sample size.

To identify counterfactual outcome means, we first group units into subpopulations called \textit{cohorts} that share the same sets of observed outcomes. We then use each cohort's data to identify factor vectors corresponding to that cohort's observed outcomes up to cohort-specific bases, using any of the existing approaches for short, balanced panels reviewed in Section \ref{sec:theory:cohort_specific_factor_estimates}. To align these cohort-specific factor vectors so they are expressed with respect to a common basis, we use them to construct a particular matrix we call an \textit{Aggregated Projection Matrix (APM)}. Our main identification result shows that the rows of any basis matrix for the APM's null space serve as aligned factor vectors for all outcomes, provided an efficient graph algorithm we call the \textit{Observed Outcome Overlap (O\textsuperscript{3}) Algorithm} converges appropriately. The O\textsuperscript{3} Algorithm iteratively merges groups of cohorts sharing sufficient overlap in their observed outcomes until no further merges are possible; our result holds if all cohorts can be merged into a single group. Since the O\textsuperscript{3} Algorithm only depends on the known sets of observed outcomes for each cohort, this condition can be verified empirically.

Given aligned factor vectors, we regress each unit's observed outcomes on the corresponding factor vectors to obtain unbiased factor loading estimates. We then use these estimates to impute counterfactual outcomes and average them to recover our counterfactual outcome means of interest.

Relative to existing factor model estimators, our approach has several desirable properties. First, it makes no assumptions about how units select into cohorts and does not require enough observed outcomes per unit to recover unit-specific confounders exactly, unlike factor-model-based approaches for long panels, e.g. those in the ``matrix completion'' family (see Section \ref{subsec:related_work}). Thus, despite not being able to ``difference out'' unit-level confounders as one can under the TWFE model, our approach identifies cohort outcome means using only a finite number of observed outcomes per unit.

Second, our approach accommodates more general missingness patterns than other methods designed for short panels. In particular, to identify a target cohort-outcome mean, it does not require a ``reference'' cohort like a ``never-treated group'' in event study settings for whom both the target outcome is observed and a sufficient number of outcomes overlap with the target cohort's observed outcomes. Both empirical examples in Figure \ref{fig:empirical_outcome_missingness_patterns} contain many target cohort-outcome pairs for which no such reference cohort exists, as quantified in Section \ref{sec:empirical_performance}. Third, our method automatically aggregates identifying information that existing methods either use in isolation or ignore entirely, improving sample efficiency even for cohort-outcome means that existing methods can identify.

We translate this identification strategy into an analogous, computationally efficient plug-in estimator that can also accommodate outcome fixed effects and covariates that vary across both units and outcomes. In an asymptotic regime in which the number of outcomes remains fixed as the cross-sectional dimension of the panel and the sizes of all cohorts grow, we show that this estimator is consistent and asymptotically normal, and that a weighted bootstrap procedure provides valid asymptotic inference. These results rely on an exact, first-order expansion of the operator mapping a symmetric matrix into the projection matrix onto the space spanned by some subset of its eigenvectors. We derive this expansion using a result called Kato's integral \citep{kato1949convergence}.

Finally, we demonstrate the empirical performance of our estimator using the VWH dataset introduced above and visualized in Figure \ref{fig:match_outcome_missingness_pattern}. Our goal is to assess whether observed wage differences across Italian provinces are driven by causal effects of locations on wages or worker sorting, in the spirit of \citet{card2023location}.
To this end, we conduct two empirical exercises. First, we run a semi-synthetic simulation study in which we mask observed cohort outcomes and compare the accuracy of our method to a TWFE-based estimator at recovering the masked outcome means. Across many masked target cohort outcome means, our method frequently delivers estimates with lower bias and root mean squared error than the TWFE-based estimator. Second, we decompose observed wage differences between provinces into shares attributable to causal effects of locations versus worker sorting using a nonparametric generalization of the decomposition proposed in \citet{finkelstein2016sources}. In line with \citet{bonhomme2019distributional}, both exercises suggest that complementarities between workers and firms across different provinces affect wages modestly.

\subsection{Related Work}\label{subsec:related_work}

A variety of factor-model-based methods estimate cohort outcome means by aggregating individual-level imputations based on estimated factor structures.\footnote{Examples include \citet{bai2009panel,moon2015linear,gobillon2016regional,xu2017generalized,athey2021matrix,bai2021matrix,arkhangelsky2021synthetic,fernandez2021low,agarwal2021causal,chernozhukov2021inference,farias2021learning,ben2022synthetic,dwivedi2022doubly,chan2022pcdid,xiong2023large,choi2023matrix,choi2023inference,imbens2023identification,freeman2023linear,arkhangelsky2023largesample,yan2024entrywise,abadie2024doubly,deaner2025inferring,athey2025identification}.}
These methods all require the number of observed outcomes per cohort to grow with the sample size, since factor vectors are estimated using cross-unit variation while loading vectors are estimated using cross-outcome variation. In short panels, unit-specific loadings cannot be recovered using these methods without persistent bias, which can also contaminate the factor vector estimates they generate, an example of the incidental parameter problem \citep{neyman1948consistent,lancaster2000incidental}. Additionally, complicated missingness patterns often require restrictions on the missingness mechanism even in long panels \citep{xiong2023large}, such as missingness at random or known probabilities of jointly observing outcome pairs, which are often implausible in our settings.

Instead, we build on an approach proposed in \citet{imbens2021controlling}, \citet{brown2022generalized}, and \citet{agarwal2020synthetic}, which uses factor vectors expressed with respect to a common basis to recover asymptotically unbiased estimates of units' factor loadings and impute missing outcomes. This approach was developed for missingness patterns where a cohort exists for whom all outcomes are observed like block missingness patterns (defined in e.g. \citet{athey2021matrix} and visualized in Figure \ref{fig:block_missingness_figure}); the data from this cohort can be used to recover factor vectors for all outcomes with respect to the same basis. In event study settings, a ``never-treated'' cohort typically plays this role, and several other short-panel methods use a never-treated group for the same purpose, e.g. \citet{callaway2023treatment,brown2022generalized,brown2023difference,callaway2023treatment2,fry2024method,arkhangelsky2024sequential}.\footnote{\citet{athey2021matrix} also require a nontrivial share of never-treated units for their estimator to be consistent in the staggered treatment adoption setting.}

\citet{agarwal2021causal} extend this approach to more general missingness patterns by finding block missingness patterns embedded within more complicated missingness patterns. However, as discussed in Section \ref{sec:method:intuition}, even embedded block missingness patterns are often insufficient to identify all cohort outcome means in our settings of interest, and when multiple such patterns exist, it is unclear how to efficiently combine the information from each. Our approach addresses both challenges by combining information across the many possible ways to express cohort-specific factor estimates with respect to a common basis.

Our assumption that the O\textsuperscript{3} graph algorithm converges to a single super cohort resembles those in the literature on fixed-effect models of bipartite match outcomes under strict exogeneity \citep{abowd2002computing,jochmans2019fixed,bonhomme2019distributional,hull2018estimating,kline2024firm}; see \citet{bonhomme2020econometric} for a review. In Appendix \ref{sec:supplement:connections_graph_based_identification}, we show that our O\textsuperscript{3} Algorithm convergence requirement generalizes theirs to factor models with arbitrary rank.

Our consistency and asymptotic normality results relate to the literature on perturbation theory for eigenspaces of random matrices \citep{bai2010spectral}. Standard bounds like the Davis--Kahan theorem \citep{yu2015useful} are too coarse to characterize asymptotic distributions. Instead, we derive an exact, first-order eigenspace projection expansion using Kato's integral \citep{kato1949convergence}. Several other papers have also applied perturbation-theoretic results to show that eigenspaces of random matrices concentrate, e.g. \citet{onatski2009testing,simons2023inference,babii2025tensor}, as well as \citet{oliveira2010concentration} and \citet{lei2020unified}, which consider random matrices with (approximately) independent entries.

%% file: sections/intuition.tex
\subsection{Short Panel Settings}\label{sec:setup:setup}

In our settings of interest, researchers observe a large, i.i.d.\ sample of $N$ units, each with $T$ outcomes, where $Y_{it}^*$ denotes unit $i$'s outcome $t$. To describe which outcomes are observed, we group units into $C$ \textit{cohorts}, where $C_i$ denotes unit $i$'s cohort and $p_c \coloneqq \Pr(C_i = c)$ denotes the share of units in cohort $c$. For cohort $c$, let $\mathcal{T}_c \subseteq [T] \coloneqq \{1, \dotsc, T\}$ denote the indices of their observed outcomes and $T_c \coloneqq \abs{\mathcal{T}_c}$ denote how many are observed. We allow cohort membership $C_i$ to be correlated with outcomes $Y_{it}^*$ in a structured way made precise below. To distinguish observed from missing outcomes, we define $Y_{it} = Y_{it}^*$ if $t \in \mathcal{T}_{C_i}$ and $Y_{it} = \text{?}$ otherwise.\footnote{In what follows, we define $? \cdot 0 = 0$.} Our asymptotic theory in Section \ref{sec:theory:estimation_inference} focuses on short panels where $T$ remains bounded as $N$ grows. Section \ref{sec:conclusion:short_to_sparse_panels} discusses the more general ``sparse panel'' setting where $T_c$ remain bounded but $C$ (and hence $T$) can grow with $N$.

This setup encapsulates several causal panel data settings. In event study settings like Figure \ref{fig:control_outcomes_over_time_missingness_pattern}'s, $Y_{it}^*$ represents unit $i$'s period-$t$ control potential outcome---their outcome in period $t$ had they not yet received treatment by period $t$---and a unit belongs to cohort $c$ if they were first treated at time $c$. Since control potential outcomes are only observed prior to treatment, post-treatment outcomes are missing: $\mathcal{T}_c = \{t \in [T] ~\setst~ t < c\}$.

In bipartite matching settings, $Y_{it}^*$ denotes ``row-type'' unit $i$'s outcome when matched with ``column-type'' unit $t$, and row-type units belong to the same cohort if matched with the same subset of column-type units. Figure \ref{fig:match_outcome_missingness_pattern} illustrates such a setting, which serves as our main running example and the focus of Section \ref{sec:empirical_performance}'s empirical application. There, row-type units $i$ are workers, $t$ indexes pairs of a two-year period and a group of similar firms in the same Veneto province of Italy, and $Y_{it}^*$ denotes the log of worker $i$'s average weekly wage at firms in group $t$ during $t$'s two-year period. Since workers are employed by only a subset of firms, counterfactual wages at firms in groups they never worked for are missing.

Given these unbalanced panel data, researchers often seek to estimate aggregations of cohort outcome means $\mu_{ct} \coloneqq \E[Y_{it}^* ~|~ C_i = c]$. We call $\mu_{ct}$ a counterfactual outcome mean since outcome $t$ may not be observed for cohort $c$. Our approach enables estimation of aggregation parameters $\theta \coloneqq h(\mu, \eta)$, where $\mu$ collects all cohort outcome means $\mu_{ct}$ and $\eta$ is a vector of nuisance parameters estimable from observed outcomes, covariates, and additional data.

In event study settings, a common $\theta$ is a vector of average treatment effects at different relative time horizons post-treatment. As introduced in \citet{callaway2021difference} and \citet{sun2021estimating} and reviewed in Example \ref{ex:dynamic_treatment_effects} of Appendix \ref{sec:supplement:target_params_examples}, these estimands can be expressed as weighted averages of differences between each cohort's counterfactual post-treatment control potential outcome means $\mu_{ct}$ and their observed post-treatment treated potential outcome means (part of $\eta$), with weights depending on relative cohort sizes $p_c$ (also part of $\eta$).\footnote{See \citet{de2020two} for treatment effect definitions when treatment is not absorbing, and \citet{de2024difference} when a unit's entire treatment history may affect future outcomes.}

In the context of a bipartite matching setting with Medicare patients as row-type units and geographic areas as column-type units, \citet{finkelstein2016sources} propose estimands $\theta$ that decompose differences in average observed outcomes between two column-type units into components attributable to column-type unit effects versus row-type unit selection. Below, we define a nonparametric generalization of their attribution parameters that we estimate in Section \ref{sec:empirical_performance} using data from Figure \ref{fig:match_outcome_missingness_pattern}'s setting:\footnote{The continuation of Example \ref{example-match-outcome-attribution} in Appendix \ref{sec:supplement:target_params_examples} shows how our generalization reduces to \citet{finkelstein2016sources}'s parameters under a TWFE model \eqref{eq:twfe_model}. \citet{hull2018estimating} and \citet{kline2024firm} study similar parameters under TWFE-like assumptions. We leave decompositions based on higher-order moments like variances \citep{abowd1999high} to future work.}

\begin{example}[Match Outcome Attribution]\label{example-match-outcome-attribution}
    Analogous to \citet{finkelstein2016sources}, the difference in average observed match outcomes for row-type units matched to two column-type units $t_1$ and $t_2$ can be decomposed additively as follows:
    \begin{equation}
    \begin{aligned}
        &\E[Y_{it_1} ~|~ t_1 \in \mathcal{T}_{C_i}] - \E[Y_{it_2} ~|~ t_2 \in \mathcal{T}_{C_i}] \\
        &= \underbrace{\E[Y_{it_1}^*] - \E[Y_{it_2}^*]}_{\text{Average effect on outcomes of column-type unit differences under at-random matches}} \\
        &\phantom{=}+ \underbrace{(\E[Y_{it_1}^* ~|~ t_1 \in \mathcal{T}_{C_i}] - \E[Y_{it_1}^*]) - (\E[Y_{it_2}^* ~|~ t_2 \in \mathcal{T}_{C_i}] - \E[Y_{it_2}^*]).}_{\text{Average effect on outcomes of differential row-type unit selection into matches}}
    \end{aligned}
    \end{equation}
    Following \citet{finkelstein2016sources}, we define $\theta_{\text{col}, t_1, t_2}$ as the ``share'' of the difference in average observed outcomes between $t_1$ and $t_2$ attributable to how these column-type units affect outcomes and $\theta_{\text{row}, t_1, t_2}$ as the share attributable to differential row-type unit selection:
    \begin{equation}
        \theta_{\text{col}, t_1, t_2} \coloneqq \frac{\E[Y_{it_1}^*] - \E[Y_{it_2}^*]}{\E[Y_{it_1}^* ~|~ t_1 \in \mathcal{T}_{C_i}] - \E[Y_{it_2}^* ~|~ t_2 \in \mathcal{T}_{C_i}]} \eqqcolon 1 - \theta_{\text{row}, t_1, t_2}.
    \end{equation}
    We put ``share'' in quotes because, as in \citet{finkelstein2016sources}, $\theta_{\text{col}}$ and $\theta_{\text{row}}$ sum to one but need not lie between zero and one.

    To express $\theta_{\text{col}, t_1, t_2}$ and $\theta_{\text{row}, t_1, t_2}$ as functions of $\mu$ and $\eta$, note that both depend on unconditional outcome means $\E[Y_{it}^*]$ and observed outcome means $\E[Y_{it}^* ~|~ t \in \mathcal{T}_{C_i}]$, which are both different cohort size-weighted averages of cohort outcome means:
    \begin{equation}
        \E[Y_{it}^*] = \sum_{c = 1}^C p_c\mu_{ct}, \quad \E[Y_{it}^* ~|~ t \in \mathcal{T}_{C_i}] = \frac{\E[\ind{t \in \mathcal{T}_{C_i}}Y_{it}^*]}{\Pr(t \in \mathcal{T}_{C_i})}
        = \sum_{c=1}^{C}\frac{\ind{t\in \mathcal{T}_{c}}p_c}{\sum_{c'=1}^{C}\ind{t\in \mathcal{T}_{c'}}p_{c'}}\mu_{ct}.
    \end{equation}
    Thus, the nuisance parameters $\eta$ in this example are simply the shares of units in each cohort $p_c$.
\end{example}

\subsection{Factor Model of Outcomes}\label{sec:setup:factor_model}

To estimate averages of counterfactual outcomes not observed for some cohorts, we impose additional structure on how units' observed and latent characteristics affect their outcomes. Specifically, we assume outcomes $Y_{it}^*$ follow a rank-$r$ linear factor model: outcomes are determined by the inner product of an outcome-specific factor vector $\gamma_t \in \R^r$ and a unit-specific loading vector $\lambda_i \in \R^r$, plus an optional outcome fixed effect $\gamma_{t0}$, an optional term involving an always-observed covariate vector $X_{it} \in \R^q$, and a mean-zero error $\epsilon_{it}$:
\begin{equation}\label{eq:factor_model}
\begin{aligned}
    Y_{it}^* = \gamma_t'\lambda_i + \gamma_{t0} + X_{it}'\beta + \epsilon_{it}, \quad \E[\epsilon_{it} ~|~ 
    \lambda_i, X_i, 
    C_i] = 0,
\end{aligned}
\end{equation}
where $X_i$ is a $T \times q$ matrix with $t$-th row $X_{it}$. Mean-independence of the residuals $\epsilon_{it}$ from cohort membership $C_i$ conditional on $\lambda_i$ and $X_i$ 
implies that only the coordinates of $\lambda_i$ can serve as unobserved confounders affecting both cohort membership and outcomes.\footnote{We treat factor vectors $\gamma_t$ as fixed, conditioning inference on them. If $\gamma_t$ and $\gamma_{t0}$ were random, the residual mean condition in \eqref{eq:factor_model} would read $\E[\epsilon_{it} ~|~ 
\lambda_i, X_{i}, 
C_i, \{(\gamma_t, \gamma_{t0}): t \in [T]\}] = 0$. As discussed in Section \ref{sec:theory:cohort_specific_factor_estimates}, mean independence of $\epsilon_{it}$ from $\lambda_i$ is sometimes stronger than necessary, but it makes clear the role that factor loadings $\lambda_i$ play as unobserved confounders in our settings of interest. Section \ref{sec:theory:cohort_specific_factor_estimates} also discusses additional restrictions on the residuals' covariance structure (e.g.\ independence across outcomes and homoskedasticity) and covariate exogeneity conditions enabling estimation of cohort-specific factor vectors.}
We assume the rank $r < T$ is known throughout, though Section \ref{sec:conclusion:choosing_factor_model_rank} discusses principled ways to empirically evaluate different values of $r$.

For intuition, consider Figure \ref{fig:match_outcome_missingness_pattern}'s worker-firm setting. There, $\lambda_i$'s coordinates could measure worker $i$'s skills (e.g., cognitive and interpersonal), and $\gamma_t$'s coordinates could measure how firms in $t$'s group value each skill, as in \citet{lazear2009firm}. The outcome fixed effect $\gamma_{0t}$, if present, could capture firm-specific wage determinants independent of worker characteristics, and $X_{it}$ could include $(i,t)$-specific wage determinants like worker $i$'s age at the beginning of $t$'s period or the interaction of $i$'s education with the industry of $t$'s firm group. To the extent workers sort into firms that value their skills, $\lambda_i$ and $X_{it}$ determine both which firms they work at ($\mathcal{T}_{C_i}$) and the wages they earn ($Y_{it}^*$) \citep{lazear2009firm}.

Before continuing, it is worth contrasting the factor model \eqref{eq:factor_model} with the canonical Two-Way Fixed Effects (TWFE) model,\footnote{Variants of the TWFE model can be derived from and imply ``parallel trends'' assumptions in both event study and bipartite matching settings \citep{borusyak2021revisiting,ghanem2022selection,kline2024firm,hull2018estimating}.} which instead specifies outcomes as the sum of unidimensional unit and outcome fixed effects $\lambda_i, \gamma_{t0} \in \R$ plus covariate and error terms $X_{it}'\beta$ and $\epsilon_{it}$ interpreted as in \eqref{eq:factor_model}:
\begin{equation}\label{eq:twfe_model}
    Y_{it}^* = \lambda_i + \gamma_{t0} + X_{it}'\beta + \epsilon_{it}, \quad \E[\epsilon_{it} ~|~ \lambda_i, X_{i}, C_i] = 0.
\end{equation}
As discussed in the event study and bipartite matching literatures \citep{athey2021matrix,arkhangelsky2021synthetic,bonhomme2019distributional}, when outcome fixed effects are included, the factor model \eqref{eq:factor_model} generalizes the TWFE model \eqref{eq:twfe_model} in two key ways.\footnote{Even without outcome fixed effects, the TWFE model \eqref{eq:twfe_model} is a special case of \eqref{eq:factor_model} with $r = 2$, $\lambda_{i1} = 1$, and $\gamma_{t2} = 1$.} First, even with $r = 1$, the factor model allows the latent characteristic $\lambda_i$ to affect different outcomes differently via interaction with $\gamma_t$, whereas the TWFE model requires $\lambda_i$ to affect all outcomes identically. In Figure \ref{fig:match_outcome_missingness_pattern}'s setting, where $\lambda_i$ is often interpreted as worker $i$'s human capital, the TWFE model's additive separability rules out complementarity between worker ability and firm productivity in determining wages.\footnote{Even with log-wages, the TWFE model rules out super-multiplicative complementarity between worker ability and firm productivity.} Second, while the TWFE model requires confounders to affect outcomes through a unidimensional index $\lambda_i$, the factor model with $r \geq 2$ accommodates multidimensional confounding effects, e.g. different firms valuing workers' skill bundles differently in Figure \ref{fig:match_outcome_missingness_pattern}'s setting.

\subsection{Intuition for Factor Model Identification}\label{sec:method:intuition}

Our identification approach proceeds in two steps: first, we recover factor vectors for all outcomes in a common basis; second, we use these to extrapolate from each cohort's observed outcome means to their counterfactual outcome means. The second step is conceptually straightforward: given factor vectors $\tilde{\gamma}_t = Q\gamma_t$ rotated by some common $r \times r$ basis matrix $Q$, regressing unit $i$'s observed outcomes $Y_{it}$ for $t \in \mathcal{T}_c$ on the corresponding $\tilde{\gamma}_t$ yields an unbiased estimate $\tilde{\lambda}_i$ of $(Q')^{-1}\lambda_i$ \citep{brown2022generalized,agarwal2021causal,imbens2021controlling}, assuming no outcome fixed effects or covariates for simplicity. The inner product $\tilde{\gamma}_t'\tilde{\lambda}_i = \gamma_t'\lambda_i$ is then an unbiased estimate of unit $i$'s outcome $t$, and averaging over the units in cohort $c$ yields $\mu_{ct}$. Crucially, factor vectors need only be recovered up to a common basis $Q$; even without missing data, only rotations $\tilde{\gamma}_t = Q'\gamma_t$ are identified, reflecting the inherent arbitrariness of $\lambda_i$'s coordinate system.\footnote{Formally, for any invertible $Q$, $Y_{it}^*$ and $\gamma_t'QQ^{-1}\lambda_i + \gamma_{t0} + X_{it}'\alpha + \epsilon_{it}$ have the same distribution, as is well-known in the factor model literature \citep{anderson1956statistical, anderson2009introduction}.} The main challenge is the first step: recovering the factor vectors for all outcomes expressed with respect to a common basis when different outcome subsets across cohorts.

For concreteness, consider Figure \ref{fig:match_outcome_missingness_pattern}'s worker-firm setting with $r = 2$. The two coordinates of $\lambda_i$ could represent worker $i$'s cognitive and interpersonal skills or any linear reparameterization thereof; the observed wage distribution is identical regardless of this choice. What can be recovered are the relative values of these skills to different firm groups in different years, and these relative values are exactly what we need to impute counterfactual wages across all firm groups in all years.\footnote{We are grateful to Jonathan Roth for suggesting this intuition as our discussant.}

\begin{figure}[t]
    \centering
    \begin{subfigure}{0.3\textwidth}
        \centering
        \includegraphics[width=0.7\textwidth,page=1]{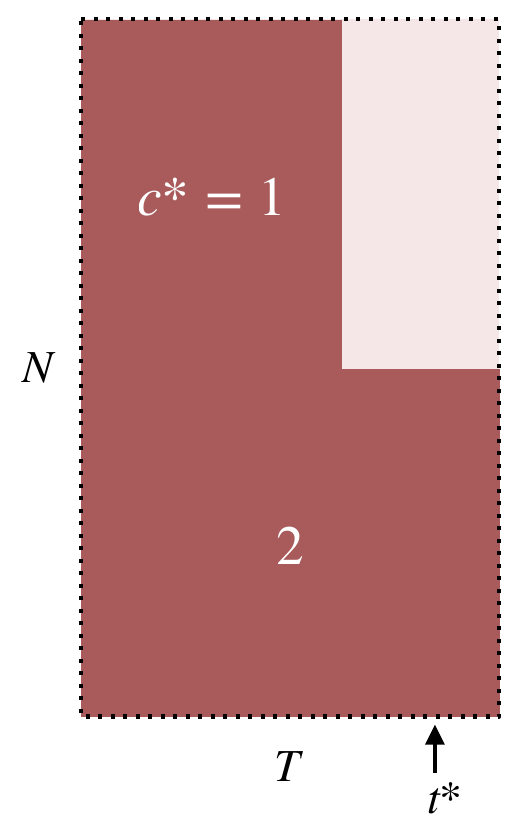}
        \caption{Block Missingness Pattern}
        \label{fig:block_missingness_figure}
    \end{subfigure}
    \hfill
    \begin{subfigure}{0.3\textwidth}
        \centering
        \includegraphics[width=0.7\textwidth,page=10]{\figurespath simple_cohort_figure_paper.pdf}
        \caption{Staircase Pattern}
        \label{fig:three_cohort_figure}
    \end{subfigure}
    \hfill
    \begin{subfigure}{0.3\textwidth}
        \centering
        \includegraphics[width=0.7\textwidth,page=12]{\figurespath simple_cohort_figure_paper.pdf}
        \caption{Higher-Order Chains}
        \label{fig:higher_order_chains_figure}
    \end{subfigure}
    \caption{Three stylized outcome missingness patterns. As in Figure \ref{fig:empirical_outcome_missingness_patterns}, a pixel is dark red if the outcome (column) is observed for that unit (row) and light red otherwise. Numbers indicate the cohorts to which blocks of units belong.}
    \label{fig:simple_cohort_figures}
\end{figure}

In certain structured missingness patterns, recovering aligned factor vectors $\tilde{\gamma}_t$ with respect to a common basis is straightforward. In the block missingness pattern of Figure \ref{fig:block_missingness_figure}, all outcomes are observed for cohort 2, so we can recover every $\tilde{\gamma}_t$ using cohort 2's data alone via any factor model identification approach for short, balanced panels (see Section \ref{sec:theory:cohort_specific_factor_estimates}). In event study settings, a ``never-treated group'' serves a similar purpose \citep{imbens2021controlling,brown2022generalized}, and in Figure \ref{fig:match_outcome_missingness_pattern}'s setting, a cohort of workers who worked at all groups of firms in all years would function in this way, enabling recovery of the relative values of those workers' skills to every firm in every year.

In the more complex staircase pattern of Figure \ref{fig:three_cohort_figure}, if our target is cohort 1's mean of outcome $t_1$, we can subset to the units and outcomes inside the dashed rectangle, reducing the problem to block missingness \citep{agarwal2021causal}. However, this idea fails for cohort 1's mean of outcome $t_2$: $t_2$ is only observed for cohort 3, which shares no observed outcomes with cohort 1. In bipartite matching settings like Figure \ref{fig:match_outcome_missingness_pattern}'s, these embedded block missingness patterns are typically insufficient since only a small fraction of possible cohorts are observed \citep{andrews2008high,jochmans2019fixed,bonhomme2023much}.

Our approach first recovers cohort-specific factor vectors $\tilde{\gamma}_{ct} \coloneqq Q_c\gamma_t$ for each cohort $c$'s observed outcomes $t \in \mathcal{T}_c$, where $Q_c$ are unknown basis matrices that generally differ across cohorts.\footnote{Imposing a common normalization as in \citet{bai2013principal} is not possible since, in many settings, fewer than $r$ outcomes (and often none) are observed across all cohorts.}
The key insight is that if cohorts $c_1$ and $c_2$ share at least $r$ observed outcomes, we can find a unique alignment matrix $R_{c_1c_2}$ satisfying $R_{c_1c_2}\tilde{\gamma}_{c_1t} = \tilde{\gamma}_{c_2t}$ by solving a system of $r \cdot \abs{\mathcal{T}_{c_1} \cap \mathcal{T}_{c_2}} \geq r^2$ equations with $r^2$ unknowns \citep{bai2021matrix}. This alignment matrix then translates all of cohort $c_1$'s factor vectors into cohort $c_2$'s basis, including those for outcomes not observed by cohort $c_2$.

Returning to Figure \ref{fig:match_outcome_missingness_pattern}'s setting for intuition, data from each worker cohort yields estimates of relative skill values across the firm groups they worked for, but skill definitions may differ across cohorts. When two cohorts share overlapping firm groups, we can translate one cohort's skill definitions into the other's and recover common measures of relative skill values for all firm groups observed for either cohort.

Importantly, these pairwise alignments can be chained to express multiple cohorts' factor vectors with respect to a common basis. For example, in Figure \ref{fig:three_cohort_figure}, we can align cohorts 1 and 2, then cohorts 2 and 3, as indicated by the black arcs. Chaining through cohort 2 expresses $\tilde{\gamma}_t$ for every outcome with respect to a common basis.
When $r \geq 2$, higher-order alignment chains enable identification in even more settings. For example, in Figure \ref{fig:higher_order_chains_figure}, cohorts 1 and 2 each share only $r/2$ outcomes with cohort 3, insufficient for direct alignment. However, first aligning cohorts 1 and 2 yields a combined set of factor vectors sharing $r$ outcomes with cohort 3, enabling subsequent alignment as indicated by the arc connecting cohorts 1 and 2 and the arc connecting that first arc to cohort 3. In Figure \ref{fig:match_outcome_missingness_pattern}'s setting, constructing these chains amounts to extending the cohort-pair-specific translation of relative skills to sequences of cohorts with overlapping firm groups, progressively building a unified picture of relative skill values.

While this intuition clarifies how these alignment chains express cohort-specific factor vectors in a common basis, how to find such chains in a computationally efficient manner---let alone enumerating all of them---is unclear.\footnote{\citet{hull2018estimating} suggests an identification and estimation strategy that enumerates analogous chains under a TWFE-like model of outcomes.} Even if enumeration were feasible, how to combine information across chains is not obvious. Our method sidesteps these difficulties via a linear-algebraic construction that automatically aggregates information from all available alignment chains.

\subsection{The Aggregated Projection Matrix}

To describe our automatic solution to the factor misalignment problem, we introduce additional notation. Let $\Gamma$ be the $T \times r$ matrix with $t$-th row equalling $\gamma_t$, and let $\tilde{\Gamma}_c$ be the $T \times r$ matrix whose $t$-th row equals $\tilde{\gamma}_{ct} = Q_c\gamma_t$ if $t \in \mathcal{T}_c$ and zero otherwise. For a subset of outcomes $\mathcal{T} \subseteq [T]$, let $E_{\mathcal{T}}$ be the $T \times T$ diagonal matrix with $t$-th entry equalling one if $t \in \mathcal{T}$ and zero otherwise, and define $E_c \coloneqq E_{\mathcal{T}_c}$. For any matrix $M$, let $\Pi(M) \coloneqq M(M'M)^{+} M'$ denote the projection onto $M$'s column space, where $M^+$ denotes the Moore-Penrose pseudo-inverse of $M$.

Two observations underpin our solution. First, since the column space of $\tilde{\Gamma}_c = E_c \Gamma Q_c'$ does not depend on $Q_c$, we have $\Pi(\tilde{\Gamma}_c) = \Pi(E_c\Gamma)$---the projection onto the column space of the matrix $E_c\Gamma$ whose non-zero rows are the true factor vectors for cohort $c$'s observed outcomes.\footnote{By definition, $\Pi(\tilde{\Gamma}_c) = E_c \Gamma Q_c' (Q_c \Gamma' E_c \Gamma Q_c)^+ Q_c \Gamma' E_c = E_c \Gamma (\Gamma' E_c' \cdot E_c \Gamma)^+ \Gamma' E_c = \Pi(E_c \Gamma)$.} Thus, $\Pi(\tilde{\Gamma}_c)$ provides a basis-free representation of information about cohort $c$'s factor vectors, uncontaminated by the cohort-specific $Q_c$.

Our second observation is that $\Gamma$'s column space lies in the null space of $E_c - \Pi(E_c\Gamma)$ for each cohort $c$.\footnote{\label{foot:col_space_Gamma_in_null_space_APM} $\Pi(E_c\Gamma) = E_c\Pi(E_c\Gamma) = \Pi(E_c\Gamma)E_c$ and $E_c^2 = E_c$, so $(E_c - \Pi(E_c\Gamma))\Gamma = E_c(I - \Pi(E_c\Gamma))E_c\Gamma = \zeros_{T \times r}$.} This fact motivates defining a particular matrix we call an \textit{Aggregated Projection Matrix} (APM):\footnote{Despite its name, the APM is not itself a projection matrix.}
\begin{equation}\label{eq:aggregated_projection_matrix_def}
 A \coloneqq \frac{1}{C}\sum_{c = 1}^C\left[E_c - \Pi(E_c\Gamma)\right].
\end{equation}
Since $A$ is the mean of $E_c - \Pi(E_c\Gamma) = E_c - \Pi(\tilde{\Gamma}_c)$ across cohorts, $\Gamma$'s column space lies in $A$'s null space. In Section \ref{sec:theory:identification}, we show that, with sufficient overlap between cohorts' observed outcomes, $A$'s null space is \textit{exactly} $\Gamma$'s column space. Thus, the rows of any basis matrix for $A$'s null space provide valid factor vectors for the downstream regression used to impute counterfactual outcomes.

%% file: sections/method.tex
Having explained the conceptual 
underpinnings of our approach, we now describe our estimation and inference procedure, which has four steps and is summarized in Algorithm \ref{alg:estimation}. A software package implementing the procedure is available at \url{https://github.com/brad-ross/apm}.

\begin{algorithm}[t]
    \SetAlgoLined
    \DontPrintSemicolon
    \caption{Estimation}
    \label{alg:estimation}
    \KwData{$\{(C_i, Y_i, X_i)\}_{i = 1}^N$}
    \BlankLine
    \nl \For{$c \in \{1, \dotsc, C\}$}{
        Compute cohort-specific factor vector estimates $\hat{\gamma}_{ct}$ for $t \in \mathcal{T}_c$ (see Section \ref{sec:theory:cohort_specific_factor_estimates})\;
        Construct a $T \times r$ matrix $\hat{\Gamma}_c$ with row $t$ equalling $\hat{\gamma}_{ct}$ if $t \in \mathcal{T}_c$, and $\zeros_r$ otherwise\;
    }
    \nl Construct estimated APM $\hat{A}$ as in \eqref{eq:estimated_APM_def}\;
    \nl Compute $T \times r$ matrix $\hat{\Gamma}$ of eigenvectors for $\hat{A}$'s $r$ smallest eigenvalues\;
    \nl Estimate the outcome imputation ingredients $(\hat{\gamma}_0, \hat{\beta}, \{\hat{\lambda}_i\}_{i = 1}^N)$ via the regression \eqref{eq:regression_imputation_estimator_def}\;
    \nl \For{$c \in \{1, \dotsc, C\}$}{
        Compute outcome mean estimate vector $\hat{\mu}_c$ for cohort $c$ as in \eqref{eq:estimator_def}\;
    }
    \nl Compute estimate $\hat{\eta}$ of nuisance parameters $\eta$ necessary for estimating $\theta$\;
    \nl Compute plug-in target parameter estimate $\hat{\theta} = h(\hat{\mu}, \hat{\eta})$, where $\hat{\mu} = (\hat{\mu}_1', \dotsc, \hat{\mu}_C')'$\;
\end{algorithm}

In step one, we use one of the methods discussed in Section \ref{sec:theory:cohort_specific_factor_estimates}---each imposing different additional assumptions---to construct estimates $\hat{\gamma}_{ct}$ of the cohort-specific factor vectors $\tilde{\gamma}_{ct}$ for each observed outcome $t \in \mathcal{T}_c$ and cohort $c$. 
We stack these estimates into a $T \times r$ matrix $\hat{\Gamma}_c$ whose $t$-th row is $\hat{\gamma}_{ct}$ if $t \in \mathcal{T}_c$ and $\zeros_r$ otherwise. The cohort-specific factor estimators we consider use only data from each cohort and thus can be computed efficiently in parallel.\footnote{In Section \ref{sec:conclusion:nonuniqueness}, we discuss sharing data across cohort-specific factor estimators, leaving details to future work.}

In step two, we use the cohort-specific estimated factor matrices $\{\hat{\Gamma}_c ~\setst~ c \in [C]\}$ to construct an estimated APM:
\begin{equation}\label{eq:estimated_APM_def}
    \hat{A} \coloneqq \frac{1}{C}\sum_{c = 1}^C [E_c - \Pi(\hat{\Gamma}_c)].
\end{equation}
Let $\hat{\Gamma}$ be the $T \times r$ matrix whose columns are eigenvectors corresponding to the $r$ smallest eigenvalues of $\hat{A}$.\footnote{Recall that eigenvectors are unique up to magnitudes, signs, and permutations between indices corresponding to repeated eigenvalues.} Row $t$ of $\hat{\Gamma}$, denoted $\hat{\gamma}_t$, consistently estimates the factor vector $\gamma_t$ in a sense made precise in Section \ref{sec:theory:estimation_inference}.

Step three treats the estimated factor vectors $\{\hat{\gamma}_t ~\setst~ t \in [T]\}$ as data to construct regression-based imputations of missing outcomes. We regress observed outcomes on the estimated factor vectors with unit-specific slopes $\hat{\lambda}_i$, outcome fixed effects $\hat{\gamma}_{t0}$, and covariates $X_{it}$:\footnote{To implement this constrained regression using standard software, the outcome fixed effect $\hat{\gamma}_{t0}$ can be replaced by including the $t$-th row of any $T \times (T - r)$ matrix of eigenvectors corresponding to $\hat{A}$'s $T - r$ largest eigenvalues as additional regressors; the inner product of this vector with its coefficient vector will equal $\hat{\gamma}_{t0}$.}
\begin{equation}\label{eq:regression_imputation_estimator_def}
    \left(\hat{\gamma}_0, \hat{\beta}, \{\hat{\lambda}_i\}_{i = 1}^N\right) \coloneqq \argmin_{\check{\gamma}_0 \in \R^T,\; \check{\beta} \in \R^q,\; \check{\lambda}_i \in \R^r} \left\{\frac{1}{N}\sum_{i = 1}^N \sum_{t \in \mathcal{T}_{C_i}}\left(Y_{it} - \left[\hat{\gamma}_t'\check{\lambda}_i + \check{\gamma}_{t0} + X_{it}'\check{\beta}\right]\right)^2 ~\setst~ \hat{\Gamma}'\check{\gamma}_0 = \zeros_r \right\}.
\end{equation}
We then estimate the mean $\mu_{ct}$ of any observed or counterfactual outcome $t$ for cohort $c$ by averaging asymptotically unbiased imputations of that outcome for the units in that cohort, where $N_c \coloneqq \sum_{i = 1}^N \ind{C_i = c}$ denotes the number of units in cohort $c$:\footnote{These regression-based imputations can also be used for downstream analyses that accommodate estimation error \citep{arkhangelsky2025causal,kline2024firm,kline2020leave,borusyak2021revisiting}.}
\begin{equation}\label{eq:estimator_def}
    \hat{\mu}_{ct} \coloneqq \frac{1}{N_c}\sum_{i = 1}^N \ind{C_i = c} \left[\hat{\gamma}_t'\hat{\lambda}_i + \hat{\gamma}_{t0} + X_{it}'\hat{\beta}\right], \quad \hat{\mu}_c \coloneqq (\hat{\mu}_{c1}, \dotsc, \hat{\mu}_{cT})'.
\end{equation}
The constraint $\sum_{t = 1}^T \hat{\gamma}_{t0}\hat{\gamma}_t = \zeros_r$ in \eqref{eq:regression_imputation_estimator_def} is one of many equivalent normalizations that ensure $\hat{\gamma}_0$ is uniquely defined without affecting $\hat{\mu}_{ct}$. When $\hat{\gamma}_t = 1$, the regression \eqref{eq:regression_imputation_estimator_def} and estimator \eqref{eq:estimator_def} reduce to a TWFE model-based imputation estimator like those studied in \citet{jochmans2019fixed} and \citet{borusyak2021revisiting}, as discussed in Section \ref{sec:setup:factor_model}.\footnote{This normalization simplifies our theoretical results in Section \ref{sec:theory}. When $\hat{\gamma}_t = 1$, the normalization of $\hat{\gamma}_0$ in \eqref{eq:regression_imputation_estimator_def} is equivalent to the commonly imposed normalization that outcome fixed effect estimates sum to zero in TWFE regressions.}

As discussed in Section \ref{sec:setup:setup}, estimands of interest are more often known functions $\theta = h(\mu, \eta)$ of cohort outcome means $\mu \coloneqq (\mu_1', \dotsc, \mu_C')'$ and estimable nuisance parameters $\eta$. Examples include the dynamic treatment effects for event studies in Example \ref{ex:dynamic_treatment_effects} of Appendix \ref{sec:supplement:target_params} and the match outcome attribution statistics in Example \ref{example-match-outcome-attribution}. Extending step three, we compute a plug-in estimate $\hat{\theta} \coloneqq h(\hat{\mu}, \hat{\eta})$, where $\hat{\mu} \coloneqq (\hat{\mu}_1', \dotsc, \hat{\mu}_C')'$ collects the estimates from \eqref{eq:estimator_def} across cohorts and $\hat{\eta}$ estimates $\eta$.

In step four, given a size $\alpha$, we construct simultaneous $1-\alpha$ confidence intervals for the $p$ coordinates of $\theta$ using a Bayesian bootstrap \citep{rubin1981bayesian}.\footnote{Simultaneous coverage $1-\alpha$ means all coordinates of $\theta$ lie inside their respective intervals with probability at least $1-\alpha$. Any bootstrap procedure that weights or resamples units could be used, but we prefer weighted procedures with continuously distributed weights to avoid cases where no units from some cohort are sampled.} Since our bootstrap algorithm is similar to other simultaneous inference procedures (see \citet{chernozhukov2013inference}, for example), we defer a detailed description to Appendix \ref{sec:target_params:inference_procedure} for brevity.

%% file: sections/theory.tex
\subsection{Factor Identification Given Cohort-Specific Factors}\label{sec:theory:identification}

To highlight our contributions, this section takes the cohort-specific factor projection matrices $\Pi(E_c \Gamma)$ as given. Section \ref{sec:theory:cohort_specific_factor_estimates} discusses assumptions from the factor model literature on short, balanced panels under which $\Pi(E_c\Gamma)$ are identified, meaning they can be expressed as functions of each cohort $c$'s observed data distribution.

\begin{algorithm}[t]
    \SetAlgoLined
    \DontPrintSemicolon
    \caption{Observed Outcome Overlap (O\textsuperscript{3}) Algorithm}
    \label{alg:o3_algorithm}
    \KwIn{observed outcomes per cohort $\{\mathcal{T}_c\}_{c = 1}^C$, factor model rank $r$}
    \BlankLine
    \nl Initialize initial super cohorts $\mathcal{S}_c^{(0)} \coloneqq \{c\}$ for $c \in [C]$, set $M^{(0)} \coloneqq C$\;
    \Repeat( \textbf{for} $k = 1, 2, \ldots$){$M^{(k)} = M^{(k - 1)}$}{
        \nl New undirected O\textsuperscript{3} Graph $\mathcal{G}_r^{(k)}$ with $M^{(k-1)}$ vertices, one per super cohort\;
        \nl\label{cond:observed_outcome_overlap_graph_edge_def}Add edge in $\mathcal{G}_r^{(k)}$ between vertices $m_1$ and $m_2$ if $\abs{\mathcal{T}_{\mathcal{S}_{m_1}^{(k - 1)}} \cap \mathcal{T}_{\mathcal{S}_{m_2}^{(k - 1)}}} \geq r$\;
        \nl Compute $M^{(k)}$ connected components (CCs) of $\mathcal{G}_r^{(k)}$: $\{\mathcal{M}_m^{(k)}\}_{m = 1}^{M^{(k)}} \subseteq [M^{(k - 1)}]$\;
        \nl $k$ super cohorts union same-CC $k-1$ super cohorts: $\mathcal{S}_m^{(k)} \coloneqq \cup_{m' \in \mathcal{M}_m^{(k)}} \mathcal{S}_{m'}^{(k - 1)}$ \;
    }(, where $K$ denotes the final iteration count)
\end{algorithm}

To characterize when factor vectors for all outcomes can be recovered with respect to a common basis given the cohort-specific factor projection matrices $\Pi(E_c \Gamma)$, we introduce the \textit{Observed Outcome Overlap (O\textsuperscript{3}) Algorithm}, defined formally in Algorithm \ref{alg:o3_algorithm}. The algorithm first constructs an \textit{O\textsuperscript{3} Graph} with one vertex per cohort and an undirected edge between cohorts sharing at least $r$ observed outcomes.
It then computes the O\textsuperscript{3} Graph's connected components and creates a \textit{super cohort} for each, defined as the union of cohorts in that connected component. 

The algorithm then repeats this process with super cohorts playing the role of the original cohorts: it constructs a new O\textsuperscript{3} Graph with super cohorts as vertices, where a super cohort $\mathcal{S} \subseteq [C]$'s observed outcome set $\mathcal{T}_{\mathcal{S}} \coloneqq \cup_{c \in \mathcal{S}} \mathcal{T}_c$ is the union of its constituent cohorts $c \in \mathcal{S}$; then it computes connected components and merges super cohorts accordingly. The algorithm terminates when no further merging occurs, which is guaranteed within a finite number of iterations $K$.\footnote{Across iterations, the number of connected components weakly decreases and is always at least one; it cannot decrease further once super cohorts stop changing.} Let $M^{(k)}$ denote the number of super cohorts after iteration $k$.

Our key requirement for recovering all factor vectors with respect to a common basis is that the O\textsuperscript{3} Algorithm converges to a single super cohort containing all cohorts. Since the algorithm depends only on the observed outcome sets $\mathcal{T}_c$ and is computationally efficient,\footnote{A crude time complexity bound is $\mathcal{O}(C^3)$: there are $K = \mathcal{O}(C)$ iterations since at least one pair of super cohorts merges in each non-final iteration, and computing connected components is linear in vertices and edges \citep{cormen2022introduction}, with at most $C$ vertices and $\mathcal{O}(C^2)$ edges per graph.} researchers can verify this condition in practice.
This single-super-cohort requirement implies every cohort has at least $r$ observed outcomes: if $T_c < r$ for some cohort $c$, no other cohort can share $r$ or more outcomes with $c$. Consequently, our identification results hold only for units with at least $r$ observed outcomes, i.e., conditional on $T_{C_i} \geq r$.\footnote{This is analogous to restricting samples in bipartite matching settings to row-type units matched with at least two column-type units.} When the algorithm converges to multiple super cohorts, our result below instead applies separately within each final super cohort and the subsets of outcomes observed for at least one cohort in those super cohorts. In Appendix \ref{sec:supplement:connections_graph_based_identification}, we show that when $r = 1$, single-super-cohort convergence is equivalent to standard graph connectedness assumptions in the literature on fixed-effect models of bipartite network data under strict exogeneity \citep{abowd2002computing,jochmans2019fixed,bonhomme2019distributional,hull2018estimating,kline2024firm,bonhomme2020econometric}.

We can now state our main identification result as follows:
\begin{theorem}\label{thm:factor_identification}
    Suppose the factor vectors $\{\gamma_t \,\setst\, t \in [T]\}$ are in general position and the O\textsuperscript{3} Algorithm converges to a single super cohort ($\mathcal{S}_1^{(K)} = [C]$) for whom all outcomes are observed ($\mathcal{T}_{\mathcal{S}_1^{(K)}} = [T]$).\footnote{A collection of vectors $\{v_j \in \R^d \,\setst\, j \in [J]\}$ is in general position if every size-$d$ subset is linearly independent.} Then $\Gamma$'s column space is exactly the null space of the population APM $A$.
\end{theorem}
\noindent That $\Gamma$'s column space lies in $A$'s null space follows directly without assumptions on factor vectors (see Footnote \ref{foot:col_space_Gamma_in_null_space_APM}). The converse---that $A$'s null space lies in $\Gamma$'s column space given single-super-cohort O\textsuperscript{3} Algorithm convergence---requires more involved arguments provided in Appendix \ref{proof:thm:factor_identification}. The general position assumption ensures each factor affects at least one observed outcome for every cohort. While this assumption simplifies exposition and makes the O\textsuperscript{3} Algorithm depend only on observed quantities, we can relax it by instead defining edges based directly on the factor vectors corresponding to those super cohorts' overlapping observed outcomes.\footnote{\label{foot:oracle_o3_algorithm} Specifically, we add an edge between super cohorts if the span of factor vectors for their overlapping outcomes has dimension $r$. Formally, replace Line \ref{cond:observed_outcome_overlap_graph_edge_def} of Algorithm \ref{alg:o3_algorithm} with: add an edge between $m_1$ and $m_2$ if $\mathrm{rank}(E_{\mathcal{T}_{\mathcal{S}_{m_1}^{(k)}}}E_{\mathcal{T}_{\mathcal{S}_{m_2}^{(k)}}}\Gamma) = r$. Single-super-cohort convergence of this ``oracle'' algorithm implies $\mathrm{rank}(E_c\Gamma) = r$; see Appendix \ref{proof:lemma:E_c_Gamma_full_rank}.}

\subsection{Consistency, Asymptotic Linearity Given Cohort-Specific Factors}\label{sec:theory:estimation_inference}

Having identified the column space of $\Gamma$ given cohort-specific factor projection matrices $\Pi(E_c\Gamma)$, we now show that the plug-in estimator from Section \ref{sec:method:procedure} is asymptotically linear. We first introduce some additional notation. Let $\hat{\E}_N$ denote expectation under the empirical measure (i.e., $\hat{\E}_N[V_i] \coloneqq \sum_{i = 1}^N V_i/N$). For square matrices $M_1$, $M_2$, and $M_3$, define the multilinear function $g(M_1, M_2, M_3) \coloneqq M_1M_2M_3 + M_3M_2M_1$. Let the $T$-vector $Y_i^* \coloneqq (Y_{i1}^*, \dotsc, Y_{iT}^*)'$ collect unit $i$'s outcomes, let $Y_i \coloneqq E_{C_i}Y_i^*$ denote observed outcomes,\footnote{Although $Y_i$ has undefined entries for unobserved outcomes, all entries of $E_{C_i} Y_i$ are well-defined since those corresponding to unobserved outcomes are zero.} and let $\epsilon_i \coloneqq (\epsilon_{i1}, \dotsc, \epsilon_{iT})'$ collect $i$'s outcome residuals. Finally, let $\mu_{X,\, c} \coloneqq \E[X_i ~|~ C_i = c]$ and $\hat{\mu}_{X,\, c} \coloneqq \hat{\E}_N[X_i ~|~ C_i = c]$ denote cohort $c$'s population and sample mean covariate matrices.

As in Section \ref{sec:theory:identification}, we assume access to consistent estimators $\Pi(\hat{\Gamma}_c)$ of $\Pi(E_c\Gamma)$ satisfying the following asymptotic linearity condition:
\begin{assumption}\label{assump:cohort_specific_factor_proj_mat_inf_fn}
    For each cohort $c \in [C]$, there exists a $(T \times T)$-matrix-valued function $\phi_{\Gamma, c}$ of $C_i$, $Y_i$, and $X_i$ such that $\E[\phi_{\Gamma, c}(C_i, Y_i, X_i)] = \zeros_{T \times T}$, $\E[\normnofit{\phi_{\Gamma, c}(C_i, Y_i, X_i)}_{\mathrm{op}}^2] < \infty$, and
    \begin{equation}\label{cond:cohort_specific_proj_mat_asymptotic_normality}
        \norm{\left[\Pi(\hat{\Gamma}_c) - \Pi(E_c \Gamma)\right] - \hat{\E}_N[\phi_{\Gamma, c}(C_i, Y_i, X_i)]}_\text{op} = \littleop{N^{-1/2}}.
    \end{equation}
\end{assumption}
\noindent Section \ref{sec:theory:cohort_specific_factor_estimates} discusses low-level assumptions from the literature on short, balanced panel factor models under which estimators satisfying Assumption \ref{assump:cohort_specific_factor_proj_mat_inf_fn} exist. For example, with no covariates ($X_i = \zeros_{T \times q}$), the Principal Components (PC) estimator applied to cohort $c$'s data satisfies Assumption \ref{assump:cohort_specific_factor_proj_mat_inf_fn}, as shown in Appendix \ref{sec:supplement:pc_estimator_consistency_asymptotic_linear}.

Under this assumption, the column space of our plug-in estimator $\hat{\Gamma}$ as represented by $\Pi(\hat{\Gamma})$ is consistent for the column space of $\Gamma$ as represented by $\Pi(\Gamma)$ and asymptotically linear:
\begin{theorem}\label{thm:factor_consistency_asymptotic_normality}
Suppose that Theorem \ref{thm:factor_identification} and Assumption \ref{assump:cohort_specific_factor_proj_mat_inf_fn} hold. Then as $N \rightarrow \infty$,
\begin{equation}\label{eq:factor_proj_mat_asymptotic_normality}
\begin{aligned}
    &\norm{[\Pi(\hat{\Gamma}) - \Pi(\Gamma)] - \hat{\E}_N\left[\phi_A(C_i, Y_i, X_i)\right]}_\text{op} = \littleop{N^{-1/2}},
\end{aligned}
\end{equation}
where $\Pi(\hat{\Gamma})$'s influence function
\begin{equation}\label{eq:apm_inf_fn_def}
    \phi_A(C_i, Y_i, X_i) \coloneqq g\left(A^+,\, \frac{1}{C}\sum_{c = 1}^C\phi_{\Gamma, c}(C_i, Y_i, X_i),\, \Pi(\Gamma)\right)
\end{equation}
has zero mean and $\E[\normnofit{\phi_A(C_i, Y_i, X_i)}_{\mathrm{op}}^2] < \infty$.
\end{theorem}
\noindent We prove Theorem \ref{thm:factor_consistency_asymptotic_normality} in Appendix \ref{proof:thm:factor_consistency_asymptotic_normality}. The proof uses an exact, first-order expansion of the operator mapping a symmetric matrix to the projection onto a subset of its eigenspaces, derived via Kato's integral \citep{kato1949convergence}. Since projection matrices uniquely represent column spaces, this expansion directly bounds the second-order error of $\Pi(\hat{\Gamma})$ under minimal assumptions on $A$'s eigenvalues. This expansion appears to be new and may be of independent interest, so we provide a self-contained description in Appendix \ref{sec:eigspace_perturbation}.

Next, we show that $\hat{\mu}_c$ is asymptotically linear, which requires additional regularity conditions. First, we impose the following moment bounds on our model's primitives:
\begin{assumption}\label{assump:cohort_sizes_and_potential_outcome_variance_bound}
$\E[\norm{\lambda_i}_2^2] < \infty$, $\E[\norm{X_i}_\mathrm{op}^4] < \infty$, $\E[\norm{\epsilon_i}_2^2] < \infty$, and for all $c \in [C]$, $p_c > 0$.
\end{assumption}
\noindent These moment bounds in turn imply that $\E[\norm{Y_{it}^*}_2^2] < \infty$, which is sufficient for the proof of the result below.
In addition, we require an assumption to rule out collinearity between the factor vectors, outcome fixed effects, and covariates:

\begin{assumption}\label{assump:A0_invertibility}
The $(r+1)$-th smallest eigenvalue of the following matrix is positive:\footnote{It is easy to verify this matrix is positive semi-definite.} $$\E\left[[I - \Pi(\Gamma)\ \ X_i]' \{E_{C_i} - \Pi(E_{C_i}\Gamma)\} [I - \Pi(\Gamma)\ \ X_i]\right].$$
\end{assumption}

\noindent In Appendix \ref{sec:A0_invertibility}, we show that, when there are no covariates ($X_i = \zeros_{T \times q}$), Assumption \ref{assump:A0_invertibility} holds if and only if the column space of $\Gamma$ equals the null space of $A$. As such, Assumption \ref{assump:A0_invertibility} is not necessary in this case if we assume Theorem \ref{thm:factor_identification} holds.
We now state our main consistency and asymptotic linearity result for $\hat{\mu}_c$:

\begin{theorem}\label{thm:outcome_mean_consistency_asymptotic_normality}
Suppose that Theorem \ref{thm:factor_identification} and Assumptions \ref{assump:cohort_specific_factor_proj_mat_inf_fn}, \ref{assump:cohort_sizes_and_potential_outcome_variance_bound}, and \ref{assump:A0_invertibility} hold.\footnote{In Appendix \ref{sec:estimation_proofs}, we prove Theorem \ref{thm:outcome_mean_consistency_asymptotic_normality} under a weaker assumption than the model \eqref{eq:factor_model}, namely that $\E[\epsilon_{it} ~|~ C_i, X_{it}] = 0$ . However, an assumption like mean-independence of $\epsilon_{it}$ from $C_i$, $\lambda_i$, and the full collection of covariates $X_i$ in addition is often necessary for cohort-specific factor estimators to satisfy Assumption \ref{assump:cohort_specific_factor_proj_mat_inf_fn}; see Section \ref{sec:theory:cohort_specific_factor_estimates}.} Then as $N \rightarrow \infty$,
\begin{equation}\label{eq:counterfactual_outcome_mean_vector_inf_fn}
\begin{aligned}
    \hat{\mu}_c - \mu_c &= \hat{\E}_N\left[\psi_c(C_i, Y_i, X_i)\right] + \littleop{N^{-1/2}}, \\
    \psi_c(C_i, Y_i, X_i) &\coloneqq \psi_{c,\, \mathrm{reg}}(C_i, Y_i, X_i) + \psi_{c,\, \mathrm{fac}}(C_i, Y_i, X_i)
\end{aligned}
\end{equation}
where the influence function $\psi_c$ decomposes into two zero-mean, finite-second-moment components: $\psi_{c,\, \mathrm{reg}}$ given in \eqref{eq:mu_c_inf_fn_reg}, the influence function if $\Pi(\Gamma)$ were known, and $\psi_{c,\, \mathrm{fac}}$ given in \eqref{eq:mu_c_inf_fn_fac}, an additional term due to factor estimation error. Both are continuous in $\Pi(\Gamma), E_c\mu_c, \mu_{X,\, c}$, and $(\gamma_0, \beta)$.
\end{theorem}

\noindent We defer full definitions of $\psi_{c,\, \mathrm{reg}}$ and $\psi_{c,\, \mathrm{fac}}$ to Theorem \ref{thm:muc_hat_influence} and prove the result in Appendix \ref{subapp:influence_function}. 

Given $\hat{\mu}_c$'s asymptotic linearity, the asymptotic properties of the plug-in estimator $\hat{\theta}$ of a target estimand $\theta = h(\mu, \eta)$ follow easily as corollaries, where $h$ is smooth and $\eta$ are nuisance parameters that are consistently estimable at parametric rates. In particular, we prove that $\hat{\theta}$ is asymptotically normal in Appendix \ref{sec:theory:target_parameters}.

When the cohort-specific factor estimator influence functions $\phi_{\Gamma, c}$ are known, $\psi_c$ and thus the influence function for $\hat{\theta}$ from \eqref{eq:target_parameter_est_asymptotic_linear} can be consistently estimated by plugging the consistent estimators $\Pi(\hat{\Gamma}), E_c\hat{\mu}_c, \hat{\mu}_{X,\, c}$, and $(\hat{\gamma}_0, \hat{\beta})$ into $\phi_{\Gamma, c}$ for their corresponding population versions, since $\psi_c$ is smooth in these quantities. As such, one could use a plug-in procedure for valid inference in principle. However, deriving $\phi_{\Gamma, c}$ for every cohort-specific factor estimator in Section \ref{sec:theory:cohort_specific_factor_estimates} is prohibitively laborious.
To maintain flexibility in the choice of cohort-specific factor estimators, we opt for our Bayesian bootstrap approach. The resulting simultaneous confidence intervals from Section \ref{sec:target_params:inference_procedure} achieve valid coverage of $\theta$ under minimal regularity conditions without requiring knowledge of the cohort-specific influence functions $\phi_{\Gamma, c}$, as proved in Appendix \ref{sec:theory:target_parameters}.

\subsection{Identifying and Estimating Cohort-Specific Factors}\label{sec:theory:cohort_specific_factor_estimates}

We now turn to discussing how to identify and estimate the cohort-specific factor projection matrices $\Pi(E_c\Gamma)$ since, without additional assumptions, $\Pi(E_c\Gamma)$ cannot be identified \citep{anderson1956statistical}. Below, we discuss several common assumptions from the literature on estimating factor models with no missing data that enable identification and estimation of $\Pi(E_c\Gamma)$ when $T_c$ remains fixed as $N_c$ grows. Our APM approach can wrap around any of these cohort-specific factor vector estimators.

\paragraph*{Models with uncorrelated errors across outcomes.} When errors $\epsilon_{it}$ are uncorrelated across outcomes $t$---a common assumption in the bipartite match outcome literature \citep{kline2020leave}---the cohort-specific factor matrix can be identified without auxiliary data. In the simplest case with no covariates and homoskedastic errors across outcomes (but not necessarily across units), the canonical Principal Components (PC) estimator $\Pi(\hat{\Gamma}_{c, \text{PC}})$ yields $N_c^{1/2}$-consistent estimates of $\Pi(E_c\Gamma)$ provided $T_c$ remains finite as $N_c$ grows and the loadings $\lambda_i$ for units in cohort $c$ have a full-rank covariance matrix \citep{connor1986performance,bai2003inferential,westerlund2020cross}. In Appendix \ref{sec:supplement:pc_estimator_consistency_asymptotic_linear}, we prove that $\Pi(\hat{\Gamma}_{c, \text{PC}})$ is asymptotically linear, as required by Assumption \ref{assump:cohort_specific_factor_proj_mat_inf_fn}, under slightly weaker conditions than typically imposed, using our first-order expansion of the eigenspace projection operator from Appendix \ref{sec:eigspace_perturbation}.

When errors are heteroskedastic, the PC estimator is generally inconsistent in short panels due to the incidental parameter problem \citep{chamberlain1983arbitrage, connor1986performance, ahn2001gmm, bai2003inferential, neyman1948consistent, lancaster2000incidental}. Unlike the homoskedastic case where $T \ge r+1$ suffices for identification, \citet{anderson1956statistical} show that $(T - r)^2 \ge T + r$ is necessary. They propose estimators requiring maximization of highly nonconcave objectives (see \citet{anderson1963asymptotic} and Chapter 14 of \citet{anderson2009introduction}). While global maximizers yield asymptotically linear estimators of $\Pi(E_c\Gamma)$, off-the-shelf algorithms are not guaranteed to converge to them, and algorithms with global optimality guarantees are difficult to scale even for moderate $r$ and $T_c$ \citep{bertsimas2017certifiably, khamaru2019computation}.

Another line of work develops computationally efficient estimators based on internal instruments, proposed by \citet{madansky1964instrumental} and generalized by \citet{heckman1987importance, freyberger2018non, harding2022estimation}. These methods assume $T \ge 2r+1$ and accommodate covariates, but typically only produce aligned factor estimates for subsets of factors. An exception is \citet{hagglund1982factor}, which yields an $N_c^{1/2}$-consistent and asymptotically linear estimate of the entire factor matrix column space under the stronger assumption that $T \ge 3r$. 

\paragraph*{Models with correlated errors and auxiliary data.} With auxiliary data, we can relax the assumption of uncorrelated errors across outcomes. One approach uses quasi-differencing to project out the factor component \citep{chamberlain1984panel}. With sufficiently many time-varying covariates, the GMM estimators of \citet{chamberlain1984panel, chamberlain1992efficiency, ahn2001gmm, ahn2013panel, phillips2024simple, robertson2015iv} are $N_c^{1/2}$-consistent and asymptotically linear in short panels with heteroskedastic and correlated errors. Other $N_c^{1/2}$-consistent methods using covariates include Projected Principal Component Analysis \citep{fan2016projected} and Diversified Projections \citep{duan2023target}. Alternatively, if some covariates share a low-rank structure with the outcomes,\footnote{In this case, one could alternatively treat the covariates as additional, always-observed outcomes, creating overlapping observed ``outcomes'' between every pair of cohorts, strengthening identification of $\Pi(\Gamma)$.} one can apply Common Correlated Effects \citep{pesaran2006estimation,westerlund2019cce}, Essential Regression \citep{bing2022inference}, or transfer learning \citep{duan2023target} to estimate $\Pi(E_c\Gamma)$.

%% file: sections/empirical.tex
We evaluate our procedure as an alternative to TWFE model-based methods for decomposing observed wage differences across locations into causal location effects versus worker sorting, in the spirit of \citet{card2023location}. As discussed in Example \ref{example-match-outcome-attribution}, such analyses require predicting counterfactual wages for workers at locations where they are not observed. Unlike TWFE-based imputations, which rule out complementarities and multidimensional match effects (Section \ref{sec:setup:factor_model}), our factor model-based method accommodates these features.\footnote{\citet{bonhomme2019distributional} note that typical empirical tests claiming no match effects have no power under models with worker-firm complementarities (see \citet{card2013workplace} for an example). \citet{card2013workplace} and \citet{woodcock2015match} find evidence for match effects using repeated measurements of the same matches over time, though such estimates are noisy and/or require random-effects-like assumptions. \citet{bonhomme2019distributional} estimate a model with discrete worker and firm types and also find evidence of complementarities.}

\subsection{Setting and Sample Construction}\label{sec:empirical_performance:setting}

Our empirical illustration uses the Veneto Worker Histories (VWH) dataset derived from Italian social security administration data on the weekly wages earned across different firms of every person who ever lived or worked in any of the seven Veneto provinces from 1975 to 2001.\footnote{The VWH dataset was developed by the Economics Department in Universit\`{a} Ca' Foscari Venezia under the supervision of Giuseppe Tattara, and can be accessed at \url{https://www.frdb.org/en/dati/dati-inps-carriere-lavorative-in-veneto/}. The code for our empirical illustration is available at \url{https://github.com/brad-ross/apm-paper}.} Each observation records a worker-firm-year triple, including weeks worked in the year and total pay.

To construct our panel, we restrict to firms existing in Veneto from 1998 through 2001 and workers employed by those firms during that period. We then cluster firms within each province into three types using the $k$-means procedure proposed in \citet{bonhomme2019distributional}.\footnote{Clustering firms within each province avoids understating the role of location in wage determination by ignoring within-province firm heterogeneity \citep{card2023location}.} Following \citet{lachowska2023firm} and \citet{chandar2025shifts}, an outcome $Y_{it}^*$ is the log of worker $i$'s average weekly wage for the triple $t$ of a two-year period (1998--1999 or 2000--2001), a province, and a firm type within that province, yielding $T = \input{\figurespath result_summaries/snippets/num_outcomes.txt}$ outcomes.\footnote{In education contexts, \citet{chetty2014measuring} and \citet{kwon2023optimal} study TWFE models with teacher-by-year fixed effects, allowing teachers' effects to drift over time.} Dropping cohorts with fewer than $\input{\figurespath result_summaries/snippets/min_cohort_size.txt}$ workers leaves $N = \input{\figurespath result_summaries/snippets/num_workers.txt}$ workers in $C = \input{\figurespath result_summaries/snippets/num_cohorts.txt}$ cohorts with $\input{\figurespath result_summaries/snippets/num_obs.txt}$ observed outcomes, an average of 2.18 observed outcomes per worker. See Appendix \ref{sec:empirical_appendix:dataset_construction} for further details.

Figure \ref{fig:match_outcome_missingness_pattern} illustrates the outcome missingness pattern in our sample. Our sample restrictions ensure at least two observed wages per worker, enabling estimation of average outcomes for both the typical ``mover'' population and ``stayers'' who remain within firm groups. For $r \in \{1, 2\}$, the O\textsuperscript{3} Algorithm converges after two iterations to a single super cohort comprising all cohorts with at least $\input{\figurespath result_summaries/snippets/min_cohort_size.txt}$ workers, so the entire $C \times T$ matrix of cohort outcome means is identified under both the TWFE model and factor models with $r \leq 2$.\footnote{Figures \ref{fig:largest_super_cohort_share_plot} and \ref{fig:num_o3_iterations_needed_plot} visualize the shares of units and firms whose cohort outcome means are identified, and the number of O\textsuperscript{3} Algorithm iterations needed, across different numbers of firm clusters per province.} In contrast, Figure \ref{fig:block_missingness_id_firm_weighted_plot} shows that no embedded block missingness patterns exist with which other factor model estimation methods could identify most cohorts' outcome means.

\subsection{Semi-Synthetic Simulation Study of Estimator Performance}\label{sec:empirical_performance:semi_synthetic}

To evaluate our method's performance at predicting counterfactual outcome means without taking a stand on the true data-generating process, we conduct a semi-synthetic simulation study. In particular, for each target cohort $c^*$ with at least three observed outcomes, we ``mask'' one outcome $t^*$, treating it as if it were unobserved. Using the Bayesian bootstrap procedure from Section \ref{sec:target_params:inference_procedure}, we construct 1,000 bootstrap replicates of five estimates of $\mu_{c^*t^*}$: one TWFE model-based estimate and four variants of our method with $r \in \{1, 2\}$ and with and without outcome fixed effects. All variants use the PC estimator from Section \ref{sec:theory:cohort_specific_factor_estimates} with no covariates for simplicity. We evaluate accuracy via absolute bias, standard error, and root mean squared error (RMSE) across bootstrap replicates, treating the actual sample mean for the masked cohort outcome as ground truth. We repeat this procedure for every observed outcome across the 15 largest cohorts with at least three observed outcomes.\footnote{For each cohort-outcome pair, we verify via the O\textsuperscript{3} Algorithm that all cohort outcome means remain identified after masking.}

\begin{table}[t]
    \centering
    \begin{subtable}[b]{0.48\textwidth}
        \centering
        \input{\figurespath result_summaries/tables/leave_cohort_outcome_out_evals_results.tex}
        \caption{Estimator Performance Evaluation}
        \label{tab:leave_cohort_outcome_out_evals_results}
    \end{subtable}
    \hfill
    \begin{subtable}[b]{0.48\textwidth}
        \centering
        \input{\figurespath result_summaries/tables/match_outcome_decomp_results.tex}
        \caption{Match Outcome Decomposition}
        \label{tab:match_outcome_decomp_results}
    \end{subtable}
    \caption{Table \ref{tab:leave_cohort_outcome_out_evals_results} summarizes the performance of the four variants of our method relative to the TWFE model-based estimator in the semi-synthetic simulation study described in Section \ref{sec:empirical_performance:semi_synthetic}; bolded values indicate the best-performing variant of our method for each error metric. Table \ref{tab:match_outcome_decomp_results} reports the share of observed wage differences we attribute to the causal effects of locations on workers' wages using the approach described in Section \ref{sec:empirical_performance:match_outcome_decomposition}. The APM estimate is computed using our method with $r = 1$ and no outcome fixed effects. Inference is based on the Bayesian bootstrap procedure described in Section \ref{sec:target_params:inference_procedure}.}
    \label{fig:rel_err_metrics_dists_orig_panel}
\end{table}

Table \ref{tab:leave_cohort_outcome_out_evals_results} summarizes the performance of our method's variants relative to the TWFE estimator. For each estimator and metric, we report the geometric mean ratio of our method's error metric to the TWFE estimator's as well as the share of cohort-outcome pairs where each estimator outperforms the TWFE estimator.\footnote{Both statistics weight cohorts by worker count and outcomes by firm count in the corresponding province-cluster.} The best-performing variant ($r = 1$, no outcome fixed effects) has lower bias than the TWFE estimator for $\input{\figurespath result_summaries/snippets/pc_1_no_fes_equal_weights_bias_share_better.txt}$ of cohort-outcome pairs and lower RMSE for $\input{\figurespath result_summaries/snippets/pc_1_no_fes_equal_weights_rmse_share_better.txt}$ despite higher variance, with average bias and RMSE reductions of $\input{\figurespath result_summaries/snippets/pc_1_no_fes_equal_weights_bias_avg_pct_better.txt}$ and $\input{\figurespath result_summaries/snippets/pc_1_no_fes_equal_weights_rmse_avg_pct_better.txt}$, respectively. In contrast, variants with $r = 2$ exhibit much higher variance and nearly double the bias. These results suggest that accommodating complementarities via $r = 1$ improves accuracy, while the multidimensional confounders allowed by $r = 2$ are less important in this setting.

\subsection{Match Outcome Decomposition}\label{sec:empirical_performance:match_outcome_decomposition}

Lastly, we estimate a variant of the decomposition from Example \ref{example-match-outcome-attribution} adapted to groups of outcomes. Specifically, we replace single-outcome means with firm-count-weighted averages across firm clusters within each province, yielding estimates of the share of observed wage differences between province pairs attributable to causal location effects; see Example \ref{example-match-outcome-attribution} in Appendix \ref{sec:supplement:target_params} for details.

Table \ref{tab:match_outcome_decomp_results} reports the average attribution statistics across all province pairs using both our method (with $r = 1$ and no outcome fixed effects) and the TWFE model-based estimator to impute counterfactual outcome means. Both estimators yield similarly imprecise estimates, with our method suggesting provinces causally contribute $\input{\figurespath result_summaries/snippets/fgw_decomp_province_difference_estimate.txt}$ percentage points more to observed wage differences than the TWFE model-based estimator implies. However, the 95\% bootstrap confidence interval does not rule out the TWFE model-based estimator implying a slightly lower province share. These results, together with those from Section \ref{sec:empirical_performance:semi_synthetic}, are consistent with \citet{bonhomme2019distributional}, who find complementarity between worker and firm heterogeneity with a modest effect on wage dispersion.

%% file: figures/result_summaries/snippets/num_outcomes.txt
42

%% file: figures/result_summaries/snippets/min_cohort_size.txt
75

%% file: figures/result_summaries/snippets/num_workers.txt
1,032,650

%% file: figures/result_summaries/snippets/num_cohorts.txt
450

%% file: figures/result_summaries/snippets/num_obs.txt
2,249,271

%% file: figures/result_summaries/tables/leave_cohort_outcome_out_evals_results.tex
\begin{tabular}[t]{llrrr}
\toprule
\multicolumn{1}{c}{ } & \multicolumn{2}{c}{FEs} & \multicolumn{2}{c}{No FEs} \\
\cmidrule(l{3pt}r{3pt}){2-3} \cmidrule(l{3pt}r{3pt}){4-5}
Stat. & $r = 1$ & $r = 2$ & $r = 1$ & $r = 2$\\
\midrule
\addlinespace[0.3em]
\multicolumn{5}{l}{\textbf{Geometric mean of statistic ratio}}\\
\hspace{1em}Bias & 0.89 & 1.72 & \textbf{0.78} & 1.41\\
\hspace{1em}Std. Err. & 1.97 & 173.42 & \textbf{1.35} & 139.53\\
\hspace{1em}RMSE & 1.12 & 33.39 & \textbf{0.89} & 26.53\\
\addlinespace[0.3em]
\multicolumn{5}{l}{\textbf{Share with statistic ratio $\leq 1$}}\\
\hspace{1em}Bias & 0.59 & 0.20 & \textbf{0.67} & 0.40\\
\hspace{1em}Std. Err. & \textbf{0.02} & 0.00 & 0.00 & 0.00\\
\hspace{1em}RMSE & 0.52 & 0.00 & \textbf{0.64} & 0.00\\
\bottomrule
\end{tabular}

%% file: figures/result_summaries/tables/match_outcome_decomp_results.tex
\begin{tabular}[t]{lccc}
\toprule
Stat. & APM & TWFE & Diff.\\
\midrule
\makecell{Province \\ Share} & 0.55 & 0.49 & 0.06\\
Std. Err. & 0.10 & 0.10 & 0.05\\
$p$-value & 0.00 & 0.00 & 0.25\\
95\% CI & \makecell{[0.34, \\ \;0.77]} & \makecell{[0.29, \\ \;0.69]} & \makecell{[-0.04, \\ \;0.16]}\\
\bottomrule
\end{tabular}

%% file: figures/result_summaries/snippets/pc_1_no_fes_equal_weights_bias_share_better.txt
67.2\%

%% file: figures/result_summaries/snippets/pc_1_no_fes_equal_weights_rmse_share_better.txt
63.7\%

%% file: figures/result_summaries/snippets/pc_1_no_fes_equal_weights_bias_avg_pct_better.txt
21.5\%

%% file: figures/result_summaries/snippets/pc_1_no_fes_equal_weights_rmse_avg_pct_better.txt
11.2\%

%% file: figures/result_summaries/snippets/fgw_decomp_province_difference_estimate.txt
6.0

%% file: sections/conclusion.tex
\subsection{Choosing the Factor Model Rank}\label{sec:conclusion:choosing_factor_model_rank}

Throughout the previous sections we assumed the factor model rank $r$ was known. Unlike the well-established literature on determining $r$ when both $N$ and $T$ are large \citep{bai2002determining,onatski2009testing,ahn2013eigenvalue,gagliardini2019diagnostic,dobriban2019deterministic,chernozhukov2021inference,fan2022estimating}, rank determination from short panel data without parametric distributional assumptions on residuals $\epsilon_{it}$ has only recently been studied \citep{ahn2013panel, fortin2022eigenvalue,fortin2023latent}.

Rather than rely on analytical results like these that require stronger assumptions, we suggest a data-driven approach based on the leave-cohort-outcome-out exercise described in Section \ref{sec:empirical_performance:semi_synthetic}. For each candidate value of $r$, we randomly select $(c^*, t^*)$ pairs, mask the corresponding observations, apply our method to construct bootstrap-replicated estimates of $\mu_{c^* t^*}$, and compute the bootstrap RMSE using the sample mean of outcome $t^*$ in cohort $c^*$ as ground truth. We then choose $r$ to minimize the average RMSE across all cohort-outcome pairs $(c^*, t^*)$. This rank selection method is analogous to network cross-validation \citep{li2020network}, enabling model comparison without ``double-dipping.''

\subsection{From Short to Sparse Panels}\label{sec:conclusion:short_to_sparse_panels}

While we focus on short panels in this paper, our proof techniques readily extend to more general settings since all of our steps are effectively non-asymptotic. Here, we generalize some of our results to \emph{sparse} panel settings, where the numbers of observed outcomes per unit $T_c$ remain uniformly bounded but the number of cohorts $C$ (and thus $T$) may grow with $N$. In these settings, cohort sizes $p_c$ mechanically decrease with $C$ and cohort connectedness as determined by the O\textsuperscript{3} Algorithm can weaken with $C$. Even with moderate $T$, there are $\sum_{t = r}^T\binom{T}{t}$ possible cohorts, so sparse panel settings with many small cohorts can be realistic even when only a small fraction of potential cohorts are observed.

For example, the matched employer-employee panel from Section \ref{sec:empirical_performance:setting} has $T = \input{\figurespath result_summaries/snippets/num_outcomes.txt}$ outcomes but $C = \input{\figurespath result_summaries/snippets/num_cohorts.txt}$ cohorts, with only approximately 10\% of workers in the largest cohort (Figure \ref{fig:largest_super_cohort_share_workers_plot}) and over 100 cohorts with fewer than 100 workers (Figure \ref{fig:cohort_size_dist_plot}). To demonstrate the effects of increasing the number of cohorts and outcomes, we can increase the number of firm clusters per province beyond the three used in Section \ref{sec:empirical_performance:setting}; as Figure \ref{fig:largest_super_cohort_share_plot} shows, the largest cohort shrinks rapidly, and the shares of workers and firms in the largest terminal super cohort of the O\textsuperscript{3} Algorithm decreases, indicating weakening cohort connectedness.
Similarly, Figure \ref{fig:control_outcomes_over_time_missingness_pattern}'s event study would become a sparse panel if new driver cohorts continued to be recruited, since all drivers are monitored for a fixed 20 weeks regardless of the growing time horizon.

To extend our theoretical results to sparse panels, we generalize Assumption \ref{assump:cohort_specific_factor_proj_mat_inf_fn} to explicitly account for how problem parameters like absolute cohort size $p_cN$ affect the quality of first-order approximations of the cohort-specific factor estimators:

\begin{assumption}\label{assump:general_cohort_specific_factor_proj_mat_inf_fn}
    For each cohort $c \in [C]$, there exists a $(T \times T)$-matrix-valued function $\phi_{\Gamma, c}$ of $C_i$, $Y_i$, and $X_i$ such that $\E[\phi_{\Gamma, c}(C_i, Y_i, X_i)] = \zeros_{T \times T}$ and
    \begin{equation}\label{cond:general_cohort_specific_proj_mat_asymptotic_linearity}
    \begin{aligned}
        \norm{\left[\Pi(\hat{\Gamma}_c) - \Pi(E_c \Gamma)\right] - \hat{\E}_N[\phi_{\Gamma, c}(C_i, Y_i, X_i)]}_\text{op} &= \bigop{rT_c \cdot (p_cN)^{-1}}, \\
        \E[\norm{\phi_{\Gamma, c}(C_i, Y_i, X_i)}_{\mathrm{op}}^2] &= \bigo{rT_cp_c^{-1}}.
    \end{aligned}
    \end{equation}
\end{assumption}
\noindent The convergence rates in Assumption \ref{assump:general_cohort_specific_factor_proj_mat_inf_fn} reflect typical parametric convergence rates of factor model estimators without missing data. For example, the PC estimator of $\Pi(E_c\Gamma)$ applied to cohort $c$'s data satisfies this assumption, as shown in Proposition \ref{proposition:cohort_specific_factor_inf_fn_homoskedastic_outcomes} in Appendix \ref{sec:supplement:pc_estimator_consistency_asymptotic_linear}. The framework easily extends to more general rates.

We now state a generalization of Theorem \ref{thm:factor_consistency_asymptotic_normality} showing that our plug-in estimator $\Pi(\hat{\Gamma})$ is consistent and asymptotically linear, where $d_{r+1}$ denotes the $(r+1)$-th smallest eigenvalue of the APM $A$:\footnote{All eigenvalues of $A$ are real and non-negative: since $E_c - \Pi(E_c\Gamma) = E_c[I - \Pi(E_c\Gamma)]E_c$ is the product of projection matrices and hence positive semidefinite, $A$ is a sum of these matrices and is thus also positive semidefinite.}
\begin{theorem}\label{thm:general_factor_consistency_asymptotic_normality}
Suppose that Theorem \ref{thm:factor_identification} and Assumption \ref{assump:general_cohort_specific_factor_proj_mat_inf_fn} hold. If 
\begin{equation}\label{cond:apm_est_err_eigenvalue_bound}
    \frac{1}{C}\sum_{c = 1}^C \sqrt{\frac{rT_c}{p_cN}} = o(d_{r+1}),
\end{equation}
then $\Pi(\hat{\Gamma})$ is consistent and asymptotically linear in the following sense:
\begin{equation}\label{eq:general_factor_proj_mat_asymptotic_linearity}
\begin{aligned}
    &\norm{\left[\Pi(\hat{\Gamma}) - \Pi(\Gamma)\right] - \hat{\E}_N\left[\phi_A(C_i, Y_i, X_i)\right]}_\text{op} = \bigop{d_{r+1}^{-1} \cdot \frac{1}{C}\sum_{c = 1}^C \frac{rT_c}{p_cN}},
\end{aligned}
\end{equation}
where $\Pi(\hat{\Gamma})$'s influence function $\phi_A(C_i, Y_i, X_i)$ defined in \eqref{eq:apm_inf_fn_def} has zero mean and
\begin{equation}\label{eq:sparse_factor_proj_mat_asymptotic_error_squared}
    \E[\normnofit{\phi_A(C_i, Y_i, X_i)}_{\mathrm{op}}^2] = \bigo{d_{r+1}^{-2} \cdot \frac{1}{C}\sum_{c = 1}^C \frac{rT_c}{p_c}}.
\end{equation}
\end{theorem}
\noindent The proof follows that of Theorem \ref{thm:factor_consistency_asymptotic_normality} closely, with additional details at the end of Appendix \ref{proof:thm:factor_consistency_asymptotic_normality}. While \eqref{eq:general_factor_proj_mat_asymptotic_linearity} is analogous to \eqref{eq:factor_proj_mat_asymptotic_normality}, we must now be explicit about the condition \eqref{cond:apm_est_err_eigenvalue_bound}: it ensures the estimation error of $\hat{A}$ (bounded by the left-hand side) is small relative to the eigenspace linearization error governed by $d_{r+1}$. Moreover, finite influence function variability for fixed $C$ as in Theorem \ref{thm:factor_consistency_asymptotic_normality} no longer suffices; the explicit bound \eqref{eq:sparse_factor_proj_mat_asymptotic_error_squared} ensures the influence function variance shrinks slowly enough to provide guarantees of downstream estimators' asymptotic normality.

Theorem \ref{thm:general_factor_consistency_asymptotic_normality} formalizes how panel sparsity affects the asymptotic properties of $\Pi(\hat{\Gamma})$ by shrinking cohorts and weakening cohort connectedness. Shrinking cohort sizes $p_c$ increase $p_c^{-1}$, which increases the left-hand side of \eqref{cond:apm_est_err_eigenvalue_bound}, the error rate in \eqref{eq:general_factor_proj_mat_asymptotic_linearity}, and the variability bound in \eqref{eq:sparse_factor_proj_mat_asymptotic_error_squared}. Moreover, it turns out that $d_{r+1}$ serves as a summary statistic for cohort connectedness as determined by the O\textsuperscript{3} Algorithm, so decreasing cohort connectedness increases the stringency of \eqref{cond:apm_est_err_eigenvalue_bound} and the bounds in \eqref{eq:general_factor_proj_mat_asymptotic_linearity} and \eqref{eq:sparse_factor_proj_mat_asymptotic_error_squared}. To see why $d_{r+1}$ proxies for cohort connectedness, we prove the following theorem that provides a lower bound on $d_{r+1}$ in terms of the number of cohort sequences with sufficient overlap in observed outcomes:\footnote{This result is analogous to Cheeger's inequality, which relates a graph's Laplacian matrix spectrum to a particular notion of its connectedness (see e.g. \citet{spielman2025spectral} for a textbook treatment). In network data settings in which observations correspond to the edges of a graph, \citet{jochmans2019fixed} shows how the spectrum of the Laplacian matrix of that graph affects the asymptotic properties of TWFE regression estimators.}
\begin{theorem}\label{thm:graphical_bound_lambda_r+1}
For any positive integer $K$, we call $(c_1, \ldots, c_K)\in [C]^K$ a \emph{nice path} of length $K$ if $K = 1$ or if there exists $L \in [K-1]$ such that
\begin{equation}\label{eq:nice_path}
\absfit{\left(\bigcup_{k=1}^{L}\mathcal{T}_{c_k}\right) \cap \left(\bigcup_{k=L+1}^{K}\mathcal{T}_{c_k}\right)} \geq r
\end{equation}
and both $(c_1, \dotsc, c_L)$ and $(c_{L+1}, \dotsc, c_K)$ are nice paths. Furthermore, we call $(c_1, \ldots, c_K)\in [C]^K$ a \emph{complete path} if 
\[\bigcup_{k=1}^{K}\mathcal{T}_{c_k} = [T].\]
Let $N_K$ be the number of nice and complete paths of length $K$. 
Assume that, for any subset $\mathcal{\mathcal{T}}\subset [T]$ with $\abs{\mathcal{\mathcal{T}}} = r$, 
\begin{equation}\label{eq:general_position_quantitative}
\sigma_{\min}(\Gamma_\mathcal{\mathcal{T}})\ge \sigma_0,
\end{equation}
for some constant $\sigma_0 > 0$ where $\sigma_{\min}(M)$ denotes the minimal singular value of a matrix $M$. Then 
\[d_{r+1} \geq 1 - \inf_{K \geq 1}\left(1 - \frac{N_K\sigma_0^{2(K-1)}}{C^K}\right)^{1/K} \geq \sup_{K \geq 1}\frac{N_K}{C^K} \frac{\sigma_0^{2(K-1)}}{K}.\]
\end{theorem}
\noindent We provide a proof of Theorem \ref{thm:graphical_bound_lambda_r+1} in Appendix \ref{proof:thm:graphical_bound_lambda_r+1}.

As it happens, the O\textsuperscript{3} Algorithm can be viewed as searching for nice and complete paths: any path in the first-iteration O\textsuperscript{3} Graph is nice, and for any path in a subsequent iteration's O\textsuperscript{3} Graph, a nice path from each of the constituent super cohorts can be concatenated to form a longer nice path. If the O\textsuperscript{3} Algorithm converges to a single super cohort, then there exists a nice path containing every cohort, which must then also be complete. As such, we can view Theorem \ref{thm:graphical_bound_lambda_r+1} as a quantitative extension of Theorem \ref{thm:factor_identification}: $d_{r+1}$ is bounded away from zero if the O\textsuperscript{3} Algorithm converges to a single super cohort, with the bound additionally characterizing $d_{r+1}$'s magnitude in terms of the number of nice and complete paths.

\subsection{Non-Uniqueness of the APM}\label{sec:conclusion:nonuniqueness}

Finally, we note that the APM $A$ defined in \eqref{eq:aggregated_projection_matrix_def} is not unique: several APM variants described below share the same identification properties as the APM studied in this paper and may suggest estimators with improved statistical efficiency, though we leave fully developing such refinements to future work.

One straightforward extension assigns more general positive weights $w_c$ to cohorts and constructs a weighted APM:
\[A(w) \coloneqq \sum_{c=1}^C w_c (E_c - \Pi(E_c\Gamma)),\]
where the original APM is recovered by setting $w_c = 1/C$ for all $c$. By an argument similar to the proof of Theorem \ref{thm:factor_identification}, the null space of any weighted APM equals the column space of $\Gamma$ provided the O\textsuperscript{3} Algorithm converges to a single super cohort. If empirical weight analogs $\hat{w}_c$ are $N^{1/2}$-consistent and asymptotically linear, we can modify the proof of Theorem \ref{thm:outcome_mean_consistency_asymptotic_normality} to derive an influence function for the weighted analog of $\hat{\mu}_c$ that additionally depends on the influence functions of the weights.

We can also expand the class of APMs substantially by considering polynomials of $E_c - \Pi(\hat{\Gamma}_c)$:
\[
A_{\mathrm{poly}}(w) = \sum_{k\ge 1}\sum_{(c_1, \dotsc, c_k) \in [C]^k}w_{c_1,\, \dotsc,\, c_k}\prod_{\ell = 1}^k (E_{c_\ell} - \Pi(E_{c_\ell}\Gamma)), \quad w_{c} > 0, \; c\in [C].
\]
for non-negative weights $w_{c_1,\, \dotsc,\, c_k}$ that can be estimated at parametric rates. Since $(E_c - \Pi(E_c\Gamma))\Gamma = 0$ for all $c$, $\Gamma$'s column space lies in $A_{\mathrm{poly}}(w)$'s null space. Moreover, $A_{\mathrm{poly}}(w)$ is lower-bounded by $A(w_{k = 1})$ in the positive semi-definite order (where $w_{k=1}$ is the subset of $w$ for $k = 1$), so its $(r+1)$-th eigenvalue is bounded away from zero so long as $A(w_{k=1})$'s is. As such, its null space must be exactly $\Gamma$'s column space, as implied by Theorem \ref{thm:factor_identification}.  Asymptotic linearity of polynomial APM-based estimators should follow via the Delta method, though the resultant estimators will have more complex influence functions.

Finally, we can add additional cohorts to the APM formed by merging existing cohorts and retaining their overlapping outcomes. For example, if cohort 1 observes outcomes $\{1, 2, 3\}$ and cohort 2 observes $\{2, 3, 4\}$, we can define a combined cohort $(1, 2)$ containing all units from both cohorts with observed outcomes $\{2, 3\}$. Adding these cohorts preserves identification since it can only strengthen cohort connectedness. The resulting estimators should be consistent and asymptotically linear, potentially with improved efficiency from larger combined cohorts enabling more precise cohort-specific factor estimates. However, one would have to carefully account for the statistical dependence among cohort-specific factor estimators these additional cohorts introduce.

%% file: appendix_body.tex
\section{Proof of Theorem \ref{thm:factor_identification}}\label{proof:thm:factor_identification}

\input{\paperpath sections/identification_proofs}

\section{An Eigenspace Operator Exact First-Order Expansion}\label{sec:eigspace_perturbation}

\input{\paperpath sections/eigenspace}

%% file: sections/identification_proofs.tex
First, as in Footnote \ref{foot:col_space_Gamma_in_null_space_APM}, we show that for any cohort $c$, the columns of $\Gamma$ lie in the null space of $E_c - \Pi(E_c\Gamma)$, in which case $\mathrm{col}(\Gamma) \subseteq \mathrm{null}(A)$, where $\mathrm{col}(\Gamma)$ denotes the column space of $\Gamma$. Since $E_c^2 = E_c$ and $E_c\Pi(E_c\Gamma) = \Pi(E_c\Gamma)E_c = \Pi(E_c\Gamma)$, we have that
\begin{equation}
    (E_c - \Pi(E_c\Gamma))\Gamma = (E_c^2 - E_c\Pi(E_c\Gamma)E_c)\Gamma = E_c \cdot (I - \Pi(E_c\Gamma))E_c\Gamma = \zeros_{T \times r},
\end{equation}
as required.

Next, we show that $\mathrm{null}(A) \subseteq \mathrm{col}(\Gamma)$ also holds. Define the matrix
\begin{equation}
    \tilde{A} \coloneqq \frac{1}{\sqrt{C}}\bmat{E_1 - \Pi(E_1\Gamma) \\ \vdots \\ E_C - \Pi(E_C\Gamma)},
\end{equation}
and note that since
\begin{equation}
    (E_c - \Pi(E_c\Gamma))^2 = E_c^2 - E_c\Pi(E_c\Gamma) - \Pi(E_c\Gamma)E_c - \Pi(E_c\Gamma)^2 = E_c - \Pi(E_c\Gamma)
\end{equation}
and $E_c - \Pi(E_c\Gamma)$ is symmetric, $A = \tilde{A}'\tilde{A}$, implying that $\mathrm{null}(A) = \mathrm{null}(\tilde{A})$.\footnote{\label{foot:null_space_A_equals_null_space_tilde_A} It is straightforward to check that if $v \in \mathrm{null}(\tilde{A})$, then $v \in \mathrm{null}(A) = \mathrm{null}(\tilde{A}'\tilde{A})$. To check the other direction of inclusion, for any $v \in \mathrm{null}(A)$, we have that $\zeros = Av = \tilde{A}'\tilde{A}v$. Further, we have that $0 = v'\zeros = v'\tilde{A}'\tilde{A}v = (\tilde{A}v)'(\tilde{A}v)$, so it must be that $\tilde{A}v = \zeros$, as required.}

Consider any $v \in \mathrm{null}(\tilde{A})$. We will show by induction on iterations $k \geq 0$ of the O\textsuperscript{3} Algorithm (Algorithm \ref{alg:o3_algorithm}) that for any super cohort $m \in \{1, \dotsc, M^{(k)}\}$, $E_{\mathcal{T}_{\mathcal{S}_m^{(k)}}} v \in \mathrm{col}(E_{\mathcal{T}_{\mathcal{S}_m^{(k)}}}\Gamma)$ (recall from Section \ref{sec:method:intuition} that, for any subset of outcomes $\mathcal{T} \subseteq [T]$, $E_{\mathcal{T}}$ is defined as the diagonal matrix whose $t$-th diagonal entry is one if $t \in \mathcal{T}$ and zero otherwise). As the base case, take $k = 0$. By the definitions of $v$ and $\tilde{A}$, for any cohort $c$, $(E_c - \Pi(E_c\Gamma))v = \zeros_T$, which in turn implies that, since the initial super cohorts $\mathcal{S}_c^{(0)}$ are singletons corresponding to cohorts $c$:
\begin{equation}
    E_{\mathcal{T}_{\mathcal{S}_c^{(0)}}} v = E_c v = \Pi(E_c\Gamma)v = \Pi(E_{\mathcal{T}_{\mathcal{S}_c^{(0)}}}\Gamma)E_{\mathcal{T}_{\mathcal{S}_c^{(0)}}}v.
\end{equation}
Therefore, $E_{\mathcal{T}_{\mathcal{S}_c^{(0)}}} v \in \mathrm{col}(E_{\mathcal{T}_{\mathcal{S}_c^{(0)}}}\Gamma)$, as required.

For the inductive step, consider $k \geq 1$ and suppose that for all super cohorts $m \in \{1, \dotsc, M^{(k - 1)}\}$, we have that $E_{\mathcal{T}_{\mathcal{S}_m^{(k - 1)}}} v \in \mathrm{col}(E_{\mathcal{T}_{\mathcal{S}_m^{(k - 1)}}}\Gamma)$. Then for every iteration-$(k-1)$ super cohort $m$, there exists a vector $\omega_m \in \R^r$ such that $E_{\mathcal{T}_{\mathcal{S}_m^{(k - 1)}}} v = E_{\mathcal{T}_{\mathcal{S}_m^{(k - 1)}}}\Gamma \omega_m$. Consider any edge $(m_1, m_2)$ in the iteration-$k$ O\textsuperscript{3} Graph $\mathcal{G}_r^{(k)}$. Since diagonal matrices commute, we have that
\begin{equation}
\begin{aligned}
    &E_{\mathcal{T}_{\mathcal{S}_{m_1}^{(k - 1)}}}E_{\mathcal{T}_{\mathcal{S}_{m_2}^{(k - 1)}}}v = E_{\mathcal{T}_{\mathcal{S}_{m_2}^{(k - 1)}}}E_{\mathcal{T}_{\mathcal{S}_{m_1}^{(k - 1)}}}v = E_{\mathcal{T}_{\mathcal{S}_{m_2}^{(k - 1)}}}E_{\mathcal{T}_{\mathcal{S}_{m_1}^{(k - 1)}}}\Gamma\omega_{m_1} = E_{\mathcal{T}_{\mathcal{S}_{m_1}^{(k - 1)}}}E_{\mathcal{T}_{\mathcal{S}_{m_2}^{(k - 1)}}}\Gamma\omega_{m_1}, \\
    &E_{\mathcal{T}_{\mathcal{S}_{m_1}^{(k - 1)}}}E_{\mathcal{T}_{\mathcal{S}_{m_2}^{(k - 1)}}}v = E_{\mathcal{T}_{\mathcal{S}_{m_1}^{(k - 1)}}}E_{\mathcal{T}_{\mathcal{S}_{m_2}^{(k - 1)}}}\Gamma\omega_{m_2},
\end{aligned}
\end{equation}
implying that $E_{\mathcal{T}_{\mathcal{S}_{m_1}^{(k - 1)}}}E_{\mathcal{T}_{\mathcal{S}_{m_2}^{(k - 1)}}}\Gamma(\omega_{m_1} - \omega_{m_2}) = \zeros_T$. By Line \ref{cond:observed_outcome_overlap_graph_edge_def} in Algorithm \ref{alg:o3_algorithm}, there must be at least $r$ overlapping observed outcomes between super cohorts $m_1$ and $m_2$, so since the factor vectors $\gamma_t$ are in general position, $E_{\mathcal{T}_{\mathcal{S}_{m_1}^{(k - 1)}}}E_{\mathcal{T}_{\mathcal{S}_{m_2}^{(k - 1)}}}\Gamma$ must have full rank. This observation in turn implies that the null space of $E_{\mathcal{T}_{\mathcal{S}_{m_1}^{(k - 1)}}}E_{\mathcal{T}_{\mathcal{S}_{m_2}^{(k - 1)}}}\Gamma$ is trivial and $\omega_{m_1} - \omega_{m_2}$ lies in it, meaning $\omega_{m_1} = \omega_{m_2}$. Now consider any connected component $\mathcal{M}_{m'}^{(k)}$ of $\mathcal{G}_r^{(k)}$, which corresponds to super cohort $\mathcal{S}_{m'}^{(k)}$. Since all iteration-$(k-1)$ super cohorts in the connected component $\mathcal{M}_{m'}^{(k)}$ of $\mathcal{G}_r^{(k)}$ must be connected by some sequence of edges in $\mathcal{G}_r^{(k)}$, chaining the edge-wise $\omega_m$-equality logic above along that sequence implies that there must exist some common vector $\omega \in \R^r$ such that $\omega_m = \omega$ for all $m \in \mathcal{M}_{m'}^{(k)}$, meaning that $E_{\mathcal{T}_{\mathcal{S}_m^{(k - 1)}}} v = E_{\mathcal{T}_{\mathcal{S}_m^{(k - 1)}}}\Gamma \omega$ for all prior-iteration super cohorts $q \in \mathcal{Q}_{q'}^{(k)}$.

Now consider any observed outcome $t$ for the iteration-$k$ super cohort $\mathcal{S}_{m'}^{(k)}$, i.e. $t \in \mathcal{T}_{\mathcal{S}_{m'}^{(k)}}$. Note that this outcome must also be observed for at least one iteration-$(k-1)$ super cohort $m_t$ belonging to the connected component $\mathcal{M}_{m'}^{(k)}$, i.e. $t \in \mathcal{T}_{\mathcal{S}_{m_t}^{(k - 1)}}$. Letting $v_t$ denote the $t$-th entry of $v$ and $e_t$ denote the $t$-th standard basis vector, we have that
\begin{equation}\label{eq:v_t_in_col_Gamma}
    v_t = e_t'E_{\mathcal{T}_{\mathcal{S}_{m_t}^{(k - 1)}}}\Gamma \omega = \gamma_t'\omega.
\end{equation}
Stacking \eqref{eq:v_t_in_col_Gamma} across $t \in \tilde{\mathcal{T}}_{q'}^{(k)}$, we have that $E_{\mathcal{T}_{\mathcal{S}_{m'}^{(k)}}} v = E_{\mathcal{T}_{\mathcal{S}_{m'}^{(k)}}}\Gamma \omega$, meaning that $E_{\mathcal{T}_{\mathcal{S}_{m'}^{(k)}}} v \in \mathrm{col}(E_{\mathcal{T}_{\mathcal{S}_{m'}^{(k)}}}\Gamma)$, as required.

Finally, recall that $K$ denotes the number of iterations required for the O\textsuperscript{3} Algorithm to converge and that we assumed the final number of super cohorts was $M^{(K)} = 1$. Since all super cohorts must be unions of prior-iteration super cohorts and every outcome is observed for some initial super cohort (since every outcome is observed for at least one cohort), it must be that this single super cohort is the set of all cohorts, i.e. $\mathcal{S}_1^{(K)} = [C]$, and therefore that all outcomes are observed for this final super cohort, i.e. $\mathcal{T}_{\mathcal{S}_1^{(K)}} = [T]$. As such, the inductive logic above implies that
\begin{equation}
    v = E_{\mathcal{T}_{\mathcal{S}_1^{(K)}}} v \in \mathrm{col}(E_{\mathcal{T}_{\mathcal{S}_1^{(K)}}}\Gamma) = \mathrm{col}(\Gamma).
\end{equation}
Therefore, $\mathrm{null}(\tilde{A}) = \mathrm{null}(A) \subseteq \mathrm{col}(\Gamma)$, as required.

%% file: sections/eigenspace.tex
In this section, we derive an exact, first-order expansion of the operator mapping a symmetric matrix into the projection matrix onto the space spanned by some subset of its eigenvectors. Our expansion is based on Kato's integral, which characterizes the projection matrix onto the space spanned by some subset of a matrix's eigenvectors as a contour integral of that matrix's resolvent \citet{kato1949convergence}. We explicitly characterize the approximation error in our expansion including exact constants, unlike the asymptotic expansions given in e.g. \citet{kato2013perturbation} and \citet{sun1991perturbation} (which is applied in \citet{simons2023inference} to construct hypothesis tests concerning eigenspaces). As such, our result below may be of independent interest.

To describe our expansion, we first introduce some notation. Let $\mathbb{S}^d$ denote the set of real-valued, $d$-dimensional symmetric matrices, and for any $M \in \mathbb{S}^d$, let $\lambda_j(M)$ be the $j$-th smallest eigenvalue of $A$, where %
\begin{equation}
    \lambda_0(M) \coloneqq -\infty < \lambda_1(M) \leq \cdots \leq \lambda_d(M) < \lambda_{d+1}(M) \coloneqq \infty,\footnote{The eigenvalues of any symmetric matrix are all real-valued.}
\end{equation}
and we denote a generic eigendecomposition of $M$ as follows:
\begin{equation}
    M = U(M)\Lambda(M)U(M)' = \sum_{j = 1}^d \lambda_j(M) \Pi(u_j(M)),
\end{equation}
where $U(M)$ is any matrix with $j$th column $u_j(M)$ being an eigenvector corresponding to the $j$th eigenvalue $\lambda_j(M)$, and $\Pi(u_j(M))$ is the projection onto the span of $u_j(M)$.\footnote{Even though $U(M)$ is required to be orthonormal, the columns of $U(M)$ are only unique up to signs and permutations any eigenvectors corresponding to repeated eigenvalues.} For notational convenience, for any integers $1 \leq j, k \leq d$, we let 
\begin{equation}
U_{j:k}(M) \coloneqq \bmat{u_j(M) & \cdots & u_k(M)}
\end{equation}
denote the matrix whose columns are the eigenvectors corresponding to $\lambda_j(M)$ through $\lambda_k(M)$. Recall also that we defined
$g(A, B, C) \coloneqq ABC + CBA$
for square matrices $A$, $B$, and $C$.

Armed with this notation, we construct an exact bound on the error incurred by a first-order expansion of the difference between the projection matrices onto eigenspaces of two matrices $M$ and $\hat{M}$ in terms of the magnitude of the difference between $M$ and $\hat{M}$:
\begin{theorem}\label{thm:eigenspace_perturbation_expansion}
    Consider any integers $s, r$ such that $1 \leq s + 1 \leq s + r \leq d$ and any real-valued, $d$-dimensional symmetric matrix $M$ satisfying the following eigen-gap condition:\begin{equation}\label{cond:well_sep_eigvals}
    \Delta(M) > 0, \quad \Delta(M) \coloneqq 4^{-1}\min\left\{\lambda_{s+1}(M) - \lambda_s(M), \lambda_{s+r+1}(M) - \lambda_{s+r}(M)\right\},
    \end{equation}
    and define the following neighborhood of $M$:
    \begin{equation}
        \mathcal{B}(M) \coloneqq \left\{\hat{M} \in \mathbb{S}^d ~\setst~ \normnofit{\hat{M} - M}_\mathrm{op} \leq \Delta(M)\right\}.
    \end{equation}
    Then for any $\hat{M} \in \mathcal{B}(M)$, the following first-order approximation holds:
    \begin{equation}\label{eq:eigenvec_first_order_approx}
    \begin{aligned}
        &\Bigg\lVert\left[\Pi(U_{(s+1):(s+r)}(\hat{M})) - \Pi(U_{(s+1):(s+r)}(M))\right] \\
    &\phantom{\Bigg\lVert}- \sum_{j = s + 1}^{s + r}\sum_{k \not\in [s + 1, s + r]} \frac{1}{\lambda_k(M) - \lambda_j(M)} g\left(\Pi(u_j(M)), (\hat{M}-M), \Pi(u_k(M))\right)\Bigg\rVert_\text{op} \\
    &\leq \frac{2}{\pi\Delta(M)^2}\normnofit{\hat{M} - M}_\text{op}^2.
    \end{aligned}
    \end{equation}
\end{theorem}
\noindent We provide a proof of Theorem \ref{thm:eigenspace_perturbation_expansion} in Appendix \ref{proof:thm:eigenspace_perturbation_expansion}.

%% file: online_appendix_body.tex
\section{Proofs of Estimation Results}\label{sec:estimation_proofs}

\input{\paperpath sections/estimation_proofs}

\section{Target Parameters: Examples, Estimation, and Inference}\label{sec:supplement:target_params}

\input{\paperpath sections/target_params}

\section{Proofs of Intermediate Results in Appendices}\label{sec:supplement:appendix_proofs}

\input{\paperpath sections/intermediate_proofs}

%% file: sections/estimation_proofs.tex
\subsection{Proof of Theorem \ref{thm:factor_consistency_asymptotic_normality}}\label{proof:thm:factor_consistency_asymptotic_normality}

Let $d_j$ be the $j$th smallest eigenvalue of the population APM $A$, and denote a generic eigendecomposition of $A$ as follows:
\begin{equation}
    A = \sum_{j = 1}^T d_j \Pi(u_j),
\end{equation}
where $u_j$ is an eigenvector corresponding to the $j$th eigenvalue $d_j$, and $\Pi(u_j)$ is the projection onto the span of $u_j$.\footnote{Even though the matrix of eigenvectors whose columns are $u_j$ is required to be orthonormal, its columns are only unique up to signs. Furthermore, if $A$ has repeated eigenvalues, then any of the eigenvectors corresponding to those repeated eigenvalues are interchangeable.} We note that all of $A$'s eigenvalues must be non-negative.\footnote{$E_c - \Pi(E_c\Gamma) = E_c[I - \Pi(E_c\Gamma)]E_c$, so since $E_c$ and $I - \Pi(E_c\Gamma)$ are both projection matrices, their product must be positive semidefinite. The sum of positive semidefinite matrices must also be positive semidefinite, so then $A$ must be positive semidefinite as well by \eqref{eq:aggregated_projection_matrix_def}.} %

Suppose $\normnofit{\hat{A} - A}_\mathrm{op} \leq 4^{-1}d_{r+1}$ holds (we will show it does with high probability later); then we can apply Theorem \ref{thm:eigenspace_perturbation_expansion} with $M = A$, $\hat{M} = \hat{A}$ as defined in Section \ref{sec:method:procedure}, $s = 0$, and $r = r$. We also note that $d_j = \lambda_j(A)$ using the notation defined in Section \ref{sec:theory:estimation_inference}. Since Theorem \ref{thm:factor_identification} implies that
\begin{equation}\label{eq:apm_eigval_zero_and_positive}
0 = d_1 = \dotsc = d_r < d_{r+1},
\end{equation}
we have that $\Delta(A) = 4^{-1}d_{r+1} > 0$, satisfying \eqref{cond:well_sep_eigvals}. Further, Theorem \ref{thm:factor_identification} implies that $\Pi(\Gamma) = \Pi(U_{1:r}(A))$, and we define $\hat{\Gamma}$ in Section \ref{sec:method:procedure} such that $\Pi(\hat{\Gamma}) = \Pi(U_{1:r}(\hat{A}))$.
Then, under our conditioning event above, Theorem \ref{thm:eigenspace_perturbation_expansion} implies that
\begin{align}
    &\norm{\Pi(\hat{\Gamma}) - \Pi(\Gamma) - \sum_{j = 1}^r\sum_{k = r + 1}^T d_k^{-1} g\left(\Pi(u_j), (\hat{A}-A), \Pi(u_k)\right)}_\mathrm{op} %
    \leq \frac{32}{\pi d_{r+1}^2}\normnofit{\hat{A} - A}_\mathrm{op}^2. \label{eq:apm_first_step_finite_sample_approximate_linearity} %
\end{align}
From \eqref{eq:apm_first_step_finite_sample_approximate_linearity}, we have that
\begin{align}
    &\sum_{j = r + 1}^T \frac{1}{d_j}\sum_{k = 1}^r \bigg\{\Pi(u_j)(\hat{A} - A)\Pi(u_k) + \Pi(u_k)(\hat{A} - A)\Pi(u_j)\bigg\}. \\
    &= \left(\sum_{j = r + 1}^T\frac{1}{d_j}\Pi(u_j)\right)(\hat{A} - A)\left(\sum_{k = 1}^r\Pi(u_k)\right) + \left(\sum_{k = 1}^r\Pi(u_k)\right)(\hat{A} - A)\left(\sum_{j = r + 1}^T\frac{1}{d_j}\Pi(u_j)\right). \label{eq:factor_proj_mat_expansion_nonsimplified}
\end{align}
Next, since $d_1 = \dotsc = d_r = 0$, it must be that
\begin{equation}
    \sum_{j = r + 1}^T\frac{1}{d_j}\Pi(u_j) = \sum_{j = 1}^T\frac{\ind{j > r}}{d_j}\Pi(u_j) = A^+,
\end{equation}
and since $\mathrm{rank}(\Gamma) = r$ and $\mathrm{null}(A) = \mathrm{col}(\Gamma)$,
\begin{equation}
    \sum_{k = 1}^r\Pi(u_k) = \Pi(\Gamma).
\end{equation}
Substituting these two simplifications back into \eqref{eq:factor_proj_mat_expansion_nonsimplified} and applying the definitions of $\hat{A}$ and $A$, we have
\begin{equation}\label{eq:factor_proj_mat_consistency}
\sum_{j = 1}^r\sum_{k = r + 1}^T d_k^{-1} g\left(\Pi(u_j), (\hat{A}-A), \Pi(u_k)\right) = g\left(A^+, \frac{1}{C}\sum_{c = 1}^C[\Pi(\hat{\Gamma}_c) - \Pi(E_c\Gamma)], \Pi(\Gamma)\right).
\end{equation}
Thus, we can show the following:
\begin{align}
    &\norm{\left[\Pi(\hat{\Gamma}_c) - \Pi(E_c\Gamma)\right] - \hat{\E}_N[\phi_A(C_i, Y_i, X_i)]}_\mathrm{op} \\
    &\leq \Bigg\lVert\left[\Pi(\hat{\Gamma}_c) - \Pi(E_c\Gamma)\right] - g\Bigg(A^+, \underbrace{\frac{1}{C}\sum_{c = 1}^C[\Pi(\hat{\Gamma}_c) - \Pi(E_c\Gamma)]}_{= \hat{A} - A}, \Pi(\Gamma)\Bigg)\Bigg\rVert_\mathrm{op} \\
    &\phantom{\leq} + \norm{g\left(A^+, \frac{1}{C}\sum_{c = 1}^C\left\{[\Pi(\hat{\Gamma}_c) - \Pi(E_c\Gamma)] - \hat{\E}_N[\phi_{\Gamma, c}(C_i, Y_i, X_i)]\right\}, \Pi(\Gamma)\right)}_\mathrm{op} && \text{($\Delta$ Ineq.)} \\
    &\leq \frac{32}{\pi d_{r+1}^2} \normnofit{\hat{A} - A}_\text{op}^2 && \text{(By \eqref{eq:apm_first_step_finite_sample_approximate_linearity}, \eqref{eq:factor_proj_mat_consistency})} \\
    &\phantom{\leq} + 2\underbrace{\norm{A^+}_\mathrm{op}}_{\text{$\leq d_{r+1}^{-1}$ by \eqref{eq:apm_eigval_zero_and_positive}}} \underbrace{\normnofit{\Pi(\Gamma)}_\mathrm{op}}_{\leq 1} \cdot \frac{1}{C}\sum_{c = 1}^C \underbrace{\norm{[\Pi(\hat{\Gamma}_c) - \Pi(E_c\Gamma)] - \hat{\E}_N[\phi_{\Gamma, c}(C_i, Y_i, X_i)]}_\mathrm{op}}_{\coloneqq \hat{R}_c} \\
    &= \frac{32}{\pi d_{r+1}^2} \normnofit{\hat{A} - A}_\text{op}^2 + \frac{2}{d_{r+1}C}\sum_{c = 1}^C \hat{R}_c. \label{eq:apm_finite_sample_approximate_linearity}
\end{align}

Next, we bound the APM estimation error $\normnofit{\hat{A} - A}_\mathrm{op}$. To do so, we first state the following lemma that we prove in Appendix \ref{proof:lemma:chebyshev_tightness}:
\begin{lemma}\label{lemma:chebyshev_tightness}
    Let $\{V_i\}_{i = 1}^\infty$ be a sequence of i.i.d. random elements taking values in a measurable, normed vector space with norm $\norm{\cdot}$. Suppose also that $\sigma^2 \coloneqq \E[\norm{V_i}^2] < \infty$, and that $\norm{\cdot}$ is equivalent to some other norm $\norm{\cdot}_\text{H}$ induced by an inner product $\inner{\cdot, \cdot}$ in the sense that there exist constants $C_1, C_2 > 0$ such that, for all $v$,
    \begin{equation}\label{cond:inner_prod_norm_equiv}
        C_1\sqrt{\inner{v, v}} = C_1\norm{v}_\text{H} \leq \norm{v} \leq C_2\norm{v}_\text{H} = C_2 \sqrt{\inner{v, v}}.
    \end{equation}
    Then $\normnofit{\hat{\E}_N[V_i] - \E[V_i]}_\text{op} = \bigop{N^{-1/2} \kappa \sigma}$ where $\kappa \coloneqq C_1^{-1} \cdot C_2$.
\end{lemma}
\noindent We can then apply Lemma \ref{lemma:chebyshev_tightness} to derive the following asymptotic bound on $\normnofit{\hat{A} - A}_\mathrm{op}$ since the operator norm is equivalent to the Frobenius norm and $\E[\phi_{\Gamma, c}(C_i, Y_i, X_i)] = \zeros_{T \times T}$:
\begin{align}
    \normnofit{\hat{A} - A}_\mathrm{op} &\leq \frac{1}{C}\sum_{c = 1}^C \normnofit{\Pi(\hat{\Gamma}_c) - \Pi(E_c\Gamma)}_\mathrm{op} && \text{(By \eqref{eq:aggregated_projection_matrix_def} and $\Delta$ Ineq.)} \\
    &\leq \frac{1}{C}\sum_{c = 1}^C \left[\norm{\hat{\E}_N[\phi_{\Gamma, c}(C_i, Y_i, X_i)]}_\mathrm{op} + \hat{R}_c\right] && \text{($\Delta$ Ineq.)} \\
    &= \frac{1}{C}\sum_{c = 1}^C \bigg\{\hat{R}_c + \mathcal{O}_{\mathbb{P}}\bigg(N^{-1/2} \cdot \underbrace{\E\left[\norm{\phi_{\Gamma, c}(C_i, Y_i, X_i)}_{\mathrm{op}}^2\right]^{1/2}}_{\coloneqq \sigma_{\Gamma, c}}\bigg)\bigg\} && \text{(By Lemma \ref{lemma:chebyshev_tightness})} \\
    &= \bigop{\frac{1}{C}\sum_{c = 1}^C \left\{\hat{R}_c + N^{-1/2} \sigma_{\Gamma, c}\right\}}. \label{eq:apm_op_norm_err_bound_general}
\end{align}
Substituting \eqref{eq:apm_op_norm_err_bound_general} into \eqref{eq:apm_finite_sample_approximate_linearity}, we obtain a general bound on the linearization error incurred by our first-order approximation of $\Pi(\hat{\Gamma}) - \Pi(\Gamma)$ so long as $\normnofit{\hat{A} - A}_\mathrm{op} \leq 4^{-1}d_{r+1}$ holds:
\begin{align}
    &\norm{\left[\Pi(\hat{\Gamma}_c) - \Pi(E_c\Gamma)\right] - \hat{\E}_N[\phi_A(C_i, Y_i, X_i)]}_\mathrm{op} \\
    &\leq \bigop{\frac{64}{\pi d_{r+1}^2C}\sum_{c = 1}^C \left\{\hat{R}_c^2 + N^{-1}\sigma_{\Gamma, c}^2\right\}} + \frac{2}{d_{r+1}C}\sum_{c = 1}^C \hat{R}_c && \text{(By \eqref{eq:apm_finite_sample_approximate_linearity}, \eqref{eq:apm_op_norm_err_bound_general}, QM-AM Ineq.)} \\
    &= \bigop{d_{r+1}^{-1} \cdot \frac{1}{C}\sum_{c = 1}^C \left\{d_{r+1}^{-1}\hat{R}_c^2 + \hat{R}_c + N^{-1}\sigma_{\Gamma, c}^2\right\}}. \label{eq:apm_inf_fn_general_asymptotic_error}
\end{align}

We can show that the random matrix $\phi_A(C_i, Y_i, X_i)$ has zero mean using the multilinearity of $g$ and the definition of $\phi_{\Gamma, c}$ in Assumption \ref{assump:cohort_specific_factor_proj_mat_inf_fn}, and we can upper-bound the expected squared operator norm of $\phi_A(C_i, Y_i, X_i)$ as follows:
\begin{align}
    \E\left[\norm{\phi_A(C_i, Y_i, X_i)}_{\mathrm{op}}^2\right] &\leq \frac{4}{C^2}\underbrace{\norm{A^+}_\mathrm{op}^2}_{\leq d_{r+1}^{-2}}\underbrace{\norm{\Pi(\Gamma)}_{\mathrm{op}}^2}_{\leq 1} \E\left[\left(\sum_{c = 1}^C \norm{\phi_{\Gamma, c}(C_i, Y_i, X_i)}_{\mathrm{op}}\right)^2\right] \\
    &\leq \frac{4}{d_{r+1}^2 C} \sum_{c = 1}^C \E\left[\norm{\phi_{\Gamma, c}(C_i, Y_i, X_i)}_{\mathrm{op}}^2\right] && \text{(QM-AM Ineq.)} \\
    &= \frac{4}{d_{r+1}^2 C}\sum_{c = 1}^C \sigma_{\Gamma, c}^2. \label{eq:apm_inf_fn_general_asymptotic_error_squared}
\end{align}

To finish the proof of Theorem \ref{thm:factor_consistency_asymptotic_normality}, we must show that $\normnofit{\hat{A} - A}_\mathrm{op} \leq 4^{-1}d_{r+1}$ holds with probability approaching one as $N \rightarrow \infty$. Since $C$ is fixed, Assumption \ref{assump:cohort_specific_factor_proj_mat_inf_fn} and \eqref{eq:apm_op_norm_err_bound_general} imply that
\begin{equation}
    \normnofit{\hat{A} - A}_\mathrm{op} \leq \mathcal{O}_{\mathbb{P}}\bigg(\frac{1}{C}\sum_{c = 1}^C \bigg\{\underbrace{\hat{R}_c}_{\text{$=\littleop{N^{-1/2}}$ by Assump. \ref{assump:cohort_specific_factor_proj_mat_inf_fn}}} + N^{-1/2} \cdot \underbrace{\sigma_{\Gamma, c}}_{\text{$< \infty$ by Assump. \ref{assump:cohort_specific_factor_proj_mat_inf_fn}}}\bigg\}\bigg) = \bigop{N^{-1/2}},
\end{equation}
meaning $\normnofit{\hat{A} - A}_\mathrm{op} \leq 4^{-1}d_{r+1}$ with high probability. Since $d_{r+1}$ remains constant in Theorem \ref{thm:factor_consistency_asymptotic_normality}'s regime, \eqref{eq:apm_inf_fn_general_asymptotic_error} and the fact that $\hat{R}_c = \littleop{N^{-1/2}}$ by Assumption \ref{assump:cohort_specific_factor_proj_mat_inf_fn} imply \eqref{eq:factor_proj_mat_asymptotic_normality}. Finally, boundedness of the expected squared operator norm of $\phi_A(C_i, Y_i, X_i)$ follows from \eqref{eq:apm_inf_fn_general_asymptotic_error_squared} and that $\sigma_{\Gamma, c}^2 < \infty$ under Assumption \ref{assump:cohort_specific_factor_proj_mat_inf_fn}.

To prove the more general result in Theorem \ref{thm:general_factor_consistency_asymptotic_normality}, we must show that $\normnofit{\hat{A} - A}_\mathrm{op} \leq 4^{-1}d_{r+1}$ holds with high probability first. By Assumption \ref{assump:general_cohort_specific_factor_proj_mat_inf_fn} and \eqref{eq:apm_op_norm_err_bound_general}, we have that
\begin{align}
    \normnofit{\hat{A} - A}_\mathrm{op} &\leq \mathcal{O}_{\mathbb{P}}\bigg(\frac{1}{C}\sum_{c = 1}^C \bigg\{\underbrace{\hat{R}_c}_{\text{$=\bigop{rT_c p_c^{-1}N^{-1}}$ by Assump. \ref{assump:general_cohort_specific_factor_proj_mat_inf_fn}}} + N^{-1/2} \cdot \underbrace{\sigma_{\Gamma, c}}_{\text{$\leq \bigo{(rT_c)^{1/2} p_c^{-1/2}}$ by Assump. \ref{assump:general_cohort_specific_factor_proj_mat_inf_fn}}}\bigg\}\bigg)
    \\
    &= \bigop{r^{1/2}N^{-1/2} \cdot \frac{1}{C}\sum_{c = 1}^C T_c^{1/2}p_c^{-1/2}},
\end{align}
meaning $\normnofit{\hat{A} - A}_\mathrm{op} \leq 4^{-1}d_{r+1}$ with high probability by \eqref{cond:apm_est_err_eigenvalue_bound}. Specializing \eqref{eq:apm_inf_fn_general_asymptotic_error} to this setting, Assumption \ref{assump:general_cohort_specific_factor_proj_mat_inf_fn} implies $\hat{R}_c = \bigop{rT_cp_c^{-1}N^{-1}}$ and $\sigma_{\Gamma, c}^2 = \bigo{rT_cp_c^{-1}}$, which in turn imply \eqref{eq:general_factor_proj_mat_asymptotic_linearity}:
\begin{align}
    &\norm{\left[\Pi(\hat{\Gamma}_c) - \Pi(E_c\Gamma)\right] - \hat{\E}_N[\phi_A(C_i, Y_i, X_i)]}_\mathrm{op} \\
    &= \mathcal{O}_{\mathbb{P}}\bigg(d_{r+1}^{-1} \cdot \frac{1}{C}\sum_{c = 1}^C \bigg\{d_{r+1}^{-1}\underbrace{\hat{R}_c^2}_{\bigop{r^2T_c^2p_c^{-2}N^{-2}}} + \underbrace{\hat{R}_c}_{\bigop{rT_cp_c^{-1}N^{-1}}} + N^{-1}\underbrace{\sigma_{\Gamma, c}^2}_{\bigo{rT_cp_c^{-1}}}\bigg\}\bigg) && \text{(By \eqref{eq:apm_inf_fn_general_asymptotic_error})} \\
    &= \bigop{\frac{1}{C}\sum_{c = 1}^C \left\{\frac{r^2T_c^2}{d_{r+1}^2p_c^2N^2} + \frac{rT_c}{d_{r+1}p_cN}\right\}}.
\end{align}
Finally, the bound \eqref{eq:sparse_factor_proj_mat_asymptotic_error_squared} on the expected squared operator norm of $\phi_A(C_i, Y_i, X_i)$ follows from \eqref{eq:apm_inf_fn_general_asymptotic_error_squared} and that $\sigma_{\Gamma, c}^2 = \bigo{rT_cp_c^{-1}}$ under Assumption \ref{assump:general_cohort_specific_factor_proj_mat_inf_fn}.

\subsection{Discussion of Assumption \ref{assump:A0_invertibility}}\label{sec:A0_invertibility}

We first prove a useful lemma that gives an equivalent form of Assumption \ref{assump:A0_invertibility}. 
Define 
\begin{equation}\label{eq:A0}
A_0 \coloneqq \E\left[[I - \Pi(\Gamma)\ \ X_i]'\{E_{C_i} - \Pi(E_{C_i}\Gamma)\} [I - \Pi(\Gamma)\ \ X_i]\right].
\end{equation}

\begin{lemma}\label{lem:A0_invertibility_equivalent}
Let 
\begin{equation}\label{eq:bar_Gamma}
\bar{\Gamma} = [\Gamma'\ \ \mathbf{0}_{T\times q}]'
\end{equation}
and, for any $\omega\in \R$,
\begin{equation}\label{eq:A0_omega}
A_{0,\omega} \coloneqq A_0 + \omega\bar{\Gamma}\bar{\Gamma}',
\end{equation}
Assumption \ref{assump:A0_invertibility} holds if and only if $A_{0,\omega}$ is invertible for any $\omega > 0$. In addition, under Assumption \ref{assump:A0_invertibility}, for any $\Lambda\in \R^{T\times q}$,
\begin{equation}\label{eq:H_A0+}
[I - \Pi(\Gamma)\ \ \Lambda]A_{0,\omega}^{-1} = [I - \Pi(\Gamma)\ \ \Lambda]A_{0}^{+},
\end{equation}
where $+$ denotes the Moore-Penrose pseudo-inverse. 
\end{lemma}

\noindent In the absence of covariates ($X_i = \zeros_{T \times q}$), we show that Assumption \ref{assump:A0_invertibility} is equivalent to the identification condition. 

\begin{proposition}\label{prop:A0_invertibility_no_covariate}
Under Assumption \ref{assump:cohort_sizes_and_potential_outcome_variance_bound}, without covariates, Assumption \ref{assump:A0_invertibility} holds if and only if $\Gamma$'s column space is exactly the null space of the APM $A$. 
\end{proposition}

\subsection{Proof of Theorem \ref{thm:outcome_mean_consistency_asymptotic_normality}}\label{subapp:influence_function}

We start by defining several useful quantities.
For any projection matrix  $\Pi\in\R^{T\times T}$ and any cohort $c\in [C]$, define the cohort-specific parameters
\[G_c(\Pi)\coloneqq \left(I-(I - E_c)\Pi(I - E_c)\right)^{+}, \]
and
\[P_c(\Pi)\coloneqq \bigl(I+E_c\Pi (I - E_c)\bigr)G_c(\Pi)\Pi E_c, \quad Q_c(\Pi) \coloneqq E_c(I - P_c(\Pi)).\]
Further, let 
\[H(\Pi, \mu)\coloneqq \bigl[I-\Pi\ \ \mu\bigr]\in \R^{T\times (T+q)}, \,\, \text{where }\mu\in \R^{T\times q},
\]
and for each unit $i\in [N]$,
\[D_i(\Pi)\coloneqq \bigl[I-\Pi\ \ X_i\bigr]\in\R^{T\times (T+q)}.\]
By definition,  \eqref{eq:A0} can be rewritten as
\[A_0 = \E[D_i(\Pi(\Gamma))'Q_{C_i}(\Pi(\Gamma))D_i(\Pi(\Gamma))].\]

Now, we restate Theorem \ref{thm:outcome_mean_consistency_asymptotic_normality} with a formal definition of the influence function $\psi_c$ for $\hat{\mu}_c$:
\begin{theorem}\label{thm:muc_hat_influence}
 Assume $\Gamma'\gamma_0 = 0$ without loss of generality.\footnote{Otherwise, we can redefine $\gamma$ as $(I - \Pi(\Gamma))\gamma_0$ and $\lambda_i$ as $\lambda_i + (\Gamma'\Gamma)^{-1}\Gamma'\gamma_0$ without changing the model \eqref{eq:factor_model}. Lemma \ref{lem:alpha_uniqueness} below shows that $\alpha$ is uniquely identified under this constraint. } We further assume the weaker condition $\E[\epsilon_{it} ~|~ C_i, X_i] = 0$ in the model \eqref{eq:factor_model}. Define $\alpha\coloneqq (\gamma_0',\beta')', \phi_i\coloneqq \phi_A(C_i, Y_i, X_i)$, and 
\begin{align}
\psi_{c,i,\mathrm{fac}}^\mathcal{P} &= E_c\phi_i (I - E_c) G_c(\Pi(\Gamma))\Pi(\Gamma) E_c  \\
& \quad +\bigl(I+E_c\Pi(\Gamma)(1-E_c)\bigr)G_c(\Pi(\Gamma))\bigl\{(I-E_c)\phi_i (I-E_c)G_c(\Pi(\Gamma))\Pi(\Gamma) + \phi_i\bigr\} E_c\label{eq:psicP_fac}\\
\psi_{c,i,\mathrm{reg}}^\mathcal{H} &= \left[\mathbf{0}_{T\times T}\ \ \frac{1\{C_i = c\}}{p_c}(X_i - \mu_{X,c})\right]\label{eq:psicH_reg}\\
\psi_{i,\mathrm{fac}}^\mathcal{H} &= \left[-\phi_i\ \ \mathbf{0}_{T\times q}\right]\label{eq:psicH_fac}\\
\psi_{i,\mathrm{reg}}^\mathcal{A} &= D_i\left(\Pi(\Gamma)\right)' Q_{C_i}(\Pi(\Gamma))D_i\left(\Pi(\Gamma)\right) - \E[D_i\left(\Pi(\Gamma)\right)' Q_{C_i}(\Pi(\Gamma)) D_i\left(\Pi(\Gamma)\right)]\label{eq:psiA_reg}\\\psi_{i,\mathrm{fac}}^\mathcal{A} &= -D_i(\Pi(\Gamma))'\psi_{C_i,i}^\mathcal{P} D_i(\Pi(\Gamma))\\
& \quad +(\psi_{i,\mathrm{fac}}^{\mathcal{H}})'Q_{C_i}(\Pi(\Gamma))D_i(\Pi(\Gamma))  + D_i(\Pi(\Gamma))'Q_{C_i}(\Pi(\Gamma))\psi_{i,\mathrm{fac}}^\mathcal{H} \label{eq:psiA_fac}\\
\psi_{i,\mathrm{reg}}^\mathcal{B} &= D_i\left(\Pi(\Gamma)\right)' Q_{C_i}(\Pi(\Gamma)) Y_i - \E[D_i\left(\Pi(\Gamma)\right)' Q_{C_i}(\Pi(\Gamma)) Y_i]\label{eq:psiB_reg}\\
\psi_{i,\mathrm{fac}}^\mathcal{B} &= -D_i(\Pi(\Gamma))'\psi_{C_i, i}^\mathcal{P} Y_i +(\psi_{i,\mathrm{fac}}^{\mathcal{H}})' Q_{C_i}(\Pi(\Gamma))Y_i.\label{eq:psiB_fac}
\end{align}
Under Assumptions 1-3, 
\[\hat{\mu}_c - \mu_c = \hat{\E}_N[\psi_{c,i,\mathrm{reg}} + \psi_{c,i,\mathrm{fac}}] + o_\mathbb{P}\left(N^{-1/2}\right),\]
where
\begin{align}
\psi_{c,i, \mathrm{reg}} &\coloneqq \psi_{c,\mathrm{reg}}(C_i, Y_i, X_i) \label{eq:mu_c_inf_fn_reg} \\
& \coloneqq \frac{1\{C_i = c\}}{p_c}P_{c}(\Pi(\Gamma))(Y_i - \mu_c) + (I-P_{c}(\Pi(\Gamma)))\psi_{c,i,\mathrm{reg}}^{\mathcal{H}}\alpha\\
&  \quad +(I - P_{c}(\Pi(\Gamma)))H(\Pi(\Gamma), \mu_{X,c}) A_0^{+}\psi_{i,\mathrm{reg}}^{\mathcal{A}} \alpha + (I - P_{c}(\Pi(\Gamma)))H(\Pi(\Gamma), \mu_{X,c}) A_0^{+}\psi_{i,\mathrm{reg}}^{\mathcal{B}}\label{eq:psii_reg}\\
\psi_{c,i, \mathrm{fac}} &\coloneqq \psi_{c,\mathrm{fac}}(C_i, Y_i, X_i) \label{eq:mu_c_inf_fn_fac} \\
& \coloneqq \psi_{c,i,\mathrm{fac}}^\mathcal{P}\mu_c -\psi_{c,i,\mathrm{fac}}^\mathcal{P}H(\Pi(\Gamma), \mu_{X,c})\alpha + (I-P_{c}(\Pi(\Gamma))) \psi_{i,\mathrm{fac}}^\mathcal{H}\alpha\\
&\quad + (I - P_{c}(\Pi(\Gamma)))H(\Pi(\Gamma), \mu_{X,c})A_0^{+}\psi_{i,\mathrm{fac}}^{\mathcal{A}}\alpha + (I - P_{c}(\Pi(\Gamma)))H(\Pi(\Gamma), \mu_{X,c}) A_0^{+} \psi_{i,\mathrm{fac}}^{\mathcal{B}},
\end{align}
where $A_0$ is defined in \eqref{eq:A0}.
The expansion continues to hold if $A_0^{+}$ is replaced by $A_{0,\omega}^{-1}$ for any $\omega > 0$. Moreover, 
\[\E[\psi_{c,i,\mathrm{reg}}] = \E[\psi_{c,i,\mathrm{fac}}] = \mathbf{0}_T, \quad \E[\|\psi_{c,i,\mathrm{reg}}\|_{\mathrm{op}}^2],\; \E[\|\psi_{c,i,\mathrm{fac}}\|_{\mathrm{op}}^2] < \infty,\]
and $\psi_{c,i,\mathrm{reg}}$ and $\psi_{c,i,\mathrm{fac}}$ are both continuous functions of $\Pi(\Gamma), \mu_c, \mu_{X, c}$, and $\alpha$. 
\end{theorem}

To prove Theorem \ref{thm:muc_hat_influence}, we first present several useful lemmas and relegate their proofs to Appendix \ref{sec:supplement:appendix_proofs}:

\begin{lemma}\label{lem:alpha_uniqueness}
Assume the weaker assumption $\E[\epsilon_{it}\mid C_i, X_{i}] = 0$ in \eqref{eq:factor_model}. Under Assumption \ref{assump:A0_invertibility}, the parameter vector $\alpha$ is uniquely identified given the constraint $\Gamma'\gamma_0 = 0$.
\end{lemma}

\begin{lemma}\label{lem:muc}
Assume the weaker assumption $\E[\epsilon_{it}\mid C_i] = 0$ in \eqref{eq:factor_model}; for any cohort $c\in [C]$, 
\[\mu_c = P_c(\Pi(\Gamma))\mu_c + \{I - P_c(\Pi(\Gamma))\} H(\Pi(\Gamma), \mu_{X,c})\alpha.\]
\end{lemma}

\begin{lemma}\label{lem:equivalent_constraint}
Fix any $\omega > 0$.  Consider the modified estimator:\footnote{This is an infeasible estimator because $\Gamma$ is unknown. We only use this formulation in the proof. } 
\begin{equation}\label{eq:modified_estimator}
\left(\tilde{\gamma}_0, \tilde{\beta}, \{\tilde{\lambda}_i\}_{i = 1}^N\right) \in  \argmin_{\check{\gamma}_0 \in \R^T,\; \check{\beta} \in \R^q,\; \check{\lambda}_i \in \R^r} \frac{1}{N}\sum_{i = 1}^N \sum_{t \in \mathcal{T}_{C_i}}\left(Y_{it} - \left[\hat{\gamma}_t'\check{\lambda}_i + \check{\gamma}_{t0} + X_{it}'\check{\beta}\right]\right)^2 + \omega\|\Gamma'\check{\gamma}_0\|_2^2,
\end{equation}
and
\begin{equation*}
\tilde{\mu}_{c} \coloneqq \frac{1}{N_c}\sum_{i = 1}^N \ind{C_i = c} \left[\hat{\Gamma}\tilde{\lambda}_i + \tilde{\gamma}_{0} + X_{i}\tilde{\beta}\right].
\end{equation*}
Assume Assumption Theorem \ref{thm:factor_consistency_asymptotic_normality} holds. Then, with probability $1-o(1)$, $\left(\tilde{\gamma}_0, \tilde{\beta}, \{\tilde{\lambda}_i\}_{i = 1}^N\right)$ is uniquely defined and invariant to $\omega$, 
$\hat{\mu}_c = \tilde{\mu}_c$, and $\hat{\beta} = \tilde{\beta}.$
\end{lemma}

\begin{lemma}\label{lem:Gc_operator_norm}
In the setting of Theorem \ref{thm:factor_identification}, $I - (I-E_c)\Pi(\Gamma)(I - E_c)$ is invertible for any $c\in [C]$. Moreover, there exists $M\in (0, \infty)$, such that
$\|G_c(\Pi(\Gamma))\|_{\mathrm{op}}\le M.$
\end{lemma}

\begin{lemma}\label{lem:barXY}
Define the empirical cohort-specific outcome and covariate mean influence functions
\begin{equation}\label{eq:xi}
\psi^Y_{c,i,\mathrm{reg}}\coloneqq \frac{1\{C_i=c\}}{p_c}\bigl(Y_i-\mu_c\bigr), \quad 
\psi^X_{c,i,\mathrm{reg}}\coloneqq \frac{1\{C_i=c\}}{p_c}\bigl(X_i-\mu_{X,c}\bigr).
\end{equation}
Then 
\[\bar{Y}_c - \mu_c = \hat{\E}_N \left[\psi^Y_{c,i,\mathrm{reg}}\right] + o_\mathbb{P}\left( N^{-1/2}\right), \quad \bar{X}_c - \mu_{X,c} = \hat{\E}_N \left[ \psi^X_{c,i,\mathrm{reg}} \right]+ o_\mathbb{P}\left( N^{-1/2}\right).\]
\end{lemma}

\begin{lemma}[Equivalent expression of $P_c$]\label{lem:Pk_equivalent}
For any $T \times r$ matrix $\tilde{\Gamma}$ such that $E_c\tilde{\Gamma}$ is full-rank, 
\begin{equation*}\label{eq:least_norm_matrix_proj_expr}
P_c\bigl(\Pi(\tilde{\Gamma})\bigr) = \tilde{\Gamma}\left(\tilde{\Gamma}'E_c\tilde{\Gamma}\right)^{-1}\tilde{\Gamma}'E_c
\end{equation*}
\end{lemma}

\begin{lemma}\label{lem:Qk_equivalent}
In the setting of Lemma \ref{lem:Pk_equivalent}, 
\[Q_c(\Pi(\tilde{\Gamma})) = E_c - \Pi(E_c \tilde{\Gamma}), \quad Q_c^2(\Pi(\tilde{\Gamma})) = Q_c(\Pi(\tilde{\Gamma})).\]
\end{lemma}

\begin{lemma}[Directional derivative of $P_c$]\label{lem:Pk_deriv}
Let $\Pi_0\in \R^{T\times T}$ be a projection matrix and $\Delta\in\R^{T\times T}$ be a matrix symmetric such that $\Pi_0 + \Delta$ is also a projection matrix. Assume $G_c(\Pi_0)$ is invertible with 
\[2\|\Delta\|_{\mathrm{op}} < \|G_c(\Pi_0)\|_{\mathrm{op}}^{-1}.\]
Then the directional derivative of $P_c(\Pi)$ at $\Pi=\Pi_0$ in direction $\Delta$ exists and equals
\begin{align} \mathcal{P}_c(\Pi_0)[\Delta] & \coloneqq  \frac{d}{dt}P_c(\Pi_0 + t\Delta)\bigg|_{t=0}\\
&=
E_c\Delta (I - E_c) G_c(\Pi_0)\Pi_0 E_c  \\
& \qquad +\bigl(I+E_c\Pi_0(1-E_c)\bigr)G_c(\Pi_0)\bigl\{(I-E_c)\Delta (I-E_c)G_c(\Pi_0)\Pi_0 + \Delta\bigr\} E_c.
\end{align}
Moreover, 
\begin{align*}
\left\|\mathcal{P}_c(\Pi_0)[\Delta]\right\|_{\mathrm{op}}\le 5\|G_c(\Pi_0)\|_{\mathrm{op}}^2\|\Delta\|_{\mathrm{op}},
\end{align*}
and 
\begin{align*}
\left\|P_c(\Pi_0 + \Delta) - P_c(\Pi_0) - \mathcal{P}_c(\Pi_0)[\Delta]\right\|_{\mathrm{op}} \le 8 \|G_c(\Pi_0)\|_{\mathrm{op}}^3 \|\Delta\|_{\mathrm{op}}^2.
\end{align*}

\end{lemma}

\begin{lemma}\label{lem:ABP_influence}
Let 
\[A_i(\Pi) = D_i\left(\Pi\right)' Q_{C_i}(\Pi) D_i\left(\Pi\right), \quad \hat{A}(\Pi) = \hat{\E}_N\left[A_i(\Pi)\right],\quad A(\Pi) = \E[A_i(\Pi)], \quad A_0 = A(\Pi(\Gamma)),\]
and 
\[B_i(\Pi) = D_i\left(\Pi\right)' Q_{C_i}(\Pi) Y_i, \quad \hat{B}(\Pi) = \hat{\E}_N\left[B_i(\Pi)\right], \quad B(\Pi) = \E[B_i(\Pi)], \quad B_0 = B(\Pi(\Gamma)).\]
Further, let 
\begin{align*}
\mathcal{A}_i(\Pi_0)[\Delta] &= -D_i(\Pi_0)'\mathcal{P}_{C_i}(\Pi_0)[\Delta]D_i(\Pi_0)\\
& \quad - [\Delta\ \ \mathbf{0}_{T\times q}]'Q_{C_i}(\Pi_0)D_i(\Pi_0) - D_i(\Pi_0)'Q_{C_i}(\Pi_0)[\Delta\ \ \mathbf{0}_{T\times q}]\\
\mathcal{B}_i(\Pi_0)[\Delta] &= -D_i(\Pi_0)'\mathcal{P}_{C_i}(\Pi_0)[\Delta]Y_i - [\Delta\ \ \mathbf{0}_{T\times q}]'Q_{C_i}(\Pi_0)Y_i.
\end{align*}
In the setting of Lemma \ref{lem:Ab_deriv}, under Assumptions 1-4, 
\begin{align*}
\hat{A}(\Pi(\hat{\Gamma})) - A_0 &= \hat{\E}_N[\psi_{i, \mathrm{reg}}^{\mathcal{A}} + \psi_{i, \mathrm{fac}}^{\mathcal{A}}] + o_{\mathbb{P}}\left(N^{-1/2}\right),
\\
\hat{B}(\Pi(\hat{\Gamma})) - B_0 & = \hat{\E}_N[\psi_{i, \mathrm{reg}}^{\mathcal{B}} + \psi_{i, \mathrm{fac}}^{\mathcal{B}}] + o_{\mathbb{P}}\left(N^{-1/2}\right),\\
P_c(\Pi(\hat{\Gamma})) - P_c(\Pi(\Gamma)) & = \hat{\E}_N[\psi_{c,i, \mathrm{fac}}^{\mathcal{P}}] + o_{\mathbb{P}}\left(N^{-1/2}\right), \\
H(\Pi(\hat{\Gamma}), \bar{X}_c) - H(\Pi(\Gamma), \mu_{X,c}) & = \hat{\E}_N\left[\psi_{c,i, \mathrm{reg}}^{\mathcal{H}} + \psi_{i,\mathrm{fac}}^\mathcal{H}\right] + o_\mathbb{P}\left(N^{-1/2}\right),
\end{align*}
where
\begin{align*}
&\psi_{i, \mathrm{reg}}^{\mathcal{A}} = A_i(\Pi(\Gamma)) - A_0, \,\, \psi_{i, \mathrm{fac}}^{\mathcal{A}} = \mathcal{A}(\Pi(\Gamma))[\phi_i],\\
& \psi_{i, \mathrm{reg}}^{\mathcal{B}} = B_i(\Pi(\Gamma)) - B_0, \,\, \psi_{i, \mathrm{fac}}^{\mathcal{B}} = \mathcal{B}(\Pi(\Gamma))[\phi_i],\\
& \psi_{c,i, \mathrm{fac}}^{\mathcal{P}} = \mathcal{P}_c(\Pi(\Gamma))[\phi_i],\\
& \psi_{c,i, \mathrm{reg}}^{\mathcal{H}} = [\mathbf{0}_{T\times T}\ \ \psi_{c,i, \mathrm{reg}}^X], \,\, \psi_{i,\mathrm{fac}}^\mathcal{H} = [-\phi_i\ \ \mathbf{0}_{T\times q}].
\end{align*}
In particular, the above definitions of $\psi_{i,\mathrm{reg}}^\mathcal{A}, \psi_{i,\mathrm{fac}}^\mathcal{A}, \psi_{i,\mathrm{reg}}^\mathcal{B}, \psi_{i,\mathrm{fac}}^\mathcal{B},\psi_{c,i,\mathrm{fac}}^\mathcal{P}, \psi_{c,i,\mathrm{reg}}^\mathcal{H}$ and $ \psi_{i,\mathrm{fac}}^\mathcal{H}$ are equivalent to \eqref{eq:psicP_fac}, \eqref{eq:psicH_reg}, \eqref{eq:psicH_fac},
\eqref{eq:psiA_reg}, \eqref{eq:psiA_fac},
\eqref{eq:psiB_reg}, \eqref{eq:psiB_fac},respectively. Moreover, 
\[\E[\psi_{i,\mathrm{reg}}^{\mathcal{A}}] = \E[\psi_{i,\mathrm{fac}}^{\mathcal{A}}] = \mathbf{0}_{(T+q)\times (T+q)}, \,\,  \E[\psi_{i,\mathrm{reg}}^{\mathcal{B}}] = \E[\psi_{i,\mathrm{fac}}^{\mathcal{B}}] = \mathbf{0}_{T+q},\] 
\[\E[\psi_{c,i,\mathrm{fac}}^{\mathcal{P}}] = \mathbf{0}_{T\times T}, \,\, \E[\psi_{c,i, \mathrm{reg}}^{\mathcal{H}}] = \E[\psi_{i,\mathrm{fac}}^\mathcal{H}] = \mathbf{0}_{T\times (T+q)},\]
and
\[\E[\|\psi_{i, \mathrm{reg}}^{\mathcal{A}}\|_{\mathrm{op}}^2], \E[\|\psi_{i, \mathrm{fac}}^{\mathcal{A}}\|_{\mathrm{op}}^2] < \infty,\quad \E[\|\psi_{i, \mathrm{reg}}^{\mathcal{B}}\|_{\mathrm{op}}^2], \E[\|\psi_{i, \mathrm{fac}}^{\mathcal{B}}\|_{\mathrm{op}}^2] < \infty,\]
\[\E[\|\psi_{c,i, \mathrm{fac}}^{\mathcal{P}}\|_{\mathrm{op}}^2] < \infty, \quad \E[\|\psi_{c,i, \mathrm{reg}}^{\mathcal{H}}\|_{\mathrm{op}}^2], \E[\|\psi_{i,\mathrm{fac}}^\mathcal{H}\|_{\mathrm{op}}^2] < \infty.\]
\end{lemma}
\par
\noindent \paragraph*{Proof of Theorem \ref{thm:muc_hat_influence}}

\noindent By Lemma \ref{lem:equivalent_constraint}, we consider the modified estimator defined in Lemma \ref{lem:equivalent_constraint}. With some abuse of notation, we write $(\hat{\gamma}_0, \{\hat{\lambda}_i\}_{i=1}^N)$ for $(\tilde{\gamma}_0, \{\tilde{\lambda}_i\}_{i=1}^N)$ throughout the rest of the proof. We split the remaining proof into four steps:

\noindent \paragraph*{Step 1: partialling out $\lambda_i$.}
Define cohort-specific averages:
\[
N_c\coloneqq \sum_{i=1}^N 1\{C_i=c\},\quad
\bar Y_c\coloneqq \frac1{N_c}\sum_{i:C_i=c}Y_i,\quad
\bar X_c\coloneqq \frac1{N_c}\sum_{i:C_i=c}X_i.
\]
For any $(\check{\gamma}_0,\check{\beta})$ define the residual
\[
r_i(\check{\gamma}_0,\check{\beta})\coloneqq Y_i-\check{\gamma}_0-X_i\check{\beta}\in\R^T.
\]
Fix any $i$ with $C_i = c$. Holding $(\check{\gamma}_0,\check{\beta})$ fixed, the subproblem in $\check{\lambda}_i$ is
\begin{equation}
\min_{\check{\lambda}\in\R^r}\ \left\|E_{c}\bigl(r_i(\check{\gamma}_0,\check{\beta})-\hat{\Gamma}\check{\lambda}\bigr)\right\|_2^2.
\label{eq:lambda_sub}
\end{equation}
Since $E_c$ is symmetric and idempotent, the objective equals
\[
\bigl(r_i(\check{\gamma}_0,\check{\beta})-\hat{\Gamma}\check{\lambda}\bigr)'E_c\bigl(r_i(\check{\gamma}_0,\check{\beta})-\hat{\Gamma}\check{\lambda}\bigr).
\]
The first-order condition at $(\hat{\lambda}_i, \hat{\gamma}, \hat{\beta})$ gives
\[
-2\hat{\Gamma}'E_c\bigl(r_i(\hat{\gamma}_0,\hat{\beta})-\hat{\Gamma}\hat{\lambda}_i\bigr)=0
\quad\Longrightarrow\quad
\hat{\Gamma}'E_c\hat{\Gamma}\hat{\lambda}_i=\hat{\Gamma}'E_c r_i(\hat{\gamma}_0,\hat{\beta}).
\]
Thus, 
\[
\hat{\lambda}_i
=\bigl(\hat{\Gamma}'E_c\hat{\Gamma}\bigr)^{-1}\hat{\Gamma}'E_c\,r_i(\hat{\gamma}_0,\hat{\beta}).
\]
Above, we use the fact that $\hat{\Gamma}'E_c\hat{\Gamma} = \left(E_c\hat{\Gamma}\right)'E_c\hat{\Gamma}$ is invertible under the assumption that $E_c\hat{\Gamma}$ is full-rank.

~\\
\noindent \paragraph*{Step 2: deriving $(\hat{\gamma}_0, \hat{\beta})$}. 
Multiplying by $\hat{\Gamma}$, Lemma \ref{lem:Pk_equivalent} implies that, 
\begin{equation}\label{eq:factor_estimate}
\hat{\Gamma}\hat{\lambda}_i
=
P_c\left(\Pi(\hat{\Gamma})\right)\,\bigl(Y_i-\hat{\gamma}_0-X_i\hat{\beta}\bigr).
\end{equation}
As a consequence,
\begin{align*}
(\hat{\gamma}_0, \hat{\beta}) & = \argmin_{\check{\gamma}_0\in \R^T, \check{\beta}\in \R^{q}} \ \hat{\E}_N\left\|E_{C_i}\left\{I - P_{C_i}\left( \Pi(\hat{\Gamma})\right)\right\}(Y_i - \check{\gamma}_0 - X_i\check{\beta})\right\|_2^2\\
& = \argmin_{\check{\gamma}_0\in \R^T, \check{\beta}\in \R^{q}} \ \hat{\E}_N\left\|Q_{C_i}(\Pi(\hat{\Gamma}))(Y_i - \check{\gamma}_0 - X_i\check{\beta})\right\|_2^2 + \omega \|\Gamma' \check{\gamma}_0\|_2^2.
\end{align*} 
Define 
\[\alpha\coloneqq (\gamma_0',\beta')', \quad \hat{\alpha}\coloneqq (\hat{\gamma}_0',\hat{\beta}')',\] 
and
\[\hat{\alpha} = \argmin_{\check{\alpha}\in \R^{T+q}} \ \hat{\E}_N\left\|Q_{C_i}(\Pi(\hat{\Gamma}))\left(Y_i - D_i\left(\Pi(\hat{\Gamma})\right)\check{\alpha}\right)\right\|_2^2  + \omega \|\Gamma \check{\gamma}_0\|_2^2.\]
The first-order condition with respect to $\check{\alpha}$ and Lemma \ref{lem:Qk_equivalent} imply
\begin{align*}
&\hat{\E}_N\left[D_i\left(\Pi(\hat{\Gamma})\right)'Q_{C_i}(\Pi(\hat{\Gamma}))\left(Y_i - D_i\left(\Pi(\hat{\Gamma})\right)\hat{\alpha}\right)\right] + \omega \bar{\Gamma}\bar{\Gamma}'\hat{\alpha} = 0.
\end{align*} 
Let $A_{0,\omega}$ be defined as in \eqref{eq:A0_omega} and 
\begin{equation}\label{eq:hatA0omega}
\hat{A}_\omega\left(\Pi(\hat{\Gamma})\right) \coloneqq \hat{A}\left(\Pi(\hat{\Gamma})\right) + \omega\bar{\Gamma}\bar{\Gamma}'.
\end{equation}
By Lemma \ref{lem:A0_invertibility_equivalent}, $A_{0,\omega}$ is invertible. By Lemma \ref{lem:ABP_influence}, \[\left\|\hat{A}_\omega\left(\Pi(\hat{\Gamma})\right) - A_{0,\omega}\right\|_{\mathrm{op}} = \left\|\hat{A}\left(\Pi(\hat{\Gamma})\right) - A_{0}\right\|_{\mathrm{op}}  = O_{\mathbb{P}}(1/\sqrt{N}).\] 
Assumption \ref{assump:A0_invertibility} then implies that $\hat{A}$ is invertible with probability $1-o(1)$. Rearranging the terms in the FOC, we have 
\begin{equation}\label{eq:hattheta}
\hat{\alpha} = 
\hat{A}_\omega\left(\Pi(\hat{\Gamma})\right)^{-1}\hat{B}\left(\Pi(\hat{\Gamma})\right).
\end{equation}

~\\
\noindent \paragraph*{Step 3: deriving $\hat{\mu}_c$ as a function of $\Pi(\hat{\Gamma})$}. Substitute \eqref{eq:factor_estimate} into $\hat{\mu}_c$ and use linearity of summation:
\begin{align*}
\hat{\mu}_c
&=\frac1{N_c}\sum_{i:C_i=c}\Bigl[
P_c\left(\Pi(\hat{\Gamma})\right)\bigl(Y_i-\hat{\gamma}_0-X_i\hat{\beta}\bigr)
+\hat{\gamma}_0+X_i\hat{\beta}
\Bigr]\\
&=
P_c\left(\Pi(\hat{\Gamma})\right)\Bigl(\bar Y_c-\hat{\gamma}_0-\bar X_c\hat{\beta}\Bigr)
+\hat{\gamma}_0+\bar X_c\hat{\beta}.
\end{align*}
Rearranging terms yields
\[
\hat{\mu}_c
=
P_c\left(\Pi(\hat{\Gamma})\right)\,\bar Y_c
+\left\{I-P_c\left(\Pi(\hat{\Gamma})\right)\right\}\,\bigl(\hat{\gamma}_0+\bar X_c\hat{\beta}\bigr).
\]
By definition of $H$, we can rewrite
\[
\hat{\gamma}_0+\bar X_c\hat{\beta}=H(\Pi(\hat{\Gamma}), \bar{X}_c)\hat{\alpha},
\]
and thus
\begin{align}
\hat{\mu}_c
&=
P_c\left(\Pi(\hat{\Gamma})\right)\,\bar Y_c
\\
& +\left\{I-P_c\left(\Pi(\hat{\Gamma})\right)\right\}H(\Pi(\hat{\Gamma}), \bar{X}_c)\hat{A}_\omega\left(\Pi(\hat{\Gamma})\right)^{-1}\hat{B}\left(\Pi(\hat{\Gamma})\right).
\label{eq:muc_hat}
\end{align}

~\\
\noindent \paragraph*{Step 4: deriving the influence function $\psi_{i}$}.  Write $P_{c,0}$ for $P_c(\Pi(\Gamma))$ and $H_0$ for $H(\Pi(\Gamma), \mu_
{X,c})$. By Lemma \ref{lem:equivalent_constraint}, $\alpha$ is invariant to $\omega$. Thus, 
\begin{equation}\label{eq:alpha_invariance}
A_{0,\omega}^{-1}B_0 = \alpha. 
\end{equation}
By Lemma \ref{lem:muc},
\begin{equation}
\mu_c
=P_{c,0}\mu_{c}
+(I-P_{c,0})H_0 A_{0,\omega}^{-1}B_0.
\label{eq:muc}
\end{equation}
This holds for every $\omega > 0$ as a result of Lemma \ref{lem:equivalent_constraint}.
By Lemma \ref{lem:barXY}, Lemma \ref{lem:ABP_influence}, and \eqref{eq:muc},
\begin{align*}
\hat{\mu}_c & = \left\{P_{c,0} + \hat{\E}_N[\psi_{c,i,\mathrm{fac}}^\mathcal{P}] + o_\mathbb{P}\left(N^{-1/2}\right)\right\}\left\{\mu_c + \hat{\E}_N[\psi_{c,i, \mathrm{reg}}^Y] + o_\mathbb{P}\left(N^{-1/2}\right)\right\}\\
& \quad + \left\{I - P_{c,0} - \hat{\E}_N[\psi_{c,i,\mathrm{fac}}^\mathcal{P}] + o_\mathbb{P}\left(N^{-1/2}\right)\right\}\left\{H_0 + \hat{\E}_N[\psi_{c,i,\mathrm{reg}}^{\mathcal{H}} + \psi_{i,\mathrm{fac}}^\mathcal{H}] + o_\mathbb{P}\left(N^{-1/2}\right)\right\}\\
& \qquad \cdot \left\{A_{0,\omega} + \hat{\E}_N[\psi_{i,\mathrm{reg}}^{\mathcal{A}} + \psi_{i,\mathrm{fac}}^{\mathcal{A}}] + o_\mathbb{P}\left(N^{-1/2}\right)\right\}^{-1}\left\{B_0 + \hat{\E}_N[\psi_{i,\mathrm{reg}}^{\mathcal{B}} + \psi_{i,\mathrm{fac}}^{\mathcal{B}}] + o_\mathbb{P}\left(N^{-1/2}\right)\right\}.
\end{align*}
By Assumption \ref{assump:cohort_sizes_and_potential_outcome_variance_bound} and Lemma \ref{lem:Pk_deriv}, the first term is given by 
\begin{equation}\label{eq:psii_piece1}
P_{c,0}\mu_c + \hat{\E}_N\left[P_{c,0}\psi_{c,i, \mathrm{reg}}^Y + \psi_{c,i,\mathrm{fac}}^\mathcal{P}\mu_c\right] + o_\mathbb{P}\left(N^{-1/2}\right).
\end{equation}
By Lemma \ref{lem:ABP_influence}, $\E[\psi_{i,\mathrm{reg}}^{\mathcal{A}} + \psi_{i,\mathrm{fac}}^{\mathcal{A}}] = 0$ and hence $\hat{\E}_N[\psi_{i,\mathrm{reg}}^{\mathcal{A}} + \psi_{i,\mathrm{fac}}^{\mathcal{A}}] = o_\mathbb{P}(1)$. By Assumption \ref{assump:A0_invertibility} and Lemma \ref{lem:A-B_inv},
\[\left\{A_{0,\omega} + \hat{\E}_N[\psi_{i,\mathrm{reg}}^{\mathcal{A}} + \psi_{i,\mathrm{fac}}^{\mathcal{A}}] + o_\mathbb{P}\left(N^{-1/2}\right)\right\}^{-1} = A_{0,\omega}^{-1} - \hat{\E}_N\left[A_{0,\omega}^{-1}(\psi_{i,\mathrm{reg}}^{\mathcal{A}} + \psi_{i,\mathrm{fac}}^{\mathcal{A}})A_{0,\omega}^{-1}\right] + o_\mathbb{P}\left(N^{-1/2}\right).\]
By Lemma \ref{lem:ABP_influence}, the second term of $\hat{\mu}_c$ is given by 
\begin{align}
& (I - P_{c,0})H_0A_{0,\omega}^{-1}B_0 \\
& + \hat{\E}_N\Big[-\psi_{c,i,\mathrm{fac}}^\mathcal{P}H_0A_{0,\omega}^{-1}B_0 + (I-P_{c,0})(\psi_{c,i,\mathrm{reg}}^{\mathcal{H}} + \psi_{i,\mathrm{fac}}^\mathcal{H})A_{0,\omega}^{-1}B_0\\
& - (I - P_{c,0})H_0A_{0,\omega}^{-1}(\psi_{i,\mathrm{reg}}^{\mathcal{A}} + \psi_{i,\mathrm{fac}}^{\mathcal{A}})A_{0,\omega}^{-1}B_0 + (I - P_{c,0})H_0 A_{0,\omega}^{-1}(\psi_{i,\mathrm{reg}}^{\mathcal{B}} + \psi_{i,\mathrm{fac}}^{\mathcal{B}})\Big]\label{eq:psii_piece2}\\
& + o_\mathbb{P}\left(N^{-1/2}\right).
\end{align}
Combining \eqref{eq:alpha_invariance}, \eqref{eq:muc}, \eqref{eq:psii_piece1} and \eqref{eq:psii_piece2}, 
\begin{align*}
&\hat{\mu}_c - \mu_c\\
& = \hat{\E}_N\Big[P_{c,0}\psi_{c,i, \mathrm{reg}}^Y + (I-P_{c,0})\psi_{c,i,\mathrm{reg}}^{\mathcal{H}}\alpha\\
& \qquad \quad +(I - P_{c,0})H_0A_{0,\omega}^{-1}\psi_{i,\mathrm{reg}}^{\mathcal{A}} \alpha + (I - P_{c,0})H_0 A_{0,\omega}^{-1}\psi_{i,\mathrm{reg}}^{\mathcal{B}} \Big] \\
& \quad + \hat{\E}_N\Big[\psi_{c,i,\mathrm{fac}}^\mathcal{P}\mu_c -\psi_{c,i,\mathrm{fac}}^\mathcal{P}H_0\alpha + (I-P_{c,0}) \psi_{i,\mathrm{fac}}^\mathcal{H}\alpha\\
&\qquad \quad + (I - P_{c,0})H_0A_{0,\omega}^{-1}\psi_{i,\mathrm{fac}}^{\mathcal{A}}\alpha + (I - P_{c,0})H_0 A_{0,\omega}^{-1} \psi_{i,\mathrm{fac}}^{\mathcal{B}}\Big] \\
& \quad + o_\mathbb{P}\left(N^{-1/2}\right)\\
& = \hat{\E}_N[\psi_{c,i,\mathrm{reg}}] + \hat{\E}_N[\psi_{c,i,\mathrm{fac}}] + o_\mathbb{P}\left(N^{-1/2}\right)
\end{align*}
The moment conditions on $\psi_i$ are directly consequences of Lemma \ref{lem:ABP_influence} and Assumption \ref{assump:cohort_sizes_and_potential_outcome_variance_bound}. 

By \eqref{eq:H_A0+} in Lemma \ref{lem:A0_invertibility_equivalent}, 
\[H_0 A_{0, \omega}^{-1} = H_0 A_0^+. \]
Thus, we can also replace the term $H_0 A_{0, \omega}^{-1}$ by $H_0 A_{0}^{+}$ in the above expression. For any fixed $\omega > 0$, Lemma \ref{lem:A0_invertibility_equivalent} implies $A_{0,\omega}$ is invertible. Since $A_{0,\omega}$ is continuous in $\Pi(\Gamma)$, $A_{0,\omega}^{-1}$ is also continuous in $\Pi(\Gamma)$.

%% file: sections/target_params.tex
In this appendix, we provide several examples of empirically relevant target parameters $\theta$ that researchers can estimate and conduct inference on using our estimator. Then, we describe our Bayesian-bootstrap-based simultaneous inference procedure for $\theta$ in detail. Finally, we establish that our plug-in estimator $\hat{\theta}$ described in Section \ref{sec:method:procedure} is a consistent and asymptotically normal estimator of $\theta$, and that the Bayesian-bootstrap-based simultaneous confidence intervals described in Section \ref{sec:target_params:inference_procedure} have valid simultaneous coverage.

\subsection{Examples of Target Parameters}\label{sec:supplement:target_params_examples}

Here, we introduce two examples of target parameters $\theta = h(\mu, \eta)$ that aggregate cohort outcome means $\mu$ and nuisance parameters $\eta$ nonlinearly through $h$ as introduced in Section \ref{sec:setup:setup}:

\begin{example}[Dynamic Treatment Effects]\label{ex:dynamic_treatment_effects}
    In event study settings, researchers are often interested in reporting dynamic treatment effect paths, i.e. average effects of a treatment across different numbers of time periods relative to units' treatment times. Recall that, in our notation, outcome $Y_{it}^*$ refers to unit $i$'s potential outcome in period $t$ had they not yet been treated by period $t$ (often denoted by $Y_{it}(\infty)$ in this literature), $C_i$ refers to unit $i$'s first treatment period, and the set of observed outcomes for the units treated in period $c$ is $\mathcal{T}_c \subset \{1, \dotsc, c - 1\}$. To express dynamic treatment effects in the form $h(\mu, \eta)$, we first define $Z_{it}^*$ as unit $i$'s potential outcome in period $t$ had they been treated in period $C_i$, and, to distinguish between observed and missing outcomes, we let $Z_{it} = Z_{it}^*$ if unit $i$ was treated by period $t$ (i.e. $C_i \leq t$) and $Z_{it} = \emptyset$ otherwise.
    
    Assuming we observe either $Y_{it}^*$ or $Z_{it}^*$ in every period for simplicity,\footnote{In many settings like Figure \ref{fig:control_outcomes_over_time_missingness_pattern}'s, we do not observe control or treated potential outcome in every period. \citet{callaway2021difference} and \citet{sun2021estimating} discuss several ways of defining estimands that take this additional unbalancedness into account; for brevity, we simply note that these estimands can also be written in the form $h(\mu, \eta)$ and refer the interested reader to Section 3.1.1 of \citet{callaway2021difference} for details.} we can write a coordinates of the $p$-dimensional parameter vector whose entries correspond to the dynamic treatment effects from $b$ periods before treatment through $p - b$ treated periods in the form of Equation (3.4) in \citet{callaway2021difference} or Equation (26) in \citet{sun2021estimating}:
    \begin{equation}\label{eq:dynamic_treatment_effect_def}
    \begin{aligned}
        \theta_{\text{dyn},j} &\coloneqq \sum_{t = 1}^T \sum_{c = 1}^C \ind{t - c = j - b - 1}p_c\\
        &\phantom{\coloneqq \sum_{t = 1}^T \sum_{c = 1}^C}\cdot\left(\underbrace{\E[\ind{t \geq c}Z_{it} + \ind{t < c}Y_{it} ~|~ C_i = c]}_{m_{ct}} - \mu_{ct}\right),
    \end{aligned}
    \end{equation}
    where $j \in \{1, \dotsc, p\}$ indexes the coordinates of $\theta_{\text{dyn}}$, $t - c$ reflects the index of the current period relative to cohort $c$'s treatment time (with zero corresponding to a cohort's first treated period in relative time), and $m_{ct}$ denotes the average observed outcome for units in cohort $c$ in period $t$; when period $t$ is before cohort $c$'s treatment time $c$, the observed outcome is $Y_{it}$ which equals the control potential outcome $Y_{it}^*$, and when period $t$ is after cohort $c$'s treatment time $c$, the observed outcome is $Z_{it}$ which equals the treated potential outcome $Z_{it}^*$. We note that $\theta_{\text{dyn},j}$ can be defined without any restrictions on heterogeneity in treatment effects $Z_{it}^* - Y_{it}^*$ across units or time periods \citep{callaway2021difference,sun2021estimating}.
\end{example}

\begin{examplecont}{example-match-outcome-attribution}
    We continue our discussion of the nonparametric decomposition of differences in observed outcome means introduced in Example \ref{example-match-outcome-attribution}. We note that the definitions of $\theta_{\text{col}, t_1, t_2}$ and $\theta_{\text{row}, t_1, t_2}$ from Example \ref{example-match-outcome-attribution} are only defined in terms of outcomes, which is beneficial for interpretation \citep{hull2018estimating,kline2024firm}, but they coincide exactly with $S_\text{place}$ and $S_\text{pat}$ in \citet{finkelstein2016sources} when $Y_{it}^*$ is determined by the TWFE model \eqref{eq:twfe_model}:
    \begin{equation}
    \begin{aligned}
        \theta_{\text{col}, t_1, t_2} &\propto \gamma_{t_1} - \gamma_{t_2} \\
        \theta_{\text{row}, t_1, t_2} &\propto \E[\lambda_i ~|~ t_1 \in \mathcal{T}_{C_i}] - \E[\lambda_i ~|~ t_2 \in \mathcal{T}_{C_i}].
    \end{aligned}
    \end{equation}

    As discussed in \citet{finkelstein2016sources}, one can also generalize this decomposition by studying differences in average observed match outcomes $\overline{\mu}_{\tilde{\mathcal{T}}_1}  - \overline{\mu}_{\tilde{\mathcal{T}}_2}$ between two \textit{groups} of outcomes $\tilde{\mathcal{T}}_1, \tilde{\mathcal{T}}_2 \subset [T]$ with outcomes assigned potentially unequal weights $w_t > 0$, i.e.
    \begin{equation}
    \begin{aligned}
        \overline{\mu}_{\tilde{\mathcal{T}}_1}  - \overline{\mu}_{\tilde{\mathcal{T}}_2}, \quad \overline{\mu}_{\tilde{\mathcal{T}}} \coloneqq \left(\sum_{c = 1}^C \Pr(C_i = c) \sum_{t \in \tilde{\mathcal{T}} \cap \mathcal{T}_c}w_t\right)^{-1}\sum_{c = 1}^C \Pr(C_i = c) \sum_{t \in \tilde{\mathcal{T}} \cap \mathcal{T}_c}w_t\E[Y_{it} ~|~ t \in \mathcal{T}_c].
    \end{aligned}
    \end{equation}
    In particular, one can define $\theta_{\text{col}, \tilde{\mathcal{T}}_1, \tilde{\mathcal{T}}_2}$ and $\theta_{\text{row}, \tilde{\mathcal{T}}_1, \tilde{\mathcal{T}}_2}$ analogously to $\theta_{\text{col}, t_1, t_2}$ and $\theta_{\text{row}, t_1, t_2}$:
    \begin{equation}
    \begin{aligned}
        \theta_{\text{col}, \tilde{\mathcal{T}}_1, \tilde{\mathcal{T}}_2} &\coloneqq \left[\overline{\mu}_{\tilde{\mathcal{T}}_1}  - \overline{\mu}_{\tilde{\mathcal{T}}_2}\right]^{-1}\left[\overline{\mu}_{\tilde{\mathcal{T}}_1}^* - \overline{\mu}_{\tilde{\mathcal{T}}_2}^*\right] \eqqcolon 1 - \theta_{\text{row}, \tilde{\mathcal{T}}_1, \tilde{\mathcal{T}}_2}, \quad \overline{\mu}_{\tilde{\mathcal{T}}}^* \coloneqq \left(\sum_{t \in \tilde{\mathcal{T}}}w_t\right)^{-1}\sum_{t \in \tilde{\mathcal{T}}}w_t\E[Y_{it}^*].
    \end{aligned}
    \end{equation}

    Given a collection of disjoint subsets of outcomes $\tilde{\mathcal{T}}_1, \ldots, \tilde{\mathcal{T}}_K$, we can summarize the pairwise differences in average observed match outcomes across all pairs of these outcome groups by averaging $\theta_{\text{col}, \tilde{\mathcal{T}}_i, \tilde{\mathcal{T}}_j}$ over all pairs $(i, j) \in [K] \times [K]$, weighted by the total outcome weight corresponding to the union of each pair of outcome groups:
    \begin{equation}
        \overline{\theta}_\text{col} \coloneqq \left(\sum_{i = 1}^K \sum_{j = i + 1}^K \sum_{t \in \tilde{\mathcal{T}}_i \cup \tilde{\mathcal{T}}_j}w_t\right)^{-1}\sum_{i = 1}^K \sum_{j = i + 1}^K \left(\sum_{t \in \tilde{\mathcal{T}}_i \cap \tilde{\mathcal{T}}_j}w_t\right)\theta_{\text{col}, \tilde{\mathcal{T}}_i, \tilde{\mathcal{T}}_j} \eqqcolon 1 - \overline{\theta}_\text{row}.
    \end{equation}
    In our empirical illustration in Section \ref{sec:empirical_performance:match_outcome_decomposition}, we estimate $\overline{\theta}_\text{col}$ for subsets of outcomes corresponding to clusters of firms within the same provinces in the two-year window 2000 to 2001. For simplicity, we restrict our attention to $\theta_{\text{col}, t_1, t_2}$ and $\theta_{\text{row}, t_1, t_2}$ in our remaining discussion of the theory of conducting valid inference on target parameters.
\end{examplecont}

\subsection{A Bootstrap-Based Simultaneous Inference Procedure}\label{sec:target_params:inference_procedure}

To describe how we construct our Bayesian-bootstrap-based simultaneous confidence intervals, we introduce more notation. Given a vector of $N$ non-negative weights $W \coloneqq (W_1, \dotsc, W_N)'$ that sum to one, we will assume that our cohort-specific factor matrix estimators $\hat{\Gamma}_c$ as well as our nuisance parameter estimator $\hat{\eta}$ can be adapted to accommodate non-uniform sampling weights $W$, which we denote $\hat{\Gamma}_c(W)$ and $\hat{\eta}(W)$, respectively. Typically, when $W_1 = \dotsc = W_N = \frac{1}{N}$, we have that $\hat{\Gamma}_c(W) = \hat{\Gamma}_c$ and $\hat{\eta}(W) = \hat{\eta}$.

Next, we construct an APM using these weighted, estimated cohort-specific factor matrices $\hat{\Gamma}_1(W), \dotsc, \hat{\Gamma}_C(W)$:
\begin{equation}
    \hat{A}(W) \coloneqq \frac{1}{C}\sum_{c = 1}^C \left[E_c - \Pi(E_c\hat{\Gamma}_c(W))\right],
\end{equation}
and we let $\hat{\Gamma}(W)$ denote the eigenvectors of $\hat{A}(W)$ corresponding to its $r$ smallest eigenvalues. We can then define a weighted analog of the regression-based imputation estimator \eqref{eq:regression_imputation_estimator_def}:
\begin{equation}\label{eq:weighted_regression_imputation_estimator_def}
\begin{aligned}
    &\left(\hat{\gamma}_0(W), \hat{\beta}(W), \{\hat{\lambda}_i(W)\}_{i = 1}^N\right) \\
    &\coloneqq \argmin_{\check{\gamma}_0 \in \R^T,\; \check{\beta} \in \R^q,\; \check{\lambda}_i \in \R^r} \left\{\sum_{i = 1}^N W_i\sum_{t \in \mathcal{T}_{C_i}}\left(Y_{it} - \left[\hat{\gamma}_t(W)'\check{\lambda}_i + \check{\gamma}_{t0} + X_{it}'\check{\beta}\right]\right)^2 ~\setst~ \hat{\Gamma}'(W)\check{\gamma}_0 = \zeros_r \right\},
\end{aligned}
\end{equation}
so we can construct a weighted analog of the regression-based imputation estimator \eqref{eq:regression_imputation_estimator_def} as follows:
\begin{equation}\label{eq:weighted_estimator_def}
\begin{aligned}
    \hat{\mu}_{ct}(W) &\coloneqq \sum_{i = 1}^N \frac{\ind{C_i = c}W_i}{\sum_{j = 1}^N \ind{C_j = c}W_j}\left[\hat{\gamma}_t(W)'\hat{\lambda}_i(W) + \hat{\gamma}_{t0}(W) + X_{it}'\hat{\beta}(W)\right], \\
    \hat{\mu}_c(W) &\coloneqq (\hat{\mu}_{c1}(W), \dotsc, \hat{\mu}_{cT}(W))'.
\end{aligned}
\end{equation}
A weighted version of our plug-in estimator for $\hat{\theta}$ is $\hat{\theta}(W) \coloneqq h(\hat{\mu}(W), \hat{\eta}(W))$, where $\hat{\mu}(W)$ is the weighted analog of $\hat{\mu}$. Before continuing, we note that, again, under uniform weights $W_1 = \dotsc = W_N = w > 0$, all of the quantities defined previously in this paragraph equal their unweighted counterparts.

Given this notation, our inference procedure proceeds as follows. First, for each iteration $m$ of a large number $M$ of repetitions,\footnote{We recommend $M \geq 500$.} we take $N$ i.i.d. draws $\xi_{m1}, \dotsc, \xi_{mN}$ from the $\mathrm{Exponential}(1)$ distribution,\footnote{Other non-negative distributions are also possible as long as they satisfy regularity conditions; see Section 3.6.2 in \citet{vaart1996empirical} for more examples.} construct a vector $W_m \coloneqq (W_{m1}, \dotsc, W_{mN})'$ of $N$ normalized weights $W_{mi} \coloneqq \xi_{mi}/\sum_{j = 1}^N \xi_{mj}$, and compute a weighted target parameter estimate $\hat{\theta}^*_m \coloneqq \hat{\theta}(W_m)$.\footnote{Since $T$ is small, each computation of $\hat{\theta}(W_m)$ should be quite fast, as discussed when describing Algorithm \ref{alg:estimation} above. Further, $\hat{\theta}(W_m)$ can be computed in parallel across iterations $m$, boosting computational efficiency further. However, in settings where $T \ll N$ but $N$ and $T$ are both very large, the simplicity of this weighted bootstrap procedure may be outweighed its computational burden. In such cases, one could instead construct a multiplier bootstrap inference procedure like the one proposed in \citet{belloni2017program} using the influence function expressions given in Theorem \ref{thm:factor_consistency_asymptotic_normality}, Theorem \ref{thm:outcome_mean_consistency_asymptotic_normality}, and Corollary \ref{cor:target_parameter_asymptotic_linear_normal}.} Next, for each coordinate $j \in \{1, \dotsc, p\}$ of $\theta$, we compute an estimate $\hat{\sigma}_j$ of the standard error of $\hat{\theta}_j$, e.g. 
\begin{equation}\label{eq:bootstrap_standard_err_est_def}
    \hat{\sigma}_j \coloneqq \frac{q_{0.75}(\hat{\theta}^*_{1j}, \dotsc, \hat{\theta}^*_{Mj}) - q_{0.25}(\hat{\theta}^*_{1j}, \dotsc, \hat{\theta}^*_{Mj})}{q_{0.75}(Z) - q_{0.25}(Z)},
\end{equation}
where we let $q_\zeta(x_1, \dotsc, x_M)$ denote the $\zeta$th quantile across scalars $x_1, \dotsc, x_M$, and we let $q_\zeta(Z)$ denote the $\zeta$th quantile of the standard Gaussian distribution.\footnote{We suggest this interquartile-range-based estimate of estimator standard errors because it is more robust to outliers than other standard error estimators like the standard deviation over weighted bootstrap draws; see the discussion in Remark 3.2 in \citet{chernozhukov2013inference} for details.} Finally, we let $\hat{q}_{1-\alpha}$ denote the following estimated critical value:
\begin{equation}\label{eq:estimated_critical_value}
    \hat{q}_{1-\alpha} \coloneqq q_{1-\alpha}(z^*_1, \dotsc, z^*_M), \quad z^*_m \coloneqq \max_{j \in \{1, \dotsc, p\}} \absfit{\hat{\theta}^*_{mj} \big/ \hat{\sigma}_j},
\end{equation}
and we define our simultaneous $1-\alpha$ confidence intervals $\hat{\mathcal{C}}_j$ for $\theta_j$ across $j = 1, \dotsc, p$ as follows:
\begin{equation}\label{eq:simultaneous_confidence_interval}
    \hat{\mathcal{C}}_j \coloneqq \left[\hat{\theta}_j - \hat{q}_{1-\alpha}\hat{\sigma}_j, \hat{\theta}_j + \hat{q}_{1-\alpha}\hat{\sigma}_j\right].
\end{equation}
For convenience, we summarize the steps of our inference procedure in Algorithm \ref{alg:inference}.

\begin{algorithm}[t]
\SetAlgoLined
\DontPrintSemicolon
\caption{Bayesian Bootstrap Inference Procedure}
\label{alg:inference}
\KwData{$\{(C_i, Y_i)\}_{i = 1}^N$, number of bootstrap samples $M$.}
\BlankLine
\nl \For{$m \in \{1, \dotsc, M\}$}{
    Sample $\xi_{m1}, \dotsc, \xi_{mN} \distiid \mathrm{Exp}(1)$\;
    Construct weight vector $W_m \coloneqq (W_{m1}, \dotsc, W_{mN})'$ with $W_{mi} \coloneqq \xi_{mi}/\sum_{j = 1}^N \xi_{mj}$\;
    \For{$c \in \{1, \dotsc, C\}$}{
        Compute weighted cohort outcome mean estimate vector $\hat{\mu}_c(W_m)$ as in \eqref{eq:weighted_estimator_def} (using Algorithm \ref{alg:estimation} with weighted analogs)\;
    }
    Compute weighted nuisance parameter estimates $\hat{\eta}(W_m)$\;
    Compute weighted target parameter estimate $\hat{\theta}^*_m = h(\hat{\mu}(W_m), \hat{\eta}(W_m))$\;
}
\nl \For{$j \in \{1, \dotsc, p\}$}{
    Compute estimate $\hat{\sigma}_j$ of $\hat{\theta}$'s standard error, e.g. as in \eqref{eq:bootstrap_standard_err_est_def}\;
}
\nl Compute estimated critical value $\hat{q}_{1-\alpha}$ as in \eqref{eq:estimated_critical_value}\;
\nl Compute simultaneous $1-\alpha$ confidence intervals $\hat{\mathcal{C}}_j$ as in \eqref{eq:simultaneous_confidence_interval}\;
\end{algorithm}

\subsection{Asymptotic Linearity, Normality, Confidence Interval Validity}\label{sec:theory:target_parameters}

Having established in Section \ref{sec:theory:estimation_inference} that $\hat{\mu}_c$ is an asymptotically linear estimator of the vector of cohort outcome means $\mu_c$, we now establish asymptotic linearity of our plug-in estimator $\hat{\theta}$ of our target estimand $\theta = h(\mu, \eta)$ and validity of our simultaneous confidence intervals $\hat{\mathcal{C}}_j$. To do so, we assume that nuisance parameter estimator $\hat{\eta}$ is also asymptotically linear:
\begin{assumption}\label{assump:nuisance_param_est_asymptotic_linear}
    There exists an additional observed random vector $\varrho_i \in \R^m$ measurable with respect to the same probability space as $(C_i, Y_i^*, X_i)$ that satisfies $\E[\varrho_i] = \zeros_m$ and $\E[\normnofit{\varrho_i}_2^2] < \infty$ such that the following expansion holds as $N \rightarrow \infty$:
    \begin{equation}
        \hat{\eta} - \eta = \hat{\E}_N\left[\varrho_i\right] + \littleop{N^{-1/2}}.
    \end{equation}
\end{assumption}
\noindent In addition, we assume that the function $h$ that defines $\theta = h(\mu, \eta)$ is sufficiently smooth:
\begin{assumption}\label{assump:h_smooth}
    $h \colon \R^{CT} \times \R^m \to \R^p$ is differentiable at $(\mu, \eta)$.
\end{assumption}

Under the additional Assumptions \ref{assump:nuisance_param_est_asymptotic_linear} and \ref{assump:h_smooth}, our desired result holds:
\begin{corollary}\label{cor:target_parameter_asymptotic_linear_normal}
    Suppose that Theorem \ref{thm:factor_identification} and Assumptions \ref{assump:cohort_specific_factor_proj_mat_inf_fn}, \ref{assump:cohort_sizes_and_potential_outcome_variance_bound}, \ref{assump:nuisance_param_est_asymptotic_linear}, and \ref{assump:h_smooth} hold. Then
    \begin{equation}\label{eq:target_parameter_est_asymptotic_linear}
    \begin{aligned}
        \hat{\theta} - \theta &= \hat{\E}_N\left[\varphi(C_i, Y_i, X_i, \varrho_i)\right] + \littleop{N^{-1/2}}, \\
        \varphi(C_i, Y_i, X_i, \varrho_i) &\coloneqq \sum_{c = 1}^C\pfrac{h}{\mu_c'}\bigg|_{(\mu, \eta) = (\mu, \eta)}\psi_c(C_i, Y_i, X_i) + \pfrac{h}{\eta'}\bigg|_{(\mu, \eta) = (\mu, \eta)}\varrho_i,
    \end{aligned}
    \end{equation}
    where $\varphi$ satisfies $\E[\varphi(C_i, Y_i, X_i, \varrho_i)] = \zeros_p$ and $\E[\normnofit{\varphi(C_i, Y_i, X_i, \varrho_i)}_2^2] < \infty$. Further, as $N \rightarrow \infty$,
    \begin{equation}\label{eq:target_parameter_asymptotic_normality}
        \sqrt{N}(\hat{\theta} - \theta) \convin{\mathcal{D}} \mathcal{N}\left(\zeros_p, \Sigma_\theta\right), \quad \Sigma_\theta \coloneqq \E[\varphi(C_i, Y_i, X_i, \varrho_i)\varphi(C_i, Y_i, X_i, \varrho_i)'].
    \end{equation}
\end{corollary}
\noindent Corollary \ref{cor:target_parameter_asymptotic_linear_normal} follows from a straightforward application of the Delta Method and a classical multivariate Central Limit Theorem.\footnote{See e.g. Theorem 3.1 in \citet{van2000asymptotic}.}

Before continuing, we note that depending on the choice of $h$, $\Sigma_\theta$ may have zeros in its diagonal entries, in which case $\hat{\theta}$ will have zero asymptotic variance. An implication of this limiting distribution degeneracy is that the confidence intervals $\hat{\mathcal{C}}_j$ defined in \eqref{eq:simultaneous_confidence_interval} based on the limiting Gaussian distribution \eqref{eq:target_parameter_asymptotic_normality} will have zero width for some coordinates and thus zero coverage of those coordinates. The following assumption rules out such knife-edge cases:
\begin{assumption}\label{assump:rule_out_pathological_dgps_for_inference}
    The diagonal entries of $\Sigma_\theta$ are all strictly positive.
\end{assumption}

To state our result on the validity of the bootstrap-based inference procedure described in Section \ref{sec:method:procedure} under this assumption, we define the random vector $\hat{\theta}^* = \hat{\theta}(W)$, where $\hat{\theta}(\cdot)$ is defined in \eqref{eq:weighted_estimator_def}, $W$ is a random vector of weights with $i$th coordinate given by $W_i = \xi_i/\sum_{j = 1}^N \xi_j$, and $\xi_1, \dotsc, \xi_N$ are draws from $\mathrm{Exponential}(1)$ independent of both each other and $\hat{\theta}(\cdot)$. Then, under some slight generalizations of Assumptions \ref{assump:cohort_specific_factor_proj_mat_inf_fn} and \ref{assump:nuisance_param_est_asymptotic_linear} stated in Appendix \ref{proof:thm:confidence_interval_validity} for brevity, our inference procedure satisfies the following validity guarantee:
\begin{theorem}\label{thm:confidence_interval_validity}
    Suppose that Theorem \ref{thm:factor_identification} and Assumptions \ref{assump:cohort_specific_factor_proj_mat_inf_fn}, \ref{assump:cohort_sizes_and_potential_outcome_variance_bound}, \ref{assump:h_smooth},  and \ref{assump:rule_out_pathological_dgps_for_inference} hold, along with Assumptions \ref{assump:cohort_specific_factor_proj_mat_hadamard_diff} and \ref{assump:nuisance_param_est_hadamard_diff} stated in Appendix \ref{proof:thm:confidence_interval_validity}. Then the confidence intervals $\hat{\mathcal{C}}_j$ defined in \eqref{eq:simultaneous_confidence_interval} for the coordinates of $\theta$ have asymptotic simultaneous coverage at least $1-\alpha$:
    \begin{equation}
        \liminf_{N \rightarrow \infty} \Pr\left(\theta \in \bigtimes_{j = 1}^p \hat{\mathcal{C}}_j\right) \geq 1 - \alpha.
    \end{equation}
\end{theorem}
\noindent We provide a proof of Theorem \ref{thm:confidence_interval_validity} in Appendix \ref{proof:thm:confidence_interval_validity}.

We now return to Examples \ref{ex:dynamic_treatment_effects} and \ref{example-match-outcome-attribution} and verify that Assumptions \ref{assump:h_smooth} and \ref{assump:nuisance_param_est_hadamard_diff} (and therefore Assumption \ref{assump:nuisance_param_est_asymptotic_linear}) hold:

\begin{examplecont}{ex:dynamic_treatment_effects}
    From \eqref{eq:dynamic_treatment_effect_def}, we can see that the relative cohort sizes $p_c$ and cohort-specific observed outcome means $m_{ct}$ are not included in $\mu$ and thus form the components of $\eta$ we must also estimate. Of course, $p_c$ and $m_{ct}$ are identified and consistently estimable via simple averages of the observables $\ind{C_i = c}$, $\ind{t \geq C_i}Z_{it}$, and $\ind{t < C_i}Y_{it}$. Thus, by standard arguments, the components of $\eta$ expressed as a map from distributions to nuisance parameter values (see Appendix \ref{proof:thm:confidence_interval_validity} for details) must be Hadamard differentiable with respect to the distribution over which the expectations in their definitions are taken, implying Assumption \ref{assump:nuisance_param_est_hadamard_diff} and, by extension, the weaker Assumption \ref{assump:nuisance_param_est_asymptotic_linear} hold.
    Since $\theta_{\text{dyn},j}$ is linear in products of the components of $\mu$ and $\eta$, Assumption \ref{assump:h_smooth} immediately holds as well.
\end{examplecont}

\begin{examplecont}{example-match-outcome-attribution}
    To verify that our target parameters $\theta_{\text{col}, t_1, t_2}$ and $\theta_{\text{row}, t_1, t_2}$ satisfy the assumptions required for Theorem \ref{thm:confidence_interval_validity} to hold, we note that, as shown in Example \ref{example-match-outcome-attribution}, the components of the nuisance parameter vector $\eta$ simply consist of the shares of units in each cohort $p_c$. By standard arguments, these components of $\eta$ expressed as a map from distributions to nuisance parameter values (again, see Appendix \ref{proof:thm:confidence_interval_validity} for details) are Hadamard differentiable with respect to the distributions over which the means are taken, so Assumption \ref{assump:nuisance_param_est_hadamard_diff} and the weaker Assumption \ref{assump:nuisance_param_est_asymptotic_linear} hold. Further since $\theta_{\text{col}, t_1, t_2}$ and $\theta_{\text{row}, t_1, t_2}$ can be expressed as ratios of linear combinations of products of the components of $\mu$ and $\eta$, it is clearly differentiable so long as
    \begin{equation}
        \E[Y_{it_1} ~|~ t_1 \in \mathcal{T}_{C_i}] - \E[Y_{it_1} ~|~ t_2 \in \mathcal{T}_{C_i}] \neq 0.
    \end{equation}
    Thus, under this additional condition, Assumption \ref{assump:h_smooth} holds as well.
\end{examplecont}

%% file: sections/intermediate_proofs.tex
\subsection{Proof of Theorem \ref{thm:eigenspace_perturbation_expansion}}\label{proof:thm:eigenspace_perturbation_expansion}

Consider any $\hat{M} \in \mathcal{B}(M)$. By Weyl's inequality (see e.g. \citet[Theorem 4.5.3]{vershynin2018high}),
\begin{equation}\label{eq:eigval_diff_A_B_bound}
\max_j \abs{\lambda_j(M) - \lambda_j(\hat{M})} \leq \normnofit{M - \hat{M}}_\text{op} \leq \Delta(M).
\end{equation}
Let 
\begin{equation}
a(M) \coloneqq \begin{cases}
\frac{\lambda_{s+1}(M) + \lambda_s(M)}{2}, & s > 0 \\
\lambda_{1}(M) - 2\Delta(M), & s = 0
\end{cases}, \quad b(M) \coloneqq \begin{cases}
\frac{\lambda_{s+r+1}(M) + \lambda_{s+r}(M)}{2}, & s+r < d\\
\lambda_{d}(M) + 2\Delta(M), & s+r = d
\end{cases}.
\end{equation}
When $s > 0$, it must be that
\begin{equation}
    a(M) - \lambda_s(\hat{M}) = \underbrace{\frac{\lambda_{s+1}(M) - \lambda_s(M)}{2}}_{\geq 2\Delta(M)} + \underbrace{\lambda_s(M) - \lambda_s(\hat{M})}_{\geq -\Delta(M) \text{ by \eqref{eq:eigval_diff_A_B_bound}}} \geq \Delta(M),
\end{equation}
and
\begin{equation}
    \lambda_{s+1}(\hat{M}) - a(M) = \underbrace{\lambda_{s+1}(\hat{M})
 - \lambda_{s+1}(M)}_{\geq -\Delta(M) \text{ by \eqref{eq:eigval_diff_A_B_bound}}} + \underbrace{\frac{\lambda_{s+1}(M) - \lambda_s(M)}{2}}_{\geq 2\Delta(M)} \geq \Delta(M).
\end{equation}
When $s = 0$, $a(M) - \lambda_{s}(\hat{M}) = \infty\ge \Delta(M)$, and 
\begin{equation}
\lambda_{s+1}(\hat{M}) - a(M) = \underbrace{\lambda_1(\hat{M}) - \lambda_1(M)}_{\geq -\Delta(M) \text{ by \eqref{eq:eigval_diff_A_B_bound}}} + 2\Delta(M)\ge \Delta(M).
\end{equation}
Thus, in both cases, by the definition of $\Delta(M)$ in \eqref{cond:well_sep_eigvals},
\begin{equation}\label{eq:avg_eigvals_on_lower_gap_sep}
\begin{aligned}
    \min\Bigg\{&a(M) - \max\{\lambda_s(M), \lambda_s(\hat{M})\}, \min\{\lambda_{s+1}(M), \lambda_{s+1}(\hat{M})\} - a(M)\Bigg\} \geq \Delta(M) > 0.
\end{aligned}
\end{equation}
Using similar logic, we can also show that
\begin{equation}\label{eq:avg_eigvals_on_upper_gap_sep}
\begin{aligned}
    \min\Bigg\{& b(M) - \max\{\lambda_{s+r}(M), \lambda_{s+r}(\hat{M})\}, \\
    &\min\{\lambda_{s+r+1}(M), \lambda_{s+r+1}(\hat{M})\} - b(M)\Bigg\} \geq \Delta(M) > 0.
\end{aligned}
\end{equation}

Next, let $\mathcal{C}$ be some closed, bounded, positively oriented curve in the complex plane $\mathbb{C}$ that intersects the real line only at $(\lambda_{s+1}(M) + \lambda_s(M))/2$ and $(\lambda_{s+r}(M) + \lambda_{s+r+1}(M))/2$. By \eqref{eq:avg_eigvals_on_lower_gap_sep} and \eqref{eq:avg_eigvals_on_upper_gap_sep}, we have that the $(s+1)$th through $(s+r)$th eigenvalues of both $M$ and $\hat{M}$ are strictly inside $\mathcal{C}$, while all other eigenvalues are strictly outside $\mathcal{C}$. Problem I-5.9 in \citet{kato2013perturbation} (a result first shown in \citet{kato1949convergence}) then dictates that for $\tilde{M} \in \{M, \hat{M}\}$,
\begin{equation}
    \Pi(U_{(s+1):(s+r)}(\tilde{M})) = -\frac{1}{2\pi\sqrt{-1}} \oint_{\mathcal{C}} \left(\tilde{M} - \zeta I_d\right)^{-1} d\zeta,
\end{equation}
so
\begin{equation}\label{eq:eigspace_A_B_diff}
\begin{aligned}
    &\Pi(U_{(s+1):(s+r)}(\hat{M})) - \Pi(U_{(s+1):(s+r)}(M)) \\
    &= \frac{1}{2\pi\sqrt{-1}}\oint_{\mathcal{C}} \left[\left(M - \zeta I_d\right)^{-1} - \left(\hat{M} - \zeta I_d\right)^{-1}\right] d\zeta.
\end{aligned}
\end{equation}
Considering the integrand in \eqref{eq:eigspace_A_B_diff} in more detail, we can apply the following lemma:
\begin{lemma}\label{lemma:matrix_inverse_diff_lemma}
    For any two invertible matrices $M, \hat{M} \in \R^{d \times d}$, we have that
    \begin{equation}
        M^{-1} - \hat{M}^{-1} = M^{-1}(\hat{M} - M)M^{-1} - M^{-1}(\hat{M} - M)M^{-1}(\hat{M} - M)\hat{M}^{-1}.
    \end{equation}
\end{lemma}
\noindent Then
\begin{equation}\label{eq:integrand_expansion}
\begin{aligned}
    &\left(M - \zeta I_d\right)^{-1} - \left(\hat{M} - \zeta I_d\right)^{-1} \\
    &=\left(M - \zeta I_d\right)^{-1}(\hat{M} - M)\left(M - \zeta I_d\right)^{-1} \\
    &\phantom{=} - \left(M - \zeta I_d\right)^{-1}(\hat{M} - M)\left(M - \zeta I_d\right)^{-1}(\hat{M} - M)\left(\hat{M} - \zeta I_d\right)^{-1}.
\end{aligned}
\end{equation}
Returning to the expression \eqref{eq:eigspace_A_B_diff}, expanding the integrand via \eqref{eq:integrand_expansion}, rearranging terms, and taking the operator norm, we have that
\begin{equation}\label{eq:op_norm_first_order_expansion_contour_integral}
\begin{aligned}
    &\bigg\lVert\Pi(U_{(s+1):(s+r)}(\hat{M})) - \Pi(U_{(s+1):(s+r)}(M)) \\
    &\phantom{\big\lVert}- \frac{1}{2\pi\sqrt{-1}}\oint_{\mathcal{C}} \left(M - \zeta I_d\right)^{-1}(\hat{M} - M)\left(M - \zeta I_d\right)^{-1} d\zeta\bigg\rVert_\text{op} \\
    &= \norm{\frac{1}{2\pi\sqrt{-1}}\oint_{\mathcal{C}} \left(M - \zeta I_d\right)^{-1}(\hat{M} - M)\left(M - \zeta I_d\right)^{-1}(\hat{M} - M)\left(\hat{M} - \zeta I_d\right)^{-1} d\zeta}_\text{op}. 
\end{aligned}
\end{equation}
Take $\mathcal{C}$ to be the boundary of a positively oriented rectangular contour on the complex plane with the following corners for some $\upsilon > 0$:
\begin{equation}
    a(M) \pm \upsilon \sqrt{-1}, \quad b(M) \pm \upsilon\sqrt{-1}.
\end{equation}
Then we can bound the operator norm of the integral in \eqref{eq:op_norm_first_order_expansion_contour_integral} along each side of $\mathcal{C}$ separately.

To do so, for any $z \in \mathbb{C}$, we define the following decomposition of $z$ into its real and imaginary parts: $z = \mathrm{re}(z) + \mathrm{im}(z)\sqrt{-1}$. We then state the following convenient lemma. 

\begin{lemma}\label{lemma:resolvent_op_norm_expr}
    For any real, symmetric matrix $B$,
    \begin{equation}
    \norm{\left(B - \zeta I_d\right)^{-1}}_\text{op} = \max_j \left(\left(\lambda_j(B) - \mathrm{re}(\zeta)\right)^2 + \mathrm{im}(\zeta)^2\right)^{-1/2}.
    \end{equation}
\end{lemma}
\begin{proof}
Since $B$ is symmetric and real, $\lambda_j(B)$ must be real, in which case, letting $z^*$ denote the complex conjugate of $z \in \mathbb{C}$,
\begin{align}
    & \norm{\left(B - \zeta I_d\right)^{-1}}_\text{op}\\
    &= \max_j \left((\lambda_j(B) - \zeta)^*(\lambda_j(B) - \zeta)\right)^{-1/2} \\
    &= \max_j \left(\left(\lambda_j(B) - \mathrm{re}(\zeta) - \mathrm{im}(\zeta)\sqrt{-1})\right)\left(\lambda_j(B) - \mathrm{re}(\zeta) + \mathrm{im}(\zeta)\sqrt{-1}\right)\right)^{-1/2} \\
    &= \max_j \left(\left(\lambda_j(B) - \mathrm{re}(\zeta)\right)^2 - \mathrm{im}(\zeta)^2 \cdot (-1)\right)^{-1/2}.
\end{align}
\end{proof}
\noindent For notational convenience, we write $a$ and $b$ for $a(M)$ and $b(M)$, respectively. On the horizontal segment $\{x + v\sqrt{-1}: x\in [a, b]\}$,
\begin{align}
& \bigg\|\frac{1}{2\pi\sqrt{-1}}\int_{a}^{b} \left(M - (x + v\sqrt{-1}) I_d\right)^{-1}(\hat{M} - M)\left(M - (x + v\sqrt{-1}) I_d\right)^{-1}\\
& \phantom{\bigg\|\frac{1}{2\pi\sqrt{-1}}\int_{a}^{b}} \cdot (\hat{M} - M)\left(\hat{M} - (x + v\sqrt{-1}) I_d\right)^{-1} dx\bigg\|_\text{op} \\
& \le \frac{\normnofit{\hat{M} - M}_\text{op}^2}{2\pi}\cdot \int_{a}^{b}\norm{\left(M - (x + v\sqrt{-1}) I_d\right)^{-1}}_{\text{op}}^2 \norm{\left(\hat{M} - (x + v\sqrt{-1}) I_d\right)^{-1}}_{\text{op}} dx \\
& \le \frac{\normnofit{\hat{M} - M}_\text{op}^2}{2\pi}\cdot \int_{a}^{b}\underbrace{\left((\lambda_j(M) - x)^2 + v^2\right)^{-1}}_{\leq v^{-2}} \underbrace{\left((\lambda_j(\hat{M}) - x)^2 + v^2\right)^{-1/2}}_{\leq v^{-1}} dx \,\, \text{(by Lemma \ref{lemma:resolvent_op_norm_expr})} \\
& \le \frac{\normnofit{\hat{M} - M}_\text{op}^2}{2\pi}\cdot \frac{b - a}{v^3}. \label{eq:horizontal1}
\end{align}

Similarly, on the horizontal segment $\{x - v\sqrt{-1}: x\in [a, b]\}$,
\begin{align}
& \bigg\|\frac{1}{2\pi\sqrt{-1}}\int_{b}^{a} \left(M - (x - v\sqrt{-1}) I_d\right)^{-1}(\hat{M} - M)\left(M - (x - v\sqrt{-1}) I_d\right)^{-1}\\
& \phantom{\bigg\|\frac{1}{2\pi\sqrt{-1}}\int_{b}^{a}}  \cdot (\hat{M} - M)\left(\hat{M} - (x - v\sqrt{-1}) I_d\right)^{-1} dx\bigg\|_\text{op} \\
& \le \frac{\normnofit{\hat{M} - M}_\text{op}^2}{2\pi}\cdot \frac{b - a}{v^3}. \label{eq:horizontal2}
\end{align}
Now we turn to the vertical segment $\{a + y\sqrt{-1}: y\in [-v, v]\}$.
By \eqref{eq:avg_eigvals_on_lower_gap_sep} and \eqref{eq:avg_eigvals_on_upper_gap_sep},
\begin{equation}
\begin{aligned}
    \abs{\lambda_j(M) - a} &\geq \min\Big\{\abs{\lambda_s(M) - a}, \abs{\lambda_{s + 1}(M) - a}\Big\}\geq \Delta(M),
\end{aligned}
\end{equation}
so by Lemma \ref{lemma:resolvent_op_norm_expr},
\begin{equation}
    \norm{\left(M - (a + y\sqrt{-1})) I_d\right)^{-1}}_\text{op} = \max_j \left((\lambda_j(M) - a)^2 + y^2\right)^{-1/2} \leq \left(\Delta(M)^2 + y^2\right)^{-1/2}.
\end{equation}
Since \eqref{eq:avg_eigvals_on_lower_gap_sep} and \eqref{eq:avg_eigvals_on_upper_gap_sep} also imply the bound $\abs{\lambda_j(\hat{M}) - a} \geq \Delta(M)$, Lemma \ref{lemma:resolvent_op_norm_expr} also implies
\begin{equation}
    \norm{\left(\hat{M} - (a + y\sqrt{-1})) I_d\right)^{-1}}_\text{op} = \max_j \left((\lambda_j(\hat{M}) - a)^2 + y^2\right)^{-1/2} \leq \left(\Delta(M)^2 + y^2\right)^{-1/2}.
\end{equation}
Then, we can upper bound the operator norm of the integral on the vertical segment as follows:
\begin{align}
    & \bigg\|\frac{1}{2\pi\sqrt{-1}}\int_{-v}^{v} \left(M - (a + y\sqrt{-1}) I_d\right)^{-1}(\hat{M} - M)\left(M - (a + y\sqrt{-1}) I_d\right)^{-1}\\
& \phantom{\bigg\|\frac{1}{2\pi\sqrt{-1}}\int_{-v}^{v}}  \cdot (\hat{M} - M)\left(\hat{M} - (a + y\sqrt{-1}) I_d\right)^{-1} dy\bigg\|_\text{op} \\
    &\leq \frac{\normnofit{\hat{M} - M}_\text{op}^2}{2\pi}\int_{-v}^{v} \left(\Delta(M)^2 + y^2\right)^{-3/2} dy \\
    & = \frac{\normnofit{\hat{M} - M}_\text{op}^2}{2\pi}\frac{1}{\Delta(M)^2}\int_{-\infty}^{\infty} \left(1 + u^2\right)^{-3/2} du\\
    & =  \frac{\normnofit{\hat{M} - M}_\text{op}^2}{\pi}\cdot\frac{1}{\Delta(M)^2},\label{eq:vertical1}
\end{align}
where the equality \eqref{eq:vertical1} applies the following identity:
\[\int_{-\infty}^{\infty}\left(1 + u^2\right)^{-3/2} du = \int_{-\pi/2}^{\pi/2}\left(1 + \tan^2\theta\right)^{-3/2} d\tan\theta = \int_{-\pi/2}^{\pi/2}\cos\theta d\theta 
= 2.\]
Similarly, on the vertical segment $\{b + y\sqrt{-1}: y\in [-v, v]\}$, 
\begin{align}
    & \bigg\|\frac{1}{2\pi\sqrt{-1}}\int_{-v}^{v} \left(M - (b + y\sqrt{-1}) I_d\right)^{-1}(\hat{M} - M)\left(M - (b + y\sqrt{-1}) I_d\right)^{-1}\\
& \phantom{\bigg\|\frac{1}{2\pi\sqrt{-1}}\int_{-v}^{v}}  \cdot (\hat{M} - M)\left(\hat{M} - (b + y\sqrt{-1}) I_d\right)^{-1} dy\bigg\|_\text{op} \\
    & \le  \frac{\normnofit{\hat{M} - M}_\text{op}^2}{\pi}\cdot \frac{1}{\Delta(M)^2}. \label{eq:vertical2}
\end{align}
Combining \eqref{eq:horizontal1}, \eqref{eq:horizontal2}, \eqref{eq:vertical1}, and \eqref{eq:vertical2}, we obtain that 
\begin{align}
&\norm{\frac{1}{2\pi\sqrt{-1}}\oint_{\mathcal{C}} \left(M - \zeta I_d\right)^{-1}(\hat{M} - M)\left(M - \zeta I_d\right)^{-1}(\hat{M} - M)\left(\hat{M} - \zeta I_d\right)^{-1} d\zeta}_\text{op} \\
& \le \frac{\normnofit{\hat{M} - M}_\text{op}^2}{\pi}\left\{\frac{b-a}{v^3} + \frac{2}{\Delta(M)^2}\right\}.
\end{align}
Taking $v \rightarrow \infty$, we can combine the bound in the above display with \eqref{eq:op_norm_first_order_expansion_contour_integral} to obtain the following bound:
\begin{equation}\label{eq:approx_2nd_order_expansion}
\begin{aligned}
    &\bigg\lVert\Pi(U_{(s+1):(s+r)}(\hat{M})) - \Pi(U_{(s+1):(s+r)}(M)) \\
    &\phantom{\big\lVert}- \frac{1}{2\pi\sqrt{-1}}\oint_{\mathcal{C}} \left(M - \zeta I_d\right)^{-1}(\hat{M} - M)\left(M - \zeta I_d\right)^{-1} d\zeta\bigg\rVert_\text{op} \\
    &\leq \frac{2\normnofit{\hat{M} - M}_\text{op}^2}{\pi\Delta(M)^2}.
\end{aligned}
\end{equation}
Next, we consider the first-order error term inside the operator norm in \eqref{eq:approx_2nd_order_expansion}:
\begin{align}
    &\oint_{\mathcal{C}} \left(M - \zeta I_d\right)^{-1}(\hat{M} - M)\left(M - \zeta I_d\right)^{-1} d\zeta \\
    & = \oint_{\mathcal{C}} U(M)(\Lambda(M) - \zeta I_d)^{-1}U(M)'(\hat{M} - M)U(M)(\Lambda(M) - \zeta I_d)^{-1}U(M)'d\zeta \\
    & = \oint_{\mathcal{C}} \left[\sum_{j = 1}^d (\lambda_j(M) - \zeta)^{-1} \Pi(u_j(M))\right](\hat{M} - M)\left[\sum_{j = 1}^d (\lambda_j(M) - \zeta)^{-1} \Pi(u_j(M))\right]d\zeta \\
    & = \sum_{j = 1}^d \sum_{k = 1}^d \Pi(u_j(M))(\hat{M}-M)\Pi(u_k(M))\oint_{\mathcal{C}} (\lambda_j(M) - \zeta)^{-1}(\lambda_k(M) - \zeta)^{-1}d\zeta \label{eq:final_expansion_summand_complex_integral}
\end{align}
When $j = k$, by Cauchy's integral formula,
\begin{equation}
    \oint_{\mathcal{C}} (\lambda_j(M) - \zeta)^{-1}(\lambda_k(M) - \zeta)^{-1}d\zeta = \oint_{\mathcal{C}} \frac{1}{(\lambda_j(M) - \zeta)^2}d\zeta = 0,
\end{equation}
and when $j \neq k$, we have that
\begin{align}
    &\oint_{\mathcal{C}} (\lambda_j(M) - \zeta)^{-1}(\lambda_k(M) - \zeta)^{-1}d\zeta \\
    &= (\lambda_j(M) - \lambda_k(M))^{-1}\\
    &\phantom{=} \cdot\left(\oint_{\mathcal{C}} (\lambda_k(M) - \zeta)^{-1} d\zeta - \oint_{\mathcal{C}} (\lambda_j(M) - \zeta)^{-1} d\zeta\right) && \text{(Partial Fraction Decomposition)} \\
    &= 2\pi\sqrt{-1}(\lambda_j(M) - \lambda_k(M))^{-1}\\
    &\phantom{=} \cdot \left(\ind{k \in [s + 1, s + r]} - \ind{j \in [s + 1, s + r]}\right). && \text{(Residue Theorem)}
\end{align}
Thus, we can rewrite \eqref{eq:final_expansion_summand_complex_integral} as follows:
\begin{align}
    &\frac{1}{2\pi\sqrt{-1}}\oint_{\mathcal{C}} \left(M - \zeta I_d\right)^{-1}(\hat{M} - M)\left(M - \zeta I_d\right)^{-1} d\zeta \\
    &= \sum_{j = 1}^d \sum_{k = 1}^d \frac{\ind{k \in [s + 1, s + r]} - \ind{j \in [s + 1, s + r]}}{\lambda_j(M) - \lambda_k(M)}\Pi(u_j(M))(\hat{M}-M)\Pi(u_k(M)) \\
    &= \sum_{j \not\in [s + 1, s + r]} \sum_{k = s + 1}^{s + r} \frac{1}{\lambda_j(M) - \lambda_k(M)}\Pi(u_j(M))(\hat{M}-M)\Pi(u_k(M)) \\
    &\phantom{=} + \sum_{j = s + 1}^{s + r}\sum_{k \not\in [s + 1, s + r]} \frac{1}{\lambda_k(M) - \lambda_j(M)}\Pi(u_j(M))(\hat{M}-M)\Pi(u_k(M)) \\
    &= \sum_{j = s + 1}^{s + r}\sum_{k \not\in [s + 1, s + r]} \frac{1}{\lambda_k(M) - \lambda_j(M)} \bigg[\Pi(u_j(M))(\hat{M}-M)\Pi(u_k(M)) \\
    &\phantom{=\sum_{j = s + 1}^{s + r}\sum_{k \not\in [s + 1, s + r]} \frac{1}{\lambda_k(M) - \lambda_j(M)}\bigg[}+ \Pi(u_k(M))(\hat{M}-M)\Pi(u_j(M))\bigg].
\end{align}
Putting together \eqref{eq:approx_2nd_order_expansion} and the final expression in the display above yields \eqref{eq:eigenvec_first_order_approx}.

\newcommand{\Es}{E_c}
\newcommand{\Pis}{\Pi(\tilde{\Gamma})}
\newcommand{\Gams}{\tilde{\Gamma}}
\newcommand{\Gs}{G_c}
\newcommand{\DG}{\hat{\Delta}_G}
\newcommand{\DGam}{\hat{\Delta}_\Gamma}
\newcommand{\normop}[1]{\|#1\|_{\mathrm{op}}}
\newcommand{\Ls}{\ell_c}
\newcommand{\Ks}{K_c}
\newcommand{\Rs}{R_c}

\subsection{Proof of Lemma \ref{lemma:chebyshev_tightness}}\label{proof:lemma:chebyshev_tightness}

\noindent We begin by stating and proving the following helpful lemma:
\begin{lemma}\label{lemma:central_to_noncentral_sq_norm_bound}
    For any random element $V$ in a normed vector space equipped with norm $\norm{\cdot}$, we have that $\E\left[\norm{V - \E[V]}^2\right] \leq 4\E\left[\norm{V}^2\right].$
\end{lemma}
\begin{proof}
\begin{align}
    \E\left[\norm{V - \E[V]}^2\right] &\leq \E\left[(\norm{V} + \norm{\E[V]})^2\right] && \text{(Triangle Ineq.)} \\
    &= \E\left[\norm{V}^2\right] + 2\E\left[\norm{V}\right]\E\left[\norm{\E[V]}\right] + \E\left[\norm{\E[V]}^2\right] \\
    &\leq \E\left[\norm{V}^2\right] + 2\E\left[\norm{V}\right]\E\left[\norm{\E[V]}\right] + \E\left[\norm{\E[V]}^2\right] && \text{(Jensen's Ineq.)} \\
    &= 4\E\left[\norm{V}^2\right],
\end{align}
as required.
\end{proof}

\noindent Now, let $W_i \coloneqq V_i - \E[V_i]$. By Chebyshev's Inequality, we have that, for any $u > 0$,
\begin{align}
    &\Pr\left(N^{1/2}\kappa^{-1}\sigma^{-1}\normnofit{\hat{\E}_N[W_i]} > u\right) \\
    &\leq u^{-2} \cdot \E\left[N\kappa^{-2}\sigma^{-1}\norm{\hat{\E}_N[W_i]}^2\right] && \text{(Chebyshev's Ineq.)} \\
    &\leq u^{-2} \cdot N \kappa^{-2} \sigma^{-2} \cdot \E\left[C_2^2\norm{\hat{\E}_N[W_i]}_\text{H}^2\right] && \text{(By \eqref{cond:inner_prod_norm_equiv}, $\norm{\cdot} \leq C_2\norm{\cdot}_\text{H}$)} \\
    &= u^{-2} \cdot N C_1^2\sigma^{-2} \cdot \E\left[\inner{\hat{\E}_N[W_i], \hat{\E}_N[W_i]}\right] \\
    &= u^{-2} \cdot N^{-1} C_1^2\sigma^{-2} \cdot \sum_{i = 1}^N \sum_{j = 1}^N\E[\inner{W_i, W_j}] \\
    &= u^{-2} \cdot N^{-1} C_1^2\sigma^{-2} \cdot \sum_{i = 1}^N \E\left[\norm{W_i}_\text{H}^2\right] && \text{($W_i \indep W_j$, $\E[W_i] = \E[W_j] = \zeros$)} \\
    &= u^{-2} \cdot C_1^2\sigma^{-2} \cdot C_1^{-2}\E\left[\norm{W_i}^2\right] && \text{(By \eqref{cond:inner_prod_norm_equiv}, $\norm{\cdot}_\text{H} \leq C_1^{-1}\norm{\cdot}$)} \\
    &\leq u^{-2} \cdot \sigma^2 \cdot \underbrace{4\E\left[\norm{V_i}^2\right]}_{= 4\sigma^2} && \text{(By Lemma \ref{lemma:central_to_noncentral_sq_norm_bound})} \\
    &= 4u^{-2}
\end{align}
which satisfies the definition of $\bigop{\cdot}$, as required.

\subsection{Proof of Lemma \ref{lem:A0_invertibility_equivalent}}

Let $\Theta\in \R^{T\times (T-r)}$ be an orthogonal matrix such that 
\begin{equation}\label{eq:Theta_Gamma}
I - \Pi(\Gamma) = \Theta\Theta'.
\end{equation}
The choice of $\Theta$ is not unique and we take any representative (e.g., the one with the least sum of absolute values of all entries). Further, let 
\begin{equation}\label{eq:tilde_Theta}
\bar{\Theta} = \begin{bmatrix}
\Theta & \mathbf{0}_{T\times q}\\
\mathbf{0}_{q\times (T-r+q)} & I_q
\end{bmatrix}\in \R^{(T+q)\times (T-r+q)}, \quad U = \begin{bmatrix}
\bar{\Gamma} & \bar{\Theta}
\end{bmatrix}\in \R^{(T+q)\times (T+q)}.
\end{equation}
Since $[\Gamma\ \ \Theta]$ is an $T$-by-$T$ orthogonal matrix, $U$ is a $(T+q)$-by-$(T+q)$ orthogonal matrix.
Then 
\[[I - \Pi(\Gamma)\ \ X_i] = [\Theta \ \ X_i]\bar{\Theta}',\]
and
\[A_0 = \bar{\Theta} \E\Big[[\Theta \ \ X_i]' \{E_{C_i} - \Pi(E_{C_i}\Gamma)\}[\Theta \ \ X_i]\Big]\bar{\Theta}'\coloneqq \bar{\Theta}\bar{A}_0 \bar{\Theta}.\]
As a result, we can re-express $A_{0,\omega}$ as 
\begin{equation}\label{eq:A0omegaU}
A_{0,\omega} = \omega \bar{\Gamma}\bar{\Gamma}' + \bar{\Theta}\bar{A}_0 \bar{\Theta} = U\begin{bmatrix}
\omega I_{r} & \mathbf{0}_{r\times (T-r+q)}\\
\mathbf{0}_{r\times (T-r+q)} & \bar{A}_0
\end{bmatrix}U'.
\end{equation}
Then $A_{0,\omega}$ is invertible iff $\bar{A}_0$ is invertible. 

Note that
\begin{equation}\label{eq:A0U}
A_{0} = A_{0,0} =  U\begin{bmatrix}
\mathbf{0}_{r\times r} & \mathbf{0}_{r\times (T-r+q)}\\
\mathbf{0}_{r\times (T-r+q)} & \bar{A}_0
\end{bmatrix}U'.
\end{equation}
We have 
\[\mathrm{rank}(\bar{A}_0) = \mathrm{rank}(A_0).\]
Thus, Assumption \ref{assump:A0_invertibility} holds iff 
\[\mathrm{rank}(A_0) = T-r+q\Longleftrightarrow \bar{A}_0 \text{ is invertible}.\]
The proof of the equivalence is then completed.

Now we prove \eqref{eq:H_A0+}. By \eqref{eq:A0omegaU} and \eqref{eq:A0U}
\[A_{0,\omega}^{-1} = U\begin{bmatrix}
\omega^{-1} I_{r} & \mathbf{0}_{r\times (T-r+q)}\\
\mathbf{0}_{r\times (T-r+q)} & \bar{A}_0^{-1}
\end{bmatrix}U',\quad A_0^{+} = U\begin{bmatrix}
\mathbf{0}_{r\times r} & \mathbf{0}_{r\times (T-r+q)}\\
\mathbf{0}_{r\times (T-r+q)} & \bar{A}_0^{-1}
\end{bmatrix}U'.\]
By \eqref{eq:bar_Gamma}, 
\[[I - \Pi(\Gamma)\ \ \Lambda]\bar{\Gamma} = 0\]
By definition of $U$ in \eqref{eq:tilde_Theta}, we have 
\[[I - \Pi(\Gamma)\ \ \Lambda]U = [\mathbf{0}_{(T+q)\times r}\ \ [I - \Pi(\Gamma)\ \ \Lambda]\bar{\Theta}].\]
Therefore, 
\[[I - \Pi(\Gamma)\ \ \Lambda]A_{0,\omega}^{-1} = [I - \Pi(\Gamma)\ \ \Lambda]A_{0}^{+} =[\mathbf{0}_{(T+q)\times r}\ \ [I - \Pi(\Gamma)\ \ \Lambda]\bar{\Theta}] \begin{bmatrix}
\mathbf{0}_{r\times r} & \mathbf{0}_{r\times (T-r+q)}\\
\mathbf{0}_{r\times (T-r+q)} & \bar{A}_0^{-1}
\end{bmatrix}U'. \]

\subsection{Proof of Proposition \ref{prop:A0_invertibility_no_covariate}}
We have shown that $\Gamma$ is in the null space of $A$. Thus, $\Gamma$'s column space is exactly the null space of $A$ iff $\mathrm{rank}(A) = T - r$. Similarly, $A_0 \Gamma = \Gamma' A_0 = 0$, $\mathrm{rank}(A_0)\le T - r$. Thus, Assumption \ref{assump:A0_invertibility} is equivalent to $\mathrm{rank}(A_0) = T-r$. As a consequence, we are left to prove 
\[\mathrm{rank}(A_0) = \mathrm{rank}(A).\]
In the absence of $X_i$, 
\[A_0 = (I - \Pi(\Gamma))\E[E_{C_i} - \Pi(E_{C_i}\Gamma)](I - \Pi(\Gamma)) = (I - \Pi(\Gamma))\left\{\sum_{c=1}^C p_c (E_c - \Pi(E_c\Gamma))\right\}(I - \Pi(\Gamma)).\]
Under the positive semidefinite order, 
\[\left(\min_{c\in [C]}p_c\right) A\preceq \sum_{c=1}^C p_c (E_c - \Pi(E_c\Gamma))\preceq A.\]
Since $A\Gamma = \Gamma' A = 0$,
\[(I - \Pi(\Gamma))A (I - \Pi(\Gamma)) = A.\]
Thus, 
\[\left(\min_{c\in [C]}p_c\right) A \preceq A_0 \preceq  A.\]
By Assumption \ref{assump:cohort_sizes_and_potential_outcome_variance_bound}, $\min_{c\in [C]}p_c > 0$. Thus, 
\[\mathrm{rank}(A_0) = \mathrm{rank}(A).\]

\subsection{Proof of Lemma \ref{lem:alpha_uniqueness}}
Under model \eqref{eq:factor_model} with the weaker assumption $\E[\epsilon_{it}\mid C_i, X_{it}] = 0$, 
\[\E[Y_i^*\mid C_i, X_i] = \Gamma \E[\lambda_i \mid C_i, X_i] + (I - \Pi(\Gamma))\tilde{\gamma}_0 + X_i\beta.\]
If there exist $(\tilde{\lambda}_i, \tilde{\gamma}_0, \tilde{\beta})$ with $\Gamma'\tilde{\gamma}_0 = 0$ that yield the same model \eqref{eq:factor_estimate}, then 
\[\gamma_0 = (I - \Pi(\Gamma))\gamma_0,\quad \tilde{\gamma}_0 = (I - \Pi(\Gamma))\tilde{\gamma}_0,\] 
and 
\[\Gamma \E[\tilde{\lambda}_i\mid C_i, X_i] + (I - \Pi(\Gamma))\tilde{\gamma}_0 + X_i\tilde{\beta}  = \Gamma \E[\lambda_i\mid C_i, X_i] + (I - \Pi(\Gamma))\gamma_0 + X_i\beta \quad \text{a.s.}.\]
Write $\tilde{\alpha}$ for $(\tilde{\gamma}_0', \tilde{\beta}')'$. Multiplying both sides by $I - \Pi(\Gamma)$, we have 
\[[I - \Pi(\Gamma)\ \ X_i](\tilde{\alpha} - \alpha) = 0\Longrightarrow [I - \Pi(\Gamma)\ \ X_i]' \{E_{C_i} - \Pi(E_{C_i}\Gamma)\} [I - \Pi(\Gamma)\ \ X_i](\tilde{\alpha} - \alpha) = 0\]
Taking expectation over $(C_i, X_i)$, 
\[A_0 (\tilde{\alpha} - \alpha) = 0.\]
Recall the definition of $\bar{\Gamma}$ in \eqref{eq:bar_Gamma}. Since $\Gamma'\gamma_0 = \Gamma'\tilde{\gamma}_0 = 0$, 
\[\bar{\Gamma}'(\tilde{\alpha} - \alpha) = 0.\]
Then for any $\omega > 0$, 
\[A_{0,\omega}(\tilde{\alpha} - \alpha) = 0,\]
By Lemma \ref{lem:A0_invertibility_equivalent}, $A_{0,\omega}$ is invertible and hence $\tilde{\alpha} = \alpha$.

\subsection{Proof of Lemma \ref{lem:muc}}
The model \eqref{eq:factor_model} can be written in the vector form 
\[Y_i^{*} = \Gamma\lambda_i + \gamma_0 + X_i\beta + \epsilon_i, \quad \E[\epsilon_i \mid C_i] = 0.\]
Taking expectation of both sides conditional on $C_i = c$, we obtain that 
\[\mu_c = \Gamma \E[\lambda_i \mid C_i = c] + \gamma_0 + [I\ \ \mu_{X,c}]\alpha.\]
Multiplying both sides by $I - P_c(\Pi(\Gamma))$ yields 
\begin{align*}
& \{I - P_c(\Pi(\Gamma))\}\mu_c  = \{I - P_c(\Pi(\Gamma))\}\Gamma \E[\lambda_i \mid C_i = c] + \{I - P_c(\Pi(\Gamma))\}[I\ \ \mu_{X,c}]\alpha.
\end{align*}
By Lemma \ref{lem:Pk_equivalent}, \[\{I - P_c(\Pi(\Gamma))\}\Gamma  = \Gamma - \Gamma\left(\Gamma'E_c\Gamma\right)^{-1}\Gamma'E_c\Gamma = 0.\]
This also implies 
\[\{I - P_c(\Pi(\Gamma))\}(I - \Pi(\Gamma)) = I - P_c(\Pi(\Gamma)).\]
Thus, 
\begin{align*}
& \{I - P_c(\Pi(\Gamma))\}\mu_c  = \{I - P_c(\Pi(\Gamma))\}[I - \Pi(\Gamma)\ \ \mu_{X,c}]\alpha
\end{align*}
Rearranging the terms completes the proof.

\subsection{Proof of Lemma \ref{lem:equivalent_constraint}}
Consider any minimizer $(\bar{\gamma}_0, \bar{\beta}, \{\bar{\lambda}_i\}_{i=1}^N)$ of the least square objective in \eqref{eq:regression_imputation_estimator_def} without the constraint $\hat{\Gamma}'\check{\gamma}_0 = 0$. Then we can map it to the constrained solution $(\hat{\gamma}_0, \hat{\beta}, \{\hat{\lambda}_i\}_{i=1}^N)$ by 
\[\hat{\lambda}_i = \bar{\lambda}_i + (\hat{\Gamma}'\hat{\Gamma})^{-1}\hat{\Gamma}'\bar{\gamma}_0, \quad \hat{\gamma}_0 = (I - \Pi(\hat{\Gamma})) \bar{\gamma}_0.\]
Plugging this into \eqref{eq:estimator_def}, they yield the same $\hat{\mu}_c$. 

Next, we prove $\Gamma'\hat{\Gamma}$ is invertible with probability $1-o(1)$. Since $\Gamma'\Gamma = I$
\[\Gamma'\hat{\Gamma}\hat{\Gamma}'\Gamma - I = \Gamma'\left( \hat{\Gamma}\hat{\Gamma}' - \Gamma \Gamma'\right)\Gamma = \Gamma'\left( \Pi( \hat{\Gamma}) - \Pi\left( \Gamma \right)\right)\Gamma.\]
By Theorem \ref{thm:factor_consistency_asymptotic_normality}, 
\[\Pi(\hat{\Gamma}) - \Pi(\Gamma) = \mathcal{O}_\mathbb{P}(1/\sqrt{N}) = o_\mathbb{P}(1).\] 
Thus, 
\[\Gamma'\hat{\Gamma}\hat{\Gamma}'\Gamma - I = o_\mathbb{P}(1).\]
Thus, with probability $1 - o(1)$, 
\[\lambda_{\min}\left(\Gamma'\hat{\Gamma}\hat{\Gamma}'\Gamma\right) \ge 1 - o(1)\Longrightarrow \Gamma'\hat{\Gamma}\text{ is invertible}.\]

Lastly, we prove $\hat{\mu}_c = \tilde{\mu}_c$ conditional on $\Gamma'\hat{\Gamma}$ being invertible. Let 
\[\kappa = (\Gamma'\hat{\Gamma})^{-1}\Gamma'\hat{\gamma}_0, \quad \bar{\gamma}_0 = \hat{\gamma}_0 - \hat{\Gamma}\kappa, \quad \bar{\lambda}_i = \lambda_i + \kappa.\]
Since $(\bar{\gamma}_0, \hat{\beta}_0, \{\bar{\lambda}_i\}_{i=1}^N)$ is a linear transformation of $(\hat{\gamma}_0, \hat{\beta}_0, \{\hat{\lambda}_i\}_{i=1}^N)$ and it gives the same prediction of $Y_{it}$, it is also a minimizer of the least squares objective. Moreover, by construction, 
\[\Gamma'\bar{\gamma}_0 = \Gamma'(\hat{\gamma}_0 - \hat{\Gamma}\kappa) = 0.\]
Therefore, 
\[(\bar{\gamma}_0, \hat{\beta}_0, \{\bar{\lambda}_i\}_{i=1}^N) \in  \argmin_{\check{\gamma}_0 \in \R^T,\; \check{\beta} \in \R^q,\; \check{\lambda}_i \in \R^r} \left\{\frac{1}{N}\sum_{i = 1}^N \sum_{t \in \mathcal{T}_{C_i}}\left(Y_{it} - \left[\hat{\gamma}_t'\check{\lambda}_i + \check{\gamma}_{t0} + X_{it}'\check{\beta}\right]\right)^2 ~\setst~ \Gamma'\check{\gamma}_0 = \zeros_r \right\}.\]
Similar to the first step, we conclude that $\hat{\mu}_c$ is invariant to the choice of the  solution under the constraint $\Gamma'\check{\gamma}_0 = 0$. Moreover, since $(\bar{\gamma}_0, \hat{\beta}_0, \{\bar{\lambda}_i\}_{i=1}^N)$ minimizes the least squares objective globally, it also minimizes the penalized objective
\begin{equation}\label{eq:modified_estimator}
\frac{1}{N}\sum_{i = 1}^N \sum_{t \in \mathcal{T}_{C_i}}\left(Y_{it} - \left[\hat{\gamma}_t'\check{\lambda}_i + \check{\gamma}_{t0} + X_{it}'\check{\beta}\right]\right)^2 + \lambda\|\Gamma'\check{\gamma}_0\|_2^2.
\end{equation}

\subsection{Proof of Lemma \ref{lem:Gc_operator_norm}}
Recall that for any invertible $M$, the Sherman-Morrison-Woodbury matrix identity states that
\begin{equation}
(M - C'C)^{-1} = M^{-1} + M^{-1}C'(I - CM^{-1}C')^{-1}CM^{-1}, \text{ if }I - CM^{-1}C'\text{ is invertible}.
\end{equation}
Taking $M = I$ and $C = \Gamma(I - E_c)$, 
\[I - \Gamma'(I - E_c)(I- E_c)\Gamma = I - \Gamma'(I - E_c)\Gamma = \Gamma'E_c \Gamma = (E_c \Gamma)'(E_c \Gamma)\]
which is invertible. Thus, $I - \Gamma'(I - E_c)(I- E_c)\Gamma$ is invertible and hence 
\[G_c(\Pi(\Gamma)) = \left\{I - (I-E_c)\Pi(\Gamma)(I - E_c)\right\}^{-1}.\]
Then 
\begin{align*}
&G_c(\Pi(\Gamma))\\
& = \left\{I - (I-E_c)\Gamma \Gamma'(I - E_c)\right\}^{-1}\\
& = I + (I - E_c)\Gamma(I - \Gamma'(I - E_c)(I- E_c)\Gamma)^{-1}\Gamma'(I - E_c)\\
& = I + (I - E_c)\Gamma(\Gamma'E_c\Gamma)^{-1}\Gamma'(I - E_c).
\end{align*}
Let 
\[m = \min_{c\in [C]}\lambda_{\min}(\Gamma'E_c\Gamma).\]
In the setting of Theorem \ref{thm:factor_identification}, $m > 0$. Then 
\[\|G_c(\Pi(\Gamma)\|_{\mathrm{op}}\le 1 + (1/m)\|(I - E_c)\Gamma\|_{\mathrm{op}}^2 \le 1 + 1/m =: M.\]

\subsection{Proof of Lemma \ref{lem:barXY}}\label{proof:lemma:inf_fn_group_avg}

We prove the following more general result than Lemma \ref{lem:barXY}:
\begin{lemma}\label{lemma:inf_fn_group_avg}
    Let $\{V_i\}_{i = 1}^\infty$ be a sequence of random elements taking values in a measurable, normed vector space with norm $\norm{\cdot}$ that is equivalent to a norm induced by an inner product as in \eqref{cond:inner_prod_norm_equiv}. In addition, let $\{D_i\}_{i = 1}^\infty$ be i.i.d. Bernoulli random variables for whom $p \coloneqq \Pr(D_i = 1) > 0$; for each $i$, $D_i$ and $V_i$ can be arbitrarily dependent. Suppose also that $\sigma^2 \coloneqq \E[\normnofit{V_i}^2 ~|~ D_i = 1] < \infty$. Then, letting $\kappa \coloneqq C_1^{-1} \cdot C_2$ where $C_1$ and $C_2$ are defined in \eqref{cond:inner_prod_norm_equiv},
    \begin{equation}\label{eq:lemma:inf_fn_group_avg_sqrt_n_consistency}
        \norm{\hat{\E}_N[V_i ~|~ D_i = 1] - \E[V_i ~|~ D_i = 1]} = \bigop{N^{-1/2}p^{-1/2}\kappa\sigma},
    \end{equation}
    and
    \begin{equation}\label{eq:lemma:inf_fn_group_avg_inf_fn_bound}
    \begin{aligned}
        &\norm{\left(\hat{\E}_N[V_i ~|~ D_i = 1] - \E[V_i ~|~ D_i = 1]\right) - \hat{\E}_N\left[\frac{D_i}{\Pr(D_i = 1)}\left(V_i - \E[V_i ~|~ D_i = 1]\right)\right]} \\
        &= \bigop{N^{-1}p^{-1}\kappa\sigma}.
    \end{aligned}
    \end{equation}
\end{lemma}

\begin{proof}
First, by Lemma \ref{lemma:chebyshev_tightness} and the fact that $\E[\norm{D_iV_i}^2] = p\E[\norm{V_i}^2 ~|~ D_i = 1]$, we have that
\begin{equation}\label{eq:D_i_V_i_tightness_rates}
\begin{aligned}
    \norm{\hat{\E}_N[D_iV_i] - p\E[V_i ~|~ D_i = 1]} &= \bigop{N^{-1/2} \kappa p^{1/2}\sigma} \\
    \hat{\E}_N[D_i] &= p + \bigop{N^{-1/2}p^{1/2}}.
\end{aligned}
\end{equation}
Then, by the Delta Method (see e.g. Theorem 3.1 in \citet{van2000asymptotic}), we have that
\begin{align}
    &\norm{\hat{\E}_N[V_i ~|~ D_i = 1] - \E[V_i ~|~ D_i = 1]} \\
    &= \norm{\frac{\hat{\E}_N[D_iV_i]}{\hat{\E}_N[D_i]} - \frac{\E[D_iV_i]}{\Pr(D_i = 1)}} \\
    &= p^{-1} \cdot \bigop{N^{-1/2} \kappa p^{1/2}\sigma} \\
    &\phantom{=} - \frac{\E[\norm{V_i} ~|~ D_i = 1]}{p} \cdot \bigop{N^{-1/2}p^{1/2}} && \left(\text{$\pfrac{}{x}\frac{x}{y} = \frac{1}{y}$, $\pfrac{}{y}\frac{x}{y} = -\frac{x}{y^2}$, \eqref{eq:D_i_V_i_tightness_rates}}\right) \\
    &= \bigop{N^{-1/2}p^{-1/2}\kappa\sigma} && \text{($\E[\norm{V_i} ~|~ D_i = 1] \leq \E[\norm{V_i}^2 ~|~ D_i = 1]^{1/2}$)},
\end{align}
proving \eqref{eq:lemma:inf_fn_group_avg_sqrt_n_consistency}. To prove \eqref{eq:lemma:inf_fn_group_avg_inf_fn_bound}, we have that
\begin{align}
    &\norm{\left(\hat{\E}_N[V_i ~|~ D_i = 1] - \E[V_i ~|~ D_i = 1]\right) - \hat{\E}_N[\varphi(D_i, V_i)]} \\
    &= \norm{\left(\frac{1}{\hat{\E}_N[D_i]}\hat{\E}_N[D_iV_i] - \E[V_i ~|~ D_i = 1]\right) - \hat{\E}_N[\varphi(D_i, V_i)]} \\
    &= \Bigg\|\left(\frac{1}{\hat{\E}_N[D_i]} - \frac{1}{\Pr(D_i = 1)}\right)\hat{\E}_N[D_iV_i] + \left(\frac{\hat{\E}_N[D_iV_i]}{\Pr(D_i = 1)} - \E[V_i ~|~ D_i = 1]\right) \\
    &\phantom{= \Bigg\|}- \hat{\E}_N[\varphi(D_i, V_i)]\Bigg\| \\
    &= \Bigg\|\frac{\Pr(D_i = 1) - \hat{\E}_N[D_i]}{\hat{\E}_N[D_i]\Pr(D_i = 1)}\hat{\E}_N[D_iV_i] + \left(\frac{\hat{\E}_N[D_iV_i]}{\Pr(D_i = 1)} - \E[V_i ~|~ D_i = 1]\right) \\
    &\phantom{= \Bigg\|}- \hat{\E}_N[\varphi(D_i, V_i)]\Bigg\| \\
    &= \Bigg\|\frac{\Pr(D_i = 1) - \hat{\E}_N[D_i]}{\Pr(D_i = 1)^2}\hat{\E}_N[D_iV_i] + \left(\frac{\hat{\E}_N[D_iV_i]}{\Pr(D_i = 1)} - \E[V_i ~|~ D_i = 1]\right) \\
    &\phantom{=}+ \left(\frac{\Pr(D_i = 1) - \hat{\E}_N[D_i]}{\hat{\E}_N[D_i]\Pr(D_i = 1)} - \frac{\Pr(D_i = 1) - \hat{\E}_N[D_i]}{\Pr(D_i = 1)^2}\right)\hat{\E}_N[D_iV_i] \\
    &\phantom{= \Bigg\|}- \hat{\E}_N[\varphi(D_i, V_i)]\Bigg\| \\
    &= \Bigg\|\frac{\Pr(D_i = 1) - \hat{\E}_N[D_i]}{\Pr(D_i = 1)^2}\E[D_iV_i] + \left(\frac{\hat{\E}_N[D_iV_i]}{\Pr(D_i = 1)} - \E[V_i ~|~ D_i = 1]\right) \\
    &\phantom{=} + \underbrace{\frac{\Pr(D_i = 1) - \hat{\E}_N[D_i]}{\Pr(D_i = 1)^2}\left(\hat{\E}_N[D_iV_i] - \E[D_iV_i]\right)}_{\bigop{N^{-1}p^{-1}\kappa\sigma}} \\
    &\phantom{=}+ \left(\frac{1}{\hat{\E}_N[D_i]\Pr(D_i = 1)} - \frac{1}{\Pr(D_i = 1)^2}\right)\left(\Pr(D_i = 1) - \hat{\E}_N[D_i]\right)\hat{\E}_N[D_iV_i] \\
    &\phantom{= \Bigg\|}- \hat{\E}_N[\varphi(D_i, V_i)]\Bigg\| \\
    &\leq \Bigg\|\left(1 - \frac{\hat{\E}_N[D_i]}{\Pr(D_i = 1)}\right)\E[V_i ~|~ D_i = 1] + \left(\frac{\hat{\E}_N[D_iV_i]}{\Pr(D_i = 1)} - \E[V_i ~|~ D_i = 1]\right) \\
    & \phantom{=}+ \underbrace{\left(\Pr(D_i = 1) - \hat{\E}_N[D_i]\right)^2}_{\bigop{N^{-1}p}}\underbrace{\frac{1}{\hat{\E}_N[D_i]\Pr(D_i = 1)^2}\hat{\E}_N[D_iV_i]}_{\bigop{p^{-2}\sigma + N^{-1/2}p^{-5/2}\kappa\sigma}} \\
    &\phantom{= \Bigg\|}- \hat{\E}_N[\varphi(D_i, V_i)]\Bigg\| + \bigop{N^{-1}p^{-1}\kappa\sigma} \\
    & = \Bigg\|-\frac{\hat{\E}_N[D_i]}{\Pr(D_i = 1)}\E[V_i ~|~ D_i = 1] + \frac{\hat{\E}_N[D_iV_i]}{\Pr(D_i = 1)} - \hat{\E}_N[\varphi(D_i, V_i)]\Bigg\| \\
    &\phantom{=} + \bigop{N^{-1}p^{-1}\kappa\sigma + N^{-1}p^{-1}\sigma + N^{-3/2}p^{-3/2}\kappa\sigma} \\
    & = \bigop{N^{-1}p^{-1}\kappa\sigma + N^{-3/2}p^{-3/2}\kappa\sigma},
\end{align}
as in \eqref{eq:lemma:inf_fn_group_avg_inf_fn_bound}. %

\end{proof}

\subsection{Proof of Lemma \ref{lem:Pk_equivalent}}

We assume without loss of generality that $\tilde{\Gamma}$ is an orthogonal matrix. First, we note that for any matrix $\tilde{\Gamma} \in \R^{T \times r}$, we can divide $\Pi(\tilde{\Gamma})$ up into four disjoint sets of entries via left (to select rows) and right (to select columns) multiplications by $E_c$ and $(I-E_c)$ as follows:
\begin{equation}\label{eq:projection_expansion}
    \Pi(\tilde{\Gamma}) = (E_c\tilde{\Gamma} + (I - E_c)\tilde{\Gamma})\left(\tilde{\Gamma}'E_c\tilde{\Gamma} + \tilde{\Gamma}'(I - E_c)\tilde{\Gamma}\right)^{-1}(E_c\tilde{\Gamma} + (I - E_c)\tilde{\Gamma})'.
\end{equation}
Recall that for any invertible $M$, the Sherman-Morrison-Woodbury matrix identity states that
\begin{equation}
(M + C'C)^{-1} = M^{-1} - M^{-1}C'(I + CM^{-1}C')^{-1}CM^{-1}.
\end{equation}
Taking $M = \tilde{\Gamma}'E_c\tilde{\Gamma}$ and $C = (I - E_c)\tilde{\Gamma}$, we have that
\begin{equation}\label{eq:sherman_morrison_woodbury}
\begin{aligned}
&\left(\tilde{\Gamma}'E_c\tilde{\Gamma} + \tilde{\Gamma}'(I - E_c)\tilde{\Gamma}\right)^{-1} \\
&= \left(\tilde{\Gamma}'E_c\tilde{\Gamma}\right)^{-1} \\
&\phantom{=} - \left(\tilde{\Gamma}'E_c\tilde{\Gamma}\right)^{-1} \tilde{\Gamma}'(I - E_c) \left(I + \underbrace{(I - E_c)\tilde{\Gamma}\left(\tilde{\Gamma}'E_c\tilde{\Gamma}\right)^{-1}\tilde{\Gamma}'(I - E_c)}_{D}\right)^{-1}(I - E_c)\tilde{\Gamma}\left(\tilde{\Gamma}'E_c\tilde{\Gamma}\right)^{-1}.
\end{aligned}
\end{equation}
Then using \eqref{eq:sherman_morrison_woodbury} and \eqref{eq:projection_expansion}, we can write an equation with our quantity of interest on one side:
\begin{align}
E_c\Pi(\tilde{\Gamma})E_c &= \tilde{\Gamma}\left(\tilde{\Gamma}'E_c\tilde{\Gamma}\right)^{-1}\tilde{\Gamma}'E_c \\
&\phantom{=} - \underbrace{(I - E_c)\tilde{\Gamma}\left(\tilde{\Gamma}'E_c\tilde{\Gamma}\right)^{-1}\tilde{\Gamma}'E_c}_{K} \\
&\phantom{=} - E_c\tilde{\Gamma}\left(\tilde{\Gamma}'E_c\tilde{\Gamma}\right)^{-1} \tilde{\Gamma}'(I - E_c)\left(I + D\right)^{-1}(I - E_c)\tilde{\Gamma}\left(\tilde{\Gamma}'E_c\tilde{\Gamma}\right)^{-1}\tilde{\Gamma}'E_c \\
&= \tilde{\Gamma}\left(\tilde{\Gamma}'E_c\tilde{\Gamma}\right)^{-1}\tilde{\Gamma}'E_c  - K - K'(I + D)^{-1}K. \label{eq:least_norm_matrix_proj_expr_B_c_H}
\end{align}
Next, note that, again by \eqref{eq:sherman_morrison_woodbury} and \eqref{eq:projection_expansion},
\begin{align}
(I - E_c)\Pi(\tilde{\Gamma})(I - E_c) &= D - D(I + D)^{-1}D \\
&= (I + D)(I + D)^{-1}D - D(I + D)^{-1}D \\
&= (I + D - D)(I + D)^{-1}D \\
&= (I+D)^{-1}D \\
&= (I+D)^{-1}(I + D - I) \\
&= I - (I+D)^{-1},\label{eq:(I-E)Pi(I-E)}
\end{align}
and
\begin{equation}\label{eq:B_c_I_plus_H_{c^*}_inv}
    E_c\Pi(\tilde{\Gamma})(I - E_c) = K' - K'(I + D)^{-1}D = K'(I - (I + D)^{-1}D) = K'(I + D)^{-1}.
\end{equation}
Therefore,
\begin{equation}\label{eq:D_B_c_equation}
\begin{aligned}
    (I + D)^{-1} &= I - (I - E_c)\Pi(\tilde{\Gamma})(I - E_c) = G_c(\Pi(\tilde{\Gamma}))^{-1}, \\
    K &= (I + D)(I - E_c)\Pi(\tilde{\Gamma})E_c \\
    & = G_c(\Pi(\tilde{\Gamma})) (I - E_c)\Pi(\tilde{\Gamma})E_c.
\end{aligned}
\end{equation}
Substituting \eqref{eq:B_c_I_plus_H_{c^*}_inv} and the components of \eqref{eq:D_B_c_equation} into \eqref{eq:least_norm_matrix_proj_expr_B_c_H} and rearranging, we have that
\begin{align}
&\tilde{\Gamma}\left(\tilde{\Gamma}'E_c\tilde{\Gamma}\right)^{-1}\tilde{\Gamma}'E_c \\
&= E_c\Pi(\tilde{\Gamma})E_c + G_c(\Pi(\tilde{\Gamma})) (I - E_c)\Pi(\tilde{\Gamma})E_c + E_c\Pi(\tilde{\Gamma})(I - E_c)G_c(\Pi(\tilde{\Gamma})) (I - E_c)\Pi(\tilde{\Gamma})E_c,
\end{align}
which simplifies to \eqref{eq:least_norm_matrix_proj_expr}. By definition,
\begin{equation}\label{eq:Gs_commute_Es}
\begin{aligned}
G_c(\Pi(\tilde{\Gamma})) E_c = (I + D)E_c = E_c = E_c(I + D) = E_c G_c(\Pi(\tilde{\Gamma})).
\end{aligned}
\end{equation}
Thus,
\begin{equation}\label{eq:Gs_commute}
\begin{aligned}
G_c(\Pi(\tilde{\Gamma}))(I - E_c) &= (I - E_c)G_c(\Pi(\tilde{\Gamma})) = G_c(\Pi(\tilde{\Gamma})) - E_c, \\
(I - E_c)G_c(\Pi(\tilde{\Gamma})) (I - E_c) &= (I - E_c)G_c(\Pi(\tilde{\Gamma})).
\end{aligned}
\end{equation}
By \eqref{eq:Gs_commute},
\begin{align}
&\tilde{\Gamma}\left(\tilde{\Gamma}'E_c\tilde{\Gamma}\right)^{-1}\tilde{\Gamma}'E_c \\
&= E_c\Pi(\tilde{\Gamma})E_c +(G_c(\Pi(\tilde{\Gamma}))  - E_c)\Pi(\tilde{\Gamma})E_c + E_c\Pi(\tilde{\Gamma})(I - E_c)G_c(\Pi(\tilde{\Gamma}))\Pi(\tilde{\Gamma})E_c\\
& = \left(I + E_c \Pi(\tilde{\Gamma}) (I - E_c)\right)G_c(\Pi(\tilde{\Gamma})) \Pi(\tilde{\Gamma}) E_c.
\end{align}

\subsection{Proof of Lemma \ref{lem:Qk_equivalent}}

By Lemma \ref{lem:Pk_equivalent}, 
\[E_c\left\{I - P_c\left(\Pi(\hat{\Gamma})\right)\right\} = E_c - E_c\hat{\Gamma}(\hat{\Gamma}'E_c\hat{\Gamma})^{-1}\hat{\Gamma}'E_c.\]
Since $E^2_c = E_c$, it is easy to verify that 
\[\left\{I - P_c\left(\Pi(\hat{\Gamma})\right)'\right\}E^2_c\left\{I - P_c\left(\Pi(\hat{\Gamma})\right)\right\} = E_c\left\{I - P_c\left(\Pi(\hat{\Gamma})\right)\right\}.\]

\subsection{Proof of Lemma \ref{lem:Pk_deriv}}

We start by proving a useful lemma. 
\begin{lemma}\label{lem:A-B_inv}
For any symmetric matrices $A$ and $B$ with $2\|B\|_{\mathrm{op}} < \lambda_{\min}(A)$ where $\lambda_{\min}$ denotes the smallest eigenvalue, we have
\begin{equation}\label{eq:inv_(A-B)_remainder}
\|(A - B)^{-1} - A^{-1} - A^{-1}BA^{-1}\|_{\mathrm{op}} \le \frac{2\|B\|_{\mathrm{op}}^2}{\lambda_{\min}^3(A)}.
\end{equation}
\end{lemma}
\begin{proof}
First, we prove the following identity:
\begin{equation}\label{eq:A-B_inv_identity}
(A - B)^{-1} = A^{-1} + A^{-1}BA^{-1} + (A^{-1}B)^2(A-B)^{-1}.
\end{equation}
Multiplying both sides by $A$ on the left and $A-B$ on the right, we just need to prove
\[A = (A - B) + BA^{-1}(A - B) + B A^{-1}B = A - B + B = A.\]
Thus, we prove \eqref{eq:A-B_inv_identity}. By Weyl's inequality, 
\[\|(A-B)^{-1}\|_{\mathrm{op}} = \lambda_{\min}^{-1}(A - B)\ge \lambda_{\min}(A) - \|B\|_{\mathrm{op}}\ge \lambda_{\min}(A)/2.\]
This implies 
\begin{align}
\|(A - B)^{-1} - A^{-1} - A^{-1}BA^{-1}\|_{\mathrm{op}} \le \|A^{-1}\|_{\mathrm{op}}^2 \|B\|_{\mathrm{op}}^2\|(A-B)^{-1}\|_{\mathrm{op}}
\le \frac{2\|B\|_{\mathrm{op}}^2}{\lambda_{\min}^3(A)}
\end{align}
\end{proof}
\noindent For $t\in\R$ sufficiently small, define $\Pi_t:=\Pi_0+t\Delta$. Then
\begin{equation}
P_c(\Pi_t)=\bigl(I + E_c \Pi_0(I - E_c) + tE_c \Delta(I - E_c)\bigr)G_c(\Pi_t)\Pi_t E_c.
\label{eq:prod}
\end{equation}
Note that
\[
G_c(\Pi_t)^{-1}=I-(I-E_c)(\Pi_0+t\Delta)(I-E_c) = G_c(\Pi_0)^{-1} - t(I - E_c)\Delta (I-E_c).
\]
Note that 
\[2\|(I - E_c)\Delta (I-E_c)\|_{\mathrm{op}}\le 2\|\Delta\|_{\mathrm{op}} < \|G_c(\Pi_0)\|_{\mathrm{op}}^{-1}.\]
Applying Lemma \ref{lem:A-B_inv} with $A = G_c(\Pi_0)^{-1}$ and $B = (I - E_c)\Delta (I - E_c)$ and noting that $\lambda_{\min}(A) = \|G_c(\Pi_0)\|_{\mathrm{op}}^{-1}$, we have that, for any $t\in [0, 1]$ ,
\begin{align}
& \left\|\underbrace{\left(G_c(\Pi_0)^{-1}-t (I- E_c)\Delta (I - E_c)\right)^{-1}}_{G_c(\Pi_t)^{-1}}-G_c(\Pi_0)- tG_c(\Pi_0)(I-E_c)\Delta (I-E_c)G_c(\Pi_0)\right\|_{\mathrm{op}}\\
& \le 2t^2\|G_c(\Pi_0)\|_{\mathrm{op}}^3 \|\Delta\|_{\mathrm{op}}^2.
\label{eq:invexp}
\end{align}
In particular, as $t\rightarrow 0$, 
\begin{align*}
& P_c(\Pi_t)=\bigl(I + E_c \Pi_0(I - E_c) + tE_c \Delta(I - E_c)\bigr)\\
& \qquad \qquad \cdot \Big\{G_c(\Pi_0)+tG_c(\Pi_0)(I-E_c)\Delta (I-E_c)G_c(\Pi_0) + o(t)\Big\}\\
& \qquad \qquad \cdot (\Pi_0 + t\Delta) E_c.
\end{align*}
The expression of $\mathcal{P}_c(\Pi_0)[\Delta]$ is then obtained by reading off the coefficient of $t$. 

Recall that 
\begin{equation}\label{eq:some_conditions_op}
\|E_c\|_{\mathrm{op}}\le 1, \,\,\|I - E_c\|_{\mathrm{op}}\le 1, \,\, \|\Pi_0\|_{\mathrm{op}}\le 1, \,\, \|\Pi\|_{\mathrm{op}}\le 1,\,\, \|G_c(\Pi_0)\|_{\mathrm{op}}\ge 1.
\end{equation}
To see the last inequality, we recall that
\[G_c(\Pi_0)^{-1} = I - (I-E_c)\Pi_0(I - E_c)\preceq I\]
where $\preceq$ denotes the positive semidefinite order. Thus, 
\[\|G_c(\Pi_0)\|_{\mathrm{op}}\le 1\Longrightarrow \|G_c^{-1}(\Pi_0)\|_{\mathrm{op}}\ge 1.\]
Then 
\begin{align*}
\left\|\mathcal{P}_c(\Pi_0)[\Delta]\right\|_{\mathrm{op}}& \le \|G_c(\Pi_0)\|_{\mathrm{op}}\|\Delta\|_{\mathrm{op}} + 2\|G_c(\Pi_0)\|_{\mathrm{op}}(\|G_c(\Pi_0)\|_{\mathrm{op}} + 1)\|\Delta\|_{\mathrm{op}}\\
& \le 5\|G_c(\Pi_0)\|_{\mathrm{op}}^2\|\Delta\|_{\mathrm{op}}.
\end{align*}

To bound the remainder, we denote by $R(t)$ the expression on the LHS of \eqref{eq:invexp} inside the operator norm. Letting $t=1$, we obtain that
\begin{align*}
&P_c(\Pi_1) - P_c(\Pi_0) - \mathcal{P}_c(\Pi_0)[\Delta]\\
& = \bigl(I + E_c \Pi_0(I - E_c)\bigr)G_c(\Pi_0)(I-E_c)\Delta (I-E_c)G_c(\Pi_0)\Delta E_c \\
& \quad +  E_c \Delta(I - E_c) G_c(\Pi_0) \Delta \\
& \quad + E_c \Delta(I - E_c) G_c(\Pi_0)(I-E_c)\Delta (I-E_c)G_c(\Pi_0)\Pi_0\\
& \quad + \bigl(I + E_c \Pi(I - E_c) \bigr)R(1) \Pi E_c.
\end{align*}
By \eqref{eq:some_conditions_op},
\begin{align*}
&\bigl(I + E_c \Pi_0(I - E_c)\bigr)G_c(\Pi_0)(I-E_c)\Delta (I-E_c)G_c(\Pi_0)\Delta E_c\\
& \le \|I + E_c \Pi_0(I - E_c)\|_{\mathrm{op}}\|G_c(\Pi_0)\|_{\mathrm{op}}^2\|I-E_c\|_{\mathrm{op}}^2 \|E_c\|_{\mathrm{op}}\|\Delta\|_{\mathrm{op}}^2\\
& \le 2\|G_c(\Pi_0)\|_{\mathrm{op}}^2 \|\Delta\|_{\mathrm{op}}^2.
\end{align*}
Similarly, we can obtain the bounds on the other three terms. Together with the triangle inequality and \eqref{eq:invexp}, we have:
\begin{align*}
& \left\|P_c(\Pi_1) - P_c(\Pi_0) - \mathcal{P}_c(\Pi_0)[\Delta]\right\|_{\mathrm{op}}\\
& \le  2\|G_c(\Pi_0)\|_{\mathrm{op}}^2 \|\Delta\|_{\mathrm{op}}^2 + \|G_c(\Pi_0)\|_{\mathrm{op}} \|\Delta\|_{\mathrm{op}}^2 + \|G_c(\Pi_0)\|_{\mathrm{op}}^2 \|\Delta\|_{\mathrm{op}}^2 +4\|G_c(\Pi_0)\|_{\mathrm{op}}^3 \|\Delta\|_{\mathrm{op}}^2\\
& \le 8\|G_c(\Pi_0)\|_{\mathrm{op}}^3 \|\Delta\|_{\mathrm{op}}^2.
\end{align*}

\subsection{Proof of Lemma \ref{lem:ABP_influence}}

We first present two lemmas and relegate their proofs to the next two subsections. 

\begin{lemma}\label{lem:Ab_deriv}
In the setting of Lemma \ref{lem:Pk_deriv}, further assume that Assumption 2 hold and $\|\Delta\|_{\mathrm{op}}\le 1/2M$. Then 
\begin{align*}
\left\|\hat{A}(\Pi_0 + \Delta
) - \hat{A}(\Pi_0) - \hat{\E}_N\left[\mathcal{A}_i(\Pi_0)[\Delta]\right]\right\|_{\mathrm{op}}&\le 31M^3 \|\Delta\|_{\mathrm{op}}^2(1 + \hat{\E}_N[\|X_i\|_{\mathrm{op}}^2]),\\
\left\|\hat{\E}_N\left[\mathcal{A}_i(\Pi_0)[\Delta]\right]\right\|_{\mathrm{op}}&\le 14M^2 \|\Delta\|_{\mathrm{op}}(1 + \hat{\E}_N[\|X_i\|_{\mathrm{op}}^2]),\\
\left\|\hat{B}(\Pi_0 + \Delta
) - \hat{B}(\Pi_0) - \hat{\E}_N\left[\mathcal{B}_i(\Pi_0)[\Delta]\right]\right\|_{\mathrm{op}}&\le 13M^3 \|\Delta\|_{\mathrm{op}}^2(1 + \hat{\E}_N[(1+ \|X_i\|_{\mathrm{op}})(1 + \|Y_i\|)],\\
\left\|\hat{\E}_N\left[\mathcal{B}_i(\Pi_0)[\Delta]\right]\right\|_{\mathrm{op}}&\le 6M^2\|\Delta\|_{\mathrm{op}}\hat{\E}_N[(1+ \|X_i\|_{\mathrm{op}})\|Y_i\|].,\\
\end{align*}
\end{lemma}

\begin{lemma}\label{lem:AiBi}
Recall the definitions of  $\mathcal{A}_i$ and $\mathcal{B}_i$ in Lemma \ref{lem:Ab_deriv}. 
Under Assumptions 1 and 2, 
\[\E[\|\mathcal{A}_i(\Pi_0)\|_{\mathrm{op}}^2] \le  49 M^4 \E[(1 + \|X_i\|_{\mathrm{op}})^4],
 \]
 and
 \[\E[\|\mathcal{B}_i(\Pi_0)\|_{\mathrm{op}}^2] \le 36M^4 \E[(1 + \|X_i\|_{\mathrm{op}})^2]\E[\|Y_i\|_2^2].\]
 Above, the operator norm is defined with respect to the Frobenius norm. Let 
\[\mathcal{A}(\Pi_0)[\Delta] = \E[\mathcal{A}_i(\Pi_0)[\Delta]], \quad \mathcal{B}(\Pi_0)[\Delta] = \E[\mathcal{B}_i(\Pi_0)[\Delta]].\]
Then 
\[\left\|\hat{\E}_N[\mathcal{A}_i(\Pi_0)] - \mathcal{A}(\Pi_0)\right\|_{\mathrm{op}} = \mathcal{O}_\mathbb{P}\left(N^{-1/2}\right),\quad \left\|\hat{\E}_N[\mathcal{B}_i(\Pi_0)] - \mathcal{B}(\Pi_0)\right\|_{\mathrm{op}} = \mathcal{O}_\mathbb{P}\left(N^{-1/2}\right),\]
where the big-$\mathcal{O}_\mathbb{P}$ terms are uniform in $\Pi_0$.
\end{lemma}

By Assumption 3, 
\begin{equation}\label{eq:RPi}
\Pi(\hat{\Gamma}) - \Pi(\Gamma) = \hat{\E}_N[\phi_i] + R_\Pi, \quad \text{where }R_\Pi = o_{\mathbb{P}}\left(N^{-1/2}\right).
\end{equation}
Since $\E[\|\phi_i\|_{\mathrm{op}}^2] < \infty$, 
\begin{equation}\label{eq:Pihat_rootn}
\Pi(\hat{\Gamma}) - \Pi(\Gamma) = O_{\mathbb{P}}\left(N^{-1/2}\right).
\end{equation}
By Assumption 1, Markov's inequality and Cauchy-Schwarz
\[\hat{\E}_N[1 + \|X_i\|_{\mathrm{op}}^2] = O_{\mathbb{P}}(1), \quad \hat{\E}_N[(1+ \|X_i\|_{\mathrm{op}})(1 + \|Y_i\|_2)] = O_{\mathbb{P}}(1).\]
Applying Lemma \ref{lem:Ab_deriv} with $\Pi_0 = \Pi(\Gamma)$ and $\Delta = \Pi(\hat{\Gamma}) - \Pi(\Gamma)$, we obtain that
\begin{align}
\hat{A}(\Pi(\hat{\Gamma})) &= \hat{A}(\Pi(\Gamma)) + \hat{\E}_N\Big[\mathcal{A}_i(\Pi(\Gamma))[\Pi(\hat{\Gamma}) - \Pi(\Gamma)]\Big] + O_{\mathbb{P}}\left(\frac{1}{N}\right)\\
& = A_0 + \hat{\E}_N\Big[A_i(\Pi(\Gamma)) - A_0 + \mathcal{A}_i(\Pi(\Gamma))[\Pi(\hat{\Gamma}) - \Pi(\Gamma)]\Big] + o_{\mathbb{P}}\left(N^{-1/2}\right).\label{eq:A_piece0}
\end{align}
By Lemma \ref{lem:AiBi}, we obtain that
\begin{equation}\label{eq:A_piece1}
\hat{\E}_N[\mathcal{A}_i(\Pi(\Gamma))[\Pi(\hat{\Gamma}) - \Pi(\Gamma)]] - \mathcal{A}(\Pi(\Gamma))[\Pi(\hat{\Gamma}) - \Pi(\Gamma)] = o_\mathbb{P}\left(N^{-1/2}\right)
\end{equation}
Since $\mathcal{A}(\Pi(\Gamma))[\Delta]$ is linear in $\Delta$, we can rewrite 
\begin{align}
\mathcal{A}(\Pi(\Gamma))[\Pi(\hat{\Gamma}) - \Pi(\Gamma)] &= \mathcal{A}(\Pi(\Gamma))[\hat{\E}_N[\phi_i]] + \mathcal{A}(\Pi(\Gamma))[R_\Pi]\\
& = \hat{\E}_N[\mathcal{A}(\Pi(\Gamma))[\phi_i]] + \mathcal{A}(\Pi(\Gamma))[R_\Pi]\label{eq:A_piece2}
\end{align}
By Lemma \ref{lem:Ab_deriv},
\begin{equation}\label{eq:A_piece3}
\mathcal{A}(\Pi(\Gamma))[R_\Pi] = O_{\mathbb{P}}(\|R_\Pi\|_{\mathrm{op}}) = o_{\mathbb{P}}\left(N^{-1/2}\right).
\end{equation}
Plugging \eqref{eq:A_piece1}-\eqref{eq:A_piece3} into \eqref{eq:A_piece0}, we conclude that 
\begin{align}
\hat{A}(\Pi(\hat{\Gamma})) - A_0 &= \hat{\E}_N\Big[A_i(\Pi(\Gamma)) - A_0 + \mathcal{A}(\Pi(\Gamma))[\phi_i]\Big] + o_{\mathbb{P}}\left(N^{-1/2}\right)\\
& = \hat{\E}_N[\psi_{i,\mathrm{reg}}^{\mathcal{A}} + \psi_{i,\mathrm{fac}}^{\mathcal{A}}] + o_{\mathbb{P}}\left(N^{-1/2}\right). \label{eq:A_influence}
\end{align}
By definition of $A_0$ and Assumption 3,
\[\E[\psi_{i, \mathrm{reg}}^\mathcal{A}] =  \E[\psi_{i, \mathrm{fac}}^\mathcal{A}] = \mathbf{0}_{(T+q)\times (T+q)}.\]
To prove they both have finite second moments,  note that
\begin{align*}
\|A_i(\Pi(\Gamma))\|_{\mathrm{op}} & \le \|D_i(\Pi(\Gamma))\|_{\mathrm{op}}^2 \|Q_{C_i}\left(\Pi(\Gamma)\right)\|_{\mathrm{op}}\le \|D_i(\Pi(\Gamma))\|_{\mathrm{op}}^2\\
& \le (\|I - \Pi(\Gamma)\|_{\mathrm{op}} + \|X_i\|_{\mathrm{op}})^2 \le (1 + \|X_i\|_{\mathrm{op}})^2\\
& \le 2(1 + \|X_i\|_{\mathrm{op}}^2).
\end{align*}
By Assumption 1, we have the uniform bound on the second moments: 
\[\E[\|A_i(\Pi(\Gamma))\|_{\mathrm{op}}^2]\le 8(1 + \E[\|X_i\|_{\mathrm{op}}^4]) < \infty.\]
By the triangle inequality, Lemma \ref{lem:AiBi}, and Assumption 3,
\begin{align*}
&\E[\|\psi_{i,\mathrm{reg}}^\mathcal{A}\|_{\mathrm{op}}^2] + \E[\|\psi_{i,\mathrm{fac}}^\mathcal{A}\|_{\mathrm{op}}^2]\\
& \le \E[\|A_i(\Pi(\Gamma)) - A_0\|_{\mathrm{op}}^2] + \|\mathcal{A}(\Pi(\Gamma))\|_{\mathrm{op}}^2\E[\|\phi_i\|_F]^2\\
& \le \E[\|A_i(\Pi(\Gamma))\|_{\mathrm{op}}^2] + \|\mathcal{A}(\Pi(\Gamma))\|_{\mathrm{op}}^2\E[\|\phi_i\|_F^2]\\
& \le \E[\|A_i(\Pi(\Gamma))\|_{\mathrm{op}}^2] + \E[\|A_i(\Pi(\Gamma))\|_{\mathrm{op}}^2]\E[\|\phi_i\|_F^2]\\
& \le 49 M^4 \E[(1 + \|X_i\|_{\mathrm{op}})^4](1 + \E[\|\phi_i\|_F^2])\\
& <\infty.
\end{align*}
This completes the proof for $\hat{A}(\Pi(\hat{\Gamma}))$. The proof for $\hat{B}(\Pi(\hat{\Gamma}))$ follows from the same argument. 

Next, we prove the result for $P_c(\Pi(\hat{\Gamma}))$. By \eqref{eq:Pihat_rootn}, we know that with probability approaching one as $N \rightarrow \infty$,
\[2\|\Pi(\hat{\Gamma}) - \Pi(\Gamma))\|_{\mathrm{op}} < 1/M \le \|G_c(\Pi(\Gamma))\|_{\mathrm{op}}^{-1}.\]
By Lemma \ref{lem:Pk_deriv} with $\Pi_0 = \Pi(\Gamma)$ and $\Delta = \Pi(\hat{\Gamma}) - \Pi(\Gamma))$ and \eqref{eq:RPi}, we have 
\begin{align*}
P_c(\Pi(\hat{\Gamma})) - P_c(\Pi(\Gamma)) &= \hat{\E}_N[\mathcal{P}_c(\Pi(\Gamma))[\phi_i]] + \mathcal{P}_c(\Pi(\Gamma))[R_\Pi]\\
& = \hat{\E}_N[\psi_{c,i, \mathrm{fac}}^\mathcal{P}]
+ \mathcal{P}_c(\Pi(\Gamma))[R_\Pi].
\end{align*}
By Lemma \ref{lem:Pk_deriv} again,
\[\mathcal{P}_c(\Pi(\Gamma))[R_\Pi] = O(\|R_\Pi\|_{\mathrm{op}}) = o_\mathbb{P}\left(N^{-1/2}\right).\]
Thus, 
\[P_c(\Pi(\hat{\Gamma})) - P_c(\Pi(\Gamma)) = \hat{\E}_N[\psi_{c,i,\mathrm{fac}}^\mathcal{P}] + o_\mathbb{P}\left(N^{-1/2}\right).\]
By Assumption 3 and Lemma \ref{lem:Pk_deriv}, we conclude that 
\[\E[\psi_{c,i,\mathrm{fac}}^\mathcal{P}] = \mathbf{0}_{T\times T}, \quad \E[\|\psi_{c,i,\mathrm{fac}}^\mathcal{P}\|_{\mathrm{op}}^2] < \infty.\]

Lastly, the result for $H(\Pi(\hat{\Gamma}), \bar{X}_c)$ is a direct consequence of  Assumption 3, Lemma \ref{lem:barXY}, and Assumption 1.

\subsection{Proof of Lemma \ref{lem:Ab_deriv}}
By definition, 
\[D_i(\Pi_0+\Delta) = D_i(\Pi_0) + [-\Delta\ \ \mathbf{0}_{T\times q}].\]
By Lemma \ref{lem:Pk_deriv} and Assumption 2,
\[P_c(\Pi_0 + \Delta) = P_c(\Pi_0) + \mathcal{P}_c(\Pi_0)[\Delta] + R, \quad \|R\|_{\mathrm{op}}\le 8M^3\|\Delta\|_{\mathrm{op}}^2.\]
Then 
\begin{align}
A_i(\Pi_0 + \Delta) & = (D_i(\Pi_0) + [-\Delta\ \ \mathbf{0}_{T\times q}])'E_c(I - P_c(\Pi_0) - \mathcal{P}_c(\Pi_0)[\Delta] - R)(D_i(\Pi_0) + [-\Delta\ \ \mathbf{0}_{T\times q}])\\
& = A_i(\Pi_0) + \mathcal{A}_i(\Pi_0)[\Delta]\\
& \qquad - (D_i(\Pi_0) + [-\Delta\ \ \mathbf{0}_{T\times q}])' R (D_i(\Pi_0) + [-\Delta\ \ \mathbf{0}_{T\times q}])\\
& \qquad - [-\Delta\ \ \mathbf{0}_{T\times q}]'E_c \mathcal{P}_c(\Pi_0)[\Delta][-\Delta\ \ \mathbf{0}_{T\times q}]\\
& \qquad - [-\Delta\ \ \mathbf{0}_{T\times q}]'E_c \mathcal{P}_c(\Pi_0)[\Delta]D_i(\Pi_0) - D_i(\Pi_0)'E_c \mathcal{P}_c(\Pi_0)[\Delta][-\Delta\ \ \mathbf{0}_{T\times q}].\label{eq:Ai_expansion}
\end{align}
Note that 
\[\|D_i(\Pi_0)\|_{\mathrm{op}}\le \|I - \Pi_0\|_{\mathrm{op}} + \|X_i\|_{\mathrm{op}}\le 1 + \|X_i\|_{\mathrm{op}},\]
Similarly, since $\Pi_0 + \Delta$ is also a projection matrix
\[\|D_i(\Pi_0+\Delta)\|_{\mathrm{op}}\le 1 + \|X_i\|_{\mathrm{op}}.\]
Then 
\begin{align}\label{eq:Ai_expansion_piece1}
\left\|(D_i(\Pi_0) + [-\Delta\ \ \mathbf{0}_{T\times q}])' R (D_i(\Pi_0) + [-\Delta\ \ \mathbf{0}_{T\times q}])\right\|_{\mathrm{op}}\le \|R\|_{\mathrm{op}}(1 + \|X_i\|_{\mathrm{op}})^2.
\end{align}
By Lemma \ref{lem:Pk_deriv} and \eqref{eq:some_conditions_op},
\begin{align}
&\left\|(D_i(\Pi_0) + [-\Delta\ \ \mathbf{0}_{T\times q}])' R (D_i(\Pi_0) + [-\Delta\ \ \mathbf{0}_{T\times q}])\right\|_{\mathrm{op}}\\
& \le \|[-\Delta\ \ \mathbf{0}_{T\times q}]\|_{\mathrm{op}}^2\|E_c\|_{\mathrm{op}} \|\mathcal{P}_c(\Pi_0)\|_{\mathrm{op}}\\
& \le 5M^2 (1 + \|X_i\|_{\mathrm{op}})^2 \|\Delta\|_{\mathrm{op}}^3.\label{eq:Ai_expansion_piece2}
\end{align}
Similarly, 
\begin{align}
&\left\|[-\Delta\ \ \mathbf{0}_{T\times q}]'E_c \mathcal{P}_c(\Pi_0)[\Delta]D_i(\Pi_0) + D_i(\Pi_0)'E_c \mathcal{P}_c(\Pi_0)[\Delta][-\Delta\ \ \mathbf{0}_{T\times q}]\right\|_{\mathrm{op}}\\
& \le 2\|[-\Delta\ \ \mathbf{0}_{T\times q}]\|_{\mathrm{op}}\|E_c\|_{\mathrm{op}} \|\mathcal{P}_c(\Pi_0)[\Delta]\|_{\mathrm{op}}\|D_i(\Pi_0)\|_{\mathrm{op}}\\
& \le 10M^2(1 + \|X_i\|_{\mathrm{op}})\|\Delta\|_{\mathrm{op}}^2\label{eq:Ai_expansion_piece3}
\end{align}
Pluggin \eqref{eq:Ai_expansion_piece1} - \eqref{eq:Ai_expansion_piece3} into \eqref{eq:Ai_expansion}, we have 
\begin{align}
&\left\|A_i(\Pi_0 + \Delta) - A_i(\Pi_0) - \mathcal{A}_i(\Pi_0)[\Delta]\right\|_{\mathrm{op}}\\
& \le 8M^3\|\Delta\|_{\mathrm{op}}^2(1+ \|X_i\|_{\mathrm{op}})^2 + 5M^2 (1 + \|X_i\|_{\mathrm{op}})^2 \|\Delta\|_{\mathrm{op}}^3 + 10M^2(1 + \|X_i\|_{\mathrm{op}})\|\Delta\|_{\mathrm{op}}^2\\
& \le 8M^3\|\Delta\|_{\mathrm{op}}^2(1+ \|X_i\|_{\mathrm{op}})^2 + \frac{5}{2}M^2 (1 + \|X_i\|_{\mathrm{op}})^2 \|\Delta\|_{\mathrm{op}}^2 + 10M^3(1 + \|X_i\|_{\mathrm{op}})^2\|\Delta\|_{\mathrm{op}}^2\\
& = \frac{31}{2}M^3(1 + \|X_i\|_{\mathrm{op}})^2\|\Delta\|_{\mathrm{op}}^2\\
& \le 31 M^3(1 + \|X_i\|_{\mathrm{op}}^2)\|\Delta\|_{\mathrm{op}}^2,\label{eq:Ai_expansion_bound}
\end{align}
where the last line is implied by the AM-GM inequality. 

By Lemma \ref{lem:Qk_equivalent}, 
\begin{equation}\label{eq:E(I-P)}
\|Q_c(\Pi(\Gamma))\|_{\mathrm{op}}\le 1.
\end{equation}
Similar to the derivation of \eqref{eq:Ai_expansion_bound}, we have 
\begin{align*}
\left\|\mathcal{A}_i(\Pi_0)[\Delta]\right\|_{\mathrm{op}}&\le 5M^2\|\Delta\|_{\mathrm{op}}(1+ \|X_i\|_{\mathrm{op}})^2 + 2 \|\Delta\|_{\mathrm{op}}(1 + \|X_i\|_{\mathrm{op}}) \\
& \le 7M^2 \|\Delta\|_{\mathrm{op}}(1+ \|X_i\|_{\mathrm{op}})^2\\
& \le 14M^2 \|\Delta\|_{\mathrm{op}}(1 + \|X_i\|_{\mathrm{op}}^2),
\end{align*}
where the last line is implied by the AM-GM inequality.

Now we turn to $B_i$. we have 
\begin{align*}
B_i(\Pi_0 + \Delta) & = (D_i(\Pi_0) + [-\Delta\ \ \mathbf{0}_{T\times q}])'E_c(I - P_c(\Pi_0) - \mathcal{P}_c(\Pi_0)[\Delta] - R)Y_i\\
& = B_i(\Pi_0) + \mathcal{B}_i(\Pi_0)[\Delta]\\
& \qquad - (D_i(\Pi_0) + [-\Delta\ \ \mathbf{0}_{T\times q}])' R Y_i - [-\Delta\ \ \mathbf{0}_{T\times q}]'E_c \mathcal{P}_c(\Pi_0)[\Delta]Y_i.
\end{align*}
By Lemma \ref{lem:Pk_deriv} and \eqref{eq:some_conditions_op}, 
\begin{align*}
&\left\|B_i(\Pi_0 + \Delta) - B_i(\Pi_0) - \mathcal{B}_i(\Pi_0)[\Delta]\right\|_{\mathrm{op}}\\
& \le 8M^3\|\Delta\|_{\mathrm{op}}^2(1+ \|X_i\|_{\mathrm{op}})\|Y_i\|_2 + 5M^2 (1 + \|X_i\|_{\mathrm{op}}) \|\Delta\|_{\mathrm{op}}^2\\
& \le 13M^3 \|\Delta\|_{\mathrm{op}}^2(1+ \|X_i\|_{\mathrm{op}})(1 + \|Y_i\|_2).
\end{align*}
By Lemma \ref{lem:Pk_deriv} and \eqref{eq:E(I-P)}, 
\begin{align*}
\left\|\mathcal{B}_i(\Pi_0)[\Delta]\right\|_{\mathrm{op}}&\le 5M^2\|\Delta\|_{\mathrm{op}}(1+ \|X_i\|_{\mathrm{op}})\|Y_i\|_2 +  \|\Delta\|_{\mathrm{op}}\|Y_i\|_2 \\
& \le 6M^2\|\Delta\|_{\mathrm{op}}(1+ \|X_i\|_{\mathrm{op}})\|Y_i\|_2.
\end{align*}

\subsection{Proof of Lemma \ref{lem:AiBi}}

We start with a few well-known identities in matrix algebra. For $A\in \R^{n\times m}, B\in \R^{m\times p}, C\in \R^{p\times r}$, 
\begin{equation}\label{eq:vecABC}
\mathrm{vec}(ABC) = (C'\otimes A)\mathrm{vec}(B),
\end{equation}
and
\begin{equation}\label{eq:vecAB}
\mathrm{vec}(AB) = (B'\otimes I_n)\mathrm{vec}(A) = (I_p\otimes A)\mathrm{vec}(B).
\end{equation}
Let $K\in \{0,1\}^{nm\times nm}$ denote the commutation matrix such that $\mathrm{vec}(A') = K\mathrm{vec}(A)$. Then \eqref{eq:vecAB} implies 
\begin{equation}\label{eq:vecA'B}
\mathrm{vec}(A'B) = (B'\otimes I_n)\mathrm{vec}(A') = (B'\otimes I_n)K\mathrm{vec}(A).
\end{equation}
Lastly, we have
\begin{equation}\label{eq:otimes_op}
\|A\otimes B\|_{\mathrm{op}} = \|A\|_{\mathrm{op}}\|B\|_{\mathrm{op}}.
\end{equation}
By \eqref{eq:vecABC}, we can rewrite $\mathcal{P}_c(\Pi_0)$ as 
\begin{align*}
\mathrm{vec}\left(\mathcal{P}_c(\Pi_0)[\Delta]\right) &= \Big\{E_c\Pi_0 G_c(\Pi_0)(I - E_c)   \otimes E_c    \\
& \qquad + E_c\Pi_0 G_c(\Pi_0)(I - E_c)   \otimes \bigl(I+E_c\Pi_0(I-E_c)\bigr)G_c(\Pi_0)(I-E_c) \\
& \qquad +E_c\otimes \bigl(I+E_c\Pi_0(I-E_c)\bigr)G_c(\Pi_0)\Big\} \mathrm{vec}(\Delta)\\
& := \mathcal{M}_{c, \mathcal{P}}(\Pi_0)\mathrm{vec}(\Delta).
\end{align*}
By \eqref{eq:otimes_op} and \eqref{eq:some_conditions_op}, and Assumption 2,
\begin{equation}\label{eq:McP_op}
\|\mathcal{M}_{c, \mathcal{P}}(\Pi_0)\|_{\mathrm{op}}\le M + 2M^2 + 2M\le 5M^2.
\end{equation}
By \eqref{eq:vecABC}, \eqref{eq:vecAB}, and \eqref{eq:vecA'B}, we have 
\begin{align*}
& \mathrm{vec}(\mathcal{A}_i(\Pi_0)[\Delta])\\
&= -(D_i(\Pi_0)'\otimes D_i(\Pi_0)')\mathrm{vec}(\mathcal{P}_c(\Pi_0)[\Delta])\\
& \quad - \left\{D_i(\Pi_0)'Q_{C_i}(\Pi_0)'[I\ \ \mathbf{0}_{T\times q}]\otimes I\right\}K\mathrm{vec}(\Delta)\\
& \quad - \mathrm{vec}\left\{I\otimes D_i(\Pi_0)'Q_{C_i}(\Pi_0)[I\ \ \mathbf{0}_{T\times q}]\right\}\mathrm{vec}(\Delta)\\
& = -\bigg\{(D_i(\Pi_0)'\otimes D_i(\Pi_0)')\mathcal{M}_{c, \mathcal{P}}(\Pi_0) \\
& \qquad + \left(D_i(\Pi_0)'Q_{C_i}(\Pi_0)'[I\ \ \mathbf{0}_{T\times q}]\otimes I\right)K + I\otimes D_i(\Pi_0)'Q_{C_i}(\Pi_0)[I\ \ \mathbf{0}_{T\times q}]\bigg\}\mathrm{vec}(\Delta)\\
& := \mathcal{M}_{c, i}^{\mathcal{A}}(\Pi_0)\mathrm{vec}(\Delta),
\end{align*}
and 
\begin{align*}
& \mathrm{vec}(\mathcal{B}_i(\Pi_0)[\Delta])\\
&= -(Y_i'\otimes D_i(\Pi_0)')\mathrm{vec}(\mathcal{P}_c(\Pi_0)[\Delta]) - \left\{Y_i'(I - P_{C_i}(\Pi_0)')E_{C_i}[I\ \ \mathbf{0}_{T\times q}]\otimes I\right\}K\mathrm{vec}(\Delta)\\
& = -\bigg\{(Y_i'\otimes D_i(\Pi_0)')\mathcal{M}_{c, \mathcal{P}}(\Pi_0) + \left(Y_i'Q_{C_i}(\Pi_0)'[I\ \ \mathbf{0}_{T\times q}]\otimes I\right)K \bigg\}\mathrm{vec}(\Delta)\\
& := \mathcal{M}_{c, i}^{\mathcal{B}}(\Pi_0)\mathrm{vec}(\Delta).
\end{align*}
By \eqref{eq:some_conditions_op},  \eqref{eq:E(I-P)}, \eqref{eq:otimes_op}, and \eqref{eq:McP_op},
we have 
\begin{align*}
\|\mathcal{M}_{c,i}^\mathcal{A}(\Pi_0)\|_{\mathrm{op}} & \le \|(D_i(\Pi_0)'\otimes D_i(\Pi_0)')\mathcal{M}_{c, \mathcal{P}}(\Pi_0)\|_{\mathrm{op}} \\
& \qquad + \|\left(D_i(\Pi_0)'(I - P_{C_i}(\Pi_0)')E_{C_i}[I\ \ \mathbf{0}_{T\times q}]\otimes I\right)K\|_{\mathrm{op}} \\
& \qquad + \|I\otimes D_i(\Pi_0)'E_{C_i}(I - P_{C_i}(\Pi_0))[I\ \ \mathbf{0}_{T\times q}]\|_{\mathrm{op}}\\
& \le \|D_i(\Pi_0)\|_{\mathrm{op}}^2 \|\mathcal{M}_{c,\mathcal{P}}(\Pi_0)\|_{\mathrm{op}}\\
& \qquad + 2\|D_i(\Pi_0)\|_{\mathrm{op}}\|(I - P_{C_i}(\Pi_0)')E_{C_i}\|_{\mathrm{op}}\\
& \le 5M^2 (1 + \|X_i\|_{\mathrm{op}})^2 + 2(1 + \|X_i\|_{\mathrm{op}})^2 \\
& \le 7M^2 (1 + \|X_i\|_{\mathrm{op}})^2,
\end{align*}
where the last line is implied by the AM-GM inequality. 

Similarly, we have 
\begin{align*}
\|\mathcal{M}_{c,i}^\mathcal{B}(\Pi_0)\|_{\mathrm{op}} & \le \|(Y_i'\otimes D_i(\Pi_0)')\mathcal{M}_{c, \mathcal{P}}(\Pi_0)\|_{\mathrm{op}}\\
& \qquad + \|\left(Y_i'Q_{C_i}'[I\ \ \mathbf{0}_{T\times q}]\otimes I\right)K\|_{\mathrm{op}}\\
& \le \|Y_i\|_2 \|D_i(\Pi_0)\|_{\mathrm{op}}\|\mathcal{M}_{c,\mathcal{P}}(\Pi_0)\|_{\mathrm{op}}\\
& \qquad + \|Y_i\|_2 \|(I - P_{C_i}(\Pi_0)')E_{C_i}\|_{\mathrm{op}}\\
& \le 5M^2 (1 + \|X_i\|_{\mathrm{op}})\|Y_i\|_2 + \|Y_i\|_2\\
& \le 6M^2 (1 + \|X_i\|_{\mathrm{op}})\|Y_i\|_2.
\end{align*}
By Assumption 1, the AM-GM inequality, and the Cauchy-Schwarz inequality, 
\begin{align}
\E\left[\|\mathcal{M}_{c,i}^\mathcal{A}(\Pi_0)\|_{\mathrm{op}}^2\right] & \le 49 M^4 \E[(1 + \|X_i\|_{\mathrm{op}})^4],
 \label{eq:MciA_op}
\end{align}
and 
\begin{align}\E\left[\|\mathcal{M}_{c,i}^\mathcal{B}(\Pi_0)\|_{\mathrm{op}}^2\right] \le 36M^4 \E[(1 + \|X_i\|_{\mathrm{op}}^2 \|Y_i\|_2^2]\le 36M^4 \E[(1 + \|X_i\|_{\mathrm{op}})^2]\E[\|Y_i\|_2^2]\label{eq:MciB_op}
\end{align}
By Chebyshev's inequality, 
\[\left\|\hat{\E}_N[\mathcal{M}_{c,i}^\mathcal{A}(\Pi_0) ] - \E[\mathcal{M}_{c,i}^\mathcal{A}(\Pi_0)]\right\|_{\mathrm{op}} = \mathcal{O}_\mathbb{P}\left(N^{-1/2}\right), \,\, \left\|\hat{\E}_N[\mathcal{M}_{c,i}^\mathcal{B}(\Pi_0) ] - \E[\mathcal{M}_{c,i}^\mathcal{B}(\Pi_0)]\right\|_{\mathrm{op}} = \mathcal{O}_\mathbb{P}\left(N^{-1/2}\right).\]
By \eqref{eq:MciA_op} and \eqref{eq:MciB_op}, we also prove bounds on the operator norms of $\mathcal{A}(\Pi_0)$ and $\mathcal{B}(\Pi_0)$ that are uniform in $\Pi_0$.

%% file: supplement_body.tex
\section{Supplementary Discussions and Results}\label{sec:supplement}

\input{\paperpath sections/supplement}

\section{Empirical Illustration: More Details}\label{sec:empirical_supplement}

\input{\paperpath sections/empirical_supplement}

\section{Proofs of Supplementary Results}

\input{\paperpath sections/supplemental_proofs}

%% file: sections/supplement.tex
\subsection{Oracle O\textsuperscript{3} Algorithm Convergence Implies Full-Rank Cohort-Specific Factor Matrices}\label{proof:lemma:E_c_Gamma_full_rank}

If the oracle O\textsuperscript{3} Algorithm discussed in Footnote \ref{foot:oracle_o3_algorithm} converges to a single super cohort, i.e. $M^{(K)} = 1$, there must exist some iteration $K_c$ of the oracle O\textsuperscript{3} Algorithm where there exists a super cohort $m_1$ such that $\mathcal{T}_c = \mathcal{T}_{\mathcal{S}_{m_1}^{(K_c)}}$ and such that super cohort $m_1$ is connected to some other super cohort $m_2$ in the O\textsuperscript{3} Graph $\mathcal{G}_r^{(K_c + 1)}$ constructed in the subsequent iteration. Since $E_{\mathcal{T}_{\mathcal{S}_{m_1}^{(K_c)}}}$ is a square matrix, the column space of $\Gamma'E_{\mathcal{T}_{\mathcal{S}_{m_1}^{(K_c)}}}$ must be a subset of the column space of $\Gamma'E_{\mathcal{T}_{\mathcal{S}_{m_1}^{(K_c)}}}E_{\mathcal{T}_{\mathcal{S}_{m_2}^{(K_c)}}}$, which, along with \eqref{cond:observed_outcome_overlap_graph_edge_def}, implies that
\begin{align}
    \mathrm{rank}(\underbrace{E_c\Gamma}_{T \times r}) &= \mathrm{rank}(\Gamma'E_{\mathcal{T}_{\mathcal{S}_{m_1}^{(K_c)}}}) && \text{($T > r$, rank is $\min\{\text{row rank}, \text{col rank}\}$, $\mathcal{T}_c = \mathcal{T}_{\mathcal{S}_{m_1}^{(K_c)}}$)} \\
    &\geq \mathrm{rank}(\Gamma'E_{\mathcal{T}_{\mathcal{S}_{m_1}^{(K_c)}}}E_{\mathcal{T}_{\mathcal{S}_{m_2}^{(K_c)}}}) \\
    & = r && \text{($m_1$ and $m_2$ are neighbors in $\mathcal{G}_r^{(K_c + 1)}$, so \eqref{cond:observed_outcome_overlap_graph_edge_def} holds)}.
\end{align}
Since $E_c\Gamma$ is $T \times r$ with $T > r$, $\mathrm{rank}(E_c\Gamma) \leq r$ must also hold.

\subsection{Equivalence of Graph-Based Identification Criteria in Panel Data Models and O\textsuperscript{3} Algorithm Convergence to a Single Super Cohort}\label{sec:supplement:connections_graph_based_identification}

Several papers in the literature on bipartite match data develop approaches for identifying $\mu_{ct}$ (or functions of it) under other models of how unobserved confounders affect outcomes $Y_{it}^*$ that still impose strict exogeneity and fixed-effect-like assumptions \citep{bonhomme2020econometric}. To prove identification of $\mu_{ct}$ under their alternative models, these papers appeal to the connectedness of different graphs with nodes that represent units and/or outcomes in the nomenclature of this paper. In this section, we show that two types of graphs and accompanying assumptions about their connectedness made in this literature are both equivalent to the O\textsuperscript{3} Algorithm (Algorithm \ref{alg:o3_algorithm}) converging to a single super cohort, i.e. $M^{(K)} = 1$, when $r = 1$.

First, we consider the identification arguments in \citet{abowd2002computing} and \citet{jochmans2019fixed}, which study TWFE models of outcomes $\E[Y_{it}^* ~|~ \lambda_i, C_i] = \gamma_t + \lambda_i$, where $\gamma_t$ and $\lambda_i$ are both one-dimensional. These papers condition on a sample of units and the set of their uni-dimensional fixed effects $\{\lambda_i\}_{i = 1}^N$, and they show that under appropriate normalizations of the fixed effects, the TWFE regression estimators $\hat{\gamma}_t$ of the outcome fixed effects $\gamma_t$ converge to an appropriately normalized instance of the corresponding outcome fixed effects $\tilde{\gamma}_t$ at a $N^{-1/2}$ rate as $N \rightarrow \infty$ under two assumptions. First, they assume that a finite-population variant of Assumption \ref{assump:cohort_sizes_and_potential_outcome_variance_bound} holds so that a non-vanishing fraction of units have outcome $t$ observed. Second, they define the bipartite graph $\tilde{G}$ consisting of nodes corresponding to units and outcomes and an edge between any unit $i$ and outcome $t$ for which outcome $Y_{it}^*$ is observed and assume that $\tilde{\mathcal{G}}$ is connected.

As it turns out, this assumption is equivalent to the O\textsuperscript{3} Algorithm converging to a single super cohort when $r = 1$:
\begin{proposition}\label{prop:bipartite_graph_observed_outcome_overlap_graph_equivalence}
    When $r = 1$ and every entry of $\Gamma$ is non-zero, the bipartite graph $\tilde{\mathcal{G}}$ being connected implies that the O\textsuperscript{3} Algorithm converges to a single super cohort after a single iteration, i.e. $K = 1$ and $M^{(1)} = 1$. If, in addition, Assumption \ref{assump:cohort_sizes_and_potential_outcome_variance_bound} holds, then the O\textsuperscript{3} Algorithm converging to a single super cohort after a single iteration implies that $\tilde{\mathcal{G}}$ is connected with high probability.
\end{proposition}
\noindent We provide a proof of Proposition \ref{prop:bipartite_graph_observed_outcome_overlap_graph_equivalence} in Appendix \ref{proof:prop:bipartite_graph_observed_outcome_overlap_graph_equivalence}. We note that since when $r = 1$, $\Gamma$ is a vector, the assumption that every entry of $\Gamma$ is non-zero is just assuming that the systematic component $\gamma_t\lambda_i$ in the factor model \eqref{eq:factor_model} matters for determining every potential outcome.

Second, we consider the identification arguments in \citet{bonhomme2019distributional}, which identifies the distributions of $Y_{it}$ under a model of outcomes with discrete unobserved heterogeneity, and \citet{hull2018estimating} and Section 3.3 of \citet{kline2024firm}, which identify differences in $\mu_{ct}$ across $t$ under a TWFE-like model of outcomes. Unlike \citet{abowd2002computing} and \citet{jochmans2019fixed} and similarly to this paper, these papers model an infinite population of units and, due to their focus on bipartite match outcomes, take seriously the fact that units form one match at a time sequentially. Abstracting away from the time dimension to match the setup in this paper, at their core, \citet{bonhomme2019distributional}, \citet{hull2018estimating}, and \citet{kline2024firm} base their identification arguments off of variants of the assumption that another graph $\check{\mathcal{G}}$ is connected, where $\check{\mathcal{G}}$ is an undirected graph whose nodes correspond to outcomes and whose edges between two outcomes $t_1$ and $t_2$ exist if a positive measure of units have both outcomes $t_1$ and $t_2$ observed.

Similarly to Proposition \ref{prop:bipartite_graph_observed_outcome_overlap_graph_equivalence}, we can also show that the assumption that $\check{\mathcal{G}}$ is connected is equivalent to the O\textsuperscript{3} Algorithm converging to a single super cohort when $r = 1$:
\begin{proposition}\label{prop:potential_outcome_unit_overlap_observed_outcome_overlap_graph_equivalence}
    When $r = 1$ and every entry of $\Gamma$ is non-zero, the graph $\check{\mathcal{G}}$ being connected implies that the O\textsuperscript{3} Algorithm converges to a single super cohort after a single iteration, i.e. $K = 1$ and $M^{(1)} = 1$. If, in addition, Assumption \ref{assump:cohort_sizes_and_potential_outcome_variance_bound} holds, then the O\textsuperscript{3} Algorithm converging to a single super cohort after a single iteration implies that $\check{\mathcal{G}}$ is connected with high probability.
\end{proposition}
\noindent We provide a proof of Proposition \ref{prop:potential_outcome_unit_overlap_observed_outcome_overlap_graph_equivalence} in Appendix \ref{proof:prop:potential_outcome_unit_overlap_observed_outcome_overlap_graph_equivalence}.

\subsection{Asymptotic Linearity of the Principal Components Estimator}\label{sec:supplement:pc_estimator_consistency_asymptotic_linear}

As discussed intuitively in Section \ref{sec:theory:cohort_specific_factor_estimates}, the Principal Components (PC) estimator of cohort-specific factors can be shown to identify the column space of the cohort-specific factor matrix $E_c\Gamma$. In this section, we formalize this fact under slightly weaker assumptions than those typically imposed in the literature. To define the PC estimator formally, let $V_c$ denote the population covariance matrix of the observed outcomes $E_cY_i$ for cohort $c$:
\begin{equation}\label{eq:PCA_Vc}
    V_c \coloneqq \E[(E_cY_i - E_c\mu_c)(E_cY_i - E_c\mu_c)' ~|~ C_i = c],
\end{equation}
and let $\hat{V}_c$ denote its empirical counterpart:
\begin{equation}
    \hat{V}_c \coloneqq  \hat{\E}_N[(E_cY_i - \hat{\E}_N[E_cY_i ~|~ C_i = c])(E_cY_i - \hat{\E}_N[E_cY_i ~|~ C_i = c])' ~|~ C_i = c];
\end{equation}
the PC estimator $\Pi(\hat{\Gamma}_{c, \text{PC}})$ of $\Pi(E_c\Gamma)$ is defined as the projection matrix onto the span of the eigenvectors of $\hat{V}_c$ corresponding to $\hat{V}_c$'s $r$ largest eigenvalues.

Under two assumptions that we define next, we will show that a population analog of the PC estimator, i.e. the projection matrix onto the span of the eigenvectors of $V_c$ corresponding to $V_c$'s $r$ largest eigenvalues, identifies $\Pi(E_c\Gamma)$, as required for Theorem \ref{thm:factor_identification} to hold. The first assumption guarantees that all of the loadings' dimensions are non-trivial:
\begin{assumption}\label{assump:nontrivial_loadings}
    For every cohort $c \in [C]$, the loading covariance matrix $\Cov(\lambda_i ~|~ C_i = c)$ has finite operator norm and is positive definite, $\E[\norm{\lambda_i}_2^4 ~|~ C_i = c] < \infty$ and $\E[\norm{\epsilon_i}_2^4 ~|~ C_i = c] < \infty$.
\end{assumption}
\noindent Our other assumption is that the outcome residuals are uncorrelated across outcomes and have the same variance within each unit:
\begin{assumption}\label{assump:uncorrelated_homoskedastic_outcomes}
    There exist positive random variables $\sigma_i^2$ defined on the same probability space as $(C_i, Y_i^*)$ such that $\E[\epsilon_i\epsilon_i' ~|~ \sigma_i^2, C_i] = \sigma_i^2 I$ almost surely and $\sigma_c^2 \coloneqq \E[\sigma_i^2 ~|~ C_i = c] < \infty$ for every cohort $c \in [C]$.
\end{assumption}

Before continuing, we provide one more definition. Let
\begin{equation}\label{eq:cohort_specific_signals}
s_{1c}^2 \leq \dotsc \leq s_{rc}^2
\end{equation}
denote the smallest through largest eigenvalues of the matrix $\Gamma' E_c \Gamma \Cov(\lambda_i ~|~ C_i = c)$. We note that $s_{1c}^2 > 0$ by Assumption \ref{assump:nontrivial_loadings} and the fact that $E_c\Gamma$ is full-rank when the factor vectors are in general position. We are now equipped to state our identification result:

\begin{lemma}\label{lemma:cohort_specific_factor_identification_homoskedastic_outcomes}
    Suppose the factor vectors $\{\gamma_t \,\setst\, t \in [T]\}$ are in general position, Assumptions \ref{assump:nontrivial_loadings} and \ref{assump:uncorrelated_homoskedastic_outcomes} hold, and let $\tilde{\Gamma}_{c, \text{PC}}$ be any $T \times r$ matrix whose columns are eigenvectors corresponding to the $r$ largest eigenvalues of $V_c$, defined in \eqref{eq:PCA_Vc}. Then $\Pi(\tilde{\Gamma}_{c, \text{PC}}) = \Pi(E_c\Gamma)$, meaning $\Pi(E_c\Gamma)$ is identified as required for Theorem \ref{thm:factor_identification} to hold. Further, $V_c$'s eigenvalues ordered from smallest to largest are $T - T_c$ zeros followed by $T_c - r$ repetitions of $\sigma_c^2$ followed by $s_{c1}^2 + \sigma_c^2, \dotsc, s_{cr}^2 + \sigma_c^2$.
\end{lemma}
\noindent We provide a proof of Lemma \ref{lemma:cohort_specific_factor_identification_homoskedastic_outcomes} in Appendix \ref{proof:lemma:cohort_specific_factor_identification_homoskedastic_outcomes}.

Next, we show that $\Pi(\hat{\Gamma}_{c, \text{PC}})$ is a consistent and asymptotically linear estimator of $\Pi(E_c\Gamma)$. To introduce this result, we let $\tilde{\gamma}_{cj}$ denote an eigenvector of $V_c$ corresponding to the $j$th smallest eigenvalue of $V_c$.

\begin{proposition}\label{proposition:cohort_specific_factor_inf_fn_homoskedastic_outcomes}
    Suppose the factor vectors $\{\gamma_t \,\setst\, t \in [T]\}$ are in general position and Assumptions \ref{assump:cohort_sizes_and_potential_outcome_variance_bound}, \ref{assump:nontrivial_loadings}, and \ref{assump:uncorrelated_homoskedastic_outcomes}, hold. Then
    \begin{equation}\label{eq:cohort_specific_proj_mat_asymp_linear_homosk}
        \norm{[\Pi(\hat{\Gamma}_{c, \text{PC}}) - \Pi(E_c\Gamma)] - \hat{\E}_N\left[\phi_{c, \text{PC}}(C_i, Y_i)\right]}_\mathrm{op} = \bigop{s_{1c}^{-2}[s_{1c}^{-2} + r(T_c - r)] N^{-1}p_c^{-1}},
    \end{equation}
    where
    \begin{equation}
        \phi_{c, \mathrm{PC}}(C_i, Y_i) \coloneqq \sum_{j = T - r + 1}^T\sum_{k = T - T_c + 1}^{T - r} \frac{\ind{C_i = c}}{p_c \cdot s_{(j - (T - r))c}^2} \cdot g\left(\Pi(\tilde{\gamma}_{cj}), (V_c - E_c(Y_i - \mu_c)(Y_i - \mu_c)'E_c), \Pi(\tilde{\gamma}_{ck})\right),
    \end{equation}
    and $\phi_{c, \mathrm{PC}}$ satisfies $\E[\phi_{c, \mathrm{PC}}(C_i, Y_i)] = \zeros_{T^2}$ and 
    \begin{equation}
    \E[\normnofit{\phi_{c, \mathrm{PC}}(C_i, Y_i)}_\mathrm{op}^2] \leq \frac{64 r(T_c - r)}{p_c s_{1c}^2}\norm{\Gamma}_\text{op}^4\left(\E\left[\norm{\lambda_i}_2^4 ~|~ C_i = c\right] + \E[\norm{\epsilon_i}_2^4 ~|~ C_i = c]\right) < \infty.
    \end{equation}
    Thus, Assumption \ref{assump:cohort_specific_factor_proj_mat_inf_fn} holds.
\end{proposition}
\noindent We provide a proof of Proposition \ref{proposition:cohort_specific_factor_inf_fn_homoskedastic_outcomes} in Appendix \ref{proof:proposition:cohort_specific_factor_inf_fn_homoskedastic_outcomes}. We note that as a consequence of \eqref{eq:cohort_specific_proj_mat_homoskedastic_err_expansion_bound_2}, Assumption \ref{assump:cohort_specific_factor_proj_mat_hadamard_diff} holds as well, although for brevity, we do not introduce the additional notation necessary to state such a result formally here.

We conclude this section with a more detailed discussion of why the assumptions needed for Proposition \ref{proposition:cohort_specific_factor_inf_fn_homoskedastic_outcomes} to hold are slightly weaker than existing consistency and asymptotic linearity results for the PC estimator in the literature of which we are aware. It has been shown that the PC estimator is consistent and asymptotic linear when $T$ remains finite as $N$ grows under Assumptions \ref{assump:nontrivial_loadings} and \ref{assump:uncorrelated_homoskedastic_outcomes} in the sense that 
\begin{equation}\label{eq:PC_factor_asymptotic_linearity_basis}
    \hat{\Gamma}_{c, \text{PC}} - E_c\Gamma \hat{Q} = \hat{\E}_N\left[\tilde{\phi}_{c, \text{PC}}(C_i, Y_i)\right] + \littleop{N^{-1/2}}
\end{equation}
for some mean-zero influence function $\tilde{\phi}_{c, \text{PC}}$ and random matrix $\hat{Q}$ such that $\hat{Q} \sconvin{{\Pr}} Q$ for some deterministic matrix $Q$ (see e.g. Theorem 5 in \citet{bai2003inferential}).\footnote{Theorem 1 in \citet{bai2002determining} does show a similar result to the statement $\sqrt{N}\normnofit{\hat{\Gamma}_{c, \text{PC}} - E_c\Gamma \hat{Q}}_F^2 = \bigop{1}$ without requiring $\hat{Q}$ to have a probability limit, but they prove it in the more general case where $\epsilon_{it}$ is allowed to be heteroskedastic and weakly dependent, so in their theorem statement the $\sqrt{N}$ factor is replaced with $\min\{\sqrt{N}, \sqrt{T}\}$ and they require $\min\{N, T\} \rightarrow \infty$.} Such results require the eigenvalues $s_{1c}^2, \dotsc, s_{rc}^2$ to be distinct.\footnote{See Assumption G in \citet{bai2003inferential} and Assumption A2(iii) in \citet{bai2021approximate} for examples of such eigenvalue uniqueness conditions. After introducing Assumption G, \citet{bai2003inferential} notes that such an assumption is not necessary to show that consistent estimators exist for identifiable quantities derived from the factor model.}

However, our theory in Section \ref{sec:theory:estimation_inference} only requires assumptions that guarantee the consistency of $\Pi(\hat{\Gamma}_{c, \text{PC}})$ as an estimator of $\Pi(E_c\Gamma)$, which in turn only requires the existence of a random basis matrix $\hat{Q}$ such that $\hat{\Gamma}_{c, \text{PC}} \hat{Q}' = E_c\Gamma + \bigop{N^{-1/2}}$, not that $\hat{Q}$ has a deterministic probability limit, as in  \eqref{eq:PC_factor_asymptotic_linearity_basis}.\footnote{To see why, note that if such a $\hat{Q}$ did not exist, then for all potentially random basis matrices $\hat{Q}$, $\Pi(\hat{\Gamma}_{c, \text{PC}}) = \Pi(\hat{\Gamma}_{c, \text{PC}}\hat{Q}') \neq \Pi(E_c\Gamma) + \bigop{N^{-1/2}}$, which would be a violation of Proposition \ref{proposition:cohort_specific_factor_inf_fn_homoskedastic_outcomes}.} As such, we do not require distinctness of the eigenvalues $s_{1c}^2, \dotsc, s_{rc}^2$ to show that the population equivalent of $\hat{\Gamma}_{c, \text{PC}}$ identifies $\Pi(E_c\Gamma)$ as required for Theorem \ref{thm:factor_identification} to hold and that $\hat{\Gamma}_{c, \text{PC}}$ itself satisfies Assumptions \ref{assump:cohort_specific_factor_proj_mat_inf_fn} and \ref{assump:cohort_specific_factor_proj_mat_hadamard_diff} required for our estimator to be consistent and to be useful for valid bootstrap inference.

%% file: sections/empirical_supplement.tex
\subsection{Dataset Construction Details}\label{sec:empirical_appendix:dataset_construction}

\begin{figure}[t]
    \centering
    \includegraphics[width=\textwidth]{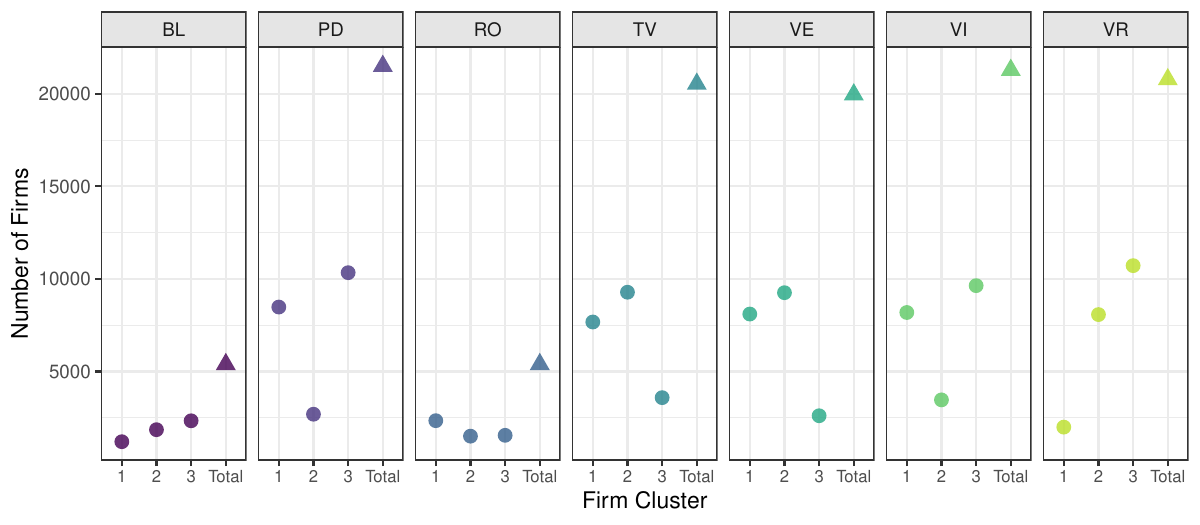}
    \caption{In this figure, we plot the number of firms from the VWH dataset in each cluster within each province as computed using the $k$-means-based procedure proposed in \protect\citet{bonhomme2019distributional} as described in Section \ref{sec:empirical_appendix:dataset_construction}. We also show the total number of firms located in each province under the ``Total'' label in the facet corresponding to each province.}
    \label{fig:cluster_firm_counts_plot}
\end{figure}

\begin{figure}[H]
    \centering
    \includegraphics[width=\textwidth]{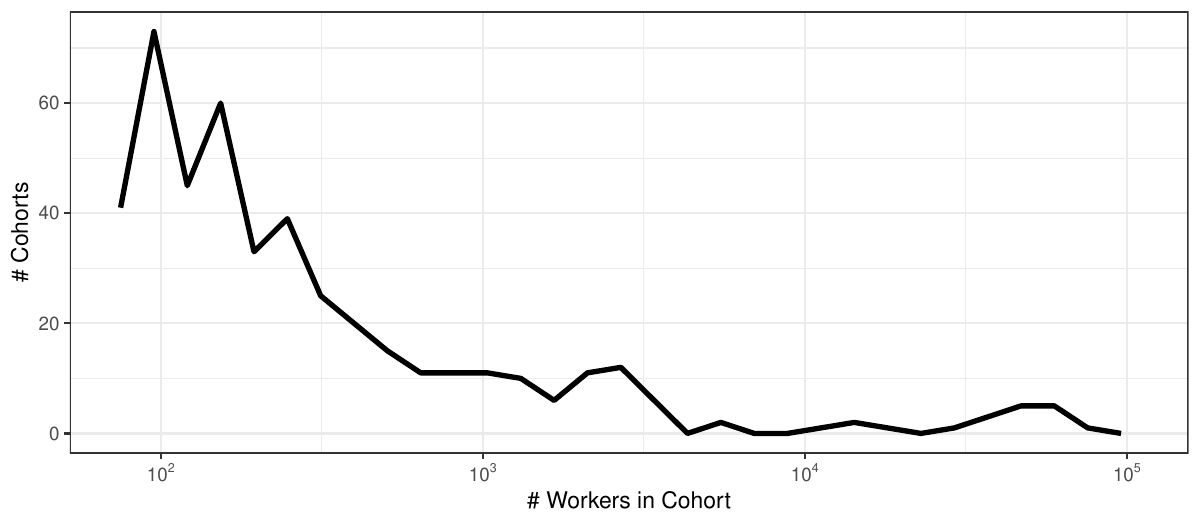}
    \caption{In this figure, we plot the distribution of cohort sizes for the clustering of firms in each province into three clusters.}
    \label{fig:cohort_size_dist_plot}
\end{figure}

As discussed in Section \ref{sec:empirical_performance:setting}, to account for within-province firm heterogeneity in a flexible way, we cluster the firms located in each province into $k = \input{\figurespath result_summaries/snippets/chosen_k.txt}$ types using the $k$-means-based procedure proposed in \citet{bonhomme2019distributional}. In particular, we let $F$ be the number of firms in our sample, we let $N_f$ denote the number of workers who ever worked for firm $f$ in 1998 and 1999, we let $\mathcal{F}_p \subset \{1, \dotsc, F\}$ denote the subset of firms located in province $p$, and we let $\hat{G}_{Y_f}$ denote the empirical cumulative distribution function of average weekly wages in 1998 and 1999 across workers that ever worked for firm $f \in \{1, \dotsc, F\}$ during 1998 and 1999. To cluster the firms $\mathcal{F}_p$ within province $p$, we solve the following clustering problem for each province $p$, as in \citet{bonhomme2019distributional}:
\begin{equation}\label{eq:firm_clustering_problem}
    \min_{\{k(f) \setst f \in \mathcal{F}_p\}, H_{p1}, \dotsc, H_{pK}} \sum_{f \in \mathcal{F}_p} N_f \int \left(\hat{G}_{Y_f}(y) - H_{pk(f)}(y)\right)^2 d\mu(y),
\end{equation}
where $\mu$ is a discrete uniform measure with mass points at 100 evenly spaced quantiles of the distribution of average weekly wages across all firms in the sample. For each province $p$, we solve \eqref{eq:firm_clustering_problem} 10 times with different initial cluster assignments and select the clustering solution with the lowest objective value. We plot the number of firms in each of the final clusters within each province in Figure \ref{fig:cluster_firm_counts_plot} and the distribution of cohort sizes under this clustering in Figure \ref{fig:cohort_size_dist_plot}. We also report summary statistics for the final panel constructed from the VWH dataset based on the sequence of sample restrictions we make to construct the final panel described in Section \ref{sec:empirical_performance:setting} in Table \ref{tab:panel_summary_stats}.

\begin{table}[H]
    \centering
    \input{\figurespath result_summaries/tables/panel_summary_stats_start_year=1998_end_year=2001_year_cluster_size=2_min_cohort_size=75_k=3.tex}
    \caption{This table provides summary statistics for panels constructed from the VWH dataset based on the sequence of sample restrictions we make to construct the final panel described in Section \ref{sec:empirical_performance:setting}. The ``Full Panel'' column refers to the panel constructed from the VWH dataset subsetted to firms located in the Veneto region's provinces and in business between 1998 and 2001 and workers who worked for at least one of those firms during that time period. The ``Always-Present Workers, Firms'' column refers to the panel constructed by subsetting the ``Full Panel'' panel to the subset of firms and workers in which every worker has at least one observed outcome at one of the firms in the subset in both the 1998 - 1999 and 2000 - 2001 two-year periods and every firm has at least one of the workers in the subset working for it during both the 1998 - 1999 and 2000 - 2001 two-year periods. The ``Cohorts Above Min. Size'' column refers to the further subsetting of the panel to just cohorts with at least 75 workers, where cohorts are defined using the final outcomes described in Section \ref{sec:empirical_performance:setting}. The ``Largest Super Cohort'' column refers to the further subsetting of the panel to just the largest super cohort constructed by the O\textsuperscript{3} Algorithm for $r = 2$; because the O\textsuperscript{3} Algorithm converges to a single super cohort, this panel is the same as the ``Cohorts Above Min. Size'' panel. The ``Num. Employment Spells'' statistic counts the total number of observed outcomes across all workers in the panel where outcomes are defined as in the final panel described in Section \ref{sec:empirical_performance:setting}, and the ``\% Workers w/ X Spells'' statistics measure the shares of workers in the panel with X observed outcomes.}
    \label{tab:panel_summary_stats}
\end{table}

\subsection{Additional Figures}\label{sec:empirical_appendix:additional_results}

\begin{figure}[H]
    \centering
    \begin{subfigure}{0.48\textwidth}
        \centering
        \includegraphics[width=\textwidth]{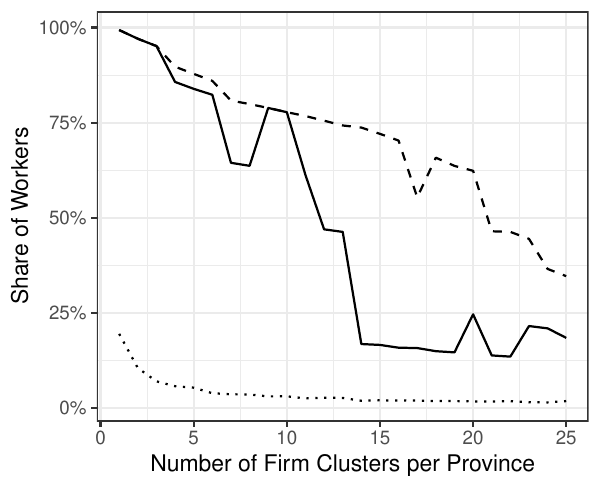}
        \caption{Share of Workers}
        \label{fig:largest_super_cohort_share_workers_plot}
    \end{subfigure}
    \begin{subfigure}{0.48\textwidth}
        \centering
        \includegraphics[width=\textwidth]{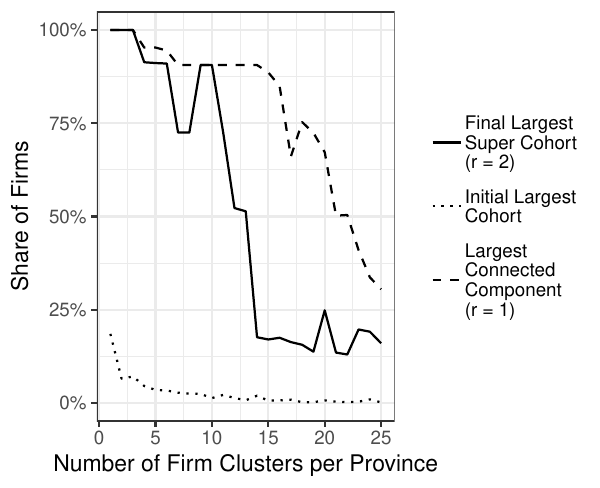}
        \caption{Share of Firms}
        \label{fig:largest_super_cohort_share_firms_plot}
    \end{subfigure}
    \caption{In this figure, we plot the shares of workers and firms whose cohort outcome means are identified using the largest super cohort constructed by the O\textsuperscript{3} Algorithm across different numbers of firm clusters per province. The solid lines represent the shares of workers and firms in the largest super cohort constructed by the O\textsuperscript{3} Algorithm for $r = 2$. The dotted line plots the share of workers in the largest cohort. The dashed lines represent the shares of workers and firms in the largest super cohort constructed by the O\textsuperscript{3} Algorithm for $r = 1$, which are equivalent to the shares of workers and firms in the largest connected component of the graph constructed by adding an edge between every pair of cohorts who share at least one observed outcome; this same graph is used to assess parameter identification under the TWFE model, as discussed in Section \ref{sec:supplement:connections_graph_based_identification}.}
    \label{fig:largest_super_cohort_share_plot}
\end{figure}

\begin{figure}[H]
    \centering
    \includegraphics[width=\textwidth]{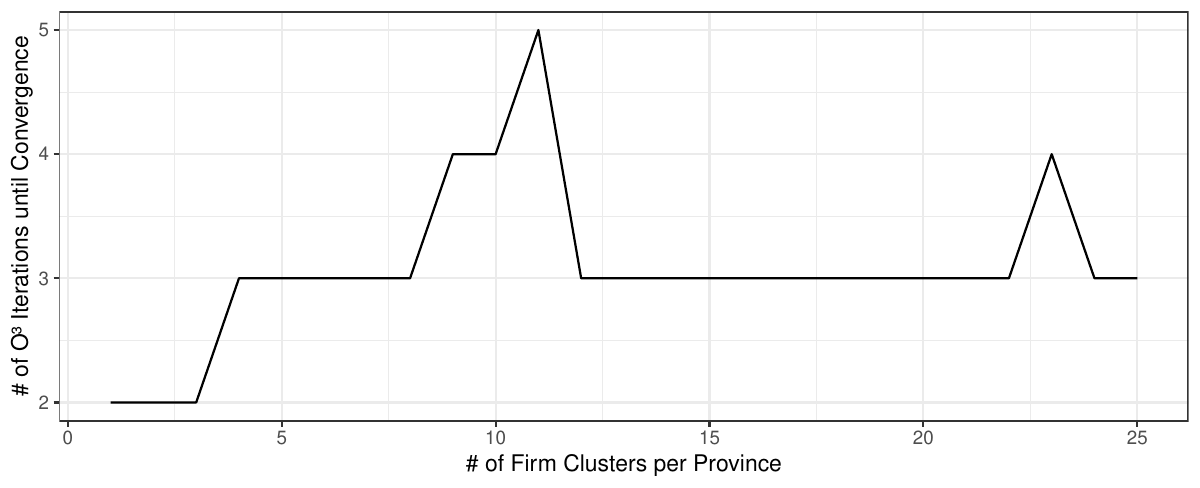}
    \caption{In this figure, we plot the number of O\textsuperscript{3} Algorithm iterations required for the O\textsuperscript{3} Algorithm for $r = 2$ to converge across different numbers of firm clusters per province.}
    \label{fig:num_o3_iterations_needed_plot}
\end{figure}

\begin{figure}[H]
    \centering
    \includegraphics[width=\textwidth]{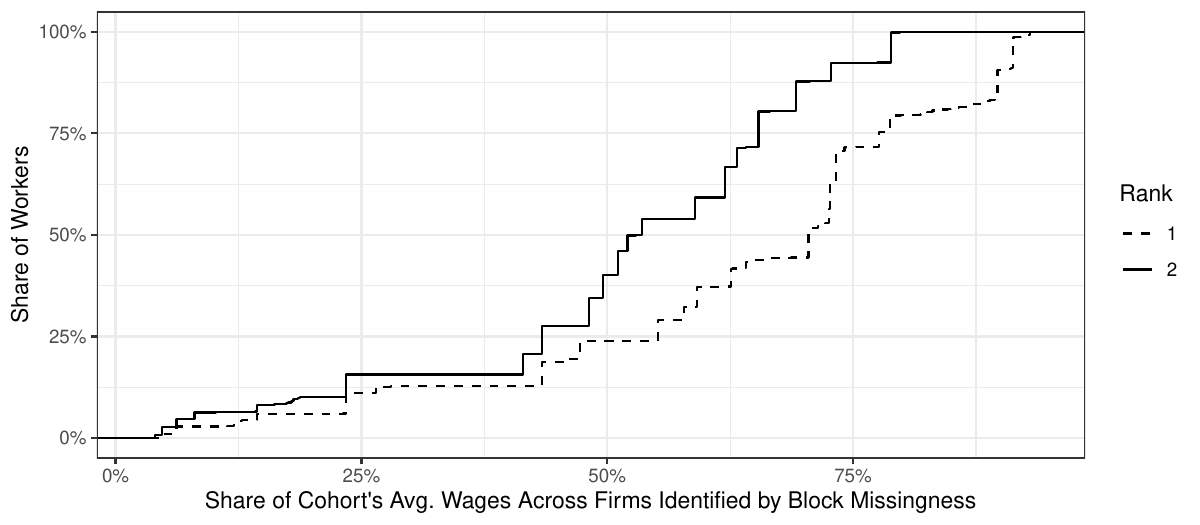}
    \caption{In this figure, we plot the cumulative distribution functions (CDFs) of the shares of workers' outcome means that can be identified using an embedded block missingness pattern for $r \in \{1, 2\}$; outcomes are weighted by the number of firms in the cluster corresponding to that outcome. For example, according to the figure, less than 50\% of workers when $r = 1$ and just over 25\% of workers when $r = 2$ have at most 50\% of their firm-cluster-size-weighted counterfactual outcome means identified using embedded block missingness patterns.}
    \label{fig:block_missingness_id_firm_weighted_plot}
\end{figure}

%% file: figures/result_summaries/snippets/chosen_k.txt
3

%% file: figures/result_summaries/tables/panel_summary_stats_start_year=1998_end_year=2001_year_cluster_size=2_min_cohort_size=75_k=3.tex
\begin{tabular}[t]{lcccc}
\toprule
Statistic & Full Panel & \makecell{Always-Present\\Workers, Firms} & \makecell{Cohorts Above\\Min. Size} & \makecell{Largest Super\\Cohort}\\
\midrule
Num. Employment Spells & 3,561,606 & 2,649,634 & 2,249,271 & 2,249,271\\
Num. Workers & 1,699,187 & 1,085,450 & 1,032,650 & 1,032,650\\
Num. Firms & 165,806 & 114,828 & 114,828 & 114,828\\
\% Workers w/ 2 Spells & 45.5\% & 71.1\% & 83.9\% & 83.9\%\\
\% Workers w/ 3 Spells & 16.1\% & 19.0\% & 14.4\% & 14.4\%\\
\% Workers w/ 4 Spells & 5.7\% & 6.5\% & 1.7\% & 1.7\%\\
\% Workers w/ 5+ Spells & 3.2\% & 3.4\% & 0.0\% & 0.0\%\\
\bottomrule
\end{tabular}

%% file: sections/supplemental_proofs.tex
\subsection{Proof of Theorem \ref{thm:graphical_bound_lambda_r+1}}\label{proof:thm:graphical_bound_lambda_r+1}

We start by presenting an alternative expression of the APM.
\begin{lemma}\label{lem:alternative_expression_APM}
 Let $\Theta\in \R^{T\times (T-r)}$ denote an orthonormal basis for the null space of $\Gamma$ and $\Theta_c \in \R^{(T-T_c)\times (T-r)}$ the submatrix with rows corresponding to $[T]\setminus\mathcal{T}_c$. Assume that $\Gamma$ is in the general position condition. Then
\[A = \Theta \left(\frac{1}{C}\sum_{c}(I_{T-r} - \Pi(\Theta_c')) \right)\Theta'.\]
\end{lemma}

Lemma \ref{lem:alternative_expression_APM} directly implies $A$ has $r$ zero eigenvalues and $\Gamma$ spans its eigenspace. Further, the $(r+1)$-th smallest eigenvalue is given by the minimal eigenvalue of 
\[\frac{1}{C}\sum_{c}(I_{T-r} - \Pi(\Theta_c')).\]

Fix any $K$ and let $\mathcal{I}_c = \mathcal{T}_c^c$. Then $\Theta_c = \Theta_{\mathcal{I}_c}$. For any nice path $(c_1, \ldots, c_K)$ of length $K \ge 1$, there exists $L\in [K-1]$ such that
\[\bigg|\left(\bigcap_{j=1}^{L}\mathcal{I}_j\right) \cup \left(\bigcap_{j=L+1}^{K}\mathcal{I}_j\right)\bigg| = T - \bigg|\left(\bigcup_{c=1}^{L}\mathcal{T}_c\right) \cap \left(\bigcup_{c=L+1}^{K}\mathcal{T}_c\right)\bigg|\le  T-r.\]
Thus, the ordered sequence $(\mathcal{T}_1, \ldots, \mathcal{T}_K)$ is ``nice'' as defined in Lemma \ref{lem:product_multiple_projection_matrices}. When $(c_1, \ldots, c_K)$ is complete, 
\[\bigcap_{k=1}^{K}\mathcal{I}_c = \left(\bigcup_{k=1}^K \mathcal{T}_c\right)^c = \emptyset.\]
By Lemma \ref{lem:product_multiple_projection_matrices} and Lemma \ref{lem:general_position}, for any nice and complete path $(c_1, \ldots, c_K)$,
\[\bigg\|\Pi(\Theta_{c_1}')\Pi(\Theta_{c_2}')\cdots \Pi(\Theta_{c_K}')\bigg\|_\text{op}\le  1 - \sigma_0^{2(K-1)}.\]
On the other hand, for any path $(c_1, \ldots, c_K)$, 
\[\bigg\|\Pi(\Theta_{c_1}')\Pi(\Theta_{c_2}')\cdots \Pi(\Theta_{c_K}')\bigg\|_\text{op}\le 1.\]
Therefore, 
\begin{align*}
&\Bigg\|\sum_c \Pi(\Theta_c')
\Bigg\|_\text{op}^K = \Bigg\|\left( \sum_c \Pi(\Theta_c')\right)
^K\Bigg\|_\text{op} = \Bigg\|\sum_{c_1, \ldots, c_K} \Pi(\Theta_{c_1}')\Pi(\Theta_{c_2}')\cdots \Pi(\Theta_{c_K}')\Bigg\|_\text{op}\\
& \le \sum_{c_1, \ldots, c_K} \bigg\|\Pi(\Theta_{c_1}')\Pi(\Theta_{c_2}')\cdots \Pi(\Theta_{c_K}')\bigg\|_\text{op}\\
& \le (C^K - N_K) + N_K (1 - \sigma_0^{2(K-1)})\\
& = C^K - N_K \sigma_0^{2(K-1)}.
\end{align*}
Thus, 
\[\lambda_{\min}\left(\frac{1}{C}\sum_{c}(I_{T-r} - \Pi(\Theta_c'))\right) \ge 1 - \Bigg\|\frac{1}{C}\sum_c \Pi(\Theta_c')
\Bigg\|_\text{op}\ge 1 - \left(1 - C^{-K} N_K \sigma_0^{2(K-1)}\right)^{1/K}.\]
The second inequalilty follows from Bernoulli's inequality, which implies
\[\left(1 - \frac{N_K\sigma_0^{2(K-1)}}{C^{K}}\right)^{1/K} \le 1 - \frac{N_K}{C^K} \frac{\sigma_0^{2(K-1)}}{K}.\]

\subsubsection{Proof of Lemma \ref{lem:alternative_expression_APM}}
By Lemma \ref{lem:general_position}, $\Theta$ is in the general position. Since $T_c \ge r$, $T - T_c \le T - r$ and $\Theta_c$ has full row rank. It remains to prove that 
\begin{equation}\label{eq:APM_component}
E_c - \Pi(E_c \Gamma) = \Theta (I_{T-r} - \Pi(\Theta_c')) \Theta'.
\end{equation}
Let $\bar{\Gamma}_c\in \R^{(T-T_c)\times r}$ denote the row complement of $\Gamma_c$ in $\Gamma$ and $\bar{\Theta}_c\in \R^{T_c\times (T-r)}$ the row complement of $\Theta_c$ in $\Theta$. Without loss of generality, we assume that $\mathcal{T}_r = \{1, \ldots, T_r\}$ and $\Gamma$ has orthonormal columns, i.e., $\Gamma' \Gamma = I_{r}$. Since $\Gamma$ and $\Theta$ jointly form a complete orthonormal basis, 
\[\begin{bmatrix}
\Gamma_c & \bar{\Theta}_c \\
\bar{\Gamma}_c & \Theta_c
\end{bmatrix}\begin{bmatrix}
\Gamma_c' & \bar{\Gamma}_c'\\
\bar{\Theta}_c' &  \Theta_c'
\end{bmatrix} = I_T\]
which implies 
\begin{align}
\Gamma_c\Gamma_c' + \bar{\Theta}_c \bar{\Theta}_c' = I_{T_c}, \quad \bar{\Gamma}_c\bar{\Gamma}_c' + \Theta_c \Theta_c' = I_{T-T_c}, \quad \Gamma_c\bar{\Gamma}_c' + \bar{\Theta}_c \Theta_c' = 0_{T_c\times (T-T_c)}.\label{eq:Gamma_Theta}
\end{align}
The LHS of \eqref{eq:APM_component} can be written as 
\[\begin{bmatrix}
I_{T_c} - \Gamma_c (\Gamma_c' \Gamma_c)^{-1}\Gamma_c & 0\\
0 & 0
\end{bmatrix},\]
and the RHS is given by
\[\begin{bmatrix}
\bar{\Theta}_c \\ \Theta_c
\end{bmatrix}(I_{T-r} - \Pi(\Theta_c'))\begin{bmatrix}
\bar{\Theta}_c' & \Theta_c'
\end{bmatrix} = \begin{bmatrix}
\bar{\Theta}_c (I_{T-r} - \Pi(\Theta_c')) \bar{\Theta}_c' & 0\\
0 & 0
\end{bmatrix}.\]
Thus, it is left to prove that 
\[\bar{\Theta}_c (I_{T-r} - \Pi(\Theta_c')) \bar{\Theta}_c' = I_{T_c} - \Gamma_c (\Gamma_c' \Gamma_c)^{-1}\Gamma_c'.\]
By \eqref{eq:Gamma_Theta}, the LHS can be reexpressed as 
\begin{align}
&\bar{\Theta}_c (I_{T-r} - \Pi(\Theta_c')) \bar{\Theta}_c'  = \bar{\Theta}_c\bar{\Theta}_c' - \bar{\Theta}_c \Theta_c' (\Theta_c \Theta_c')^{-1}\Theta_c \bar{\Theta}_c'\\
& = I_{T_c} - \Gamma_c \Gamma_c' - \Gamma_c \bar{\Gamma}_c' (I - \bar{\Gamma}_c \bar{\Gamma}_c')^{-1}\bar{\Gamma}_c \Gamma_c'\\
& = I_{T_c} - \Gamma_c \left( I_{T-T_c} -\bar{\Gamma}_c' (I - \bar{\Gamma}_c \bar{\Gamma}_c')^{-1}\bar{\Gamma}_c \right) \Gamma_c'\label{eq:intermediate_Theta_Gamma}
\end{align}
Since $\Gamma \Gamma' = I_r$, 
\[\Gamma_c' \Gamma_c + \bar{\Gamma}_c' \bar{\Gamma}_c = I_r.\]
As $\Gamma_c$ has full column rank, 
\[\lambda_{\max}(\bar{\Gamma}_c \bar{\Gamma}_c') = \lambda_{\max}(\bar{\Gamma}_c' \bar{\Gamma}_c) \le 1 - \lambda_{\min}(\Gamma_c' \Gamma_c) < 1.\]
Thus, 
\begin{align*}
&I_{T-T_c} -\bar{\Gamma}_c' (I - \bar{\Gamma}_c \bar{\Gamma}_c')^{-1}\bar{\Gamma}_c  = I_{T-T_c} -\bar{\Gamma}_c' \left\{\sum_{j\ge 0} (-1)^j(\bar{\Gamma}_c \bar{\Gamma}_c')^{j}\right\}\bar{\Gamma}_c \\
& = I_{T-T_c} + \sum_{j\ge 1}(-1)^{j}(\bar{\Gamma}_c' \bar{\Gamma}_c)^{j}= \sum_{j\ge 0}(-1)^{j}(\bar{\Gamma}_c' \bar{\Gamma}_c)^{j}\\
& = (I - \bar{\Gamma}_c' \bar{\Gamma}_c)^{-1} = (\Gamma_c' \Gamma_c)^{-1}.
\end{align*}
Plugging this back in \eqref{eq:intermediate_Theta_Gamma} yields the desired result.

\subsubsection{Technical lemmas}
Throughout this section we denote by $\lambda_{\min/\max}$ and $\sigma_{\min/\max}$ the minimal/maximal eigenvalue and singular value, respectively. 
\begin{lemma}\label{lem:general_position}
Let $[\Gamma; \Theta]$ be the matrix an orthonormal basis of $\R^{T}$ where $\Gamma\in \R^{T\times r}$ and $\Theta\in \R^{T\times (T-r)}$. 
\[\min_{\mathcal{J}: \mathcal{J}\subset [T], |\mathcal{J}| = T-r}\sigma_{\min}(\Theta_{\mathcal{J}}) = \min_{\mathcal{I}: \mathcal{I}\subset [T], |\mathcal{J}| = r}\sigma_{\min}(\Gamma_{\mathcal{I}}).\]
In particular, if $\Gamma$ is in the general position, in the sense that $\Gamma_\mathcal{I}$ for any $\mathcal{I}\subset [T]$ and $|\mathcal{I}| = r$, $\Theta$ is also in the general position.
\end{lemma}
\begin{proof}
Since $[\Gamma; \Theta]$ forms an orthonormal basis, 
\[\Gamma\Gamma' + \Theta \Theta' = I_T, \quad \Gamma'\Gamma = I_r, \quad \Theta'\Theta = I_{T-r}.\]
For any $\mathcal{I} \subset [T]$ with $|\mathcal{I}| = r$,
\[\Gamma_\mathcal{I} \Gamma_\mathcal{I}' + \Theta_\mathcal{I} \Theta_\mathcal{I}' = I_{r}.\]
This implies 
\[\lambda_{\min}(\Gamma_\mathcal{I} \Gamma_\mathcal{I}') = 1 - \lambda_{\max}(\Theta_\mathcal{I} \Theta_\mathcal{I}') = 1 - \lambda_{\max}(\Theta_\mathcal{I}'\Theta_\mathcal{I}).\]
Note that 
\[1 - \lambda_{\max}(\Theta_\mathcal{I}'\Theta_\mathcal{I}) = \lambda_{\min}(I_{T-r} - \Theta_\mathcal{I}'\Theta_\mathcal{I}) = \lambda_{\min}(\Theta_{\mathcal{I}^c}'\Theta_{\mathcal{I}^c}).\]
As a result,
\[\lambda_{\min}(\Gamma_\mathcal{I} \Gamma_\mathcal{I}') = \lambda_{\min}(\Theta_{\mathcal{I}^c}'\Theta_{\mathcal{I}^c}).\]
Taking minimum over all sets $I$ completes the proof.
\end{proof}

\begin{lemma}\label{lem:projection_matrix_B1B2}
Let $B_1\in \R^{M\times d_1}$ and $B_2\in \R^{M\times d_2}$ be two matrices with full column rank and $d_1 + d_2 \le M$. Then 
\[\Pi([B_1; B_2]) = \Pi(B_1) + \Pi\left(\left\{I - \Pi(B_1)\right\}B_2\right),\]
where $[B_1; B_2]\in \R^{M\times (d_1 + d_2)}$ denotes the row concatenation of $B_1$ and $B_2$.
\end{lemma}
\begin{proof}
By definition, 
\[[B_1; (I - \Pi(B_1))B_2] = [B_1; B_2]\begin{bmatrix}
I_{d_1} & -(B_1'B_1)^{-1}B_1'B_2\\
0_{d_2\times d_1} & I_{d_2}
\end{bmatrix}.\]
Since the projection matrix is invariant to multiplication by an invertible matrix, 
\begin{align*}
&\Pi([B_1; B_2]) = \Pi([B_1; (I - \Pi(B_1))B_2])\\
& = [B_1; (I - \Pi(B_1))B_2] 
\begin{bmatrix}
B_1'B_1 & 0_{d_1}\\
0_{d_2} & B_2'(I - \Pi(B_1))B_2
\end{bmatrix}
\begin{bmatrix}
B_1'\\
B_2'(I - \Pi(B_1))
\end{bmatrix}\\
& = \Pi(B_1) + \Pi\left(\left\{I - \Pi(B_1)\right\}B_2\right).
\end{align*}
\end{proof}

\begin{lemma}\label{lem:IAAI}
Let $A\in \R^{d_1\times d_2}$ be a matrix. Then 
$$\lambda_{\min}\left(\begin{bmatrix}
I_{d_1} & A\\
A' & I_{d_2}
\end{bmatrix}\right) = 1 - \|A\|_\text{op}$$
\end{lemma}
\begin{proof}
Without loss of generality, we assume that $d_1 \ge d_2$. Let 
$U\Sigma V'$ be a type of SVD of $A$, where $U\in \R^{d_1\times d_1}, V\in \R^{d_2\times d_2}$ are orthogonal matrices and $\Sigma\in \R^{d_1\times d_2}$ is a tall rectangular diagonal matrix in the form of 
\[\Sigma = \begin{bmatrix}
D \\ 0_{d_2\times (d_1 - d_2)}
\end{bmatrix}, \quad D = \mathrm{diag}(\sigma_1(A), \ldots, \sigma_{d_2}(A)),\] 
where $\sigma_j(\cdot)$ denotes the $j$-th largest singular value. 
Then 
\[
\begin{bmatrix}
I_{d_1} & A\\
A' & I_{d_2}
\end{bmatrix}
= \begin{bmatrix}
UU' & U\Sigma V'\\
V\Sigma' U' & VV'
\end{bmatrix}
= \begin{bmatrix}
U & 0\\
0 & V
\end{bmatrix}\begin{bmatrix}
I_{d_1} & \Sigma\\
\Sigma' & I_{d_2}
\end{bmatrix}\begin{bmatrix}
U' & 0\\
0 & V'
\end{bmatrix}.\]
Both $U$ and $V$ are orthogonal, the matrix $\begin{bmatrix}
U & 0\\
0 & V
\end{bmatrix}$ is orthogonal as well. Thus, 
\[\lambda_{\min}\left(\begin{bmatrix}
I_{d_1} & A\\
A' & I_{d_2}
\end{bmatrix}\right) = \lambda_{\min}\left(\begin{bmatrix}
I_{d_1} & \Sigma\\
\Sigma' & I_{d_2}
\end{bmatrix}\right).\]
By definition, 
\[\begin{bmatrix}
I_{d_1} & \tilde{\Sigma}\\
\tilde{\Sigma}' & I_{d_2}
\end{bmatrix} = \begin{bmatrix}
I_{d_2} & 0 &  D\\
0 & I_{d_1 - d_2} & 0 \\
D & 0 & I_{d_2}
\end{bmatrix}.\]
Thus, 
\[\lambda_{\min}\left(\begin{bmatrix}
I_{d_1} & \Sigma\\
\Sigma' & I_{d_2}
\end{bmatrix}\right) = \min\left\{1, \lambda_{\min}\left(\begin{bmatrix}
I_{d_2} & D\\
D & I_{d_2}
\end{bmatrix}\right)\right\}.\]
Rearranging the rows and columns, we can rewrite $\begin{bmatrix}
I_{d_2} & D\\
D & I_{d_2}
\end{bmatrix}$ as
\[\begin{bmatrix}
\begin{bmatrix}
1 & \sigma_1(A)\\
\sigma_1(A) & 1\\
\end{bmatrix} & & \\
& \begin{bmatrix}
1 & \sigma_2(A)\\
\sigma_2(A) & 1\\
\end{bmatrix} & & \\
& & \ddots & \\
& & & \begin{bmatrix}
1 & \sigma_{d_2}(A)\\
\sigma_{d_2}(A) & 1\\
\end{bmatrix}\\
\end{bmatrix}\]
where $\sigma_j(\cdot)$ denotes the $j$-th largest singular value. 
It is easy to check that the two eigenvalues of the $2$-by-$2$ matrix $\begin{bmatrix}
1 & a\\
a & 1
\end{bmatrix}$ are $1 \pm a$. Therefore, the eigenvalues of $\begin{bmatrix}
I_{d_2} & D\\
D & I_{d_2}
\end{bmatrix}$ are $1 \pm \sigma_j(A), j = 1,\ldots,d_2$. In particular, 
\[\lambda_{\min}\left(\begin{bmatrix}
I_{d_1} & A\\
A' & I_{d_2}
\end{bmatrix}\right) = 1 - \sigma_1(A) = 1 - \|A\|_\text{op}.\]
\end{proof}

\begin{lemma}\label{lem:product_two_projection_matrix}
Let $B_1\in \R^{M\times d_1}$ and $B_2\in \R^{M\times d_2}$ be two tall matrices (i.e., $M\ge \max\{d_1, d_2\}$). Then 
\[\|\Pi(B_1)\Pi(B_2)\|_\text{op}\le1 - \frac{\sigma_{\min}^2([B_1; B_2])}{\sigma_{\max}^2([B_1; B_2])}.\]
\end{lemma}
\begin{proof}
Let $B = [B_1; B_2]$. If $d_1 + d_2 > M$, $\sigma_{\min}^2(B) = 0$ and the result holds trivially. Throughout the rest of the proof we assume $d_1 + d_2 \le M$. Without loss of generality, we assume $d_1 \ge d_2$ and both $B_1$ and $B_2$ have full column rank. For $k=1,2$, let $U_k \Sigma_k V_k'$ be a type of SVD of $B_k$ where $U_k \in \R^{M\times d_k}, V_k \in \R^{d_k \times d_k}$ are orthogonal matrices and $\Sigma_k = \mathrm{diag}(\sigma_{1}(B_k), \ldots, \sigma_{d_k}(B_k))\in \R^{d_k\times d_k}$ is a diagonal matrix. Then 
\[\Pi(B_1) = U_1 U_1', \quad \Pi(B_2) = U_2 U_2',\]
and 
\begin{equation}\label{eq:PiB1B2_U1U2op}
\|\Pi(B_1)\Pi(B_2)\|_\text{op} = \|U_1 U_1' U_2 U_2'\|_\text{op} \le \|U_1\|_\text{op} \|U_1' U_2 \|_\text{op} \|U_2\|_\text{op} = \|U_1' U_2 \|_\text{op}.
\end{equation}
Let $U = [U_1; U_2]$. Then
\[U'U = \begin{bmatrix}
I_{d_1} & U_1'U_2\\
U_2'U_1 & I_{d_2}
\end{bmatrix}.\]
By Lemma \ref{lem:IAAI},
\begin{equation}\label{eq:sigmamin_U'U_U_1'U_2}
\sigma_{\min}^2(U) = \sigma_{\min}(U'U) = 1 - \|U_1'U_2\|_\text{op}.
\end{equation}
Lastly, by definition,
\[B = U\begin{bmatrix}
\Sigma_1 V_1' & 0\\
0 & \Sigma_2 V_2' 
\end{bmatrix}.\]
By the multiplicative Weyl's inequality \citep[e.g.][Theorem 3.3.16]{horn2012matrix}, 
\begin{align}
\sigma_{\min}(B) &\le \sigma_{\min}(U)\sigma_{\max}\left(\begin{bmatrix}
\Sigma_1 V_1' & 0\\
0 & \Sigma_2 V_2' 
\end{bmatrix}\right) \\
& \le \sigma_{\min}(U)\max\{\sigma_{\max}(B_1), \sigma_{\max}(B_2)\}\\
& \le \sigma_{\min}(U)\sigma_{\max}(B)\label{eq:sigmaminB_sigmaminU}
\end{align}
Combining \eqref{eq:PiB1B2_U1U2op}, \eqref{eq:sigmamin_U'U_U_1'U_2}, and \eqref{eq:sigmaminB_sigmaminU}, we obtain that
\[\|\Pi(B_1)\Pi(B_2)\|_\text{op} =  1 - \sigma_{\min}^2(U)\le 1 - \frac{\sigma_{\min}^2(B)}{\sigma_{\max}^2(B)}\]
\end{proof}

\begin{lemma}\label{lem:product_two_residual_projection_matrices}
Let $B_0\in \R^{M\times d_0}, B_1\in \R^{M\times d_1},$ and $B_2\in \R^{M\times d_2}$ be three tall matrices. Further, let 
\[\tilde{B}_1 = (I_{M} - \Pi(B_0))B_1, \quad \tilde{B}_2 = (I_{M} - \Pi(B_0))B_1.\]
Then 
\[\|\Pi(\tilde{B}_1)\Pi(\tilde{B}_2)\|_\text{op}\le1 - \frac{\sigma_{\min}^2([B_0; B_1; B_2])}{\sigma_{\max}^2([B_0; B_1; B_2])}\]
\end{lemma}
\begin{proof}
Similar to the proof of Lemma \ref{lem:product_two_projection_matrix}, we assume $d_0 + d_1 + d_2\le M$. Let $B = [B_0; B_1; B_2]$ and $\tilde{B} = [\tilde{B}_1, \tilde{B}_2]$. By Lemma \ref{lem:IAAI}, 
\[\|\Pi(\tilde{B}_1)\Pi(\tilde{B}_2)\|_\text{op}\le1 - \frac{\sigma_{\min}^2(\tilde{B})}{\sigma_{\max}^2(\tilde{B})}.\]
It suffices to prove that 
\[\sigma_{\min}(\tilde{B})\ge \sigma_{\min}(B), \quad \sigma_{\max}(\tilde{B})\le \sigma_{\max}(B).\]
Note that 
\[\tilde{B} = B \begin{bmatrix}
-(B_0'B_0)^{-1}B_0'B_1 & -(B_0'B_0)^{-1}B_0'B_2\\
I_{d_2} & 0_{d_2\times d_3}\\
0_{d_3\times d_2} & I_{d_3}
\end{bmatrix}\triangleq B \begin{bmatrix}
C\\
I_{d_2+d_3}
\end{bmatrix}.\]
Then 
\[\tilde{B}'\tilde{B} = \begin{bmatrix}
C' &
I_{d_2+d_3}
\end{bmatrix}B'B \begin{bmatrix}
C\\
I_{d_2+d_3}
\end{bmatrix}\succeq \sigma_{\min}^2(B)\begin{bmatrix}
C' &
I_{d_2+d_3}
\end{bmatrix}\begin{bmatrix}
C\\
I_{d_2+d_3}
\end{bmatrix} = \sigma_{\min}^2(B)(C'C + I_{d_2+d_3}).\]
Since $C'C$ is positive semidefinite, 
\[\sigma_{\min}(\tilde{B})\ge \sigma_{\min}(B).\]
Alternatively, we have 
\[[0_{M\times d_0}; \tilde{B}] = (I - \Pi(B_0)) B.\]
This implies that 
\[\sigma_{\max}([0_{M\times d_0}; \tilde{B}]) \le \|I - \Pi(B_0)\|_\text{op}\cdot \sigma_{\max}(B) = \sigma_{\max}(B).\]
The proof is completed by noting that 
\[\sigma_{\max}([0_{M\times d_0}; \tilde{B}]) = \sigma_{\max}(\tilde{B}).\]
\end{proof}

\begin{lemma}\label{lem:product_multiple_projection_matrices}
Let $\Theta\in \R^{M\times d}$ be a tall orthogonal matrix (i.e., $d\le M$) with $\Theta_\mathcal{I}\in \R^{|\mathcal{I}|\times d}$ denoting the submatrix formed by the rows of $\Theta$ corresponding to $\mathcal{I}$. Assume that there exists a constant $\sigma_0$ such that
\begin{equation}\label{eq:general_position_singular_value}
\sigma_{\min}(\Theta_\mathcal{I})\ge \sigma_0, \quad \text{for any }\mathcal{I}\subset [M] \text{ and } |\mathcal{I}| = d.
\end{equation}
Given $K\ge 1$. An ordered sequence $(\mathcal{I}_1, \ldots, \mathcal{I}_K)$ of subsets of $[M]$ is ``nice'' if $K = 1$ or there exists $L \in [K-1]$ such that
\begin{equation}\label{eq:nice_path_proof}
\bigg|\left(\bigcap_{j=1}^{L}\mathcal{I}_j\right) \cup \left(\bigcap_{j=L+1}^{K}\mathcal{I}_j\right)\bigg|\le d,
\end{equation}
and both $(\mathcal{I}_1, \ldots, \mathcal{I}_L)$ and $(\mathcal{I}_{L+1}, \ldots, \mathcal{I}_{K})$ are ``nice''.
Then 
\[\prod_{k=1}^{K}\Pi(\Theta_{\mathcal{I}_k}') = \Pi(\Theta_{\mathcal{I}_1}')\Pi(\Theta_{\mathcal{I}_2}')\cdots \Pi(\Theta_{\mathcal{I}_K}')\preceq \Pi(\Theta_{\cap_{j=1}^{K} \mathcal{I}_j}') + (I_d - \Pi(\Theta_{\cap_{j=1}^{K} \mathcal{I}_j}'))Q (I_M - \Pi(\Theta_{\cap_{j=1}^{K} \mathcal{I}_j}'))\]
for some matrix $Q\in \R^{M\times M}$ with 
\[\|Q\|_\text{op}\le \left(1 - \left\{\sigma_0^2(2 - \sigma_0^2)\right\}^{K-1}\right)^{1/2}\le 1 - \sigma_0^{2(K-1)}.\]
\end{lemma}
\begin{proof}
We prove the first inequality by induction on $K$. To prove the second inequality, we note that the AM-GM inequality implies  
\[(2 - \sigma_0^2)^{K-1} + (\sigma_0^2)^{K-1}\ge 2.\]
As a result, 
\[1 - \left\{\sigma_0^2(2 - \sigma_0^2)\right\}^{K-1}\le 1 - \sigma_0^{2(K-1)}(2 - \sigma_0^{2(K-1)}) = \left(1 - \sigma_0^{2(K-1)}\right)^2.\]

When $K= 1$, the decomposition holds with $Q = 0$. When $K = 2$, let $\mathcal{J} = \mathcal{I}_1 \cap \mathcal{I}_2$. By definition, $(\mathcal{I}_1, \mathcal{I}_2)$ is ``nice'' iff $|\mathcal{I}_1 \cup \mathcal{I}_2|\le d$. Then \eqref{eq:general_position_singular_value} implies that both $\Theta_{\mathcal{I}_1}$ and $\Theta_{\mathcal{I}_2}$ have full row rank and 
\begin{equation}\label{eq:sigmamin_I1I2}
\sigma_{\min}([\Theta_{\mathcal{I}_1\setminus \mathcal{J}}'; \Theta_{\mathcal{I}_2\setminus \mathcal{J}}'; \Theta_\mathcal{J}'])\ge \sigma_0.
\end{equation}
Since $\Theta$ is orthogonal, 
\begin{equation}\label{eq:sigmamax_I1I2}
\sigma_{\max}([\Theta_{\mathcal{I}_1\setminus \mathcal{J}}'; \Theta_{\mathcal{I}_2\setminus \mathcal{J}}'; \Theta_\mathcal{J}'])\le \sigma_{\max}(\Theta') = 1.
\end{equation}
By Lemma \ref{lem:projection_matrix_B1B2}
\[\Pi(\Theta_{\mathcal{I}_1}') = \Pi(\Theta_\mathcal{J}') + \Pi\left(\left\{I - \Pi(\Theta_\mathcal{J}')\right\}\Theta_{\mathcal{I}_1\setminus \mathcal{J}}'\right)\]
and 
\[\Pi(\Theta_{\mathcal{I}_2}') = \Pi(\Theta_\mathcal{J}') + \Pi\left(\left\{I - \Pi(\Theta_\mathcal{J}')\right\}\Theta_{\mathcal{I}_2\setminus \mathcal{J}}'\right).\]
Then 
\[\Pi(\Theta_{\mathcal{I}_1}')\Pi(\Theta_{\mathcal{I}_2}') = \Pi(\Theta_\mathcal{J}') +  \Pi\left(\left\{I - \Pi(\Theta_\mathcal{J}')\right\}\Theta_{\mathcal{I}_1\setminus \mathcal{J}}'\right)\Pi\left(\left\{I - \Pi(\Theta_\mathcal{J}')\right\}\Theta_{\mathcal{I}_2\setminus \mathcal{J}}'\right).\]
It is easy to see that 
\[\Pi\left(\left\{I - \Pi(\Theta_\mathcal{J}')\right\}\Theta_{\mathcal{I}_1\setminus \mathcal{J}}'\right) = (I - \Pi(\Theta_\mathcal{J}'))\Pi\left(\left\{I - \Pi(\Theta_\mathcal{J}')\right\}\Theta_{\mathcal{I}_1\setminus \mathcal{J}}'\right)\]
and 
\[\Pi(\left\{I - \Pi(\Theta_\mathcal{J}')\right\}\Theta_{\mathcal{I}_1\setminus \mathcal{J}}') = \Pi(\left\{I - \Pi(\Theta_\mathcal{J}')\right\}\Theta_{\mathcal{I}_1\setminus \mathcal{J}}')(I - \Pi(\Theta_\mathcal{J}')).\]
Let $Q = \Pi(\{I - \Pi(\Theta_\mathcal{J}')\}\Theta_{\mathcal{I}_1\setminus \mathcal{J}}')\Pi((I - \Pi(\Theta_\mathcal{J}'))\Theta_{\mathcal{I}_2\setminus \mathcal{J}}')$. Then 
\[\Pi(\Theta_{\mathcal{I}_1}')\Pi(\Theta_{\mathcal{I}_2}') = \Pi(\Theta_\mathcal{J}') + (I - \Pi(\Theta_\mathcal{J}'))Q (I - \Pi(\Theta_\mathcal{J}')).\]
By \eqref{eq:sigmamin_I1I2}, \eqref{eq:sigmamax_I1I2}, and Lemma \ref{lem:product_two_residual_projection_matrices}, 
\[\|Q\|_\text{op}^2\le (1 - \sigma_0^2)^2 = 1 - \{\sigma_0^2(2 - \sigma_0^2)\}.\]

Now assume the result holds for $K-1$. For notational convenience, we write 
\[\gamma_k = 1 - \{\sigma_0^2(2 - \sigma_0^2)\}^{k-1}, \quad k\ge 1.\]
The induction hypothesis implies that 
\begin{equation}\label{eq:product_left}
\prod_{j=1}^{L}\Pi(\Theta_{\mathcal{I}_j}') = \Pi(\Theta_{\mathcal{J}_1}') + (I - \Pi(\Theta_{\mathcal{J}_1}'))Q_1 (I - \Pi(\Theta_{\mathcal{J}_1}')),
\end{equation}
and 
\begin{equation}\label{eq:product_right}
\prod_{j=L+1}^{K}\Pi(\Theta_{\mathcal{I}_j}') = \Pi(\Theta_{\mathcal{J}_2}') + (I - \Pi(\Theta_{\mathcal{J}_2}'))Q_2 (I - \Pi(\Theta_{\mathcal{J}_2}')),
\end{equation}
where        
\[\mathcal{J}_1 = \cap_{j=1}^{L}\mathcal{I}_j, \quad \mathcal{J}_2 = \cap_{j=L+1}^{K}\mathcal{I}_j,\] 
and 
\[\|Q_1\|_\text{op}^2\le \gamma_L, \quad \|Q_2\|_\text{op}^2 \le \gamma_{K-L}.\] 
Write $\mathcal{J}_0$ for $\mathcal{J}_1\cap \mathcal{J}_2$. By Lemma \ref{lem:projection_matrix_B1B2}, 
\[\Pi(\Theta_{\mathcal{J}_1}') = \Pi(\Theta_{\mathcal{J}_0}') + \Pi(\{I - \Pi(\Theta_{\mathcal{J}_0}')\}\Theta_{\mathcal{J}_1\setminus \mathcal{J}_0}'),\]
and 
\[\Pi(\Theta_{\mathcal{J}_2}') = \Pi(\Theta_{\mathcal{J}_0}') + \Pi(\{I - \Pi(\Theta_{\mathcal{J}_0}')\}\Theta_{\mathcal{J}_2\setminus \mathcal{J}_0}').\]
For notational convenience, we write $P_0, P_1, P_2, \tilde{P}_1, \tilde{P}_2$ for $\Pi(\Theta_{\mathcal{J}_0}'), \Pi(\Theta_{\mathcal{J}_1}'), \Pi(\Theta_{\mathcal{J}_2}')$, \\ $\Pi(\{I - \Pi(\Theta_{\mathcal{J}_0}')\}\Theta_{\mathcal{J}_1\setminus \mathcal{J}_0}'), \Pi(\{I - \Pi(\Theta_{\mathcal{J}_0}')\}\Theta_{\mathcal{J}_2\setminus \mathcal{J}_0}')$, respectively. 
Then \eqref{eq:product_left} and \eqref{eq:product_right} can be rewritten as 
\begin{equation}\label{eq:product_left_simplify}
\prod_{j=1}^{L}\Pi(\Theta_{\mathcal{I}_j}') = P_0 + \tilde{P}_1 + (I - P_1)Q_1 (I - P_1),
\end{equation}
and 
\begin{equation}\label{eq:product_right_simplify}
\prod_{j=L+1}^{K}\Pi(\Theta_{\mathcal{I}_j}') = P_0 + \tilde{P}_2 + (I - P_2)Q_2 (I - P_2). 
\end{equation}
By definition, 
\[P_0(I - P_k) = (I - P_k)P_0 = P_0\tilde{P}_k = \tilde{P}_k P_0 =\tilde{P}_k(I - P_k) = (I-  P_k)\tilde{P}_k = 0,\quad k = 1,2.\] 
As a result, 
\begin{align*}
\prod_{j=1}^K \Pi(\Theta'_{\mathcal{I}_j}) & = P_0 + \left\{\tilde{P}_1 + (I - P_1)Q_1 (I - P_1)\right\}\left\{\tilde{P}_2 + (I - P_2)Q_2 (I - P_2)\right\}\\
& = P_0 + (I - P_0)\left\{\tilde{P}_1 + (I - P_1)Q_1 (I - P_1)\right\}\left\{\tilde{P}_2 + (I - P_2)Q_2 (I - P_2)\right\}(I - P_0)\\
& \triangleq P_0 + (I - P_0)Q_0 (I - P_0).
\end{align*}
Then 
\begin{align*}
Q_0 Q_0' = \left\{\tilde{P}_1 + (I - P_1)Q_1 (I - P_1)\right\}\left\{\tilde{P}_2 + (I - P_2)Q_2 (I - P_2)Q_2'(I - P_2)\right\}\left\{\tilde{P}_1 + (I - P_1)Q_1' (I - P_1)\right\}.
\end{align*}
Since $\sigma_{\max}(Q_2 (I - P_2)Q_2') \le \sigma_{\max}^2(Q_2)\le \gamma_{K-L}$ and $P_2 = \tilde{P}_2 + P_0\succeq \tilde{P}_2$, 
\begin{align}
Q_0 Q_0' &\preceq \left\{\tilde{P}_1 + (I - P_1)Q_1 (I - P_1)\right\}\left\{\tilde{P}_2 + \gamma_{K-L} (I - P_2)\right\}\left\{\tilde{P}_1 + (I - P_1)Q_1' (I - P_1)\right\}\\
&\preceq \left\{\tilde{P}_1 + (I - P_1)Q_1 (I - P_1)\right\}\left\{\tilde{P}_2 + \gamma_{K-L} (I - \tilde{P}_2)\right\}\left\{\tilde{P}_1 + (I - P_1)Q_1' (I - P_1)\right\}\\
& = \left\{\tilde{P}_1 + (I - P_1)Q_1 (I - P_1)\right\}\left\{\gamma_{K-L}I + (1 - \gamma_{K-L}) \tilde{P}_2\right\}\left\{\tilde{P}_1 + (I - P_1)Q_1' (I - P_1)\right\}\\
& = \gamma_{K-L}\left\{\tilde{P}_1 + (I - P_1)Q_1 (I - P_1) Q_1' (I - P_1) \right\}\\
& \quad + (1 - \gamma_{K-L})\left\{\tilde{P}_1 + (I - P_1)Q_1 (I - P_1)\right\}\tilde{P}_2\left\{\tilde{P}_1 + (I - P_1)Q_1' (I - P_1)\right\}.\label{eq:Q_0Q_0'}
\end{align}
We analyze the above two terms separately. For the first term, since $P_1 = \tilde{P}_1 + P_0 \succeq \tilde{P}_1$ and $Q_1 \preceq I$,
\begin{equation}\label{eq:Q_0Q_0'_term1}
\tilde{P}_1 + (I - P_1)Q_1 (I - P_1)Q_1' (I - P_1)\preceq P_1 + (I - P_1) = I.
\end{equation}
For the second term, 
\begin{align*}
& \lambda_{\max}\left(\left\{\tilde{P}_1 + (I - P_1)Q_1 (I - P_1)\right\}\tilde{P}_2\left\{\tilde{P}_1 + (I - P_1)Q_1' (I - P_1)\right\}\right)\\
& =\lambda_{\max}\left(\left\{\tilde{P}_1 + (I - P_1)Q_1 (I - P_1)\right\}\tilde{P}_2 \tilde{P}_2\left\{\tilde{P}_1 + (I - P_1)Q_1' (I - P_1)\right\}\right)\\
& = \lambda_{\max}\left(\tilde{P}_2\left\{\tilde{P}_1 + (I - P_1)Q_1' (I - P_1)\right\}\left\{\tilde{P}_1 + (I - P_1)Q_1 (I - P_1)\right\}\tilde{P}_2\right) \\
& = \lambda_{\max}\left(\tilde{P}_2\left\{\tilde{P}_1 + (I - P_1)Q_1' (I - P_1) Q_1 (I - P_1)\right\}\tilde{P}_2\right).
\end{align*}
Similar to the previous argument, 
\[P_1 \succeq \tilde{P}_1, \quad Q_1' (I - P_1) Q_1\preceq \gamma_L I.\]
This implies 
\begin{align*}
&\lambda_{\max}\left(\tilde{P}_2\left\{\tilde{P}_1 + (I - P_1)Q_1' (I - P_1) Q_1 (I - P_1)\right\}\tilde{P}_2\right)\\
& \le \lambda_{\max}\left(\tilde{P}_2\left\{\tilde{P}_1 + \gamma_L(I - P_1)\right\}\tilde{P}_2\right)\\
& \le \lambda_{\max}\left(\tilde{P}_2\left\{\tilde{P}_1 + \gamma_L(I - \tilde{P}_1)\right\}\tilde{P}_2\right)\\
& = \lambda_{\max}\left(\tilde{P}_2\left\{\gamma_L I + (1 - \gamma_L) \tilde{P}_1)\right\}\tilde{P}_2\right)\\
& = \lambda_{\max}\left(\gamma_L \tilde{P}_2 + (1 - \gamma_L) \tilde{P}_2\tilde{P}_1\tilde{P}_2\right)\\
& \le \lambda_{\max}\left(\gamma_L \tilde{P}_2\right) + \lambda_{\max}\left((1 - \gamma_L) \tilde{P}_2\tilde{P}_1\tilde{P}_2\right)\\
& = \gamma_L + (1 - \gamma_L)\lambda_{\max}\left(\tilde{P}_2\tilde{P}_1\tilde{P}_2\right)\\
& = \gamma_L + (1 - \gamma_L)\lambda_{\max}\left(\tilde{P}_2\tilde{P}_1 \tilde{P}_1\tilde{P}_2\right)\\
& = \gamma_L + (1 - \gamma_L)\sigma_{\max}^2\left(\tilde{P}_1\tilde{P}_2\right)
\end{align*}
By the condition \eqref{eq:nice_path_proof} and Lemma \ref{lem:product_two_residual_projection_matrices}, 
\[\sigma_{\max}\left(\tilde{P}_1\tilde{P}_2\right)\le 1 - \sigma_0^2.\]
Thus, 
\begin{equation}\label{eq:Q_0Q_0'_term2}
\lambda_{\max}\left(\left\{\tilde{P}_1 + (I - P_1)Q_1 (I - P_1)\right\}\tilde{P}_2\left\{\tilde{P}_1 + (I - P_1)Q_1' (I - P_1)\right\}\right)\le \gamma_L + (1 - \gamma_L)(1 - \sigma_0^2)^2.
\end{equation}
Putting \eqref{eq:Q_0Q_0'}, \eqref{eq:Q_0Q_0'_term1}, and \eqref{eq:Q_0Q_0'_term2} together, we obtain that
\begin{align*}
\lambda_{\max}(Q_0 Q_0')&\le \gamma_{K-L} + (1 - \gamma_{K-L})\left\{\gamma_L + (1 - \gamma_L)(1 - \sigma_0^2)^2\right\}\\
& = 1 - (1 - \gamma_{K-L})(1 - \gamma_L)\{\sigma_0^2 (2 - \sigma_0^2)\}\\
& = 1 - \{\sigma_0^2(2 - \sigma_0^2)\}^{K-L-1}\{\sigma_0^2(2 - \sigma_0^2)\}^{L-1}\{\sigma_0^2 (2 - \sigma_0^2)\}\\
& = 1-\{\sigma_0^2 (2 - \sigma_0^2)\}^{K-1}\\
& = 1 - \gamma_K.
\end{align*}
This proves the statement for $K$ and hence completes the induction.

\end{proof}

\subsection{Preliminaries for and Proof of Theorem \ref{thm:confidence_interval_validity}}\label{proof:thm:confidence_interval_validity}

\input{\paperpath sections/target_params_proofs}

\subsection{Proof of Proposition \ref{prop:bipartite_graph_observed_outcome_overlap_graph_equivalence}}\label{proof:prop:bipartite_graph_observed_outcome_overlap_graph_equivalence}

First, we will show that the bipartite graph $\tilde{\mathcal{G}}$ being connected implies that $\mathcal{G}_1^{(1)}$ is connected and thus implying that the O\textsuperscript{3} Algorithm converges to a single super cohort after a single iteration. The connectedness of the bipartite $\tilde{\mathcal{G}}$ implies there must exist a sequence of edges with some length $\ell$ denoted
\begin{equation}\label{eq:bipartite_graph_path}
    ((i_1, t_1), (i_2, t_1), \dotsc, (i_{\ell-1}, t_{\ell / 2}), (i_\ell, t_{\ell / 2}))
\end{equation}
between any two unit $i_1$ and $i_\ell$ in $\tilde{\mathcal{G}}$. Consider any pair of edges $(i_j, t), (i_{j+1}, t)$ in the path \eqref{eq:bipartite_graph_path} that connect to the same outcome $t$. If $C_{i_j} \neq C_{i_{j+1}}$, i.e. that units $i_j$ and $i_{j+1}$ belong to different cohorts. Since the edges $(i_j, t)$ and $(i_{j+1}, t)$ both belong to $\tilde{\mathcal{G}}$, by definition, outcome $t$ is observed for $i_j$ and $i_{j+1}$, in which case cohorts $C_{i_j}$ and $C_{i_{j+1}}$ must share an edge in $\mathcal{G}_1^{(1)}$. If on the other hand $C_{i_j} = C_{i_{j+1}}$, then the same logic would imply there is a self-edge connected to $C_{i_j}$ if not for the fact that $\mathcal{G}_1^{(1)}$ has no self edges.

Based on the argument above, we can iteratively construct a length $\ell$ path in $\mathcal{G}_1^{(1)}$ corresponding to the path \eqref{eq:bipartite_graph_path} in $\tilde{\mathcal{G}}$ that connects cohort $C_{i_1}$ to cohort $C_{i_\ell}$. Since connectedness of $\tilde{\mathcal{G}}$ implies that a path between any two units $i_1$ and $i_2$ exists in $\tilde{\mathcal{G}}$ and every cohort $c$ must have at least one unit belonging to it, any two cohorts in $\mathcal{G}_1^{(1)}$ must have a path between them. Thus, $\mathcal{G}_1^{(1)}$ must also be connected, so the O\textsuperscript{3} Algorithm converges to a single super cohort after a single iteration.

Next, we will show that if Assumption \ref{assump:cohort_sizes_and_potential_outcome_variance_bound} also holds, then the O\textsuperscript{3} Algorithm converging to a single super cohort after a single iteration implies that the bipartite graph $\tilde{\mathcal{G}}$ is connected with probability approaching one as $N \rightarrow \infty$. First, we note that since Assumption \ref{assump:cohort_sizes_and_potential_outcome_variance_bound} requires a unit to belong to each cohort with positive probability and there are a finite number of cohorts, it must be that $\tilde{\mathcal{G}}$ contains a unit belonging to every cohort with probability approaching one as $N \rightarrow \infty$. As such, for the remainder of the proof, we shall condition on this event.

Next, since the O\textsuperscript{3} Algorithm converges to a single super cohort after a single iteration, $\mathcal{G}_1^{(1)}$ is connected, so there exists a sequence of edges with some length $\ell$ denoted
\begin{equation}\label{eq:observed_outcome_overlap_graph_path}
    ((c_1, c_2), \dotsc, (c_{\ell - 1}, c_\ell))
\end{equation}
that connects any two cohorts $c_1 \neq c_\ell$. For a given edge $(c_j, c_{j+1})$, consider any two units $i_j$ in cohort $c_j$ and $i_{j+1}$ in cohort $c_{j+1}$. Since the edge $(c_j, c_{j+1})$ exists in $\mathcal{G}_1$, there must be at least one outcome $t_j$ that is observed for the units in both cohort $c_j$ and cohort $c_{j+1}$. As such, there must exist edges $(i_j, t_j)$ and $(i_{j+1}, t_j)$ in $\tilde{\mathcal{G}}$. Based on the argument above, we can iteratively construct a length $\ell$ path in $\tilde{\mathcal{G}}$ between any unit $i_1$ in cohort $c_1$ and any unit $i_\ell$ in cohort $c_\ell$.

To show that any two units $i_1$ and $i_2$ in the same cohort $c$ are connected in $\tilde{\mathcal{G}}$, we note that if the units are in the same cohort, the same set of outcomes is observed for both of them, meaning there exists at least one outcome $t$ such that the edges $(i_1, t)$ and $(i_2, t)$ exist. As such, there exists a length two path in $\tilde{\mathcal{G}}$ connecting any two units in the same cohort. In addition, since without loss of generality, every outcome is observed for at least one unit, for any outcome $t$, there exists at least one edge $(i, t)$ connecting it to some unit $i$. Putting everything together, we know that there exists a path in $\tilde{\mathcal{G}}$ connecting any two units, regardless of whether they belong to the same or different cohorts, and every outcome is connected to at least one unit. Thus, $\tilde{\mathcal{G}}$ must be connected, as required.

\subsection{Proof of Proposition \ref{prop:potential_outcome_unit_overlap_observed_outcome_overlap_graph_equivalence}}\label{proof:prop:potential_outcome_unit_overlap_observed_outcome_overlap_graph_equivalence}

First, we will show that the assumption that $\check{\mathcal{G}}$ is connected implies that the O\textsuperscript{3} Algorithm converges to a single super cohort after a single iteration. Since $\check{\mathcal{G}}$ is connected, there exists a path in $\check{\mathcal{G}}$ of some length $\ell$ denoted
\begin{equation}\label{eq:potential_outcome_unit_overlap_graph_path}
    ((t_1, t_2), \dotsc, (t_\ell, t_{\ell + 1}))
\end{equation}
between any two outcomes $t_1$ and $t_\ell$. If $\ell = 1$, that directly implies the existence of a cohort of units for whom outcomes $t_1$ and $t_\ell = t_2$ are both observed.

If $\ell \geq 2$ on the other hand, consider any two adjacent edges $(t_j, t_{j+1})$ and $(t_{j+1}, t_{j+2})$ in the path \eqref{eq:potential_outcome_unit_overlap_graph_path}. Since the edge $(t_j, t_{j+1})$ exists in $\check{\mathcal{G}}$, there must be some cohort of units $c_j$ for whom both outcomes $t_j$ and $t_{j+1}$ are observed. By similar logic, there must be some cohort of units $c_{j+1}$ for whom both outcomes $t_{j+1}$ and $t_{j+2}$ are observed. Since outcome $t_{j+1}$ is observed for the units in cohorts $c_j$ and $c_{j+1}$, they must then share an edge in $\mathcal{G}_1$ (see the proof of Proposition \ref{prop:bipartite_graph_observed_outcome_overlap_graph_equivalence} in Appendix \ref{proof:prop:bipartite_graph_observed_outcome_overlap_graph_equivalence} to see why we need to assume all entries of $\Gamma$ are non-zero). Applying this logic iteratively along the path in \eqref{eq:potential_outcome_unit_overlap_graph_path}, we can construct a path in $\mathcal{G}_1^{(1)}$ from any cohort for whom outcome $t_1$ is observed to any cohort for whom outcome $t_\ell$ is observed. Since without loss, at least one outcome is observed for the units in every cohort, we can therefore construct a path in $\mathcal{G}_1^{(1)}$ between any pair of cohorts, meaning $\mathcal{G}_1^{(1)}$ is connected, and thus that the O\textsuperscript{3} Algorithm must converge to a single super cohort after a single iteration.

Next, we show that the assumption that the O\textsuperscript{3} Algorithm converges to a single super cohort after a single iteration implies that the graph $\check{\mathcal{G}}$ is connected. Since $\mathcal{G}_1^{(1)}$ must therefore be connected, there exists a path in $\mathcal{G}_1^{(1)}$ of some length $\ell$ denoted
\begin{equation}\label{eq:observed_outcome_overlap_graph_2}
    ((c_1, c_2), \dotsc, (c_{\ell - 1}, c_\ell))
\end{equation}
between any two cohorts $c_1$ and $c_\ell$. For a given edge $(c_j, c_{j+1})$ in $\mathcal{G}_1^{(1)}$, by definition, there must be some outcome $t_j$ that is observed for both of the positive measures of units in cohorts $c_j$ and $c_{j+1}$ by Assumption \ref{assump:cohort_sizes_and_potential_outcome_variance_bound}. Thus, for any outcome observed for the units in $c_j$, there must be an edge in $\check{\mathcal{G}}$ between it and $t_j$, and similarly, for any outcome observed for the units in $c_{j+1}$, there must be an edge in $\check{\mathcal{G}}$ between it and $t_j$. Thus, a path of length two exists in $\check{\mathcal{G}}$ between any outcome observed for the units in $c_j$ and any outcome observed for the units in $c_{j+1}$. Applying this same logic iteratively along the path \eqref{eq:observed_outcome_overlap_graph_2}, a path can be constructed in $\check{\mathcal{G}}$ connecting any outcome observed for the units in cohort $c_1$ to any outcome observed for the units in cohort $c_\ell$. Since without loss of generality, every outcome is observed for the units in at least one cohort, a path in $\check{\mathcal{G}}$ can be constructed between any two outcomes. Thus, $\check{\mathcal{G}}$ is connected, as required.

\subsection{Proof of Lemma \ref{lemma:cohort_specific_factor_identification_homoskedastic_outcomes}}\label{proof:lemma:cohort_specific_factor_identification_homoskedastic_outcomes}

Let $U_\Gamma S_\Gamma V_\Gamma'$ be a compact singular value decomposition of $E_c\Gamma$, and recall that, since the factor vectors $\gamma_t$ are assumed to be in general position, $E_c\Gamma$ is rank $r$. Next, note that
\begin{align}
    V_c &= E_c\E[(Y_i^* - \E[Y_i^* ~|~ C_i = c])(Y_i^* - \E[Y_i^* ~|~ C_i = c])' ~|~ C_i = c]E_c \\
    &= E_c\E[(\Gamma (\lambda_i - \E[\lambda_i ~|~ C_i = c]) + \epsilon_i) \\
    &\phantom{= E_c\E[}(\Gamma (\lambda_i - \E[\lambda_i ~|~ C_i = c]) + \epsilon_i)' ~|~ C_i = c]E_c && \text{(by \eqref{eq:factor_model})} \\
    &= E_c\big(\Gamma\E[(\lambda_i - \E[\lambda_i ~|~ C_i = c])(\lambda_i - \E[\lambda_i ~|~ C_i = c])' ~|~ C_i = c]\Gamma' \\
    &\phantom{= E_c\big(} + \E[\epsilon_i\epsilon_i' ~|~ C_i = c] + \Gamma\E[\lambda_i\epsilon_i' ~|~ C_i = c] + \E[\epsilon_i\lambda_i' ~|~ C_i = c]\Gamma' && \text{(by \eqref{eq:factor_model})} \\
    &= E_c\left(\Gamma\Cov(\lambda_i ~|~ C_i = c)\Gamma' + \E[\epsilon_i\epsilon_i' ~|~ C_i = c]\right)E_c && \text{($\E[\epsilon_i ~|~ \lambda_i, C_i] = \zeros_T$ by \eqref{eq:factor_model})} \\
    &= E_c\left(\Gamma\Cov(\lambda_i ~|~ C_i = c)\Gamma' + \underbrace{\E[\sigma_i^2 ~|~ C_i = c]}_{\sigma_c^2}I\right)E_c && \text{(by Assumption \ref{assump:uncorrelated_homoskedastic_outcomes})} \\
    &= U_\Gamma \underbrace{S_\Gamma V_\Gamma' \Cov(\lambda_i ~|~ C_i = c) V_\Gamma S_\Gamma}_{M} U_\Gamma' + \sigma_c^2 E_c. && \text{($E_c\Gamma = U_\Gamma S_\Gamma V_\Gamma'$)}
\end{align}
Now, let $U_MS_M^2U_M'$ denote an eigendecomposition of the $r \times r$ matrix $M$ in the display above, and note that since the $r \times r$ matrix $\Cov(\lambda_i ~|~ C_i = c)$ is rank $r$ by Assumption \ref{assump:nontrivial_loadings} and so is $E_c\Gamma$, $S_M^2$ has $r$ non-zero eigenvalues. Then we have that
\begin{equation}\label{eq:cohort_2nd_moment_mat_chained_eigendecomp}
    V_c = U_\Gamma U_M S_M^2 U_M'U_\Gamma' + \sigma_c^2 E_c.
\end{equation}

Since $U_M$ is an orthonormal matrix and the columns of $U_\Gamma$ are orthonormal to one another, $U_\Gamma U_M$ must also have $r$ orthonormal columns. Further, we will show that $E_c U_\Gamma = U_\Gamma$, in which case $E_c U_\Gamma U_M = U_\Gamma U_M$. To see why, note that
\begin{align}
    E_c \Gamma \Gamma' E_c \cdot E_c U_\Gamma &= E_c \cdot E_c \Gamma \Gamma' E_c \cdot U_\Gamma && \text{($E_c$ is idempotent)} \\
    &= E_cU_\Gamma S_\Gamma V_\Gamma' V_\Gamma S_\Gamma U_\Gamma'U_\Gamma && \text{($E_c\Gamma = U_\Gamma S_\Gamma V_\Gamma'$)} \\
    &= E_c U_\Gamma S_\Gamma^2. && \text{($V_\Gamma'V_\Gamma = U_\Gamma'U_\Gamma = I$)}
\end{align}
Thus, $E_c U_\Gamma$ are eigenvectors corresponding to the $r$ non-zero eigenvalues of $E_c \Gamma \Gamma' E_c$, so they must also be left singular vectors of $E_c\Gamma$.

Next, let $U_{Y1} \coloneqq E_c U_\Gamma U_M$, let $U_{Y2}$ be a $T \times (T_c - r)$ matrix with orthonormal columns such that $U_{Y1}'U_{Y2} = \zeros$ and $E_c U_{Y2} = U_{Y2}$, i.e. the columns of $U_{Y2}$ are orthogonal to the columns of $U_{Y1}$, and $U_{Y2}$ only has non-zero entries in the indices $\mathcal{T}_c$, and let $U_{Y3}$ be a $T \times (T - T_c)$ matrix such that
\begin{equation}
    U_Y \coloneqq \bmat{U_{Y1} & U_{Y2} & U_{Y3}} \eqqcolon \bmat{U_{Yc} & U_{Y3}}
\end{equation}
is orthonormal. We note that the existence of the aforementioned matrices is guaranteed constructively by applications of the Gram-Schmidt process. By construction, since $U_YU_Y' = I$ and $U_{Yc}$ has orthonormal columns with non-zero entries only in the indices $\mathcal{T}_c$, it must be that $U_{Yc}U_{Yc}' = E_c$ and $U_{Y3}U_{Y3}' = I - E_c$. Then, expanding the right side of \eqref{eq:cohort_2nd_moment_mat_chained_eigendecomp} using these matrices, we have that
\begin{align}
    V_c &= \bmat{U_{Y1} & \bmat{U_{Y2} & U_{Y3}}} \bmat{S_M^2 & \zeros_{r \times (T - r)} \\ \zeros_{(T - r) \times r} & \zeros_{(T - r) \times (T - r)}} \bmat{U_{Y1}' \\ \bmat{U_{Y2}' \\ U_{Y3}'}} \\
    &\phantom{=}+ \bmat{U_{Yc} & U_{Y3}}\bmat{\sigma_c^2 I_{T_c} & \zeros_{T_c \times (T - T_c)} \\ \zeros_{(T - T_c) \times T_c} & \zeros_{(T - T_c) \times (T - T_c)}}\bmat{U_{Yc}' \\ U_{Y3}'} \\
    &= \underbrace{\bmat{U_{Y1} & U_{Y2} & U_{Y3}}}_{U_Y}\bmat{S_M^2 + \sigma_c^2 I_r & \zeros & \zeros \\ \zeros & \sigma_c^2 I_{T_c - r} & \zeros \\ \zeros & \zeros & \zeros}\underbrace{\bmat{U_{Y1}' \\ U_{Y2}' \\ U_{Y3}'}}_{U_Y'}.
\end{align}
Since $U_Y$ is an orthonormal matrix, the center matrix in the last line of the display above is diagonal, and the diagonal entries of $S_M^2 + \sigma_c^2 I_r$ are strictly larger than those of $\sigma_c^2 I_{T_c - r}$ from the fact that $S_M$ is positive definite, the expression in the display above must be an eigendecomposition of $V_c$.

Because the eigenvalues of products of nonsingular square matrices are invariant to cyclic permutations of the product terms,\footnote{For any square matrices $A, B \in \R^{d \times d}$, if $\lambda$ is an eigenvalue of $AB$ with corresponding eigenvector $v$, then since $ABv = \lambda v$, we have that $BA(Bv) = B(ABv) = B (\lambda v) = \lambda(Bv)$. Thus, $\lambda$ is also an eigenvalue of $BA$ with corresponding eigenvector $Bv$.} the eigenvalues of the matrices $S_\Gamma V_\Gamma' \Cov(\lambda_i ~|~ C_i = c)V_\Gamma S_\Gamma$ and $V_\Gamma S_\Gamma^2 V_\Gamma' \Cov(\lambda_i ~|~ C_i = c)$ are the same. Since $S_M^2$ is the diagonal matrix whose non-zero entries are the ordered eigenvalues of $S_\Gamma V_\Gamma' \Cov(\lambda_i ~|~ C_i = c)V_\Gamma S_\Gamma$, and $V_\Gamma S_\Gamma^2 V_\Gamma' = \Gamma' E_c \cdot E_c \Gamma = \Gamma' E_c \Gamma$, we equivalently have that $S_M^2$ is the diagonal matrix whose non-zero entries are the ordered eigenvalues of $\Gamma' E_c \Gamma \Cov(\lambda_i ~|~ C_i = c)$.
The results stated in the statement of the lemma then follow from the fact that $\Pi(U_{Y1}) = \Pi(E_c U_\Gamma U_M) = \Pi(U_\Gamma) = \Pi(E_c \Gamma)$ and inspection of the eigendecomposition.

\subsection{Proof of Proposition \ref{proposition:cohort_specific_factor_inf_fn_homoskedastic_outcomes}}\label{proof:proposition:cohort_specific_factor_inf_fn_homoskedastic_outcomes}

First, we note that, under Assumptions \ref{assump:cohort_sizes_and_potential_outcome_variance_bound} and \ref{assump:nontrivial_loadings}, as well as the fact that for any matrix $A$, $\mathrm{rank}(A)^{-1/2}\norm{A}_\text{F} \leq \norm{A}_\text{op} \leq \norm{A}_\text{F}$ and the rank of $\hat{V}_c - V_c$ is at most $T_c$ (since both $\hat{V}_c$ and $V_c$ have zero rows/columns with indices $t \not\in \mathcal{T}_c$), Lemma \ref{lemma:inf_fn_group_avg} implies that
\begin{align}
    \normnofit{\hat{V}_c - V_c}_\text{op} &\leq \normnofit{\hat{V}_c - \hat{\E}_N[(E_cY_i - E_c\mu_c)(E_cY_i - E_c\mu_c)' ~|~ C_i = c]}_\text{op} \\
    &\phantom{\leq} + \underbrace{\normnofit{\hat{\E}_N[(E_cY_i - E_c\mu_c)(E_cY_i - E_c\mu_c)' ~|~ C_i = c] - V_c}_\text{op}}_{\text{$=\bigop{N^{-1/2}p_c^{-1/2}T_c^{1/2}}$ by Lemma \ref{lemma:inf_fn_group_avg}}} \\
    &\leq 2\normnofit{E_c\mu_c(\hat{\E}_N[E_cY_i ~|~ C_i = c] - E_c\mu_c)'}_\text{op} \\
    &\phantom{\leq} + \normnofit{(\hat{\E}_N[E_cY_i ~|~ C_i = c] - E_c\mu_c)(\hat{\E}_N[E_cY_i ~|~ C_i = c] - E_c\mu_c)'}_\text{op} \\
    &\phantom{\leq} + \bigop{N^{-1/2}p_c^{-1/2}T_c^{1/2}} \\
    &\leq 2\normnofit{E_c\mu_c}_2\underbrace{\normnofit{\hat{\E}_N[E_cY_i ~|~ C_i = c] - E_c\mu_c}_2}_{\text{$=\bigop{N^{-1}p_c^{-1}}$ by Lemma \ref{lemma:inf_fn_group_avg}}} + \underbrace{\normnofit{\hat{\E}_N[E_cY_i ~|~ C_i = c] - E_c\mu_c}_2^2}_{\text{$=\bigop{N^{-1}p_c^{-1}}$ by Lemma \ref{lemma:inf_fn_group_avg}}} \\
    &\phantom{\leq} + \bigop{N^{-1/2}p_c^{-1/2}T_c^{1/2}} \\
    &= \bigop{N^{-1/2}p_c^{-1/2}T_c^{1/2}}. \label{eq:est_V_c_op_norm_error_bound}
\end{align}
As such, the probability that $\normnofit{\hat{V}_c - V_c}_\text{op} \leq 4^{-1}s_{1c}^2$ converges to one as long as $p_c \cdot N \to \infty$; for the rest of the proof we also condition on this event.

We can then apply Theorem \ref{thm:eigenspace_perturbation_expansion} with $M = V_c$, $\hat{M} = \hat{V}_c$, $s = T - r$, and $r = r$. Lemma \ref{lemma:cohort_specific_factor_identification_homoskedastic_outcomes} implies that, using the notation from Appendix \ref{sec:eigspace_perturbation},
\begin{equation}
    \lambda_{T -r}(V_c) = \sigma_c^2 < s_{1c}^2 + \sigma_c^2 = \lambda_{T - r + 1}(V_c) \leq \dotsc \leq s_{rc}^2 + \sigma_c^2 = \lambda_T(V_c) < \lambda_{T + 1}(V_c) = \infty.
\end{equation}
As such, we have that $\Delta(V_c) = 4^{-1}s_{1c}^2 > 0$, satisfying \eqref{cond:well_sep_eigvals}. Lemma \ref{lemma:cohort_specific_factor_identification_homoskedastic_outcomes} also implies that, for $j \in \{T - r + 1, \dotsc, T\}$ and $k \in \{1, \dotsc, T - r\}$,
\begin{equation}
    \lambda_j - \lambda_k = s_{(j - (T - r))c}^2 + \sigma_c^2\ind{k \leq T - T_c}.
\end{equation}
Further, Lemma \ref{lemma:cohort_specific_factor_identification_homoskedastic_outcomes} implies that $\Pi(E_c\Gamma) = \Pi(U_{(T - r + 1):T}(V_c))$
and we define $\hat{\Gamma}_{c, \text{PC}}$ such that $\Pi(\hat{\Gamma}_{c,\text{PC}}) = \Pi(U_{(T -r + 1):T}(\hat{V}_c))$.

Before continuing, we note that if $\tilde{\gamma}_{ck}$ lies in the null space of $\hat{V}_c - V_c$, then since projection matrices are symmetric, $(\hat{V}_c - V_c)\Pi(\tilde{\gamma}_{ck}) = \Pi(\tilde{\gamma}_{ck})(\hat{V}_c - V_c) = \zeros_{T \times T}$. Since the rows and columns $t$ of $V_c$ corresponding to $t \not\in \mathcal{T}_c$ must be zero and $V_c$ has $T - T_c$ zero eigenvalues, it must be that for any $k < T - T_c$, the eigenvector $\tilde{\gamma}_{ck}$ of $V_c$ corresponding to a zero eigenvalue has non-zero entries only in the indices $[T] \setminus \mathcal{T}_c$. Since $\hat{V}_c$ also has zero rows and columns corresponding to $t \not\in \mathcal{T}_c$, it must be that $\tilde{\gamma}_{ck}$ lies in the null space of both $V_c$ and $\hat{V}_c$. Therefore, $(\hat{V}_c - V_c)\Pi(\tilde{\gamma}_{ck}) = \Pi(\tilde{\gamma}_{ck})(\hat{V}_c - V_c) = \zeros_{T \times T}$. Further, since for all $j$, $\Pi(\tilde{\gamma}_{cj})$ is the projection operator onto a vector, $g(\Pi(\tilde{\gamma}_{cj}), (\hat{V}_c-V_c), \Pi(\tilde{\gamma}_{ck}))$ must be at most rank one.

Combining our invocation of Theorem \ref{thm:eigenspace_perturbation_expansion} with these results, then we have that
\begin{align}\label{eq:cohort_specific_factor_space_finite_sample_consistency}
    &\Bigg\lVert\left[\Pi(\hat{\Gamma}_{c,\text{PC}}) - \Pi(E_c\Gamma)\right] \\
    &\phantom{\Bigg\lVert}- \sum_{j = T - r + 1}^T\sum_{k = 1}^{T - r} \frac{-1}{s_{(j - (T - r))c}^2 + \sigma_c^2\ind{k \leq T - T_c}} g\left(\Pi(\tilde{\gamma}_{cj}), (\hat{V}_c-V_c), \Pi(\tilde{\gamma}_{ck})\right) \Bigg\rVert_\text{op} \\
    &\Bigg\lVert\left[\Pi(\hat{\Gamma}_{c,\text{PC}}) - \Pi(E_c\Gamma)\right] \\
    &\phantom{\Bigg\lVert} + \sum_{j = T - r + 1}^T\sum_{k = T - T_c + 1}^{T - r} s_{(j - (T - r))c}^{-2} \cdot g\left(\Pi(\tilde{\gamma}_{cj}), (\hat{V}_c-V_c), \Pi(\tilde{\gamma}_{ck})\right) \Bigg\rVert_\text{op} \\
&\leq \frac{64}{\pi s_{1c}^4}\normnofit{\hat{V}_c - V_c}_\text{op}^2 \label{eq:cohort_specific_proj_mat_homoskedastic_err_expansion_bound} \\
&= \bigop{N^{-1}p_c^{-1}T_c s_{1c}^{-4}}. && \text{(By \eqref{eq:est_V_c_op_norm_error_bound})} \label{eq:cohort_specific_proj_mat_homoskedastic_err_expansion}
\end{align}
Applying this result and Lemma \ref{lemma:inf_fn_group_avg}, we show \eqref{eq:cohort_specific_proj_mat_asymp_linear_homosk} as follows:
\begin{align}
    &\norm{\left[\Pi(\hat{\Gamma}_{c,\text{PC}}) - \Pi(E_c\Gamma)\right] - \hat{\E}_N\left[\phi_{c,\text{PC}}(C_i, Y_i)\right]}_\text{op} \\
    &= \Bigg\lVert\left[\Pi(\hat{\Gamma}_{c,\text{PC}}) - \Pi(E_c\Gamma)\right] \\
    &\phantom{=\Bigg\lVert}+ \sum_{j = T - r + 1}^T\sum_{k = T - T_c + 1}^{T - r} s_{(j - (T - r))c}^{-2} \cdot g\left(\Pi(\tilde{\gamma}_{cj}), (\hat{V}_c-V_c), \Pi(\tilde{\gamma}_{ck})\right) \Bigg\rVert_\text{op} \\
    &\phantom{=} + \Bigg\lVert\hat{\E}_N\left[\phi_{c,\text{PC}}(C_i, Y_i)\right] \\
    &\phantom{=+ \Bigg\lVert}- \sum_{j = T - r + 1}^T\sum_{k = T - T_c + 1}^{T - r} s_{(j - (T - r))c}^{-2} \cdot g\left(\Pi(\tilde{\gamma}_{cj}), (\hat{V}_c-V_c), \Pi(\tilde{\gamma}_{ck})\right) \Bigg\rVert_\text{op} \\
    &\leq \frac{64}{\pi s_{1c}^4}\normnofit{\hat{V}_c - V_c}_\text{op}^2 && \text{(\eqref{eq:cohort_specific_proj_mat_homoskedastic_err_expansion_bound})} \\
    &\phantom{} + \sum_{j = T - r + 1}^T\sum_{k = T - T_c + 1}^{T - r} \frac{-2}{s_{(j - (T - r))c}^2} \\
    &\phantom{+} \cdot \underbrace{\norm{g\left(\Pi(\tilde{\gamma}_{cj}), \hat{\E}_N\left[\frac{\ind{C_i = c}}{p_c}(E_c(Y_i - \mu_c)(Y_i - \mu_c)'E_c - V_c)\right] - [\hat{V}_c - V_c], \Pi(\tilde{\gamma}_{ck})\right)}_\text{op}}_{\text{$= \bigop{N^{-1}p_c^{-1}}$ by Lemma \ref{lemma:inf_fn_group_avg}}} \label{eq:cohort_specific_proj_mat_homoskedastic_err_expansion_bound_2} \\
    &= \bigop{s_{1c}^{-4}N^{-1}p_c^{-1} + r \cdot (T_c - r) \cdot s_{1c}^{-2} \cdot N^{-1}p_c^{-1}}, && \text{(\eqref{eq:cohort_specific_proj_mat_homoskedastic_err_expansion})}
\end{align}
as in \eqref{eq:cohort_specific_proj_mat_asymp_linear_homosk}.
Given the linearity of $g$, $\E[\phi_{c, \text{PC}}(C_i, Y_i)] = \zeros_{T^2}$, and we have that
\begin{align}
    &\E\left[\norm{\phi_{c, \text{PC}}(C_i, Y_i)}_\text{op}^2\right] \\
    &\leq \E\Bigg[\Bigg(\sum_{j = T - r + 1}^T\sum_{k = T - T_c + 1}^{T - r} 2 s_{(j - (T - r))c}^{-2} \underbrace{\normnofit{\Pi(\tilde{\gamma}_{cj})}_\text{op} \normnofit{\Pi(\tilde{\gamma}_{ck})}_\text{op}}_{\leq 1}\\
    &\phantom{\leq \E\Bigg[\Bigg(\sum_{j = T - r + 1}^T\sum_{k = T - T_c + 1}^{T - r}} \cdot \norm{\frac{\ind{C_i = c}}{p_c}(E_c(Y_i - \mu_c)(Y_i - \mu_c)'E_c - V_c)}_\text{op}\Bigg)^2\Bigg] && \text{(Triangle Ineq.)} \\
    &\leq \frac{4 r(T_c - r)}{p_c s_{1c}^2}\E\left[\norm{E_c(Y_i - \mu_c)(Y_i - \mu_c)'E_c - V_c}_\text{op}^2 ~\middle|~ C_i = c\right] && \text{($s_{1c} \leq s_{(j - (T - r))c}$)} \\
    &\leq \frac{16 r(T_c - r)}{p_c s_{1c}^2}\E\left[\norm{E_c(Y_i^* - \mu_c)(Y_i^* - \mu_c)'E_c}_\text{op}^2 ~\middle|~ C_i = c\right] && \text{(Lemma \ref{lemma:central_to_noncentral_sq_norm_bound})} \\
    &\leq \frac{16 r(T_c - r)}{p_c s_{1c}^2}\underbrace{\norm{E_c}_\text{op}^2}_{\leq 1}\E\left[\norm{Y_i^* - \mu_c}_2^4 ~\middle|~ C_i = c\right] && \text{($\norm{xx'}_\text{op}^2 = \norm{x}_2^4$)} \\
    &\leq \frac{16 r(T_c - r)}{p_c s_{1c}^2}\norm{\Gamma}_\text{op}^4\left(\E\left[\norm{\lambda_i - \E[\lambda_i ~|~ C_i = c]}_2^4 + \norm{\epsilon_i}_2^4 ~|~ C_i = c\right]\right) \\
    &\leq \frac{64 r(T_c - r)}{p_c s_{1c}^2}\norm{\Gamma}_\text{op}^4\left(\E\left[\norm{\lambda_i}_2^4 ~|~ C_i = c\right] + \E[\norm{\epsilon_i}_2^4 ~|~ C_i = c]\right). && \text{(Lemma \ref{lemma:central_to_noncentral_sq_norm_bound})}.
\end{align}
where the boundedness of this upper bound follows from the fact that $p_c > 0$ under Assumption \ref{assump:cohort_sizes_and_potential_outcome_variance_bound} and the boundedness of $\lambda_i$ and $\epsilon_i$'s fourth moments by Assumption \ref{assump:nontrivial_loadings}.

%% file: sections/target_params_proofs.tex
\paragraph*{Preliminaries.} Let $\tilde{P}$ denote some distribution over $(C_i, Y_i^*, X_i)$, and let $\tilde{P}_\text{obs}$ denote the distribution over $(C_i, Y_i, X_i)$ implied by $\tilde{P}$. In addition, let $\Gamma(\tilde{P})$ denote the parameter $\Gamma$ implied by the distribution $\tilde{P}$. The assumption made in Theorem \ref{thm:factor_identification} that $\Pi(E_c\Gamma)$ for each cohort $c \in [C]$ are identified, and there exist known functions $\Pi(\tilde{\Gamma}_c(\tilde{P}_\text{obs}))$ mapping distributions over observables $\tilde{P}_\text{obs}$ to $\Pi(E_c\Gamma(\tilde{P}))$. We now slightly strengthen Assumption \ref{assump:cohort_specific_factor_proj_mat_inf_fn} in the following manner:
\begin{assumption}\label{assump:cohort_specific_factor_proj_mat_hadamard_diff}
    For each cohort $c \in [C]$, the maps $\Pi(\tilde{\Gamma}_c(\cdot))$ are Hadamard differentiable and satisfy $\Pi(\tilde{\Gamma}_c(\hat{P}_N)) = \Pi(\hat{\Gamma}_c)$.
\end{assumption}
\noindent We note that Assumption \ref{assump:cohort_specific_factor_proj_mat_hadamard_diff} implies Assumption \ref{assump:cohort_specific_factor_proj_mat_inf_fn} by the Delta method (see e.g. Theorem 3.9.4 in \citet{vaart1996empirical}), the fact that $\hat{P}_N$ converges to $P$, and the fact that $\phi_{\Gamma,c}$ are the Riesz representers of the linear Hadamard derivatives of $\Pi(\tilde{\Gamma}_c(\cdot))$. As we discuss in Appendix \ref{sec:supplement:pc_estimator_consistency_asymptotic_linear}, the PC estimator satisfies Assumption \ref{assump:cohort_specific_factor_proj_mat_hadamard_diff} when $X_{it} = \zeros_q$.

Next, we let $\eta(\tilde{P})$ denote the value of the nuisance parameter implied by the distribution $\tilde{P}$, we let $\tilde{\eta}(\tilde{P}_\text{obs})$ be the representation of the consistent estimator $\hat{\eta}$ as a map from observable data distributions to estimate values. We can then also slightly strengthen Assumption \ref{assump:nuisance_param_est_asymptotic_linear} in a similar manner:
\begin{assumption}\label{assump:nuisance_param_est_hadamard_diff}
    The map $\tilde{\eta}(\cdot)$ is Hadamard differentiable and satisfies $\tilde{\eta}(\hat{P}_N) = \hat{\eta}$.
\end{assumption}

Finally, we let $\tilde{\theta}(\tilde{P}_\text{obs})$ denote the mapping from observed data distributions $\tilde{P}_\text{obs}$ to target parameter values such that $\tilde{\theta}(\hat{P}_N) = \hat{\theta}$. Given these definitions and assumptions, we are now equipped to prove Theorem \ref{thm:confidence_interval_validity}.

\paragraph*{Proof.} First, we note that Assumption \ref{assump:cohort_specific_factor_proj_mat_hadamard_diff}, \eqref{eq:apm_finite_sample_approximate_linearity} from the proof of Theorem \ref{thm:factor_consistency_asymptotic_normality}, the continuity of $\psi_c$ with respect to the population parameters guaranteed by Theorem \ref{thm:outcome_mean_consistency_asymptotic_normality}, and Assumptions \ref{assump:nuisance_param_est_hadamard_diff} and \ref{assump:h_smooth} allow us to apply the chain rule (see e.g. Lemma 3.9.3 in \citet{vaart1996empirical}) to say that the map $\tilde{\theta}(\cdot)$ is itself Hadamard differentiable.

Next, we note that the random weight vector $W$ satisfies Equation (3.6.8) in \citet{vaart1996empirical} with $c=1$ by Example 3.6.9 in the same book. Thus, Theorem 3.6.13 in \citet{vaart1996empirical} implies that the Bayesian bootstrap yields a consistent estimate $\hat{P}_N^*$ of the true data-generating distribution $P_\text{obs}$, while the bootstrap delta method given in \citet{vaart1996empirical}'s Theorem 3.9.11 implies that the asymptotic distribution of $\hat{\theta}^* = \tilde{\theta}(\hat{P}_N^*)$ converges to that of $\hat{\theta}$ in probability. Since quantile functions are also Hadamard differentiable (see Example 3.9.21 in \citet{vaart1996empirical}), we have that
\begin{equation}
    \sqrt{N}\hat{\sigma}_j = \sqrt{\Sigma_{\theta,jj}} + \littleop{1}.
\end{equation}
Given the bootstrap and standard error estimator consistency results above, we can appeal to Proposition 3 and Lemma 1 in the supplemental appendix of \citet{montiel2019simultaneous} to show that our confidence intervals have simultaneous coverage, as required.